\RequirePackage{fix-cm}
\documentclass[smallextended,referee]{svjour3}       
\smartqed  
\usepackage{graphicx}
\usepackage{subfigure}
\usepackage{float}
\usepackage{algorithm}
\usepackage{algorithmic}
\usepackage{appendix}
\usepackage[bbgreekl]{mathbbol}
\usepackage{epsfig,subfigure}
\usepackage{amsmath}
\usepackage{amsfonts}
\usepackage{amssymb} 
\usepackage[latin1]{inputenc}
\usepackage{IEEEtrantools}
\usepackage{lineno}
\usepackage{color}
\usepackage{longtable}

\begin{document}

\title{Iterative ensemble smoother as an approximate solution to a regularized minimum-average-cost problem: theory and applications 
}


\author{Xiaodong Luo \and   
	    Andreas S. Stordal \and  
	    Rolf J. Lorentzen \and
	    Geir N{\ae}vdal 
}


\institute{ Xiaodong Luo \and     
	    	Andreas S. Stordal \and
	    	Rolf J. Lorentzen \and
	    	Geir N{\ae}vdal  
	    	\at
              International Research Institute of Stavanger (IRIS), 5008 Bergen, Norway \\
              \email{xiaodong.luo@iris.no}           
}

\date{}

\maketitle

\begin{abstract}

The focus of this work is on an alternative implementation of the iterative ensemble smoother (iES). We show that iteration formulae similar to those used in \cite{chen2013-levenberg,emerick2012ensemble} can be derived by adopting a regularized Levenberg-Marquardt (RLM) algorithm \cite{jin2010regularized} to approximately solve a minimum-average-cost (MAC) problem. This not only leads to an alternative theoretical tool in understanding and analyzing the behaviour of the aforementioned iES, but also provides insights and guidelines for further developments of the smoothing algorithms. For illustration, we compare the performance of an implementation of the RLM-MAC algorithm to that of the approximate iES used in \cite{chen2013-levenberg} in three numerical examples: an initial condition estimation problem in a strongly nonlinear system, a facies estimation problem in a 2D reservoir and the history matching problem in the Brugge field case. In these three specific cases, the RLM-MAC algorithm exhibits comparable or even better performance, especially in the strongly nonlinear system.  

\end{abstract}

\section*{Introduction}
\label{sec:introduction}

For data assimilation problems there are different ways in utilizing the available observations. While certain data assimilation algorithms, for instance, the ensemble Kalman filter (EnKF, see, for example, \cite{Aanonsen-ensemble-2009,Evensen2006}), assimilate the observations sequentially in time, other data assimilation algorithms may instead collect the observations at different time instants and assimilate them simultaneously. Examples in this aspect include the ensemble smoother (ES, see, for example, \cite{Evensen2000}) and its iterative variants.

The EnKF has been widely used for reservoir data assimilation (history matching) problems since its introduction to the community of petroleum engineering \cite{naevdal2002near}. The applications of the ES to reservoir data assimilation problems are also investigated recently (see, for example, \cite{skjervheim2011ensemble}). Compared to the EnKF, the ES has certain technical advantages, including, for instance, avoiding the restarts associated with each update step in the EnKF for certain reservoir simulators (e.g., ECLIPSE$^\copyright$ \cite{eclipse2010}) and also having fewer variables to update. The formal benefit (avoiding restarts) may result in a significant reduction of simulation time in certain circumstances \cite{skjervheim2011ensemble}, while the latter (having fewer variables) can reduce the amount of computer memory in use.

To further improve the performance of the ES, some iterative ensemble smoothers (iES) are suggested in the literature, in which the iterations are carried out in the forms of certain iterative optimization algorithms, e.g., the Gaussian-Newton \cite{chen2012ensemble,Gu2007-iterative,Li2009-iterative} or the Levenberg-Marquardt method \cite{chen2013-levenberg,emerick2012ensemble,Iglesias2013regularizing,Lorentzen2011-iterative,luo2014alternative,Luo2014ensemble}\footnote{These optimization algorithms are also applied to iterative EnKFs, e.g., in some of the aforementioned works.}, or in the context of adaptive Gaussian mixture (AGM, see \cite{stordal2014Iterative}). In \cite{Iglesias2013regularizing,Luo2014ensemble}, the iteration formulae are adopted following the regularized Levenberg-Marquardt (RLM) algorithm in the deterministic inverse problem theory (see, for example, \cite{Engl2000-regularization,kaltenbacher2008iterative}). Essentially these formulae aim to find a single solution of the inverse problem, and the gradient involved in the iteration is obtained either through the adjoint model \cite{Iglesias2013regularizing}, or through a stochastic approximation method \cite{Luo2014ensemble}. 
While in \cite{emerick2012ensemble}, the iteration formula is derived based on the idea that, the final result of the iES should be equal to the estimate of the EnKF, at least for linear systems with Gaussian model and observation errors. Consequently, this algorithm is called \textit{ensemble smoother with multiple data assimilation} (ES-MDA for short). On the other hand, in \cite{chen2013-levenberg} an iteration formula is obtained based on the standard Levenberg-Marquardt (LM) algorithm. By discarding a model term in the standard LM algorithm, an approximate iteration formula is derived, which is similar to that in \cite{emerick2012ensemble}. For distinction, we call it approximate Levenberg-Marquardt ensemble randomized maximum likelihood (aLM-EnRML) algorithm. 


In this work we show that an iteration formula similar to those used in the ES-MDA and RLM-MAC can be derived by applying the RLM to find an ensemble of solutions via solving a minimum-average-cost problem \cite{luo2014alternative}. The gradient involved in the iteration is obtained in a way similar to the computation of the Kalman gain matrix in the EnKF. This derivation not only leads to an alternative theoretical tool in understanding and analyzing the behaviour of the aforementioned iES, but also provides insights and guidelines for further developments of the iES algorithm\footnote{For instance, instead of using the RLM algorithm, one could apply other types of regularized inversion methods, e.g., the regularized Gaussian-Newton or conjugate gradient algorithm \cite{Engl2000-regularization,kaltenbacher2008iterative}, to solve the minimum-average-cost problem. This will thus lead to more types of iterative ES algorithms.}.
As an example, we derive an alternative implementation of the iterative ES based on the RLM algorithm. Three numerical examples are then used to illustrate the performance of this new algorithm and compare it to the aLM-EnRML.

\section*{Methodologies}
\subsection*{ES-MDA and aLM-EnRML}
Let $\mathbf{m}$ denote an $m$-dimensional reservoir model that contains the petrophysical properties to be estimated, $\mathbf{g}$ the reservoir simulator, and $\mathbf{d}^o$ a $p$-dimensional vector that contains all available observations in a certain time interval. Here $\mathbf{d}^o$ is assumed to contain certain measurement errors, with zero mean and covariance $\mathbf{C}_d$. In the context of iterative ensemble smoothing, suppose that there is an ensemble $\mathbf{M}^i \equiv \{ \mathbf{m}_j^i \}_{j=1}^{{N_e}}$ of ${N_e}$ reservoir models available at the $i$th iteration step, then an iES updates $\mathbf{M}^i$ to its counterpart $\mathbf{M}^{i+1} \equiv \{ \mathbf{m}_j^{i+1} \}_{j=1}^{{N_e}}$ at the next iteration step via a certain iteration formula, which is the focus of our discussion below. For convenience of discussion later, let us define the following square root matrix $\mathbf{S}_m^i$ with respect to the model $\mathbf{m}$ (\textit{model square root} for short): 
{\small
\begin{linenomath*} 
\begin{IEEEeqnarray}{clc} \label{eq:model_sqrt}
  & \mathbf{S}_m^i = \frac{1}{\sqrt{{N_e}-1}}\left[\mathbf{m}_1^i - \bar{\mathbf{m}}^i,\dotsb, \mathbf{m}_{N_e}^i - \bar{\mathbf{m}}^i \right] \, , & \quad \bar{\mathbf{m}}^i = \frac{1}{{N_e}} \sum_{j=1}^{{N_e}} \mathbf{m}_j^i \, ,   
\end{IEEEeqnarray}
\end{linenomath*}   
}
in the sense that the product $\mathbf{S}_m^i \left( \mathbf{S}_m^i \right)^T$ is equal to the sample covariance of $\mathbf{M}^i$.

The idea in the ES-MDA is to make the final iES estimate equal the estimate of the EnKF, at least for linear systems with Gaussian model and observation errors (some thoughts are also provided in Appendix \ref{sec:appdendix_iES} in situations with nonlinearity and/or non-Gaussianity). To this end, an iteration formula is derived in terms of \cite{emerick2012ensemble}
{\small
\begin{linenomath*} 
\begin{IEEEeqnarray}{clc} \label{eq:linear_ls}
  & \mathbf{m}_j^{i+1} = \mathbf{m}_j^{i} + \mathbf{S}_m^i \left( \mathbf{S}_d^i \right)^T \left( \mathbf{S}_d^i \left( \mathbf{S}_d^i \right)^T + \gamma^i \, \mathbf{C}_{d} \right)^{-1} \left( \mathbf{d}_j^o - \mathbf{g} \left( \mathbf{m}_j^{i} \right) \right),~ i = 1, 2, \dotsb , I;  & \\  
  \label{eq:esmad_Sdi}&  \mathbf{S}_d^i = \dfrac{1}{\sqrt{{N_e}-1}} \left[\mathbf{g}\left(\mathbf{m}_1^{i}\right) -  \overline{\mathbf{g}\left(\mathbf{m}_j^{i}\right)}, \dotsb, \mathbf{g}\left(\mathbf{m}_{N_e}^{i}\right) - \overline{\mathbf{g}\left(\mathbf{m}_j^{i}\right)} \right] \, ; \\  
  & \overline{\mathbf{g}\left(\mathbf{m}_j^{i}\right)} = \dfrac{1}{{N_e}} \sum\limits_{j=1}^{N_e} \mathbf{g}\left(\mathbf{m}_j^{i}\right) \, . & 
\end{IEEEeqnarray}
\end{linenomath*} 
}
The total number of iteration steps $I$ is chosen before the iteration starts; $\mathbf{d}_j^o$ ($j = 1, \dotsb, {N_e}$) are the perturbations of $\mathbf{d}^o$, similar to those in the EnKF; and $\gamma^i$ ($i = 1, \dotsb , I$) satisfy 
{\small
\begin{linenomath*} 
\begin{IEEEeqnarray}{clc} \label{eq:ES-MDA_constraints}
  \gamma^i \geq 1 \text{~ and ~} \sum\limits_{i=1}^I \dfrac{1}{\gamma^i} = 1 \, . & 
\end{IEEEeqnarray}
\end{linenomath*} 
}
For convenience, we also call $\mathbf{S}_d^i$ \textit{data square root} hereafter, in the sense that $\mathbf{S}_d^i (\mathbf{S}_d^i)^T$ is equal to the sample covariance matrix of the simulated data $\{ \mathbf{g}\left(\mathbf{m}_j^{i}\right) \}_{j=1}^{{N_e}}$.  

In the Levenberg-Marquardt ensemble randomized maximum likelihood (LM-EnRML) method \cite{chen2013-levenberg}, history matching is recast as a weighted least-squares problem, in the sense that each model, say $\mathbf{m}_j^{i}$ (the $j$th ensemble model at the $i$th iteration step), is updated to the one $\mathbf{m}_j^{i+1}$ at the next iteration step by solving the following minimization problem:
{
{\small 
\begin{linenomath*} 
\begin{IEEEeqnarray}{clc} \label{eq:individual_cost}
& \underset{\mathbf{m}}{\operatorname{argmin}} \, \left( \mathbf{d}_j^o - \mathbf{g} \left( \mathbf{m} \right) \right)^T \mathbf{C}_{d}^{-1} \left( \mathbf{d}_j^o - \mathbf{g} \left( \mathbf{m} \right) \right) + \left( \mathbf{m} -  \mathbf{m}_j^{0}\right)^T \left( \mathbf{C}_m^0 \right)^{-1} \left( \mathbf{m} -  \mathbf{m}_j^{0}\right) , & 
\end{IEEEeqnarray}
\end{linenomath*} 
}
}
with ${\mathbf{C}_m^0} = \mathbf{S}_m^0 \left(\mathbf{S}_m^0\right)^T$, and $\mathbf{S}_m^0$ given by Eq. (\ref{eq:model_sqrt}). Applying the Levenberg-Marquardt algorithm \cite{marquardt1963algorithm} to (\ref{eq:individual_cost}), one has the following iteration formula to update the model \cite{chen2013-levenberg}
{\small
\begin{linenomath*} 
\begin{IEEEeqnarray}{crlc} \label{eq:lm_sln}
& \mathbf{m}_j^{i+1}  = & \, \mathbf{m}_j^{i} + \Delta\mathbf{m}_j^{i} \, , \\
& \Delta\mathbf{m}_j^{i} = & \, \left[ \left( \mathbf{C}_m^0 \right)^{-1} + \left( \mathbf{G}^i_j \right)^T \mathbf{C}_{d}^{-1}  \mathbf{G}^i_j  \right]^{-1} \left[ \left( \mathbf{G}^i_j \right)^T \mathbf{C}_{d}^{-1} \left( \mathbf{d}_j^o - \mathbf{g} \left( \mathbf{m}_j^{i} \right) \right) 
 - \left( \mathbf{C}_m^0 \right)^{-1} \left( \mathbf{m}_j^{i} - \mathbf{m}_j^{0} \right) \right]  \nonumber \\
& \approx & \, \left[ (1+\lambda^i) \left( \mathbf{C}_m^i \right)^{-1} + \left( \mathbf{G}^i_j \right)^T \mathbf{C}_{d}^{-1}  \mathbf{G}^i_j  \right]^{-1} \left[ \left( \mathbf{G}^i_j \right)^T \mathbf{C}_{d}^{-1} \left( \mathbf{d}_j^o - \mathbf{g} \left( \mathbf{m}_j^{i} \right) \right) 
 - \left( \mathbf{C}_m^0 \right)^{-1} \left( \mathbf{m}_j^{i} - \mathbf{m}_j^{0} \right) \right] \, ,  \nonumber
\end{IEEEeqnarray}
\end{linenomath*} 
}  
where ${\mathbf{C}_m^i} = \mathbf{S}_m^i \left(\mathbf{S}_m^i\right)^T$, $\mathbf{G}^i_j$ is the Jacobian of $\mathbf{g}$ evaluated at $\mathbf{m}_j^{i}$, and $\lambda^i$ a positive scalar. In the course of deriving the third line of Eq. (\ref{eq:lm_sln}), the first $\left( \mathbf{C}_m^0 \right)^{-1}$ is approximated by $(1+\lambda^i) \left( \mathbf{C}_m^i \right)^{-1}$, in accordance to the Levenberg-Marquardt algorithm \cite{chen2013-levenberg}.    
In large-scale problems, it can be expensive to evaluate the term $\left( \mathbf{C}_m^0 \right)^{-1} \left( \mathbf{m}_j^{i} - \mathbf{m}_j^{0} \right)$, {\color{black}because the model size $m$ is usually much larger than the observation size $p$ (i.e., $m \gg p$). Hence discarding the term $\left( \mathbf{C}_m^0 \right)^{-1} \left( \mathbf{m}_j^{i} - \mathbf{m}_j^{0} \right)$,} using the Sherman-Morrison-Woodbury formula \cite{sherman1950adjustment} and some algebra, one has  
{\small
\begin{linenomath*}   
\begin{IEEEeqnarray}{crlc} \label{eq:discarding_a_term_enrml}
& \mathbf{m}_j^{i+1} \approx \mathbf{m}_j^{i} + \mathbf{C}_m^i (\mathbf{G}^i_j)^T \left[ \mathbf{G}^i_j \mathbf{C}_m^i (\mathbf{G}^i_j)^T + (1+\lambda^i) \, \mathbf{C}_{d}  \right]^{-1} \left( \mathbf{d}_j^o - \mathbf{g} \left( \mathbf{m}_j^{i} \right) \right) \, . & &
\end{IEEEeqnarray}
\end{linenomath*} 
}
As to be shown later (see Eq. (\ref{eq:wls_rlm_sln_con})), with a suitable choice for $\lambda^i$, the iteration formula in Eq. (\ref{eq:discarding_a_term_enrml}) is consistent with the one derived from the RLM algorithm. Under suitable technical conditions (see, for example, \cite{jin2010regularized}), $\mathbf{m}_j^{i}$ will converge to a solution of the equation $\mathbf{g} \left( \mathbf{m} \right) = \mathbf{d}_j^o$ as $i \rightarrow +\infty$. 

Following the EnKF, one may introduce a further approximation to the gain matrix in Eq. (\ref{eq:discarding_a_term_enrml}), such that
{\small
\begin{linenomath*}  
\begin{IEEEeqnarray}{crlc} \label{eq:gain_approx_enrml}
& \mathbf{C}_m^i (\mathbf{G}^i_j)^T \left[ \mathbf{G}^i_j \mathbf{C}_m^i (\mathbf{G}^i_j)^T + (1+\lambda^i) \, \mathbf{C}_{d}  \right]^{-1} \approx \mathbf{S}_m^i \left( \mathbf{S}_d^i \right)^T \left( \mathbf{S}_d^i \left( \mathbf{S}_d^i \right)^T + (1+\lambda^i) \, \mathbf{C}_{d} \right)^{-1} \, . & & 
\end{IEEEeqnarray}  
\end{linenomath*} 
}
Therefore, with the above approximations, one has the following final iteration formula
{\small
\begin{linenomath*}  
\begin{IEEEeqnarray}{crlc} \label{eq:final_iteration_form_enrml}
& \mathbf{m}_j^{i+1} = \mathbf{m}_j^{i} + \mathbf{S}_m^i \left( \mathbf{S}_d^i \right)^T \left( \mathbf{S}_d^i \left( \mathbf{S}_d^i \right)^T + (1+\lambda^i) \, \mathbf{C}_{d} \right)^{-1} \left( \mathbf{d}_j^o - \mathbf{g} \left( \mathbf{m}_j^{i} \right) \right) \, . & &  
\end{IEEEeqnarray}
\end{linenomath*}  
}  
Comparing Eqs. (\ref{eq:final_iteration_form_enrml}) and (\ref{eq:linear_ls}), it is clear that the  iteration formulae become identical when $(1+\lambda^i)$ in Eq. (\ref{eq:final_iteration_form_enrml}) is replaced by $\gamma^i$ in Eq. (\ref{eq:linear_ls}).

For distinction, we call the algorithm in Eq. (\ref{eq:final_iteration_form_enrml}) \textit{approximate Levenberg-Marquardt ensemble randomized maximum likelihood} (aLM-EnRML) method. In the aLM-EnRML, starting from an initial value $\lambda^0$, the subsequent values of $\lambda^i$ will decrease if the average data mismatch with respect to the models is reduced, otherwise the values of $\lambda^i$ will increase instead. 
The iteration process will stop if the maximum iteration number is reached, or if the relative change of the average data mismatch in two consecutive iteration steps is lower than a given threshold.  

\subsection*{The regularized Levenberg-Marquardt (RLM) algorithm and an extension}

In the sequel we first introduce the regularized Levenberg-Marquardt (RLM, as used in \cite{Iglesias2013regularizing,Luo2014ensemble}) algorithm in the context of deterministic inverse problems theory, and then consider an extension of the RLM algorithm as an ensemble data assimilation method. In the course of deduction, the similarities between the ES-MDA/aLM-EnRML and the extended method in the current study will become clear, while some implications in the extended method will also be discussed. Note that in a recent work \cite{Iglesias2014iterative}, the author also considers using the RLM algorithm in the context of ensemble smoothing. Compared to \cite{Iglesias2014iterative}, the current study has different focuses, e.g., in terms of the way in approximating the gradient of the cost function of the minimization problem, and the parameter rule used in the RLM algorithm (see (\ref{eq:par_rule}) later)\footnote{To summarize, \cite{Iglesias2014iterative,Iglesias2013regularizing,Luo2014ensemble} and the current work are all related to the RLM algorithm. \cite{Iglesias2014iterative,Iglesias2013regularizing} use the parameter rule based on, for example, the work \cite{hanke1997regularizing}, while \cite{Luo2014ensemble} and the current work use the parameter rule based on \cite{jin2010regularized}. Although different in the concrete forms, both parameter rules are proven to lead to local convergence of the RLM algorithm \cite{hanke1997regularizing,jin2010regularized}. In addition, \cite{Iglesias2013regularizing,Luo2014ensemble} both aim to find a single solution, but employ different methods in approximating the gradient of the cost function, as discussed previously. In contrast, \cite{Iglesias2014iterative} and the current study target for multiple solutions in the context of iES, and differentiate each other in terms of the ways in constructing the data square root matrices (e.g., \cite{Iglesias2014iterative} uses the same way as in the EnKF). In Appendix \ref{sec:appdendix_iES}, we also provide an alternative point of view to interpret the iES algorithm derived in the current study.}. Readers are referred to \cite{Iglesias2014iterative} for more detail. 

In the conventional deterministic inverse problem theory (see, for example, \cite{Engl2000-regularization,kaltenbacher2008iterative}), one aims to find a single set of parameters (e.g., a reservoir model $\mathbf{m}$) whose simulated output (e.g., $\mathbf{g} (\mathbf{m})$) matches the observation $\mathbf{d}^o$. Intuitively, one can search for such a model by solving the following weighted least-squares problem 
{\small
\begin{linenomath*}  
\begin{IEEEeqnarray}{cc} \label{eq:wls_original}
\underset{\mathbf{m}}{\operatorname{argmin}} \left( \mathbf{d}^o - \mathbf{g} \left( \mathbf{m} \right) \right)^T \mathbf{C}_{d}^{-1} \left( \mathbf{d}^o - \mathbf{g} \left( \mathbf{m} \right) \right) \, ,
\end{IEEEeqnarray} 
\end{linenomath*}   
}
which aims to minimize the \textit{data mismatch} $\left( \mathbf{d}^o - \mathbf{g} \left( \mathbf{m} \right) \right)^T \mathbf{C}_{d}^{-1} \left( \mathbf{d}^o - \mathbf{g} \left( \mathbf{m} \right) \right)$, where $\mathbf{C}_{d}^{-1}$ is the weight matrix associated with the data mismatch term. For convenience of discussion later, let us define $\Vert \mathbf{d} \Vert_{2} \equiv \sqrt{\mathbf{d}^T \mathbf{d}}$ and $\Vert \mathbf{d} \Vert_{\mathbf{C}_{d}} \equiv \sqrt{\mathbf{d}^T \mathbf{C}_{d}^{-1} \mathbf{d}}$ for a vector $\mathbf{d}$, then it is clear that $\left( \mathbf{d}^o - \mathbf{g} \left( \mathbf{m} \right) \right)^T \mathbf{C}_{d}^{-1} \left( \mathbf{d}^o - \mathbf{g} \left( \mathbf{m} \right) \right) = \Vert \mathbf{d}^o - \mathbf{g} \left( \mathbf{m} \right) \Vert_{\mathbf{C}_{d}}^2 = \Vert \mathbf{C}_{d}^{-1/2}\left(\mathbf{d}^o - \mathbf{g} \left( \mathbf{m} \right) \right) \Vert_{2}^2$. 

When the dimension of the observation space is lower than that of the model space (as is often the case in practice), (\ref{eq:wls_original}) is an under-determined and ill-posed problem, which has some undesirable features, e.g., non-uniqueness of the solution and potentially large sensitivity of the solution to the observation, in the sense that even a very small change in the observation might lead to a large change in the solution \cite{Engl2000-regularization}. To mitigate the above problems, in deterministic inverse problems theory it is customary to introduce a certain regularization term to (\ref{eq:wls_original}). Here we consider Tikhonov regularization, in which a regularization term, in terms of $\left( \mathbf{m} - \mathbf{m}^{b} \right)^T \mathbf{C}_{m}^{-1} \left( \mathbf{m} - \mathbf{m}^{b} \right)$, is introduced to (\ref{eq:wls_original}), such that the weighted least-squares problem becomes 
{\small
\begin{linenomath*}  
\begin{IEEEeqnarray}{cc} \label{eq:wls_stablized}
\underset{\mathbf{m}}{\operatorname{argmin}} \left( \mathbf{d}^o - \mathbf{g} \left( \mathbf{m} \right) \right)^T \mathbf{C}_{d}^{-1} \left( \mathbf{d}^o - \mathbf{g} \left( \mathbf{m} \right) \right) +  \gamma \left( \mathbf{m} - \mathbf{m}^{b} \right)^T \mathbf{C}_{m}^{-1} \left( \mathbf{m} - \mathbf{m}^{b} \right) \, ,
\end{IEEEeqnarray} 
\end{linenomath*}   
}
where $\gamma$ is a positive scalar, $\mathbf{m}^{b}$ denotes a background model, and $\mathbf{C}_{m}^{-1}$ is the associated weight matrix. The choice of $\gamma$ affects the relative weights assigned to the data mismatch term and the regularization term, and thus influences the solution of (\ref{eq:wls_stablized}). For instance, if $\gamma$ is relatively large (e.g., tending to $\infty$ in the extreme case), then the obtained solution will approach $\mathbf{m}^{b}$. On the other hand, if $\gamma$ is relatively small (e.g., tending to $0$), then (\ref{eq:wls_original}) and (\ref{eq:wls_stablized}) tend to coincide with each other, and the solution of (\ref{eq:wls_stablized}) can be considered as an approximation to one of the solutions of (\ref{eq:wls_original}). Apart from $\gamma$, the matrices $\mathbf{C}_{m}$ and $\mathbf{C}_{d}$ would also affect the relative weights between the regularization and data mismatch terms. If, in addition, $\mathbf{C}_{m}$ and $\mathbf{C}_{d}$ are chosen to be the error covariance matrices of the background and the observations, respectively, then the correlations of the variables in the model and observation spaces, respectively, are also incorporated in the minimization problem.      

In general, one may wish to choose a $\gamma$ such that the data mismatch of the resulting solution is comparable to the noise level of the observations \cite{Engl2000-regularization,kaltenbacher2008iterative}. The true noise level is often unknown, therefore in practice it may be replaced by, for instance, a threshold value proportional to $p$ (i.e., the size of $\mathbf{d}^o$). Readers are referred to, for example, \cite{Iglesias2013regularizing,Luo2012-residual,Oliver2008} for the rationales behind this choice.   

For linear systems, there are straightforward ways for one to construct a solution that has the desired data mismatch (see, for example, \cite{Luo2012-residual}). For nonlinear systems, however, one often has to rely on a certain iterative algorithm to find the solution. One of such methods, called the regularized Levenberg-Marquardt algorithm following \cite{jin2010regularized} (or the regularizing Levenberg-Marquardt algorithm following \cite{hanke1997regularizing}), constructs a sequence of model $\{ \mathbf{m}^i \}$ ($i=1,2,\dotsb$) by solving a linearized weighted least-squares problem     
{\small
\begin{linenomath*}  
\begin{IEEEeqnarray}{ll} \label{eq:wls_rlm}
\underset{\mathbf{m}^{i+1}}{\operatorname{argmin}} & \left( \mathbf{d}^o - \mathbf{g} \left( \mathbf{m}^{i} \right) -\mathbf{G}_{i} (\mathbf{m}^{i+1} - \mathbf{m}^{i}) \right)^T \mathbf{C}_{d}^{-1} \left( \mathbf{d}^o - \mathbf{g} \left( \mathbf{m}^{i} \right) -\mathbf{G}_{i} (\mathbf{m}^{i+1} - \mathbf{m}^{i}) \right) \\
& +  \gamma^{i} \left( \mathbf{m}^{i+1} - \mathbf{m}^{i+1,b} \right)^T \mathbf{C}_{m}^{-1} \left( \mathbf{m}^{i+1} - \mathbf{m}^{i+1,b} \right) \, , \nonumber
\end{IEEEeqnarray} 
\end{linenomath*}  
}
at each iteration step. In (\ref{eq:wls_rlm}), $\mathbf{G}_{i}$ is the Jacobian of $\mathbf{g}$ evaluated at $\mathbf{m}^{i}$, such that $\mathbf{g} \left( \mathbf{m}^{i} \right) + \mathbf{G}_{i} (\mathbf{m}^{i+1} - \mathbf{m}^{i})$ represents a first order Taylor approximation to $\mathbf{g} \left( \mathbf{m}^{i+1} \right)$; $\gamma^{i}$ is a positive scalar as in (\ref{eq:wls_stablized}), but now changes over the iteration; $\mathbf{m}^{i+1,b}$ is the background model that may also be adaptive with the iteration. A convenient choice can be $\mathbf{m}^{i+1,b} = \mathbf{m}^{i}$, meaning that we want the new model $\mathbf{m}^{i+1}$ not to be too far away from the previous model, such that the approximation $ \mathbf{g} \left( \mathbf{m}^{i+1} \right) \approx \mathbf{g} \left( \mathbf{m}^{i} \right) + \mathbf{G}_{i} (\mathbf{m}^{i+1} - \mathbf{m}^{i})$ can be roughly valid \cite{Luo2014ensemble}. In addition, the choice $\mathbf{m}^{i+1,b} = \mathbf{m}^{i}$ also simplifies the iteration formula, as will be shown below.

The solution of the weighted least-squares problem (\ref{eq:wls_rlm}) is given by
{\small
\begin{linenomath*}  
\begin{IEEEeqnarray}{rll} \label{eq:wls_rlm_sln_inv}
&\mathbf{m}^{i+1}  = & \, \mathbf{m}^{i} +  \Delta\mathbf{m}^{i} \\
& \Delta\mathbf{m}^{i} = & \, \left[ \gamma^{i} \mathbf{C}_m^{-1} + \left( \mathbf{G}^i \right)^T \mathbf{C}_{d}^{-1}  \mathbf{G}^i  \right]^{-1} \left[ \left( \mathbf{G}^i \right)^T \mathbf{C}_{d}^{-1} \left( \mathbf{d}^o - \mathbf{g} \left( \mathbf{m}^{i} \right) \right) 
 - \gamma^{i} \mathbf{C}_m^{-1} \left( \mathbf{m}^{i} - \mathbf{m}^{i+1,b} \right) \right] \, , \nonumber
\end{IEEEeqnarray} 
\end{linenomath*}   
}
which is similar to Eq. (\ref{eq:lm_sln}) (e.g., by replacing $1+\lambda^i$ therein with $\gamma^i$). By letting $\mathbf{m}^{i+1,b} = \mathbf{m}^{i}$ and with some algebra, one has the simplified iteration formula   
{\small
\begin{linenomath*}  
\begin{IEEEeqnarray}{crlc} \label{eq:wls_rlm_sln_con}
& \mathbf{m}^{i+1} = \mathbf{m}^{i} + \mathbf{C}_m (\mathbf{G}^i)^T \left[ \mathbf{G}^i \mathbf{C}_m (\mathbf{G}^i)^T + \gamma^i \, \mathbf{C}_{d}  \right]^{-1} \left( \mathbf{d}^o - \mathbf{g} \left( \mathbf{m}^{i} \right) \right) \, . & &
\end{IEEEeqnarray}
\end{linenomath*}  
}
which is thus similar to Eq. (\ref{eq:final_iteration_form_enrml}). One may also let the weight matrix $\mathbf{C}_m$ be adaptive with the iteration, which will be considered later in the context of iES. Intuitively, provided that the step size $\Delta\mathbf{m}^{i}$ is small enough (which can be controlled by $\gamma^i$) such that the first order Taylor expansion of $\mathbf{g}( \mathbf{m}^{i+1} )$ around $\mathbf{m}^{i}$ is a roughly valid approximation of $\mathbf{g}( \mathbf{m}^{i+1} )$, one has
{\small
\begin{linenomath*}  
\begin{IEEEeqnarray}{ll} \label{eq:apprx_ineq}
\Vert \mathbf{g}( \mathbf{m}^{i+1} ) - \mathbf{d}^{o} \Vert_{\mathbf{C}_d}^2 & \approx \Vert \mathbf{g}\left( \mathbf{m}^{i} \right) - \mathbf{d}^{o}  + \mathbf{G}^i \left(\mathbf{m}^{i+1}  - \mathbf{m}^{i}\right) \Vert_{\mathbf{C}_d}^2 \nonumber \\
& = \Vert [ \mathbf{I} - \mathbf{G}^i \mathbf{C}_m (\mathbf{G}^i)^T \left[ \mathbf{G}^i \mathbf{C}_m (\mathbf{G}^i)^T + \gamma^i \, \mathbf{C}_{d}  \right]^{-1} ]  \left(\mathbf{g}( \mathbf{m}^{i} ) - \mathbf{d}^{o} \right) \Vert_{\mathbf{C}_d}^2  \nonumber \\
& = \Vert \gamma^i \, \mathbf{C}_{d} \left[ \mathbf{G}^i \mathbf{C}_m (\mathbf{G}^i)^T + \gamma^i \, \mathbf{C}_{d}  \right]^{-1}  \left(\mathbf{g}( \mathbf{m}^{i} ) - \mathbf{d}^{o} \right) \Vert_{\mathbf{C}_d}^2  \nonumber \\
& = \left( \mathbf{C}_{d}^{-1/2} \left(\mathbf{g}( \mathbf{m}^{i} ) - \mathbf{d}^{o} \right) \right)^T \left[ \gamma^i \left[ \left( \mathbf{C}_{d}^{-1/2} \mathbf{G}^i \right) \mathbf{C}_m (\mathbf{C}_{d}^{-1/2} \mathbf{G}^i)^T + \gamma^i \, \mathbf{I}_{p}  \right]^{-1} \right]^2 \left( \mathbf{C}_{d}^{-1/2} \left(\mathbf{g}( \mathbf{m}^{i} ) - \mathbf{d}^{o} \right) \right)  \nonumber \\
& = \Vert \gamma^i \left[ \left( \mathbf{C}_{d}^{-1/2} \mathbf{G}^i \right) \mathbf{C}_m (\mathbf{C}_{d}^{-1/2} \mathbf{G}^i)^T + \gamma^i \, \mathbf{I}_{p}  \right]^{-1}  \left( \mathbf{C}_{d}^{-1/2} \left(\mathbf{g}( \mathbf{m}^{i} ) - \mathbf{d}^{o} \right) \right) \Vert_2^2  \nonumber \\
& \leq \Vert \gamma^i \left[ \left( \mathbf{C}_{d}^{-1/2} \mathbf{G}^i \right) \mathbf{C}_m (\mathbf{C}_{d}^{-1/2} \mathbf{G}^i)^T + \gamma^i \, \mathbf{I}_{p}  \right]^{-1} \Vert_{2}^2 ~ \Vert \mathbf{g}( \mathbf{m}^{i} ) - \mathbf{d}^{o} \Vert_{\mathbf{C}_d}^2  \nonumber \\
& \leq \Vert \mathbf{g}( \mathbf{m}^{i} ) - \mathbf{d}^{o} \Vert_{\mathbf{C}_d}^2 \, , \nonumber
\end{IEEEeqnarray}
\end{linenomath*}  
since $\Vert \gamma^i \left[ \left( \mathbf{C}_{d}^{-1/2} \mathbf{G}^i \right) \mathbf{C}_m (\mathbf{C}_{d}^{-1/2} \mathbf{G}^i)^T + \gamma^i \, \mathbf{I}_{p}  \right]^{-1} \Vert_{2}^2 \leq 1$, where $\mathbf{I}_{p}$ denotes the $p$-dimensional identity matrix, and $\Vert \mathbf{A} \Vert_2$ represents the spectral norm of the matrix $\mathbf{A}$.} This implies that the data mismatch $\Vert \mathbf{g}( \mathbf{m}^{i+1} ) - \mathbf{d}^{o} \Vert_{\mathbf{C}_d}^2$ at the $(i+1)$-th iteration tends to be no larger than the one $\Vert \mathbf{g}( \mathbf{m}^{i} ) - \mathbf{d}^{o} \Vert_{\mathbf{C}_d}^2$ at the previous iteration. 

In terms of the choice of $\gamma^i$, the following parameter rule from \cite{jin2010regularized}
{\small
\begin{linenomath*}  
\begin{IEEEeqnarray}{ll} \label{eq:par_rule}
& \gamma^0 > 0 \, , \nonumber \\
& \gamma^{i+1} = \rho \, \gamma^i, \text{~with~} 1/r < \rho < 1 \text{~for some scalar~} r > 1 \, , \\
& \underset{i \rightarrow +\infty}{\lim} \gamma^i = 0 \, , \nonumber
\end{IEEEeqnarray} 
\end{linenomath*}  
}
can be used, in which the scalar sequence $\{\gamma^i\}$ gradually reduces to zero as $i$ tends to $+\infty$, while the presence of the lower bound $1/r$ for the coefficient $\rho$ aims to prevent any abrupt drop-down of $\gamma^i$ to zero. When Eq. (\ref{eq:wls_rlm_sln_con}) is used in conjunction with (\ref{eq:par_rule}), it can be analytically shown that the data mismatch of the sequence of models $\{\mathbf{m}^{i}\}$ converges to zero locally as $i \rightarrow +\infty$ and the observation noise level tends to zero, provided that the equation $\mathbf{g}(\mathbf{m}) = \mathbf{d}^{o}$ is solvable and some other conditions are satisfied \cite{jin2010regularized}. Instead of using (\ref{eq:par_rule}), one can adopt other parameter rules that also lead to local convergence of the data mismatch under similar technical conditions (see, for example,  \cite{hanke1997regularizing}).  

When the observation noise is present, however, it may not be desirable to have a too small data mismatch in order to prevent over-fitting the observations. In general, one may wish to let the iteration stop when the data mismatch of the current iterated model is comparable to the noise level, which is often referred to as the \textit{discrepancy principle} \cite{Engl2000-regularization,kaltenbacher2008iterative}. Of course, in practice the true noise level is typically unknown. As a result, one may let the iteration process Eq. (\ref{eq:wls_rlm_sln_con}) stop when either of the following three conditions are satisfied: (a) the data mismatch of the iterated model is lower than a pre-set threshold $\beta_u^2 \, p$ for the first time for a given positive scalar $\beta_u$, in light of the discussions in \cite{Iglesias2013regularizing,Luo2014ensemble,Luo2012-residual,Oliver2008}; or (b) the maximum iteration number is reached; or (c) the relative change of the data mismatch in two consecutive iteration steps is lower than a given threshold ($0.01\%$ in our implementation). Note that, with the stopping condition (a), the total number of (actual) iteration steps can be less than the maximum iteration number, for instance, when the RLM reaches the threshold $\beta_u^2 \, p$ earlier than up to the maximum iteration number. Conditions (b) and (c) are the same as those in the aLM-EnRML, and are mainly introduced to control the runtime of the iteration process. These settings will be applied to the aLM-EnRML and our extension scheme in the experiments later, in order to make the comparison of both algorithms be conducted under the same experimental settings as far as possible.  

In the practical implementation of the RLM algorithm (and our extension scheme below), there are a few issues that need to be taken into account. Firstly, in the theoretical analysis of the RLM algorithm (see, for example, \cite{Engl2000-regularization,hanke1997regularizing,jin2010regularized,kaltenbacher2008iterative}), it typically assumes that the initial model stays sufficiently close to a solution of the equation $\mathbf{g}(\mathbf{m}) = \mathbf{d}^{o}$, which is not necessarily true in reality.
In addition, as can be seen in our previous discussion, the validity of the first order Taylor approximation $ \mathbf{g} \left( \mathbf{m}^{i+1} \right) \approx \mathbf{g} \left( \mathbf{m}^{i} \right) + \mathbf{G}_{i} (\mathbf{m}^{i+1} - \mathbf{m}^{i})$ is essential to the derivation of the RLM algorithm.  
In practice, it is often not a trivial issue to obtain the Jacobian $\mathbf{G}_{i}$. If the exact Jacobian $\mathbf{G}_{i}$ is not available, then one must evaluate it based on a certain approximation scheme, which often has to struggle between the computational efficiency and accuracy. Putting these factors together, in practice it is likely that the iteration process, Eq. (\ref{eq:wls_rlm_sln_con}), may occasionally lead to higher data mismatch for the current iterated model, in comparison to that of the previous one, due to, for example, a too large step size that invalidates the first order Taylor approximation. In such cases, following \cite{chen2013-levenberg}, a simple remedy can then be to increase the parameter $\gamma^i$ to some extent and re-execute the iteration in Eq. (\ref{eq:wls_rlm_sln_con}) (which will then have a smaller step size). This strategy is essentially similar to the back-tracking line search method in optimization theory (see, for example, \cite{Nocedal-numerical}) and works well in our numerical studies.
 %


In the remainder of this section we consider an extended RLM scheme in the context of ensemble smoothing. As in the aLM-EnRML, instead of examining the data mismatch of a single model, we are more interested in studying how the average data mismatch of the ensemble models changes over the iteration step. To this end, the weighted least-squares problem in (\ref{eq:wls_stablized}) is replaced with the following minimum-average-cost (MAC) problem, namely,
{\small
\begin{linenomath*}    
\begin{IEEEeqnarray}{lll} \label{eq:wls_rlm_mac}
\underset{\{\mathbf{m}^{i+1}_j\}_{j=1}^{N_e}}{\operatorname{argmin}} & \dfrac{1}{N_e} \sum\limits_{j=1}^{N_e} & \, \left[ \left( \mathbf{d}^o_j - \mathbf{g} \left( \mathbf{m}^{i+1}_j \right) \right)^T \mathbf{C}_{d}^{-1} \left( \mathbf{d}^o_j - \mathbf{g} \left( \mathbf{m}^{i+1}_j \right) \right) \right. \\
& & \quad \left. +  \gamma^{i} \left( \mathbf{m}^{i+1}_j - \mathbf{m}^{i}_j \right)^T \left( \tilde{\mathbf{C}}_{m}^{i} \right)^{-1} \left( \mathbf{m}^{i+1}_j - \mathbf{m}^{i}_j \right) \right] \, , \nonumber
\end{IEEEeqnarray}
\end{linenomath*}  
}
where $\tilde{\mathbf{C}}_{m}^{i} = \tilde{\mathbf{S}}_{m}^{i} \left( \tilde{\mathbf{S}}_{m}^{i} \right)^T$ ($\tilde{\mathbf{S}}_{m}^{i}$ will be specified later), and $\mathbf{d}^o_j$, $\mathbf{m}^{i}_j$ and $\mathbf{m}^{i+1}_j$ are the same as those defined previously. For distinction, we call the resulting iES \textit{regularized Levenberg-Marquardt algorithm based on minimum-average-cost} (RLM-MAC for short). Appendix \ref{sec:appdendix_iES} also includes some thoughts that aim to interpret the RLM-MAC from the point of view of an expectation-maximization algorithm \cite{Dempster1977maximum}.        

To solve (\ref{eq:wls_rlm_mac}), it is necessary to linearize the forward model $\mathbf{g}$ as in the RLM algorithm. Note that in (\ref{eq:wls_rlm_mac}) there are $N_e$ simulated observation terms $\mathbf{g} \left( \mathbf{m}^{i+1}_j \right)$ ($j=1,2,\dotsb,N_e$). To reduce the computational cost in evaluating the Jacobian matrices, one may choose to linearize the forward model $\mathbf{g}$, for each ensemble member $\mathbf{m}^{i+1}_j$, around a ``common'' point, say $\mathbf{m}^{i}_c$, such that 
{\small
\begin{linenomath*}  
\begin{IEEEeqnarray}{ll} \label{eq:linearization_common}
\mathbf{g} \left( \mathbf{m}^{i+1}_j \right) \approx \mathbf{g} \left(\mathbf{m}^{i}_c \right) + \mathbf{G}^{i}_c \left( \mathbf{m}^{i+1}_j - \mathbf{m}^{i}_c \right) \, , 
\end{IEEEeqnarray}
\end{linenomath*}  
}
where $\mathbf{G}^{i}_c$ is the Jacobian of $\mathbf{g}$ at $\mathbf{m}^{i}_c$. Inserting Eq. (\ref{eq:linearization_common}) into (\ref{eq:wls_rlm_mac}) and with some algebra (see Appendix \ref{sec:appdendix_deduction}), one has the following iteration formula 
{\small
\begin{linenomath*}  
\begin{IEEEeqnarray}{lllll} \label{eq:wls_rlm_mac_orig}
& \mathbf{m}^{i+1}_j = & \, \mathbf{m}^{i}_j + \Delta\mathbf{m}^{i}_j & & \\
& \Delta\mathbf{m}^{i}_j = & \, \tilde{\mathbf{S}}_m^i (\mathbf{G}^{i}_c \tilde{\mathbf{S}}_m^i)^T \left[ (\mathbf{G}^{i}_c \tilde{\mathbf{S}}_m^i) (\mathbf{G}^{i}_c \tilde{\mathbf{S}}_m^i)^T + \gamma^i \, \mathbf{C}_{d}  \right]^{-1} \left( \mathbf{d}^o_j - \mathbf{g} \left( \mathbf{m}^{i}_c \right) - \mathbf{G}^{i}_c \left( \mathbf{m}^{i}_j - \mathbf{m}^{i}_c \right) \right) \, ,& & \nonumber 
\end{IEEEeqnarray}
\end{linenomath*}  
}
for $j = 1, \dotsb, N_e$. If one lets 
{\small  
\begin{linenomath*}  
\begin{IEEEeqnarray}{ll} \label{eq:new_Smi}
\tilde{\mathbf{S}}_m^i = \frac{1}{\sqrt{{N_e}-1}}\left[\mathbf{m}_1^i - {\mathbf{m}}^i_c,\dotsb, \mathbf{m}_{N_e}^i - {\mathbf{m}}^i_c \right], 
\end{IEEEeqnarray}
\end{linenomath*}  
}
and adopts the approximation $\mathbf{G}^{i}_c \left( \mathbf{m}^{i}_j - \mathbf{m}^{i}_c \right) \approx \mathbf{g} \left( \mathbf{m}^{i}_j \right) - \mathbf{g} \left(\mathbf{m}^{i}_c \right)$, then Eq. (\ref{eq:wls_rlm_mac_orig}) reduces to
{\small
\begin{linenomath*}  
\begin{IEEEeqnarray}{crlc} \label{eq:wls_rlm_mac_simplified}
& \mathbf{m}^{i+1}_j = \mathbf{m}^{i}_j + \tilde{\mathbf{S}}_m^i (\tilde{\mathbf{S}}_d^i)^T \left( \tilde{\mathbf{S}}_d^i (\tilde{\mathbf{S}}_d^i)^T + \gamma^i \, \mathbf{C}_{d}  \right)^{-1} \left( \mathbf{d}^o_j - \mathbf{g} \left( \mathbf{m}^{i}_j \right) \right) \, , & &
\end{IEEEeqnarray}
\end{linenomath*}  
}
where 
{\small
\begin{linenomath*}  
\begin{IEEEeqnarray}{ll} \label{eq:new_Sdi}
\tilde{\mathbf{S}}_d^i = \frac{1}{\sqrt{{N_e}-1}}\left[\mathbf{g} \left( \mathbf{m}^{i}_1 \right) - \mathbf{g} \left(\mathbf{m}^i_c \right),\dotsb, \mathbf{g} \left( \mathbf{m}^{i}_{N_e} \right) - \mathbf{g} \left(\mathbf{m}^i_c \right) \right]. 
\end{IEEEeqnarray}
\end{linenomath*}   
}   
Compared to the iteration formulae of ES-MDA and aLM-EnRML (e.g., Eq. (\ref{eq:final_iteration_form_enrml})), the iteration formula of RLM-MAC in general differs in the way of constructing the square root matrices $\tilde{\mathbf{S}}_m^i$ and $\tilde{\mathbf{S}}_d^i$. Possible choices for $\mathbf{m}^i_c$ could be, for instance, the mean $\bar{\mathbf{m}}^{i}$ or a member (e.g., the one closest to $\bar{\mathbf{m}}^{i}$ ) of the ensemble $\{ \mathbf{m}^{i}_j \}_{j=1}^{N_e}$. In the current work we let $\mathbf{m}^i_c = \bar{\mathbf{m}}^{i}$. 

Under the choice $\mathbf{m}^i_c = \bar{\mathbf{m}}^{i}$, $\tilde{\mathbf{S}}_m^i$ coincides with ${\mathbf{S}}_m^i$, therefore the resulting RLM-MAC algorithm mainly differs from the aLM-EnRML in the data square root matrix. In general, $\tilde{\mathbf{S}}_d^i$ and ${\mathbf{S}}_d^i$ may have different ``centering points'': in Eq. (\ref{eq:esmad_Sdi}), the simulated observations $\mathbf{g} \left( \mathbf{m}^{i}_j \right)$ in ${\mathbf{S}}_d^i$ center around $\overline{\mathbf{g}\left(\mathbf{m}_j^{i}\right)}$, while in $\tilde{\mathbf{S}}_d^i$ they center around $\mathbf{g} \left(\bar{\mathbf{m}}^{i} \right)$ instead. In certain circumstances, e.g., when the forward model $\mathbf{g}$ is linear or weakly nonlinear, $\overline{\mathbf{g}\left(\mathbf{m}_j^{i}\right)}$ and $\mathbf{g} \left(\bar{\mathbf{m}}^{i} \right)$ may be close to each other\footnote{In particular, when $\mathbf{g}$ is linear, $\overline{\mathbf{g}\left(\mathbf{m}_j^{i}\right)} = \mathbf{g} \left(\bar{\mathbf{m}}^{i} \right)$, such that the RLM-MAC and aLM-EnRML become identical.}. In some other cases, the distinction between $\overline{\mathbf{g}\left(\mathbf{m}_j^{i}\right)}$ and $\mathbf{g} \left(\bar{\mathbf{m}}^{i} \right)$ may be more substantial, and the behaviour of aLM-EnRML and RLM-MAC may thus become more diverged, as will be shown later. In terms of the performance comparison between the aLM-EnRML and RLM-MAC, however, we stress that, although one may see the relative superiority of one algorithm over the other in some specific cases, it is expected that the conclusion might not be valid in a broader sense, following the ``no free lunch theorems for optimization'' \cite{wolpert1997no}. Intuitively, since both the aLM-EnRML and RLM-MAC are local optimization algorithms and they may have different search directions, these two algorithms might end up with different local optima in general, and the relative superiority between these two algorithms may thus vary from case to case.    
       
Finally we note that, in the course of solving (\ref{eq:wls_rlm_mac}), we have made use of the first order Taylor approximation around a common point $\mathbf{m}^{i}_c$ in order to reduce the computational cost. In this regard, a theoretically more sound -- yet computationally more expensive -- strategy would be to linearize the forward model $\mathbf{g}$, for each ensemble model, around a certain point in its neighbourhood. A trade-off strategy may also be employed, e.g., by clustering the ensemble models into a small number of groups, and then linearizing $\mathbf{g}$ around a common point for each group. In addition, in the context of solving a MAC problem, it is also possible for one to explicitly incorporate into (\ref{eq:wls_rlm_mac}) additional terms that impose certain constraints onto the statistics of the ensemble members or the corresponding simulated observations, such that the iteration step may be in a form beyond the Kalman update formula as in Eq. (\ref{eq:wls_rlm_mac_simplified}). An investigation in this aspect will be carried out in the future.   

\section*{Numerical examples}
After discussing the similarities and differences between the ES-MDA, aLM-EnRML and RLM-MAC in the previous section, we focus on demonstrating potentially different behaviour of the ensemble smoothers that is resulted from using different ``centering points'' ($\overline{\mathbf{g}\left(\mathbf{m}_j^{i}\right)}$ or $\mathbf{g} \left(\bar{\mathbf{m}}^{i} \right)$) to construct the data square root matrix. 

As noted previously, in the ES-MDA, the total number $I$ of iteration steps is decided before the iteration process starts, and the parameters $\gamma^i$ are dependent on $I$, in light of the constraint (\ref{eq:ES-MDA_constraints}). The aLM-EnRML and RLM-MAC do not have the above features in general: Under the stopping conditions (a-c) in the previous section, the total numbers of the iteration steps of the aLM-EnRML and RLM-MAC depend on the threshold $\beta_u^2 \, p$ and the convergence speed of the algorithms in specific problems, subject to the constraint of maximum iteration number. In addition, the parameters $\gamma^i$ are chosen according to a different rule, which is independent of $I$ or the maximum iteration number in general. Therefore, for our purpose here, we confine ourselves to the comparison between the aLM-EnRML and the RLM-MAC. In the implementations, both algorithms apply the truncated singular value decomposition (TSVD) to the data square root matrices (normalized by $\mathbf{C}_d^{-1/2}$) in their iteration formulae\footnote{In doing so, the iteration formulae, e.g., that in Eq. (\ref{eq:wls_rlm_mac_simplified}), are re-expressed as $\mathbf{m}^{i+1}_j = \mathbf{m}^{i}_j + \tilde{\mathbf{S}}_m^i ( \mathbf{C}_d^{-1/2} \tilde{\mathbf{S}}_d^i)^T \left( (\mathbf{C}_d^{-1/2} \tilde{\mathbf{S}}_d^i ) ( \mathbf{C}_d^{-1/2} \tilde{\mathbf{S}}_d^i)^T + \gamma^i \, \mathbf{I}_p  \right)^{-1} \mathbf{C}_d^{-1/2} \left( \mathbf{d}^o_j - \mathbf{g} \left( \mathbf{m}^{i}_j \right) \right)$, where $\mathbf{I}_p$ is the $p$-dimensional identity matrix. The TSVD is applied to the product $(\mathbf{C}_d^{-1/2} \tilde{\mathbf{S}}_d^i ) ( \mathbf{C}_d^{-1/2} \tilde{\mathbf{S}}_d^i)^T$, which can have certain numerical benefits, e.g., mitigating the issue of different orders of amplitudes in the measurement data.}, and retain the leading eigenvalues which sum up to at least $99\%$ of the ``total energies'' (see \cite{chen2013-levenberg,tavakoli2010history} for the details). In the experiments below we first examine the performance of both algorithms in a strongly nonlinear system, and then apply both algorithms to a reservoir facies estimation problem and the Brugge field case. 

\subsection*{An initial state estimation problem in a strongly nonlinear system}
Here we adopt the strongly nonlinear $40$-dimensional Lorenz 96 (L96) model \cite{Lorenz-optimal} to demonstrate the potentially different behaviour between the aLM-EnRML and RLM-MAC. The governing equations of the L96 model are given by
\begin{linenomath*}  
\begin{equation} \label{eq:LE98}
\frac{dx_k}{dt} = \left( x_{k+1} - x_{k-2} \right) x_{k-1} - x_k + F, \, k=1, \dotsb, 40.
\end{equation}
\end{linenomath*}  
In Eq.~(\ref{eq:LE98}) the variables $x_{k}$ are defined in a cyclic way such that $x_{-1}=x_{39}$, $x_{0}=x_{40}$ and $x_{41}=x_{1}$. The L96 model is designed to mimic baroclinic turbulence in the midlatitude atmosphere. Because of its strongly nonlinear nature, this model is often employed to test the performance of data assimilation algorithms in the weather data assimilation community.

In this work we let $F=8$, with which the L96 model is chaotic. The L96 model is discretized through the fourth-order Runge-Kutta method with a constant integration step of $0.05$ (corresponding to 6 hours in real time, see \cite{Lorenz-optimal}). To initialize the model, we draw an initial state vector at random from the multivariate normal distribution $N(\mathbf{0},\mathbf{I}_{40})$, where $\mathbf{I}_{40}$ is the 40-dimensional identity matrix. To avoid the transit effect, the model states within the first 500 integration steps are discarded, and data assimilation starts after this ``spin-up'' period. For reference, hereafter we re-label the model state at time step 500 as the initial state, 
which has an impact on subsequent model states and their observations. In the experiment the initial state vector is assumed unknown, and will be estimated based on the observations at subsequent time steps. 

In the experiment the assimilation time window is 10 days, and the observations are available every 24 hours. At each observation time instant (e.g., 24h, 48h etc.), the synthetic observations are generated by first applying the nonlinear function $f(x) = x^3/5$ to the odd number state variables $x_1,x_3,\dotsb,x_{39}$ (overall 20 variables), and then adding a sample from the normal distribution $N(0,1)$ to each function value $f(x_k)$ ($k=1,3,\dotsb,39$). The total number of observation elements in the assimilation time window is thus $10 \times 20 = 200$. 

The initial background ensemble is drawn from a multivariate normal distribution whose mean and covariance are obtained from the ``climatological'' statistics of the L96 model in a long free-run with 100,000 integration steps (see, for example, \cite{Luo2014ensemble} for the details). The ensemble size of the aLM-EnRML is 100. Note that in the RLM-MAC one also needs to compute $\mathbf{g} \left(\bar{\mathbf{m}}^{i} \right)$, the simulated observation of the ensemble mean, apart from those of the ensemble members. Because of this, we let the ensemble size of the RLM-MAC be 99, such that the aLM-EnRML and RLM-MAC have the same number of simulated observations at each iteration step. The maximum number of iterations is 100, and the coefficient $\beta_u = 2$. 

In light of the requirement that the first order Taylor approximation be roughly valid, it is suggested to start with a relatively small step size at the beginning \cite{Luo2014ensemble}. As a result, the initial value $\gamma^0$ should be relatively large (but not too large in order to reduce the total number of iteration steps). In our implementation, we let $\gamma^i = \alpha^i \, \sqrt{\text{trace}(\mathbf{S}_d^i (\mathbf{S}_d^i)^T)} / N_e$, where $\text{trace}(\bullet)$ calculates the trace of a matrix, $\alpha^0 = 1$ and $\alpha^{i+1} = 0.9 \, \alpha^{i}$ if the average data mismatch of the ensemble is reduced, otherwise the $i$th iteration step is repeated, with $\alpha^{i} = 2 \, \alpha^{i} $, for maximum 5 times, following \cite{chen2013-levenberg}. In the subsequent two other case studies, the same way is applied to tune $\alpha^i$ adaptively. We stress that this should not be considered as the optimal rule. Instead, it is expected that one may improve it for specific applications in general.    

As the dimension of the L96 model is relatively low, we can afford to repeat the experiment 100 times. In each repetition, the initial background ensemble and the associated observations are drawn at random. In the same repetition run, the aLM-EnRML and RLM-MAC have the same initial background ensemble, the associated observations and the parameter rule in choosing $\gamma^i$.
  
Figures \ref{fig:L96_histogram} - \ref{fig:L96_pie_charts_rmse} report the data mismatch and RMSEs of the aLM-EnRML and RLM-MAC, obtained from 100 repetitions of the experiment\footnote{Here the data mismatch is calculated according to (\ref{eq:wls_original}), and is averaged over all ensemble members. The RMSE of a $m$-dimensional estimate $\hat{\mathbf{v}}$ with respect to its truth ${\mathbf{v}}^{tr}$ is give by $\Vert \hat{\mathbf{v}} - {\mathbf{v}}^{tr} \Vert_2 / \sqrt{m}$, where $\Vert \bullet \Vert_2$ denotes the $\ell_2$-norm.}. In terms of the data mismatch, from Figures \ref{fig:L96_histogram} and \ref{fig:L96_pie_charts_data_mismtach}, one can see that for the aLM-EnRML, its range is roughly between $10^3$ and $10^6$. Overall, $2\%$ of the data mismatch lies in the interval $[10^3,10^4]$, $51\%$ in the interval $[10^4,10^5]$, and $47\%$ in the interval $[10^5,10^6]$. For the RLM-MAC, the range of the data mismatch is roughly between $10^2$ and $10^6$. there are $11\%$ of the data mismatch contained in the interval $[10^2,10^3]$, $13\%$ in the interval $[10^3,10^4]$, $51\%$ in the interval $[10^4,10^5]$, and the remaining $25\%$ in the interval $[10^5,10^6]$. This suggests that in some cases the RLM-MAC will have data mismatch lower than the minimum one of the aLM-EnRML. Similar results can also be observed in terms of the RMSEs (Figures \ref{fig:L96_histogram} and \ref{fig:L96_pie_charts_rmse}). As can be seen in Figure \ref{fig:L96_histogram}, for the aLM-EnRML, its RMSEs seem to follow a uni-modal distribution, with its (single) peak around 3. On the other hand, for the RLM-MAC, the distribution of its RMSEs tends to be more flat, with one of its peaks being closer to zero. By counting the numbers of the RMSEs in the sub-intervals (e.g., $[0,1]$, $[1,2]$ etc) of the range $[0 , 6]$, the pie charts in Figure \ref{fig:L96_pie_charts_rmse} indicate that, the percentages of the RMSEs of the aLM-EnRML in the sub-intervals (from low to high values) are $2\%$, $18\%$, $34\%$, $31\%$, $12\%$ and $3\%$, respectively, while the percentages with respect to the RLM-MAC are $22\%$, $18\%$, $28\%$, $16\%$, $14\%$ and $2\%$. Overall, Figures \ref{fig:L96_histogram} and \ref{fig:L96_pie_charts_rmse} also suggest that in some cases the RLM-MAC will have RMSEs lower than the minimum one of the aLM-EnRML.    


Figure \ref{fig:L96_single_data_mismatch} shows the box plots of the data mismatch of the aLM-EnRML (left) and RLM-MAC (right) at different iteration steps in one repetition run. The ensemble average data mismatch in both iES decreases with the iteration. In this particular case, however, the data mismatch of the RLM-MAC tends to converge faster than that of the aLM-EnRML. In addition, its final data mismatch (up to the maximum iteration step in the figure) is substantially lower than that of the aLM-EnRML. We also note that, the top outlier in each box plot of the aLM-EnRML does not appear to follow the tendency of changes of other data mismatch values in the same box plot, which is a phenomenon not observed in the RLM-MAC. A possible explanation of this difference might be as follows: In the RLM-MAC, in order for the cost function in the MAC problem to decrease with respect to the change of each ensemble member (see the deduction in Appendix \ref{sec:appdendix_deduction}), it is natural to use the deviations
\begin{linenomath*} 
\begin{equation} \label{eq:data_sqrt_rlmmac}
\mathbf{g} \left( \mathbf{m}_j^{i} \right) - \mathbf{g} \left( \bar{\mathbf{m}}^i \right) 
\end{equation} 
\end{linenomath*} 
to construct the data square root matrix on top of the first order Taylor approximation. In the aLM-EnRML, however, (\ref{eq:data_sqrt_rlmmac}) is replaced by 
\begin{linenomath*}   
\begin{equation} \label{eq:data_sqrt_aLMEnRML}
\mathbf{g} \left( \mathbf{m}_j^{i} \right) - \overline{\mathbf{g}\left(\mathbf{m}_j^{i}\right)} \, . 
\end{equation}
\end{linenomath*} 
The difference between (\ref{eq:data_sqrt_rlmmac}) and (\ref{eq:data_sqrt_aLMEnRML}) would then depend on the degree of nonlinearity in $\mathbf{g}$. In this particular case, because of the strong nonlinearity in the L96 model, the difference between (\ref{eq:data_sqrt_aLMEnRML}) and (\ref{eq:data_sqrt_rlmmac}) may be relatively large (see Figure \ref{fig:obsDiff_boxPlot} later for a numerical investigation), such that the iteration formula based on the choice (\ref{eq:data_sqrt_aLMEnRML}) does not form gradient descent directions for all the ensemble members in the aLM-EnRML.

In accordance to the results in Figure \ref{fig:L96_single_data_mismatch}, top panels of Figure \ref{fig:L96_rmse_spread} show the box plots of RMSEs of the ensemble members with respect to the initial conditions at different iteration steps in the aLM-EnRML (left) and RLM-MAC (right), and bottom panels indicate the associated ensemble spreads\footnote{Suppose that $\{\mathbf{m}_j\}_{j=1}^{N_e}$ is an ensemble of models with the ensemble size being $N_e$, and that $m_{j,s}$ is the $s$th element of the vector $\mathbf{m}_j$. Let $\sigma^2_s$ be the sample variance of the $s$th elements of all ensemble members, then the ensemble spread is defined as $\sqrt{\sum_{s=1}^{m}\sigma^2_s / m}$, where $m$ is the size of the ensemble members $\mathbf{m}_j$.}. For both the aLM-EnRML and RLM-MAC, their RMSEs drop at the first few iteration steps, bounce back subsequently, and then drop again until they converge (although a slight oscillation can also be observed around iteration step 85 in the aLM-EnRML). In this particular case, the final RMSEs of the RLM-MAC tend to be lower than those of the aLM-EnRML. The ensemble spread of the aLM-EnRML monotonically decreases with the iteration, while that of the RLM-MAC reaches its minimum around iteration step 10, and then slightly increases afterwards. The final ensemble spreads of both algorithms are close to each other, and they are both underestimated in comparison with the corresponding RMSEs. 
One possible reason of this underestimation phenomenon may be because in both smoothers the ensemble members tend to converge to regions surrounding certain local minima, therefore the ensemble spreads are relatively small, while the RMSEs are determined by the distances between the local minima and the truth, and thus are not necessarily as small as the corresponding ensemble spreads.  
Apart from this, we also envision that the effects of finite ensemble size (e.g., sampling errors and correlations among ensemble members \cite{naevdal2011quantifying}) may also contribute to the underestimation phenomenon. 
 
Figure \ref{fig:obsDiff_boxPlot} depicts the normalized differences $\mathbf{C}_d^{-1/2} \left( \mathbf{g} \left(\bar{\mathbf{m}}^{i} \right) - \overline{\mathbf{g}\left(\mathbf{m}_j^{i}\right)} \right)$ at different iteration steps, based on the corresponding ensemble members of the RLM-MAC. Please note that in this case, the box plots indicate the spatial distribution of the elements of the vector $\mathbf{C}_d^{-1/2} \left( \mathbf{g} \left(\bar{\mathbf{m}}^{i} \right) - \overline{\mathbf{g}\left(\mathbf{m}_j^{i}\right)} \right)$, whose size is equal to the number of observations (rather than the ensemble size). In addition, in the box plots, $\mathbf{g} \left(\bar{\mathbf{m}}^{i} \right)$ and $\overline{\mathbf{g}\left(\mathbf{m}_j^{i}\right)}$ are computed from the same ensemble of the RLM-MAC at each iteration step, rather than one (e.g., $\mathbf{g} \left(\bar{\mathbf{m}}^{i} \right)$) from the ensemble of the RLM-MAC and another (e.g., $\overline{\mathbf{g}\left(\mathbf{m}_j^{i}\right)}$) from the corresponding ensemble of the aLM-EnRML. 

From Figure \ref{fig:obsDiff_boxPlot} one can see that the spreads of the normalized differences are relatively large at the first few iteration steps. This may be because the initial ensemble spreads are relatively large (see the lower right panel of Figure \ref{fig:L96_rmse_spread}), and the chaos nature of the L96 model tends to further amplify them. However, as the number of iterations increases, the ensemble members of the RLM-MAC tend to converge toward a certain local minimum, and the ensemble spread decreases rapidly. 
As a result, $\mathbf{g} \left(\bar{\mathbf{m}}^{i} \right)$ approach $ \overline{\mathbf{g}\left(\mathbf{m}_j^{i}\right)}$, such that the normalized differences decrease and tend to center around zero.     
 
\subsection*{A reservoir facies estimation problem} 

\begin{figure*}
\centering
\includegraphics[scale=0.5]{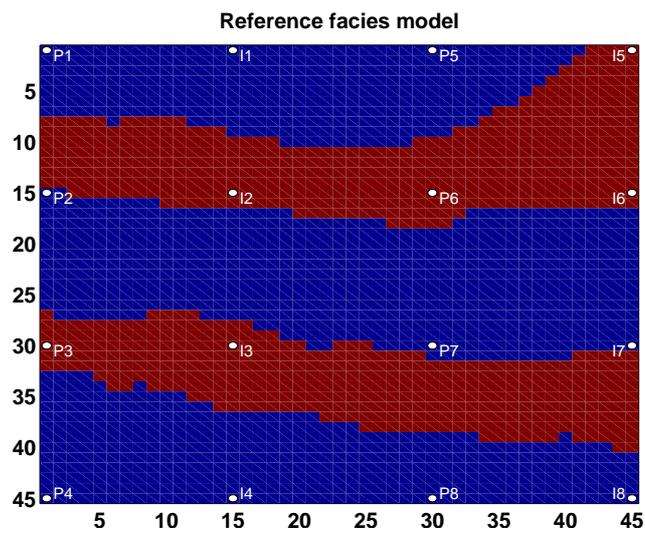}
\caption{\label{fig:true_facies} Reference facies model. The white dots indicate the locations of injectors (I1 - I8) and producers (P1 - P8).} 
\end{figure*}

Here we consider a simple 2D facies model (see Figure \ref{fig:true_facies}) previously studied in \cite{lorentzen2012history}. The reservoir consists of $45 \times 45$ gridblocks, with only oil and water inside. The porosity of the reservoir is 0.1 in each gridblock, but the permeability varies spatially: It is 10000 md in the channels, and 500 md in the background. In the field there are 8 water injectors (I1 - I8) and 8 producers (P1 - P8), and their locations are marked as white dots in Figure \ref{fig:true_facies}. The injectors are constrained by injection rates (15.9 m$^3$/day in each), while the producers are constrained by bottom hole pressures (BHP, 399 bar in each). 

The measurements used for history matching are the oil and water production rates in the producers and BHP in the injectors. To generate the synthetic measurements, we use the reference facies model in Figure \ref{fig:true_facies} as the input to ECLIPSE$^\copyright$ \cite{eclipse2010} and run the simulation for 1900 days. The measurements are then taken as the outputs of ECLIPSE$^\copyright$ every 190 days, plus certain Gaussian noise (whose variances are 0.16 m$^3$/day for production rates, and 0.07 bar for BHP). To enhance the connectivities of the estimated channels, eight statistical measures used in \cite{lorentzen2012history} are also assimilated into the model. These statistics measurements include the sample mean values of: (1) area of bodies (a body is defined as neighboring connected gridblocks with the same facies type, see \cite{lorentzen2012history}); (2) volume of bodies; (3) area-to-volume ratio of bodies; (4) maximal body extension in x-direction; (5) maximal body extension in y-direction; (6) $75$th percentile for the extensions in the x-direction; (7) number of bodies; and (8) fraction of the total volume of the facies with respect to the total volume of the field, where all the mean values are evaluated with respect to the initial background ensemble \cite{lorentzen2012history}. Accordingly, in the course of assimilation, the measurement error variances associated with these statistics measurements are taken as their sample variances that are also evaluated with respect to the initial background ensemble.  
We have also run both the aLM-EnRML and RLM-MAC when the statistics measurements are not included in assimilation. In this case, the final data mismatch values are close to what we show in Figure \ref{fig:facies_obj} later, but the corresponding estimated channels tend to be less connected (results not shown), consistent with the observation in \cite{lorentzen2012history}.    

We follow \cite{lorentzen2012history} and use the signed distance function in the level set method to characterize the facies. When using an iES for history matching, there is no need to augment the model and the corresponding state variables as in the EnKF. Therefore for the problem considered here, the initial ensemble only contains the signed distance function values at each gridblock, which are converted from the facies models generated by SNESIM \cite{strebelle2002conditional} through the MATLAB level set toolbox (version 1.1, \cite{mitchell2008flexible}). As in the previous example, the ensemble size in the aLM-EnRML is 100, while that in the RLM-MAC is 99. In addition, the aforementioned parameter rule $\gamma^i = \alpha^i \, \sqrt{\text{trace}(\mathbf{S}_d^i (\mathbf{S}_d^i)^T)} / N_e$ is also adopted in this example. The maximum number of iterations is 20, and the coefficient $\beta_u = 2$.         

A few initial models are shown in the first column of Figure \ref{fig:facies_models} (a, d, g), and the corresponding models obtained at the final iteration steps of the aLM-EnRML and RLM-MAC are plotted in the second column (b, e, h) and the third one (e, f, i), respectively. As can be seen in these images, despite the clear difference among the initial models, all the updated models of the aLM-EnRML and RLM-MAC bear certain similarities to each other, and they also, to some extent, resemble the reference model in Figure \ref{fig:true_facies}.  

Figure \ref{fig:facies_scores} reports the average scores of the initial ensemble (panel (a)) and the final ensembles of the aLM-EnRML (panel (b)) and RLM-MAC (panel (c)) in matching the facies of the reference model. The score is calculated as follows: at each gridblock, if a model matches the facies, and it has a score of 1, otherwise it has a score of 0. The final score is then averaged over all the models in the ensemble. Therefore a higher score means that more models have the correct estimated facies. From Figure \ref{fig:facies_scores} it is seen that, compared with the scores in the initial ensemble, it seems that the aLM-EnRML and RLM-MAC tend to improve the facies estimation in the channels, but also have deteriorated matching in certain areas (e.g., that around the injector I2). In addition, in this particular case, it seems that the RLM-MAC tends to perform better than the aLM-EnRML in terms of matching the facies.   

Figure \ref{fig:facies_obj} indicates the data mismatch of the aLM-EnRML (left) and RLM-MAC (right) as functions of the number of iteration steps. The aLM-EnRML stops at the 10th iteration step due to the stopping condition (c) mentioned early (relative change of data mismatch less than $0.01\%$), while the RLM-MAC stops at the maximum iteration step. Despite this difference, in this case the aLM-EnRML and RLM-MAC appear to have comparable performance, in terms of the convergence speed and the final data mismatch value. For the aLM-EnRML, its final data mismatch is $2197.71 \pm 234.63$, where the number before $\pm$ is the mean, and that after $\pm$ is the standard deviations (STD). On the other hand, for the RLM-MAC, its final data mismatch is $1864.27 \pm 98.9975$, slightly lower than that of the aLM-EnRML. 

Following \cite{lorentzen2012history}, Figures \ref{fig:facies_forecasts_WOPR} and \ref{fig:facies_forecasts_WWPR} show the performance of history matching oil and water production rates in producers P5 and P7, for the initial ensemble and the ensembles of the aLM-EnRML and RLM-MAC at their final iteration steps, respectively. In both figures, compared to the results with respect to the initial ensemble, those obtained from the final ensembles of the aLM-EnRML and RLM-MAC appear to have reduced ensemble spread and improved data mismatch. The aLM-EnRML performs better than the RLM-MAC in history matching producer P5, but the situation is reverted in producer P7. {\color{black} In the initial ensemble the oil production rates in P5 and the water production rates in P7 tend to be lower than the reference values (which may be related to the bias in the facies distribution of the initial ensemble, as highlighted in Figure \ref{subfig:fieldScore0_1}), and this bias trend remains in the final ensembles obtained by the aLM-EnRML and RLM-MAC. Similarly, in the initial ensemble the oil production rates in P7 and the water production rates in P5 tend to be higher than the reference values. This bias trend is clearly visible in the final ensemble of the aLM-EnRML, but seems to be better mitigated in the final ensemble of the RLM-MAC.}

The top panels of Figure \ref{fig:facies_rmse_spread} show the box plots of RMSEs (in permeability values) of ensemble models with respect to the reference model at different iteration steps in the aLM-EnRML (left) and RLM-MAC (right), and bottom ones indicate the corresponding spreads in the ensemble. Similar to the situation in Figure \ref{fig:L96_rmse_spread}, oscillations of the RMSEs (e.g., in terms of the medians of the box plots) are observed in both iES before the RMSE values enter certain plateaus. In this particular case, the final RMSEs of the RLM-MAC tend to be lower than those of the aLM-EnRML. On the other hand, the ensemble spread of the aLM-EnRML tends to decrease with the iteration, while that of the RLM-MAC exhibits slight U-turn behaviour twice before it enters a plateau. The final ensemble spread of the RLM-MAC is higher than that of the aLM-EnRML, and both are underestimated in comparison with the corresponding RMSEs.  

Figure \ref{fig:facies_obsDiff_boxPlot} shows the normalized differences $\mathbf{C}_d^{-1/2} \left( \mathbf{g} \left(\bar{\mathbf{m}}^{i} \right) - \overline{\mathbf{g}\left(\mathbf{m}_j^{i}\right)} \right)$ at different iteration steps, calculated based on the ensemble models of the RLM-MAC. Compared to Figure \ref{fig:obsDiff_boxPlot}, it is seen that the normalized differences remain relatively small over the iterations, which may be because in this case the phase flows are non-turbulent, and they tend not to amplify the differences among the ensemble models (in contrast to the turbulence nature in the L96 model of the previous subsection).   

\subsection*{History matching study in the Brugge field case}

Here we compare the history matching performance of the aLM-EnRML and RLM-MAC in the Brugge benchmark case study \cite{peters2010results}. The simulation model of the Brugge field case has $60048$ ($139 \times 48 \times 9$) gridblocks, with $44550$ of them being active. 
In the case study there are 20 producers and 10 water injectors. The dataset we used is from TNO \cite{peters2010results} and contains 20 years' production data\footnote{The second decade's production data are provided by TNO based on IRIS' proposed production optimization strategy for that period, see, for example, \cite{peters2010results}.}. The first decade's data are collected at 127 time instants, while the second decade's at 185 time instants. Following \cite{chen2013-levenberg}, we use the production data at 20 out of 127 time instants in the first decade for history matching, and the rest of the 20 years' production data for out-of-sample cross validation, including, for example, predicting the oil production rates during the second decade based on the history-matched models. 

In the course of history matching, the producers and injectors are controlled by the liquid rate (LRAT) target. The production data consist of the historical oil production rates (WOPRH) and water cuts (WWCTH) at 20 producers, and the historical bottom hole pressure (WBHPH) at all 30 wells. Therefore the total number of production data is $1400$. The standard deviations for WOPRH, WWCTH and WBHPH are 100 stb/day, 0.05 and 50 psi, respectively. The model variables to be updated include porosity (PORO), permeability (PERMX, PERMY, PERMZ) and net-to-gross (NTG) ratio at all active gridblocks. Consequently the total number of parameters is $222,750$. In the aLM-EnRML we use the 104 model realizations from the initial ensemble provided by TNO \cite{peters2010results}, while in the RLM-MAC we only use 103 of them, after dropping an ensemble member at random (so that the aLM-EnRML and RLM-MAC have the same number of forward runs at each iteration). In both iES, the parameter rule is $\gamma^i = \alpha^i \, \text{trace}(\mathbf{S}_d^i (\mathbf{S}_d^i)^T) / N_e$, with the maximum number of iterations being 20, and the coefficient $\beta_u = 0.5$. In all experiment runs we use ECLIPSE$^\copyright$ \cite{eclipse2010} for reservoir simulation.    

Figure \ref{fig:brugge_permx_layers12} indicates the distributions of PERMX on layers 1 and 2. The top images are from a realization of the reservoir models in the initial ensemble, the middle ones from the corresponding reservoir model updated by the aLM-EnRML, and the bottom ones from the corresponding model obtained by the RLM-MAC. Although the final data mismatch values of the aLM-EnRML and RLM-MAC are close to each other (see Figure \ref{fig:brugge_data_mismatch_hm}), there do not seem to be substantial similarities between the resulting PERMX images in Figure \ref{fig:brugge_permx_layers12}. This may be due to the difference of both iES in their ways to approximate the gradient of the objective function, hence the difference in the subsequent search directions, as we have discussed previously. 

Figure \ref{fig:brugge_data_mismatch_hm} shows the box plots of data mismatch of the aLM-EnRML (left) and RLM-MAC (right) as the function of iteration steps. The aLM-EnRML stops at the 15th iteration step due to the stopping condition on the relative change of data mismatch (less than 0.01\%), and the RLM-MAC
stops at the maximum iteration step. The aLM-EnRML and RLM-MAC have close performance again, in terms of the convergence speed and the final data mismatch value. In this case, the final data mismatch of the aLM-EnRML is $869.90 \pm 23.38$, while that of the RLM-MAC is $928.96 \pm 25.76$, about 7\% higher in the average data mismatch value.

Figure \ref{fig:brugge_wopr_hm} depicts the history matching profiles of oil production rates (WOPR) at producers BR-P-9 (top row), BR-P-13 (middle row) and BR-P-19 (bottom row) in the first 10 years, with respect to the initial ensemble of reservoir models (1st column), the history matched reservoir models of the aLM-EnRML using production data at 20 time instants (2nd column), and the counterpart history matched reservoir models of the RLM-MAC (3rd column), respectively. In the pictures, the red dots represent the historical WOPRH data at the producers, and the blue curves are the forecasts with respect to the initial ensemble (1st column), and the history matching profiles of both smoothers (2nd and 3rd columns). Consistent with the results in Figure \ref{fig:brugge_data_mismatch_hm}, the aLM-EnRML and RLM-MAC exhibit close performance in matching the WOPRH data. Similar results are also obtained for history matching the WWCTH data at BR-P-9, BR-P-13 and BR-P-19, as can be seen in Figure \ref{fig:brugge_WWCT_hm}.   

Figure \ref{fig:Brugge_obsDiff_boxPlot} also shows the normalized differences $\mathbf{C}_d^{-1/2} \left( \mathbf{g} \left(\bar{\mathbf{m}}^{i} \right) - \overline{\mathbf{g}\left(\mathbf{m}_j^{i}\right)} \right)$ at different iteration steps, calculated based on the ensemble models of the RLM-MAC, in the Brugge field case. Similar to Figure \ref{fig:facies_obsDiff_boxPlot}, the normalized differences remain relatively small over the iterations, in contrast to the corresponding results in the L96 model (see Figure \ref{fig:obsDiff_boxPlot}). This may be also because in this case the reservoir fluid dynamics is non-turbulent, and tends to be less nonlinear than that in the L96 model.

Next we present some cross validation results. Figure \ref{fig:brugge_data_mismatch_validation} shows the box plots of the data mismatch values in 20 years, with respect to the reservoir models from the initial ensemble (left), the final ensemble of the aLM-EnRML using production data at 20 time instants in the first 10 years (middle), and the corresponding final ensemble of the RLM-MAC (right). In this case, the data mismatch values with respect to the initial ensemble and the final ensembles of the aLM-EnRML and RLM-MAC are $3.48 \times 10^5 \pm 2.46 \times 10^5$, $4707.93 \pm 357.83$ and $4540.31 \pm 285.34$, respectively. Therefore, the reservoir models history matched by both iES lead to substantially lower data mismatch than those of the initial ensemble. The data mismatch values of the aLM-EnRML and RLM-MAC are close to each other again, with that of the RLM-MAC being slightly lower.

Figure \ref{fig:brugge_wopr_validation} depicts the WOPR profiles at producers BR-P-9 (top row), BR-P-13 (middle row) and BR-P-19 (bottom row) in 20 years, with respect to the reservoir models from the initial ensemble (1st column), the final ensemble of the aLM-EnRML (2nd column), and the corresponding final ensemble of the RLM-MAC (3rd column), respectively. In the pictures, the red dots represent the historical WOPRH data at the producers, and the blue curves are the forecasts with respect to the initial ensemble and the final ensembles obtained by both smoothers. The pictures in the 2nd and 3rd columns also contain some vertical dashed lines, which are used to separate the production periods between the first and second decades. In this case, the aLM-EnRML and RLM-MAC again exhibit close performance in cross-validating the WOPRH data at BR-P-9. However, in the second decade, the aLM-EnRML predicts the WOPRH data at BR-P-19 better than the RLM-MAC does, which may be related to the fact that the WWCTH data at BR-P-19 is better predicted by the aLM-EnRML (see Figure \ref{fig:brugge_WWCT_validation}). The situation is reverted for the prediction of the WOPRH data at BR-P-13 in the second decade, and the RLM-MAC outperforms the aLM-EnRML instead. This may be because the WWCTH data at BR-P-13 is better predicted by the RLM-MAC instead, as can be seen in Figure \ref{fig:brugge_WWCT_validation}.

\section*{Conclusion}

In this work we showed that an iteration formula similar to those used in the ensemble smoother with multiple data assimilation (ES-MDA) and the approximate Levenberg-Marquardt ensemble randomized maximum likelihood (aLM-EnRML) method can be derived from the regularized Levenberg-Marquardt (RLM) algorithm in inverse problems theory, when history matching is recast as a minimum-average-cost (MAC) problem. 

Specifically, to reduce the computational cost in solving the MAC problem, a simple strategy is to approximate the nonlinear simulator, for all ensemble models, through a first order Taylor expansion around a single common point, e.g., the ensemble mean at the previous iteration step as considered in the current study. In addition, to avoid directly evaluating the resulting (common) Jacobian matrix, one may follow the convention in the ensemble Kalman filter (EnKF) to approximate the product between the Jacobian matrix and the model square root matrix by the corresponding data square root matrix. In this way, the resulting iteration formula is similar to those in the ES-MDA and aLM-EnRML, but differs in the way of constructing the data square root matrix: In the ES-MDA and aLM-EnRML, the simulated data center around their sample mean, while in the RLM-MAC, they center around the simulated data of the mean of the ensemble models. In general, it is expected that one may solve the MAC problem in a more accurate way -- but possibly with a higher computational cost -- by evaluating the gradients at multiple common points. 

To examine the effect of different ways in constructing the data square root matrix, we compare the performance of the aLM-EnRML and RLM-MAC in three numerical examples. In the first example we consider a strongly nonlinear system (the Lorenz 96 model). In this particular case the aLM-EnRML and RLM-MAC exhibit more diverged behaviour and the RLM-MAC tends to perform better than the aLM-EnRML. In the second and third examples we apply the aLM-EnRML and RLM-MAC to a reservoir facies estimation problem and the history matching problem in the Brugge field case, respectively. In both cases it appears that the aLM-EnRML and RLM-MAC exhibit close performance due to weaker nonlinearity in the reservoir fluid dynamics. 

\newpage
\section*{Nomenclature}
\begin{tabular}{rcl}
$\mathbf{d}$ & = & data \\
$\mathbf{d}^o$ & = & observation data \\
$\mathbf{f}$ & = & function \\
$\mathbf{g}$ & = & reservoir simulator \\
$m$ & = & dimension of reservoir model \\
$\mathbf{m}$ & = & reservoir model \\
$\mathbf{m}_c$ & = & common reservoir model in the minimum-average-cost problem \\
$\bar{\mathbf{m}}$ & = & ensemble-average reservoir model \\
$N_e$ & = & ensemble size \\
$p$ & = & number of observation data\\
$p_k$ & = & probability density function (pdf) at the $k$th iteration step  \\
$r$ & = & scalar coefficient larger than 1 \\
$t$ & = & unit-less sudo-time variable in the Lorenz 96 model \\
$\mathbf{v}$ & = & generic random vector \\
$\hat{\mathbf{v}}$ & = & estimation of $\mathbf{v}$ \\
$\mathbf{v}^{tr}$ & = & true realization value of the random vector $\mathbf{v}$ \\
$x$ & = & state variable in the Lorenz 96 model \\
$\mathbf{y}$ & = & random vector that represents observations  \\
$\mathbf{A}$ & = & generic matrix\\
$\mathbf{C}_d$ & = & observation error covariance matrix\\
$\mathbf{C}_m$ & = & model error covariance matrix as in the EnKF\\
$\tilde{\mathbf{C}}_m$ & = & modified model error covariance matrix in the RLM-MAC\\
$F$ & = & unit-less driving force in the Lorenz 96 model \\
$\mathcal{F}$ & = & functional \\
$\mathcal{G}$ & = & inverse functional of $\mathcal{F}$  \\
$\mathbf{G}$ & = & Jacobian of reservoir simulator\\
$\mathcal{G}^C$ & = & composite functional  \\
$I$ & = & number of iterations in ES-MDA\\
\end{tabular}

\begin{tabular}{rcl}
$\mathbf{I}_p$ & = & identity matrix of size $p$\\
$\mathbf{M}$ & = & ensemble of reservoir models \\
$\mathcal{O}$ & = & objective function \\
$\bar{\mathcal{O}}$ & = & expectation of $\mathcal{O}$ \\
$\mathbf{S}_d$ & = & data square root matrix as in the EnKF \\
$\tilde{\mathbf{S}}_d$ & = & modified data square root matrix in the RLM-MAC \\
$\mathbf{S}_m$ & = & model square root matrix \\
\end{tabular}

\subsection*{Greek symbols}
\begin{tabular}{ccl}
$\alpha$ & = & adaptive coefficient in the RLM-MAC that determines $\gamma$ \\
$\delta$ & = & Dirac delta function \\
$\gamma$ & = & adaptive coefficient in iterative ensemble smoothers \\
$\sigma^2$ & = & sample variances of the ensemble of reservoir models  \\
$\rho$ & = & tapering factor of $\gamma$ \\
$\Delta m$ & = & model change \\
\end{tabular}

\subsection*{Subscript}
\begin{tabular}{ccl}
$a$ & = & analysis \\
$b$ & = & background \\
$d$ & = & data\\
$e$ & = & ensemble \\
$j$ & = & index of ensemble member \\
$k$ & = & iteration index \\
$m$ & = & model \\
$s$ & = & index of sample variances of the ensemble of reservoir models  \\
\end{tabular}
\subsection*{Superscript}
\begin{tabular}{ccl}
$b$ & = & background \\
$i$ & = & iteration index \\
$o$ & = & observation\\
$tr$ & = & truth\\
\end{tabular}

\newpage
\begin{acknowledgements}
We would like to thank Drs. Yan Chen (Total), Randi Valestrand (IRIS) and three anonymous reviewers for their valuable helps, comments and suggestions, which have significantly improved the work. We also acknowledge the IRIS/CIPR cooperative research project ``Integrated Workflow and Realistic Geology'' which is funded by industry partners ConocoPhillips, Eni,  Petrobras, Statoil, and Total, as well as the Research Council of Norway (PETROMAKS) for financial support, and thank Schlumberger for ECLIPSE academic licenses and the developers of SGeSM and MATLAB level set toolbox (Ian M. Mitchell) for providing their softwares.
\end{acknowledgements}

\appendix
\numberwithin{equation}{section}

\section{Characterizing the RLM-MAC from the point of view of an expectation-maximization algorithm} \label{sec:appdendix_iES} 

{\color{black}Following a recent working paper \cite{luo2015Optimization}}, the RLM-MAC algorithms can also be described in analogy to the expectation-maximization (EM) algorithm \cite{Dempster1977maximum}, which is an iterative algorithm that consists of an expectation step (E-step) and an maximization step (M-step) at each iteration. At the E-step, one constructs an expectation-type cost function that depends on certain unknown quantities-of-interest (QOI) and their prior knowledge, while at the subsequent M-step, one solves a maximization problem to obtain an updated estimation of the QOI. The updated QOI are then passed to the next iteration as the new prior knowledge of the corresponding E-step, and so on.  

The cost function in the RLM-MAC can be considered as a Monte-Carlo approximation to the following expectation-type cost function at the E-step of the $k$th iteration:
{\color{black}
\begin{linenomath*} 
	\begin{equation} \label{eq:e-cost}
	\bar{\mathcal{O}}_k = \int \mathcal{O}_k(\mathbf{m}_a^k,\mathbf{m}_b^k,\tilde{\mathbf{y}}) \, p_k(\mathbf{m}_a^k,\mathbf{m}_b^k,\tilde{\mathbf{y}}|\mathbf{y}) \, d\mathbf{m}_b^k \, d\mathbf{m}_a^k \, d\tilde{\mathbf{y}} \, ,
	\end{equation}     
\end{linenomath*}  
where $\mathbf{m}_b^k$ is the prior knowledge of the QOI, $\mathbf{m}_a^k$ the QOI to be updated, $\mathbf{y}$ the real observation, and $\tilde{\mathbf{y}}$ a perturbation of $\mathbf{y}$; $\mathcal{O}_k(\mathbf{m}_a^k,\mathbf{m}_b^k,\tilde{\mathbf{y}})$ is a chosen function (e.g., as the quadratic function in the RLM-MAC) and may depend on $\mathbf{m}_b^k$, $\mathbf{m}_a^k$ and $\tilde{\mathbf{y}}$, and $\bar{\mathcal{O}}_k$ is the expectation of $\mathcal{O}_k(\mathbf{m}_a^k,\mathbf{m}_b^k,\tilde{\mathbf{y}})$ with respect to the (conditional) joint pdf $p_k(\mathbf{m}_a^k,\mathbf{m}_b^k,\tilde{\mathbf{y}}|\mathbf{y})$. In the context of data assimilation, the updated QOI $\mathbf{m}_a^k$ often depend on both $\mathbf{m}_b^k$ and $\tilde{\mathbf{y}}$, so that one has $\mathbf{m}_a^k = \mathcal{F}_k \left(\mathbf{m}_b^k, \tilde{\mathbf{y}}\right)$, where $\mathcal{F}_k$ is an unknown deterministic map (or functional), and is determined at the subsequent ``M-step'' under a certain criterion. In the RLM-MAC, $\mathcal{F}_k$ are determined by solving the following functional optimization problem
\begin{linenomath*} 
\begin{equation} \label{eq:e-cost_lagrangian}
\underset{\mathcal{F}_k}{\operatorname{argmax}} \, \bar{\mathcal{O}}_k \, .
\end{equation}     
\end{linenomath*}  

Note that $\mathcal{F}_k$ may also be chosen from other points of views. For instance, they may be constructed in a way such that the conditional pdf $p_k(\mathbf{m}_a^k|\mathbf{y})$ is consistent with Bayes' rule, 
in the sense of satisfying
\begin{linenomath*} 
\begin{equation} \label{eq:bayes_rule}
\begin{split}
p_k(\mathbf{m}_a^k | \mathbf{y}) = \dfrac{p(\mathbf{y}|\mathbf{m}_a^k)p_0(\mathbf{m}_a^k)}{p(\mathbf{y})}.
\end{split} 
\end{equation}     
\end{linenomath*}

On the other hand, note that $\mathbf{m}_b^{k} = \mathbf{m}_a^{k-1}$. Through recursion, one has $\mathbf{m}_a^{k} = \mathcal{F}_k \left(\mathbf{m}_b^k, \tilde{\mathbf{y}}\right) = \mathcal{F}_k \left(\mathcal{F}_{k-1} \left(\mathbf{m}_b^{k-1}, \tilde{\mathbf{y}}\right), \tilde{\mathbf{y}}\right) = \dotsb$. Let $\mathbf{m}_a^{k} \equiv \mathcal{F}_k^{C} \left(\mathbf{m}_b^0, \tilde{\mathbf{y}}\right)$, where $\mathcal{F}_k^{C}$ is the composition of the functionals $\mathcal{F}_i$, 
then by the definition of pdf of transformed random variables we have the relation between $p_0(\mathbf{m}_b^0,\tilde{\mathbf{y}}|\mathbf{y})$ and $p_k(\mathbf{m}_a^k|\mathbf{y})$ given by
\begin{linenomath*} 
\begin{equation} \label{eq:transformed_pdf}
\begin{split}
p_k(\mathbf{m}_a^k | \mathbf{y}) & = \dfrac{\partial}{\partial \mathbf{m}_a^k} \int\limits_{\footnotesize \mathcal{F}_k^{C} \left(\mathbf{m}_b^{0}, \tilde{\mathbf{y}}\right) \leq \mathbf{m}_a^k } p_0(\mathbf{m}_b^0, \tilde{\mathbf{y}}|\mathbf{y}) ~ d\mathbf{m}_b^0 d \tilde{\mathbf{y}} \\
& = \dfrac{\partial}{\partial \mathbf{m}_a^k} \int\limits_{\footnotesize \mathcal{F}_k^{C} \left(\mathbf{m}_b^{0}, \tilde{\mathbf{y}}\right) \leq \mathbf{m}_a^k } p_0(\mathbf{m}_b^0) p(\tilde{\mathbf{y}}|\mathbf{y}) ~ d\mathbf{m}_b^0 d \tilde{\mathbf{y}} \, ,
\end{split} 
\end{equation}     
\end{linenomath*} 
where $p(\tilde{\mathbf{y}}|\mathbf{y})$ is the pdf of the perturbed observation $\tilde{\mathbf{y}}$, given the real observation $\mathbf{y}$. Note that in the second line of Eq. (\ref{eq:transformed_pdf}), we have assumed that the initial prior $\mathbf{m}_b^0$ is independent of observation.   
Combining Eqs. (\ref{eq:transformed_pdf}) and (\ref{eq:bayes_rule}), we require the following identity
\begin{linenomath*} 
\begin{equation} \label{eq:general_condition2}
\begin{split}
\dfrac{\partial}{\partial \mathbf{m}_a^k} \int\limits_{\footnotesize \mathcal{F}_k^{C} \left(\mathbf{m}_b^{0}, \tilde{\mathbf{y}}\right) \leq \mathbf{m}_a^k } p_0(\mathbf{m}_b^0) p(\tilde{\mathbf{y}}|\mathbf{y}) ~ d\mathbf{m}_b^0 d \tilde{\mathbf{y}} = \dfrac{p(\mathbf{y}|\mathbf{m}_a^k)p_0(\mathbf{m}_a^k)}{p(\mathbf{y})}
\end{split} 
\end{equation}  
\end{linenomath*}
to hold in order for $p_k(\mathbf{m}_a^k|\mathbf{y})$ to satisfy Bayes' rule.

}
Eq. (\ref{eq:general_condition2}) suggests that the choice of $\mathcal{F}_i$ depends on both the prior pdf $p_0(\bullet)$ and the likelihood $p(\mathbf{y}|\bullet)$. As a side remark, we also note that Eq. (\ref{eq:general_condition2}) only imposes a constraint on the compositional functional $\mathcal{G}_k^{C}$, but does not specify how $\mathcal{F}_i$ should be chosen at individual iteration steps. This freedom allows one to incorporate additional criteria (e.g., in terms of solving optimization problems in (\ref{eq:e-cost_lagrangian})) at some of the iteration steps, subject to the condition in Eq. (\ref{eq:general_condition2}).

In linear systems with Gaussian model and observation errors, \cite{emerick2012ensemble} provides proof and examples on how to construct the functionals $\mathcal{F}_i$ to make the resulting $p_k(\mathbf{m}_a^k|\mathbf{y})$ consistent with Bayes' rule. In general situations with nonlinearity and/or non-Gaussianity, the constructions of $\mathcal{F}_i$ become more complicated and are beyond the scope of the current study.  
   
\section{Derivation of the iteration formula in RLM-MAC} \label{sec:appdendix_deduction} 

By inserting Eq. (\ref{eq:linearization_common}) into (\ref{eq:wls_rlm_mac}), one has the following approximate cost function
{\small 
\begin{linenomath*} 
\begin{IEEEeqnarray}{lll} \label{eq:cost_rlm_mac}
\tilde{J}^{i+1} = & \dfrac{1}{N_e} \sum\limits_{j=1}^{N_e} & \, \left[ \left( \mathbf{d}^o_j - \mathbf{g} \left(\mathbf{m}^{i}_c \right) - \mathbf{G}^{i}_c \left( \mathbf{m}^{i+1}_j - \mathbf{m}^{i}_c \right) \right)^T \mathbf{C}_{d}^{-1} \left( \mathbf{d}^o_j - \mathbf{g} \left(\mathbf{m}^{i}_c \right) - \mathbf{G}^{i}_c \left( \mathbf{m}^{i+1}_j - \mathbf{m}^{i}_c \right) \right) \right. \\
& & \quad \left. +  \gamma^{i} \left( \mathbf{m}^{i+1}_j - \mathbf{m}^{i}_j \right)^T \left( \tilde{\mathbf{C}}_{m}^{i} \right)^{-1} \left( \mathbf{m}^{i+1}_j - \mathbf{m}^{i}_j \right) \right] \, . \nonumber
\end{IEEEeqnarray}
\end{linenomath*}
}
A necessary condition for $\mathbf{m}^{i+1}_j$ ($j = 1, 2, \dotsb, N_e$) being the solution of the MAC problem is thus $\partial \tilde{J}^{i+1} / \partial \mathbf{m}^{i+1}_j = 0$, which leads to the following equation: 
{\small
\begin{linenomath*}   
\begin{IEEEeqnarray}{lll} \label{eq:derivation1_rlm_mac}
-2 \, \left(\mathbf{G}^{i}_c\right)^T \, \mathbf{C}_{d}^{-1} \left( \mathbf{d}^o_j - \mathbf{g} \left(\mathbf{m}^{i}_c \right) - \mathbf{G}^{i}_c \left( \mathbf{m}^{i+1}_j - \mathbf{m}^{i}_c \right) \right) + 2 \, \gamma^{i} \left( \tilde{\mathbf{C}}_{m}^{i} \right)^{-1} \left( \mathbf{m}^{i+1}_j - \mathbf{m}^{i}_j \right) = 0 \, . 
\end{IEEEeqnarray}
\end{linenomath*} 
}
With some algebraic operations, the above equation becomes
{\small  
\begin{linenomath*} 
\begin{IEEEeqnarray}{lll} \label{eq:derivation2_rlm_mac}
\left[ \left(\mathbf{G}^{i}_c\right)^T \, \mathbf{C}_{d}^{-1} \, \mathbf{G}^{i}_c + \gamma^{i} \left( \tilde{\mathbf{C}}_{m}^{i} \right)^{-1} \right] \left(\mathbf{m}^{i+1}_j - \mathbf{m}^{i}_j\right)  = \left(\mathbf{G}^{i}_c\right)^T \, \mathbf{C}_{d}^{-1} \left[ \mathbf{d}^o_j - \mathbf{g} \left(\mathbf{m}^{i}_c \right) - \mathbf{G}^{i}_c \left( \mathbf{m}^{i}_j - \mathbf{m}^{i}_c \right) \right] \, .
\end{IEEEeqnarray}
\end{linenomath*} 
} 
By the Sherman-Morrison-Woodbury formula \cite{sherman1950adjustment}, one has  
\begin{linenomath*} 
\[
\left[ \left(\mathbf{G}^{i}_c\right)^T \, \mathbf{C}_{d}^{-1} \, \mathbf{G}^{i}_c + \gamma^{i} \left( \tilde{\mathbf{C}}_{m}^{i} \right)^{-1} \right]^{-1} \left(\mathbf{G}^{i}_c\right)^T \, \mathbf{C}_{d}^{-1} =  \tilde{\mathbf{C}}_{m}^{i} \left(\mathbf{G}^{i}_c\right)^T \left[ \mathbf{G}^{i}_c \tilde{\mathbf{C}}_{m}^{i} \left(\mathbf{G}^{i}_c\right)^T + \gamma^{i} \mathbf{C}_{d} \right]^{-1} \, .
\]
\end{linenomath*} 
Inserting the above identity into Eq. (\ref{eq:derivation2_rlm_mac}) and letting $\tilde{\mathbf{C}}_{m}^{i} = \tilde{\mathbf{S }}_{m}^{i} \left(\tilde{\mathbf{S}}_{m}^{i}\right)^T$, one obtains Eq. (\ref{eq:wls_rlm_mac_orig}).

%


\bibliographystyle{spmpsci}      
\bibliography{references}

\clearpage
\listoffigures

\section*{}
\clearpage
\newcommand{\nScale}{0.45}
\begin{figure*}
\centering
\subfigure[Histogram of average data mismatch]{ \label{subfig:L96_nlnObs_obj_nstep40_ensize100}
\includegraphics[scale=\nScale]{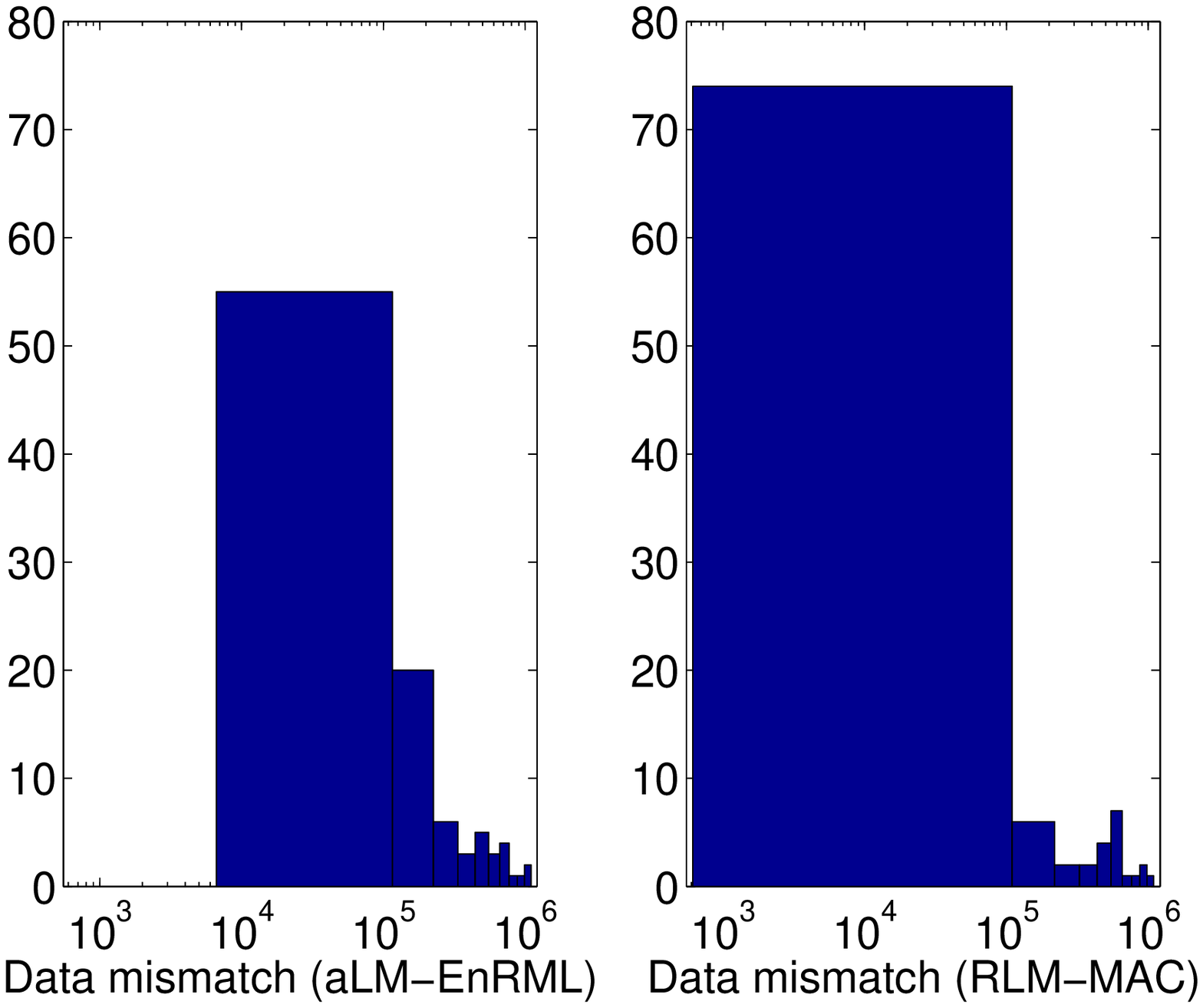}
}
\subfigure[Histogram of average RMSEs]{ \label{subfig:L96_nlnObs_mean_state_rmse_nstep40_ensize100}
\includegraphics[scale=\nScale]{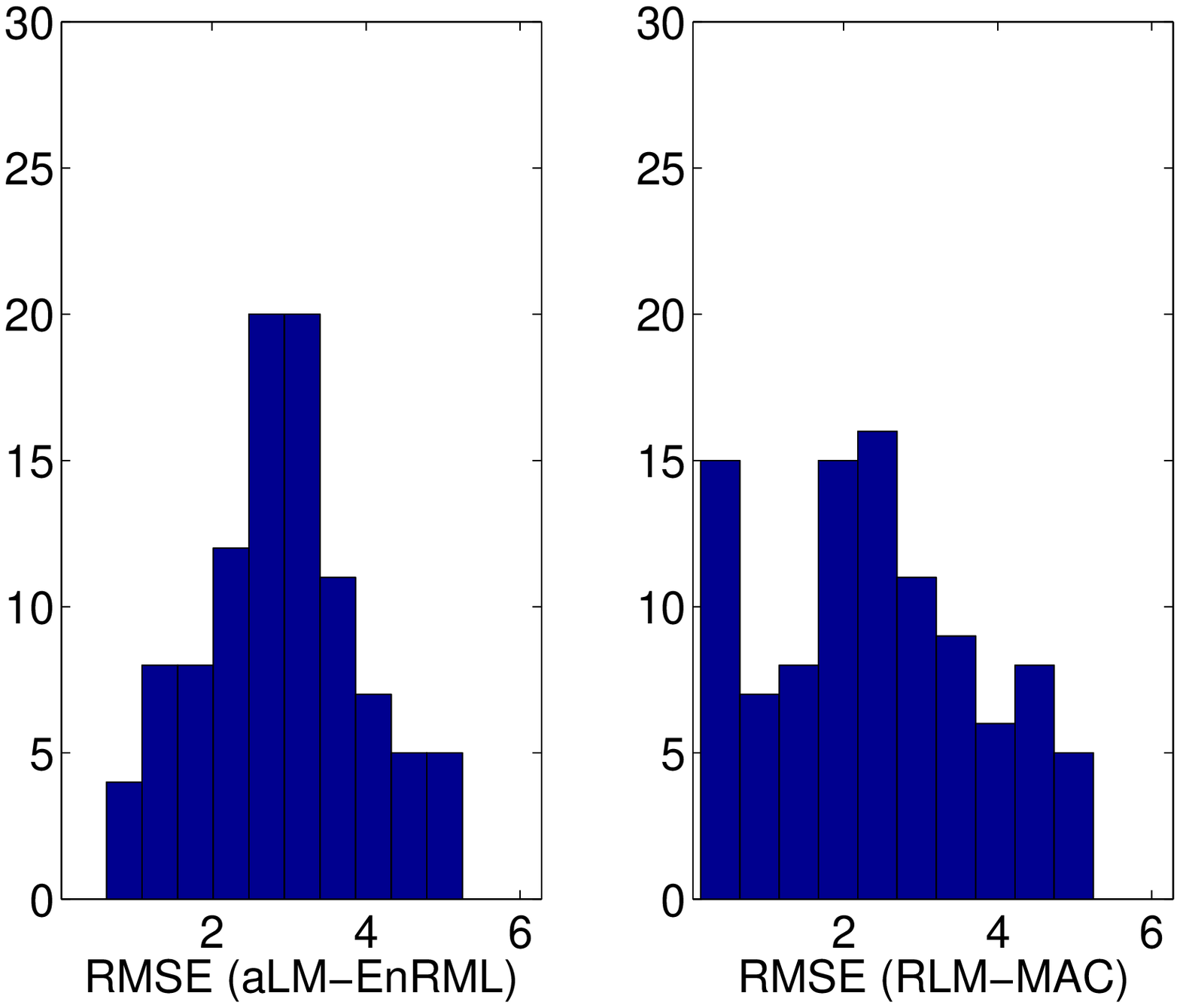}
}
\caption{\label{fig:L96_histogram} Histograms of (a) data mismatch and (b) RMSEs of the aLM-EnRML and RLM-MAC obtained from 100 repetitions of the experiment in the L96 model. For visualization, the horizontal axes of Figure \ref{subfig:L96_nlnObs_obj_nstep40_ensize100} are in the logarithmic scale. In Figure \ref{subfig:L96_nlnObs_mean_state_rmse_nstep40_ensize100}, the RMSEs are calculated based on the differences between the trajectory generated by the true initial state (in 10 days) and those generated by the estimated ones.} 
\end{figure*} 

\clearpage  
\renewcommand{\nScale}{0.45}
\begin{figure*}
\centering
\subfigure[aLM-EnRML]{ \label{L96_pie_enrml_obj}
\includegraphics[scale=\nScale]{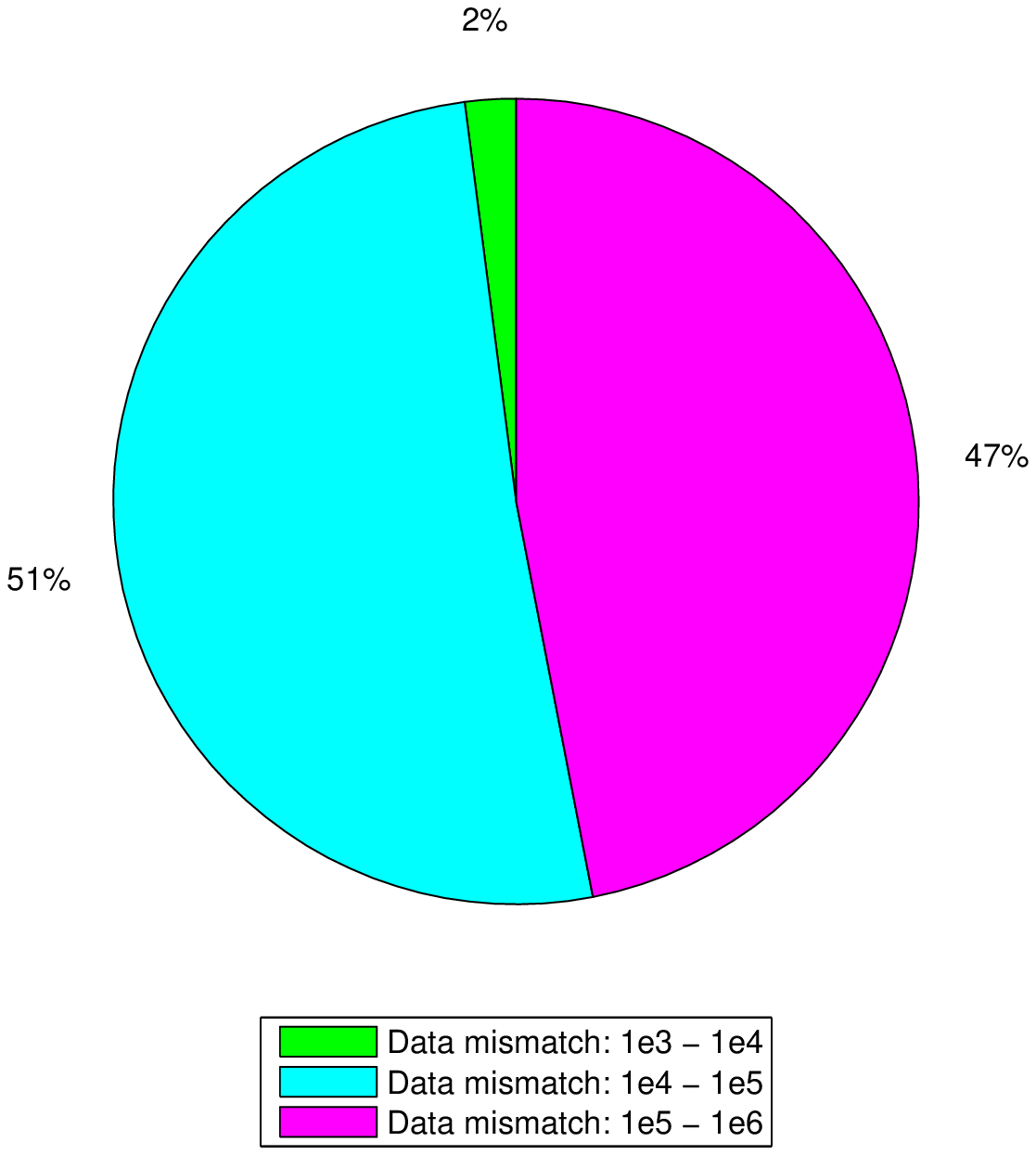}
}
\subfigure[RLM-MAC]{ \label{subfig:L96_pie_rlm_obj}
\includegraphics[scale=\nScale]{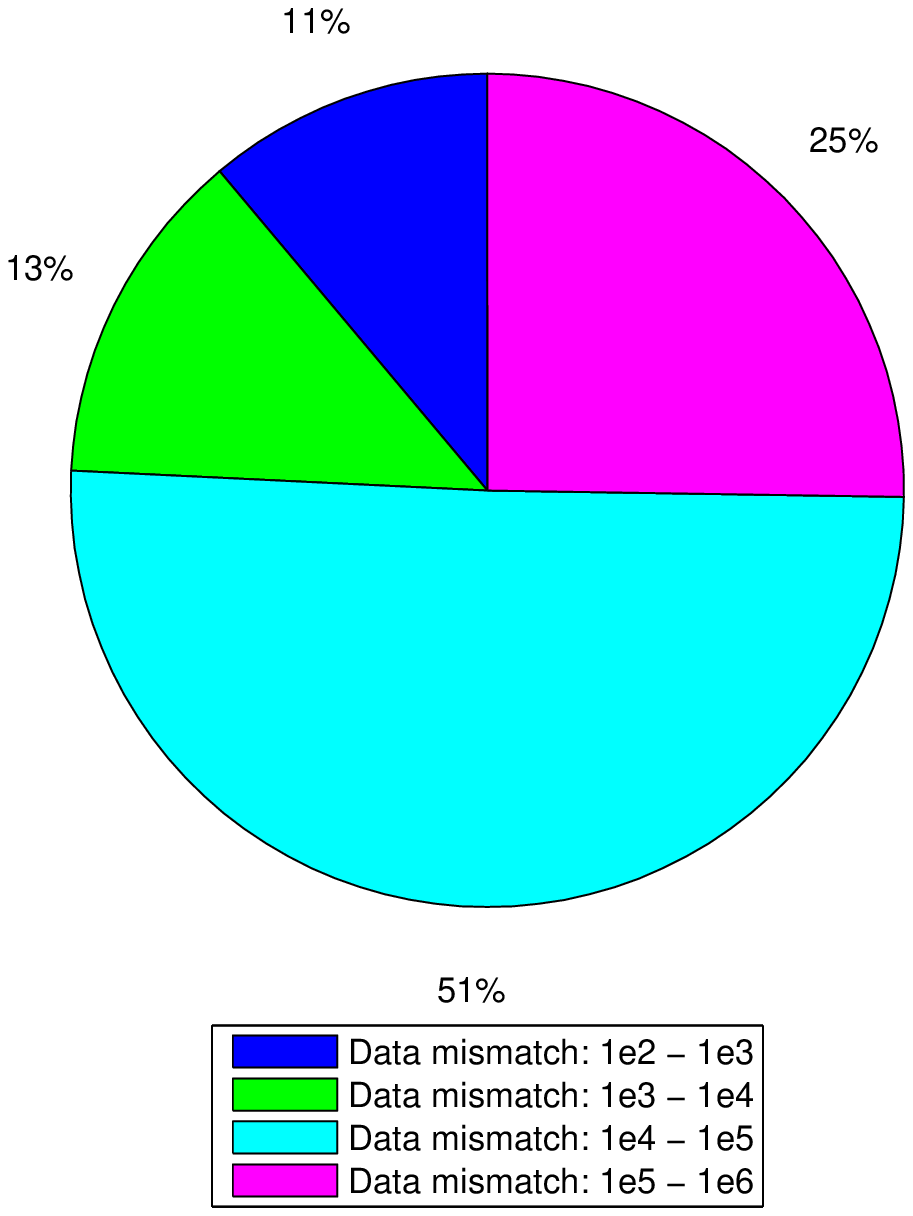}
}

\caption{\label{fig:L96_pie_charts_data_mismtach} Pie charts of the data mismatch of the aLM-EnRML (left) and RLM-MAC (right) with respect to the results in Fig. \ref{fig:L96_histogram}.} 
\end{figure*} 

\clearpage
\renewcommand{\nScale}{0.45}
\begin{figure*}
\centering

\subfigure[aLM-EnRML]{ \label{L96_pie_enrml_rmse}
\includegraphics[scale=\nScale]{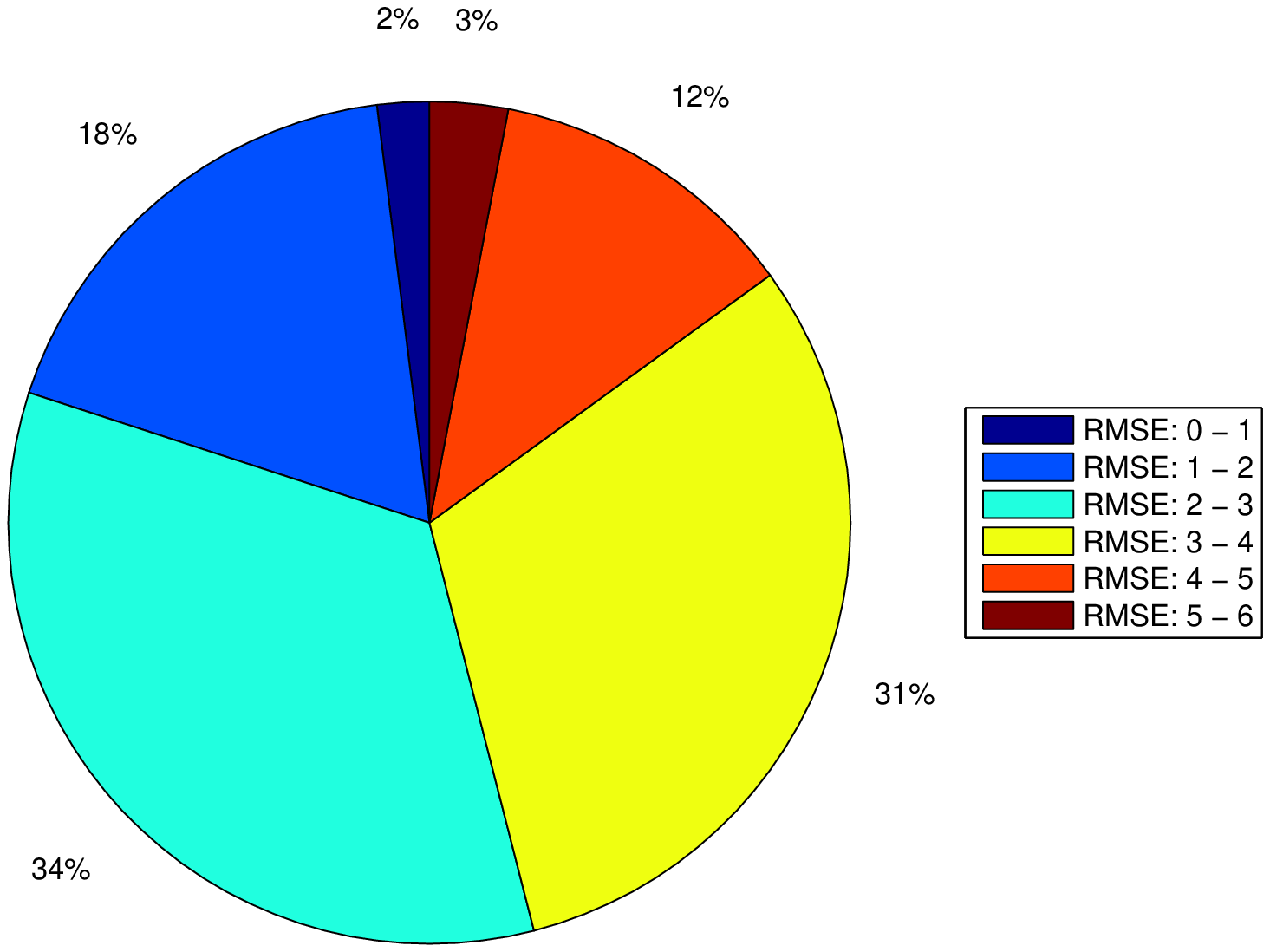}
}
\subfigure[RLM-MAC]{ \label{subfig:L96_pie_rlm_rmse}
\includegraphics[scale=\nScale]{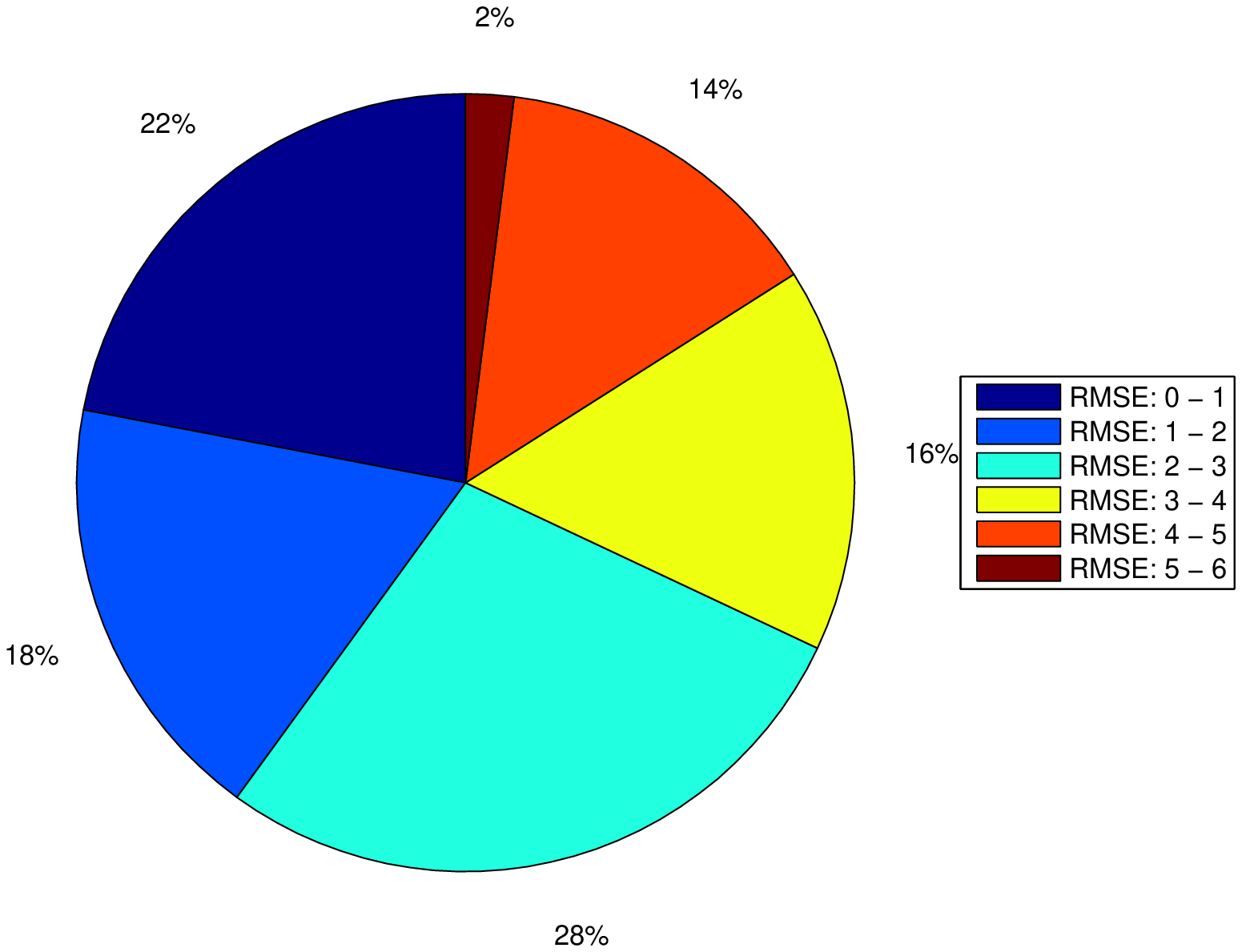}
}

\caption{\label{fig:L96_pie_charts_rmse} Pie charts of the RMSEs of the aLM-EnRML (left) and RLM-MAC (right) with respect to the results in Fig. \ref{fig:L96_histogram}.} 
\end{figure*} 

\clearpage
\renewcommand{\nScale}{0.43}
\begin{figure*}
\centering
\subfigure{ \label{subfig:L96_nlnObs_aLM-EnRML_boxplot_objReal-iter}
\includegraphics[scale=\nScale]{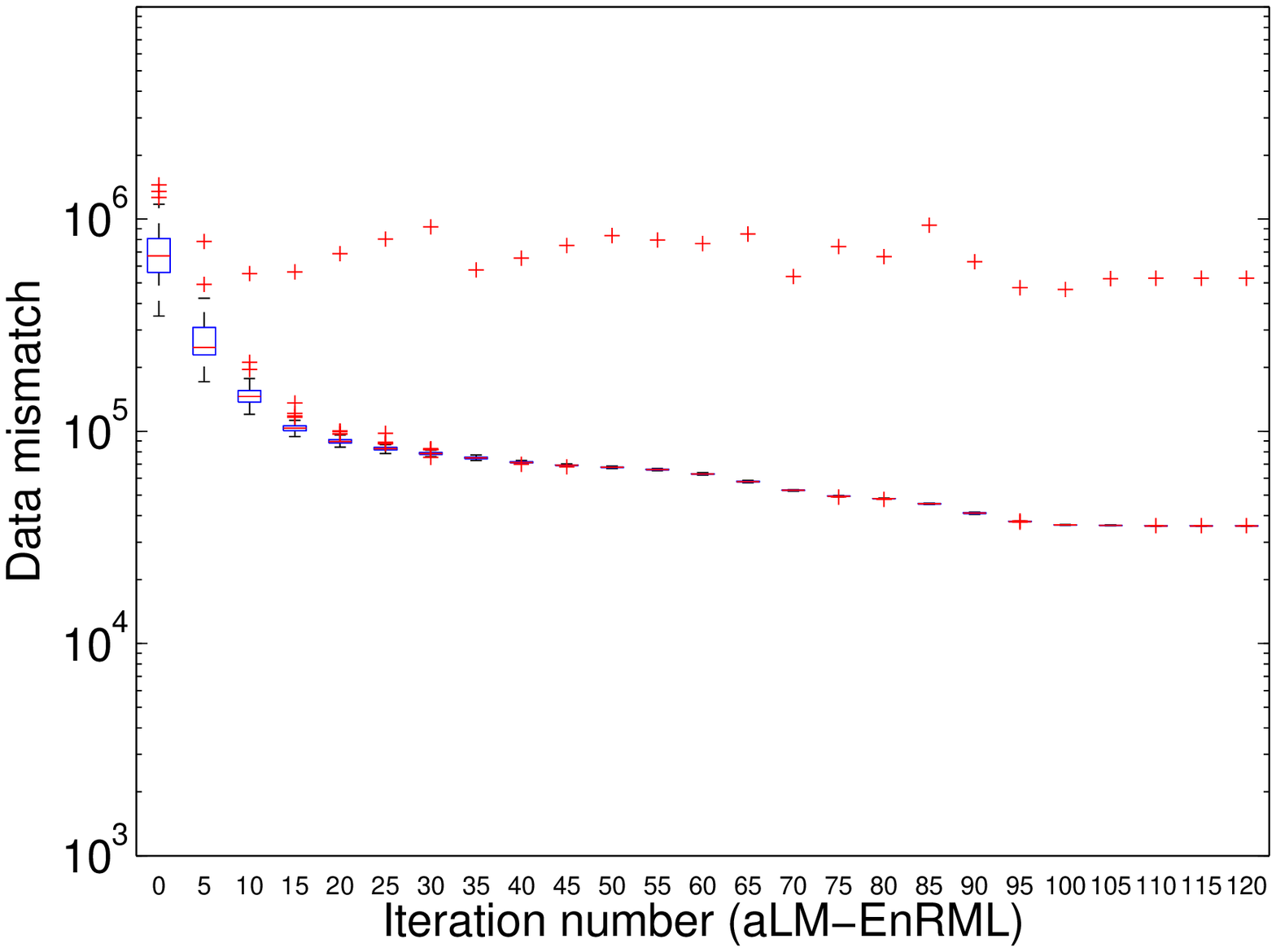}
}
\subfigure{ \label{subfig:L96_nlnObs_RLM-MAC_boxplot_objReal-iter}
\includegraphics[scale=\nScale]{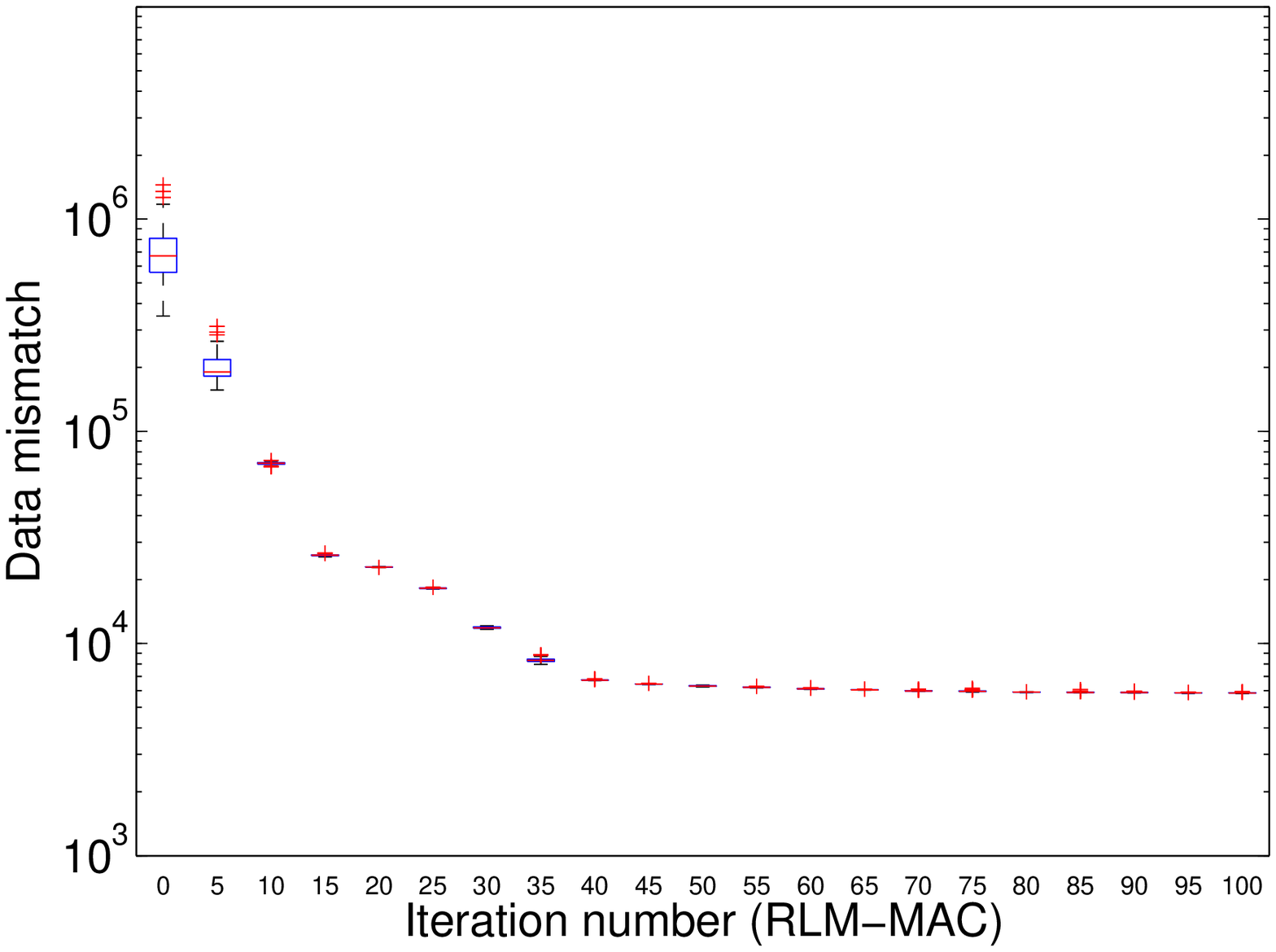}
}

\caption{\label{fig:L96_single_data_mismatch} Box plots of the data mismatch of the aLM-EnRML (left) and the RLM-MAC (right) at different iteration steps in one run in the L96 model. The maximum number of iterations for the RLM-MAC is 100, while that for the aLM-EnRML is increased to 200 and the aLM-EnRML converges after 120 iteration steps. For visualization, the vertical axes are in the logarithmic scale, and the results are plotted at iteration steps 0, 5, and every 5 iteration steps thereafter, along the horizontal axes. In each box plot, the horizontal line (in red) inside the box denotes the median; the top and bottom of the box represent the $75$th and $25th$ percentiles, respectively; the whiskers indicate the ranges beyond which the data are considered outliers, and the whiskers' positions are determined using the default MATLAB$^\copyright$ setting (please see the corresponding documentation in MATLAB$^\copyright$ R2012a), while the outliers themselves are plotted individually as plus signs (in red).} 
\end{figure*}  

\clearpage
\renewcommand{\nScale}{0.42}
\begin{figure*}
\centering
\subfigure{ \label{subfig:rmse_aLM-EnRML_boxplot_ensemble-iter}
\includegraphics[scale=\nScale]{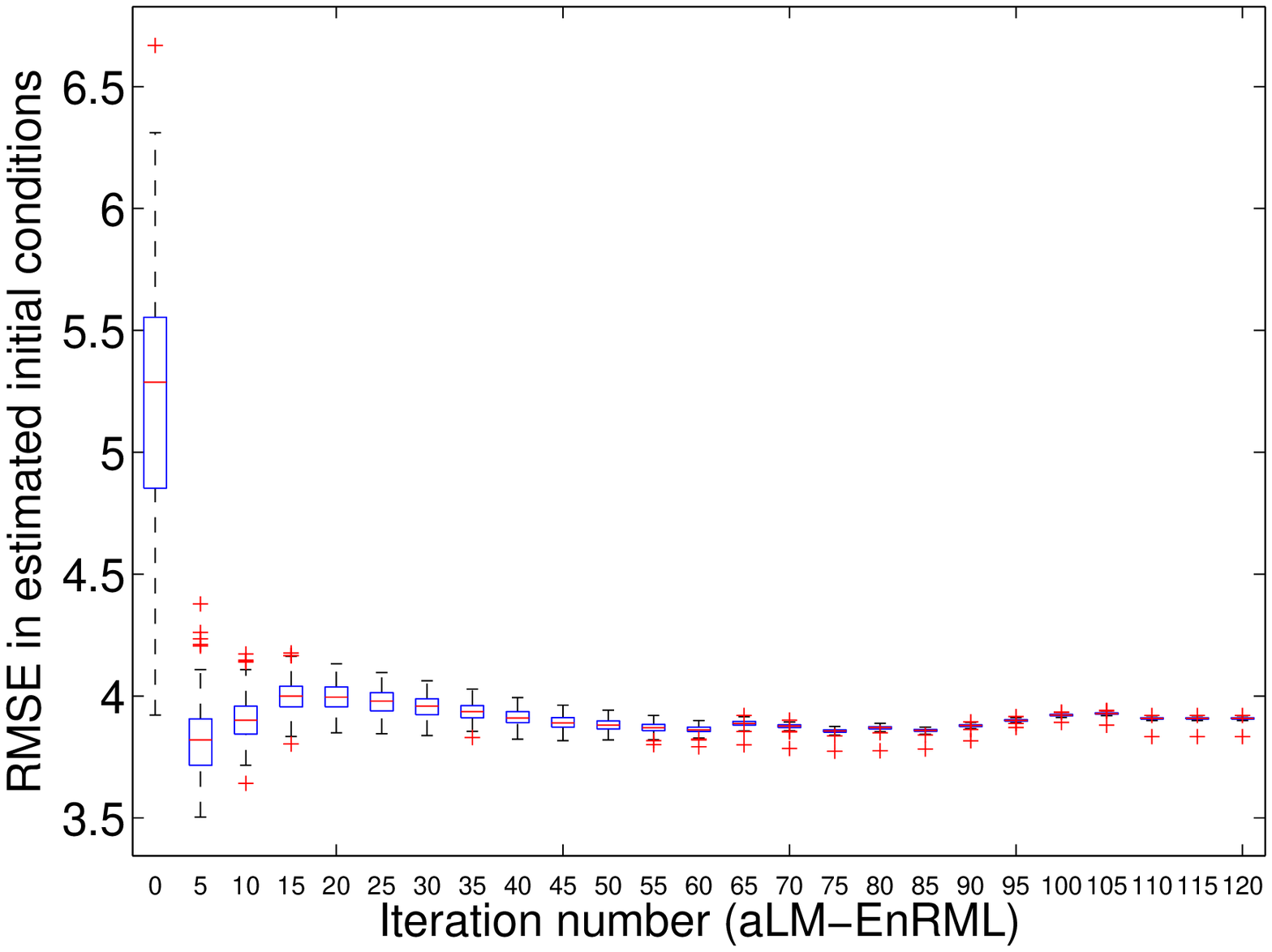}
}
\subfigure{ \label{subfig:rmse_RLM-MAC_boxplot_ensemble-iter}
\includegraphics[scale=\nScale]{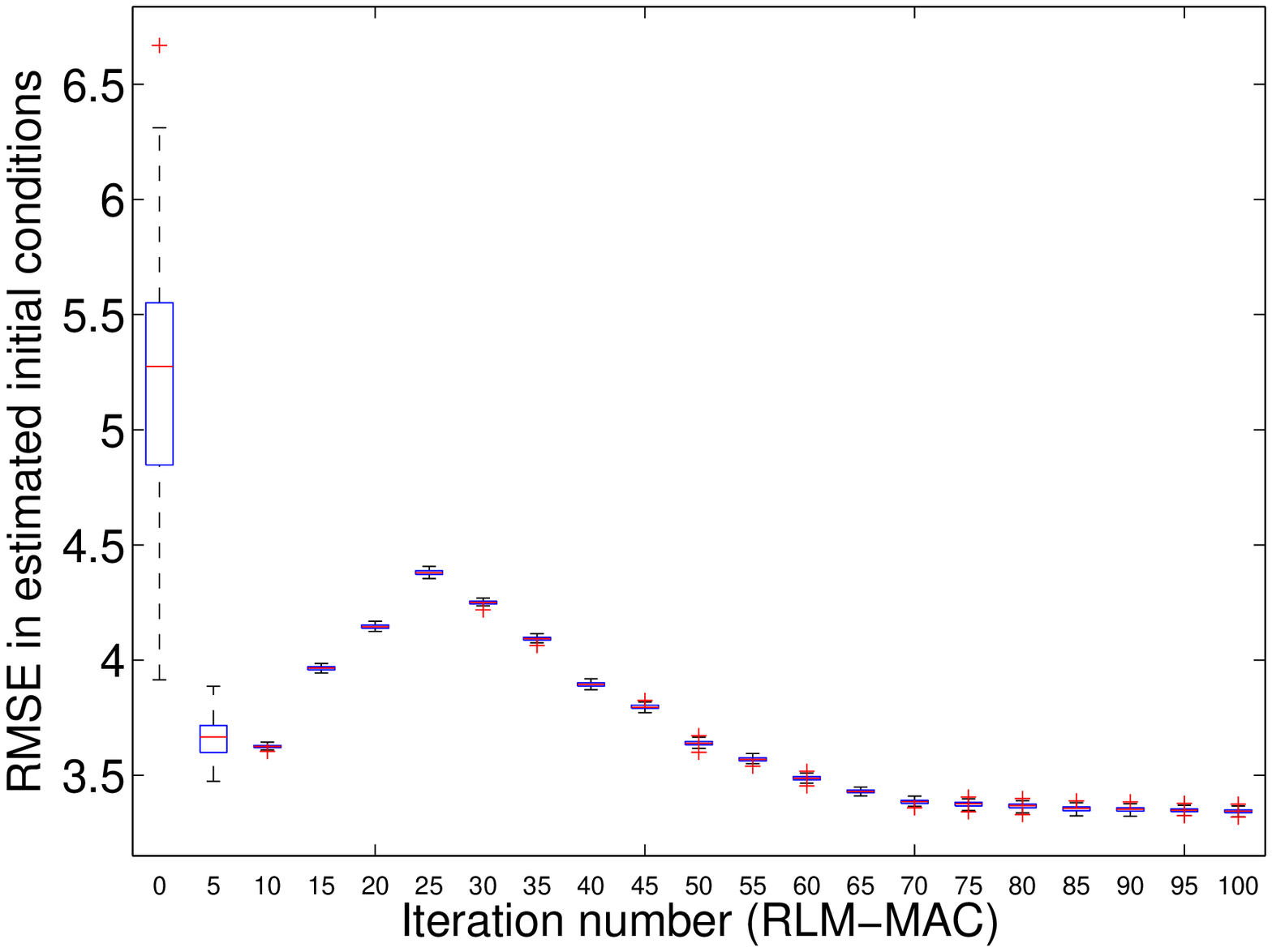}
}

\subfigure{ \label{subfig:spread_aLM-EnRML_boxplot_ensemble-iter}
\includegraphics[scale=\nScale]{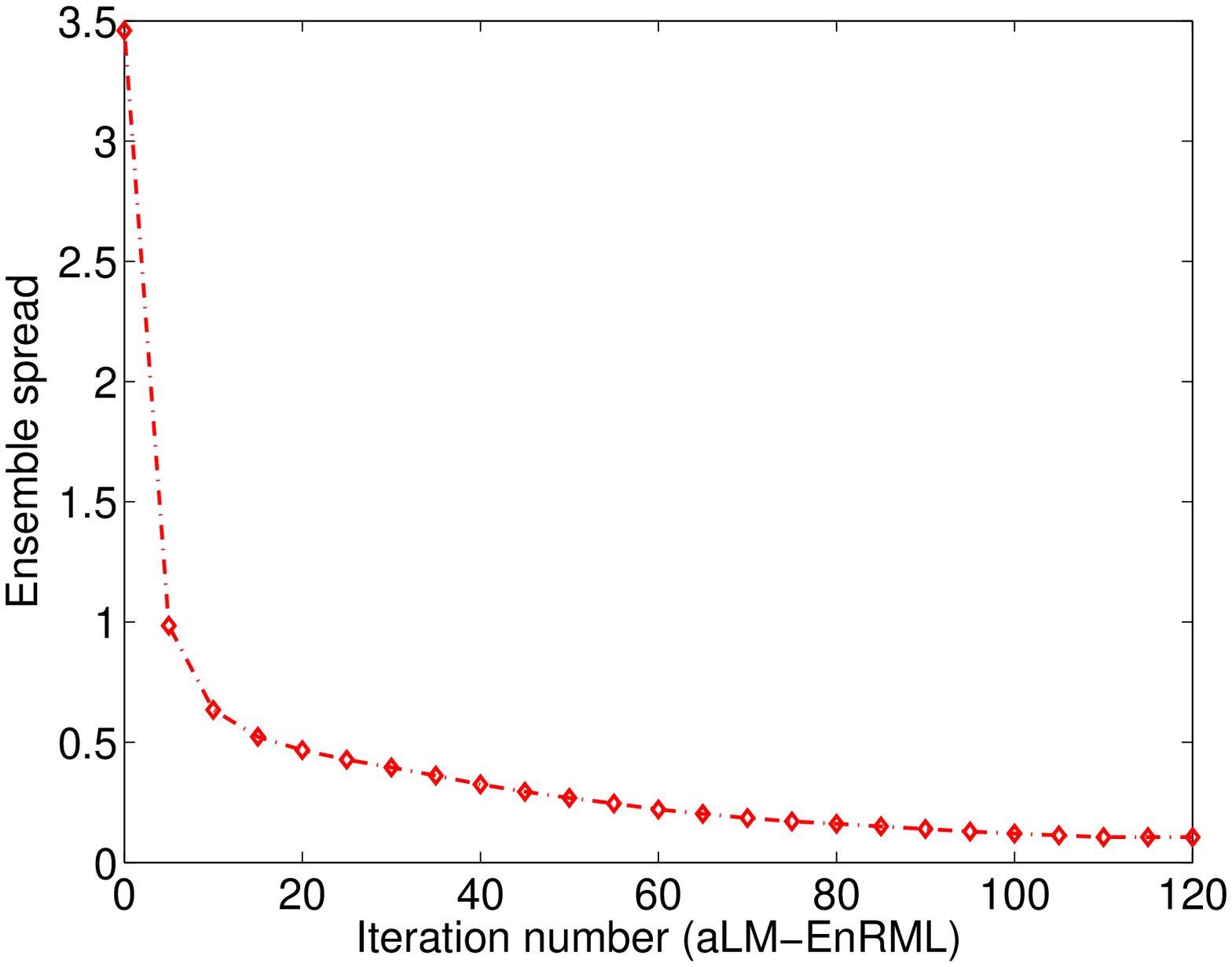}
}
\subfigure{ \label{subfig:spread_RLM-MAC_boxplot_ensemble-iter}
\includegraphics[scale=\nScale]{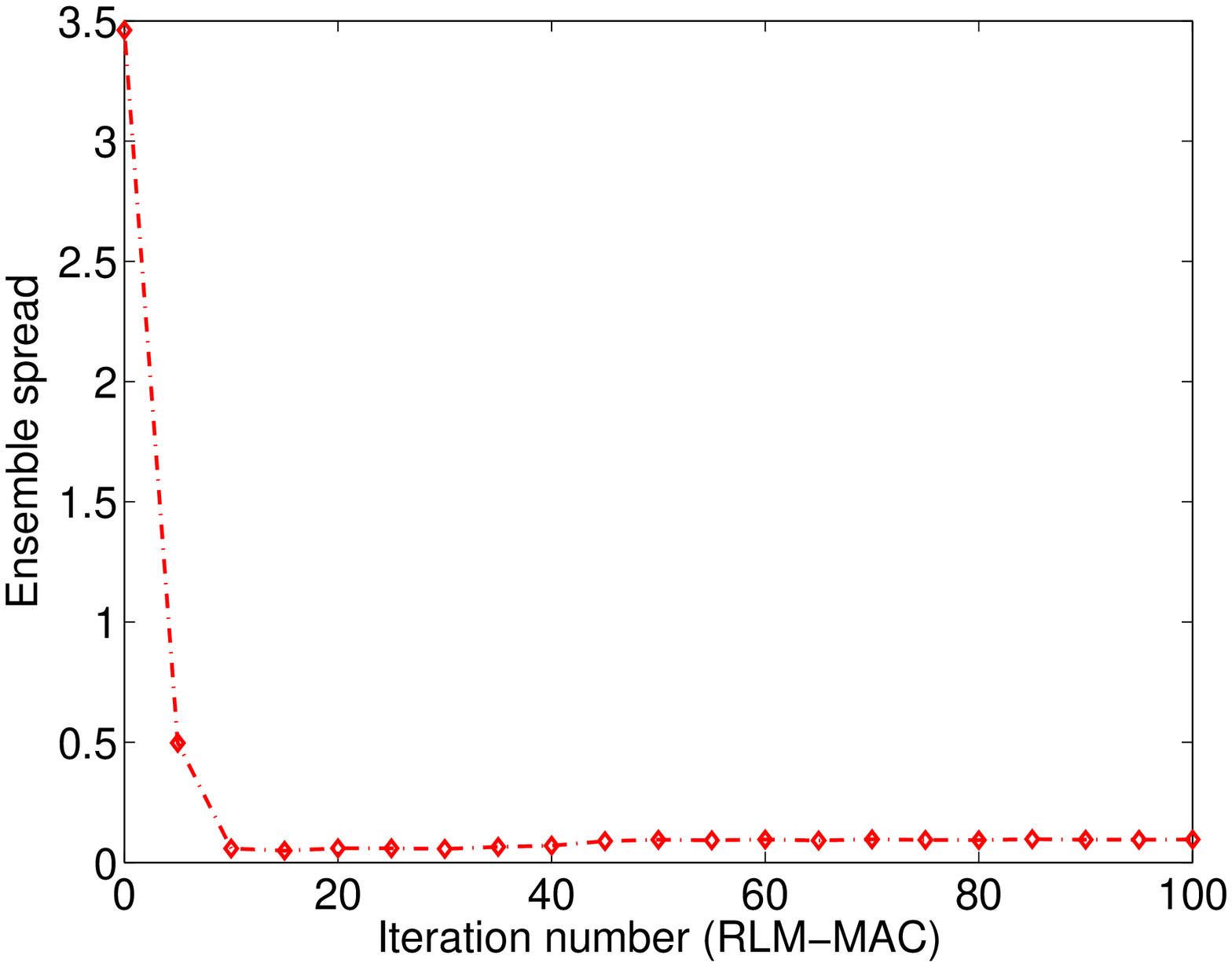}
}

\caption{\label{fig:L96_rmse_spread} Top panels: Box plots of RMSEs of the ensemble members with respect to the initial conditions at different iteration steps in the aLM-EnRML (left) and RLM-MAC (right). Bottom panels: Ensemble spreads of the aLM-EnRML (left) and RLM-MAC (right) at different iteration steps of the L96 model.} 
\end{figure*} 

\clearpage
\renewcommand{\nScale}{0.45}
\begin{figure*}
\centering
\includegraphics[scale=\nScale]{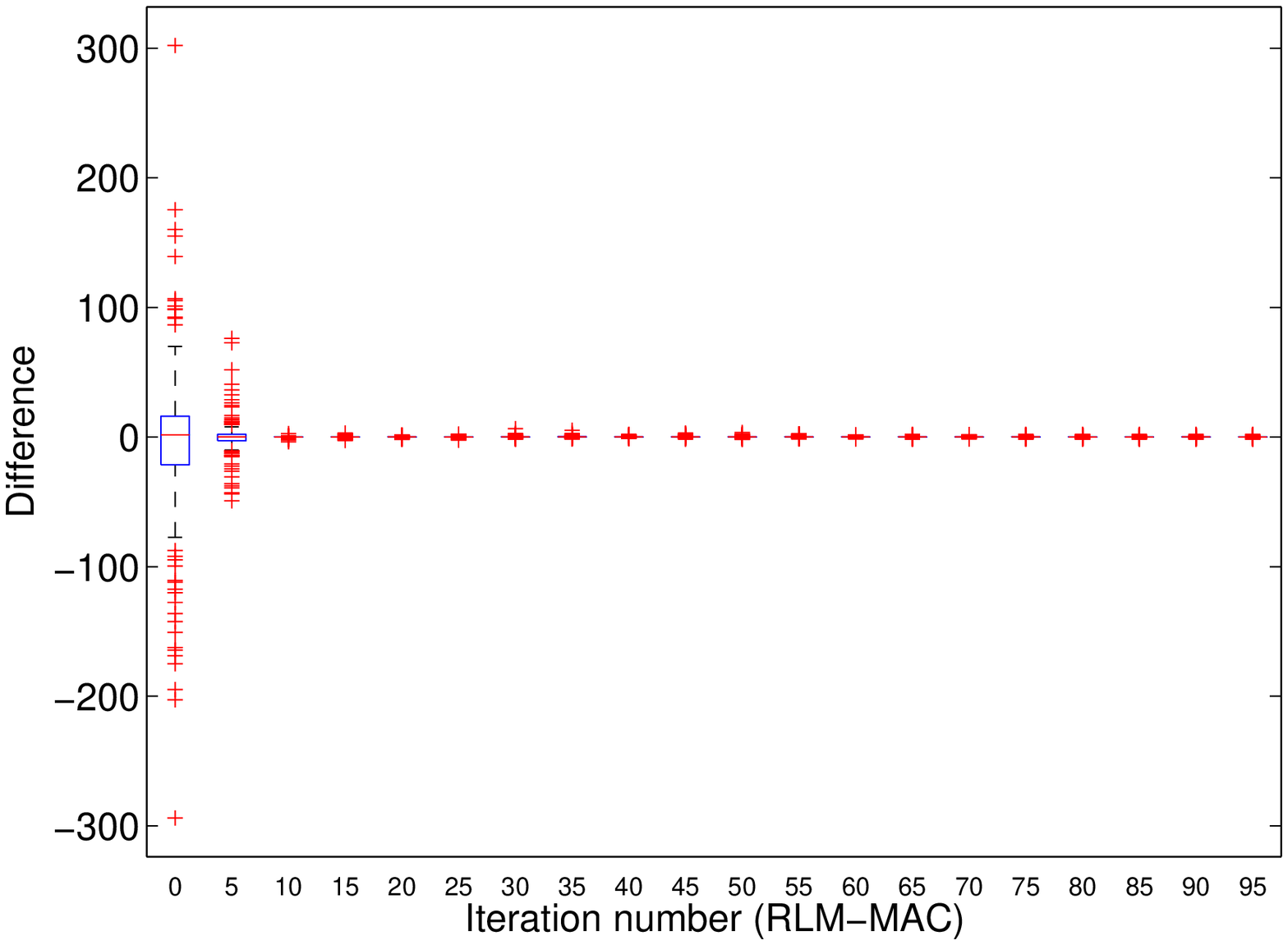}

\caption{\label{fig:obsDiff_boxPlot} Box plots of the normalized differences $\mathbf{C}_d^{-1/2} \left( \mathbf{g} \left(\bar{\mathbf{m}}^{i} \right) - \overline{\mathbf{g}\left(\mathbf{m}_j^{i}\right)} \right)$ at different iteration steps in the L96 model.} 
\end{figure*} 


\clearpage

\clearpage
\renewcommand{\nScale}{0.3}
\begin{figure*}
\centering
\subfigure[]{ \label{subfig:field_PERMX_1_1_iniEns}
\includegraphics[scale=\nScale]{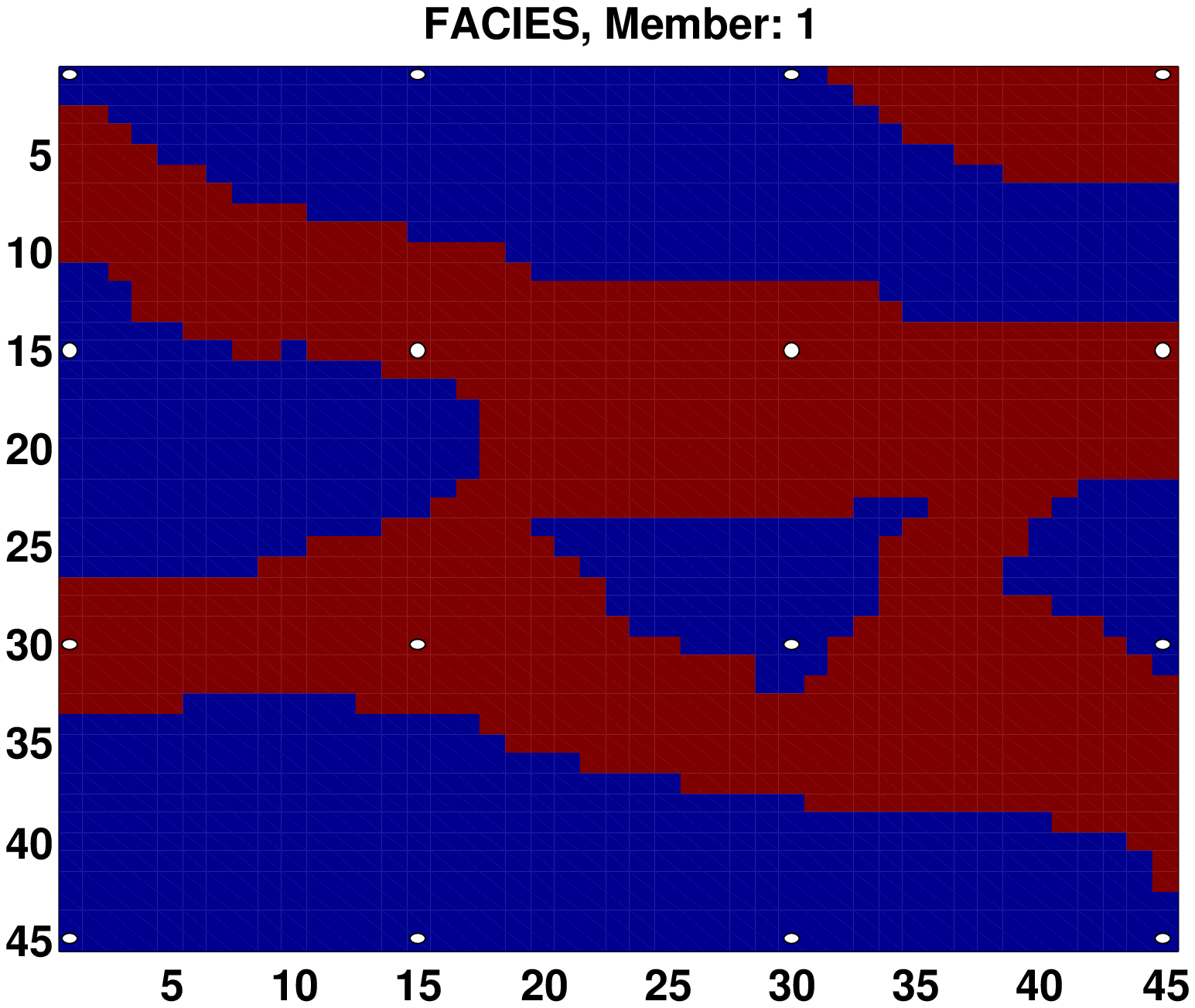}
}
\subfigure[]{ \label{subfig:aLM-EnRML_field_PERMX_1_1_ensemble10}
\includegraphics[scale=\nScale]{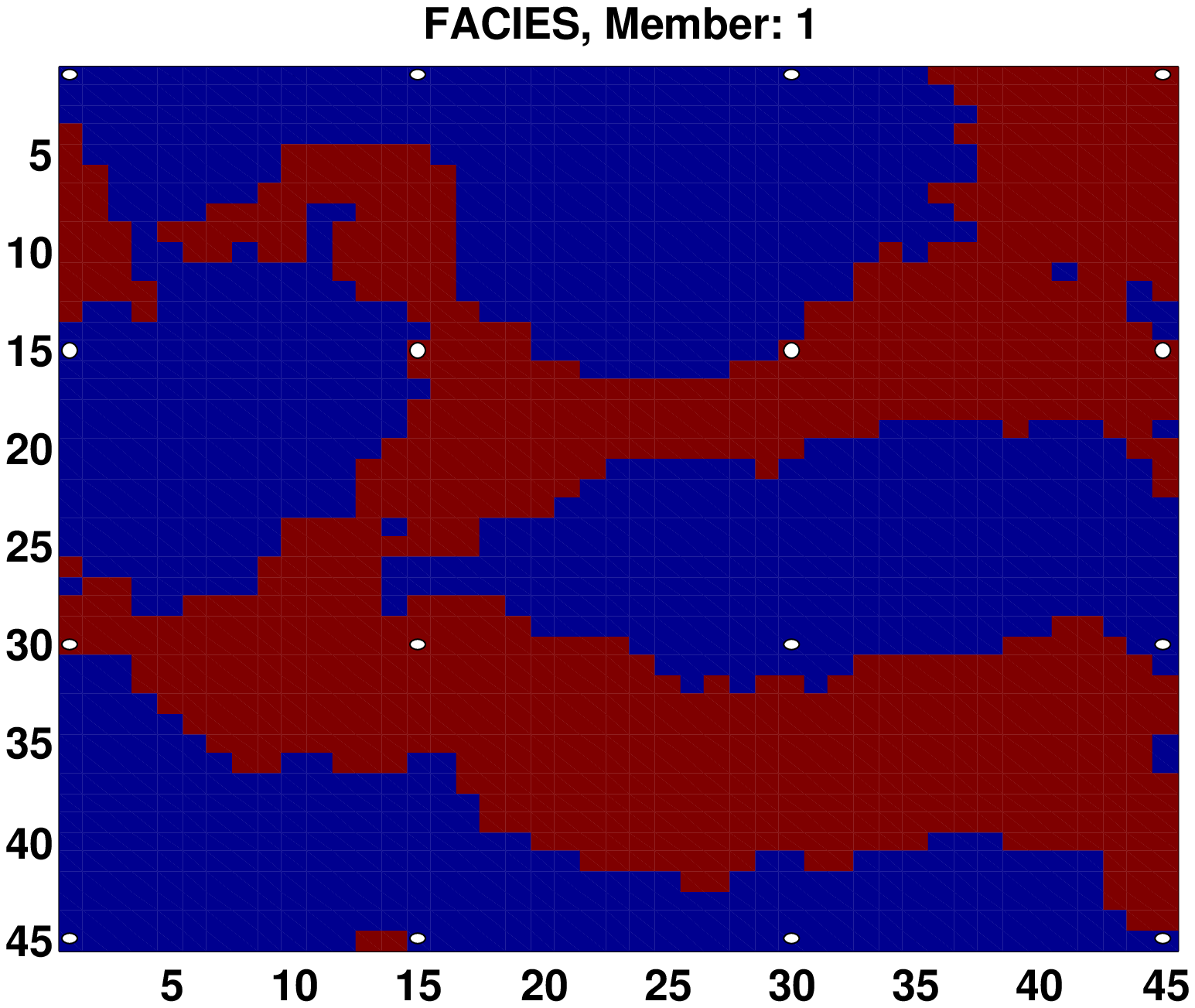}
}
\subfigure[]{ \label{subfig:RLM-MAC_field_PERMX_1_1_ensemble10}
\includegraphics[scale=\nScale]{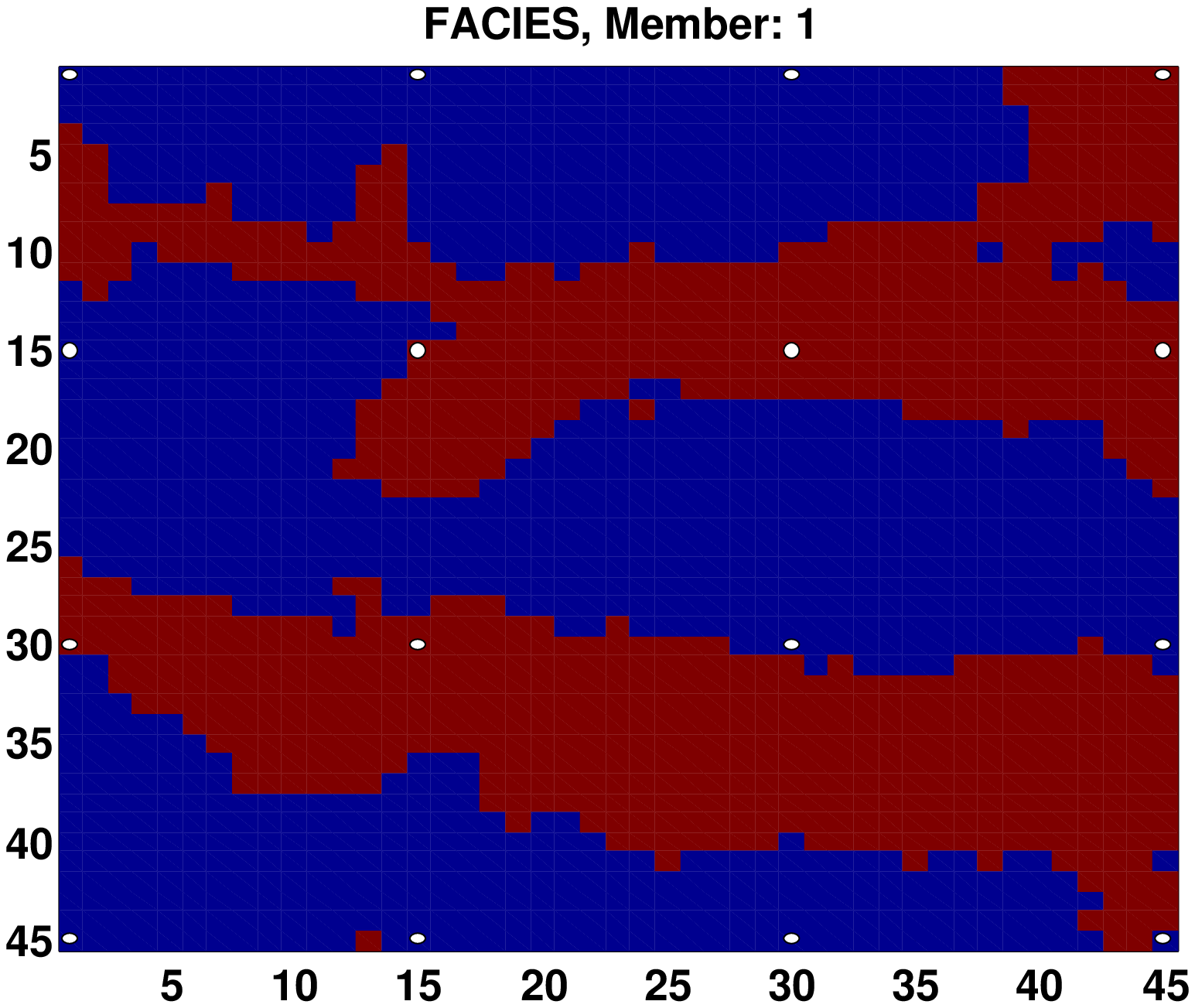}
}

\subfigure[]{ \label{subfig:field_PERMX_1_2_iniEns}
\includegraphics[scale=\nScale]{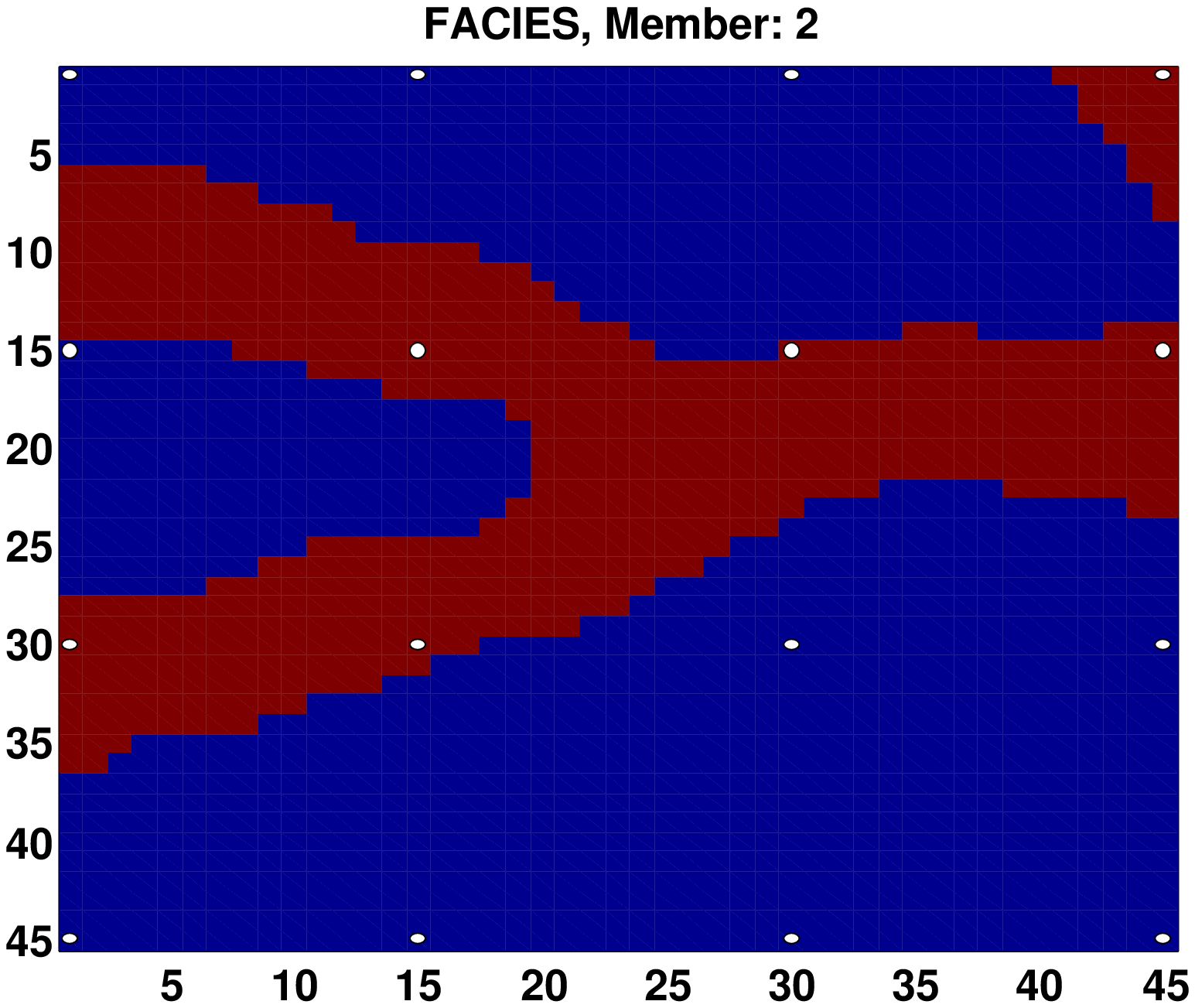}
}
\subfigure[]{ \label{subfig:aLM-EnRML_field_PERMX_1_2_ensemble10}
\includegraphics[scale=\nScale]{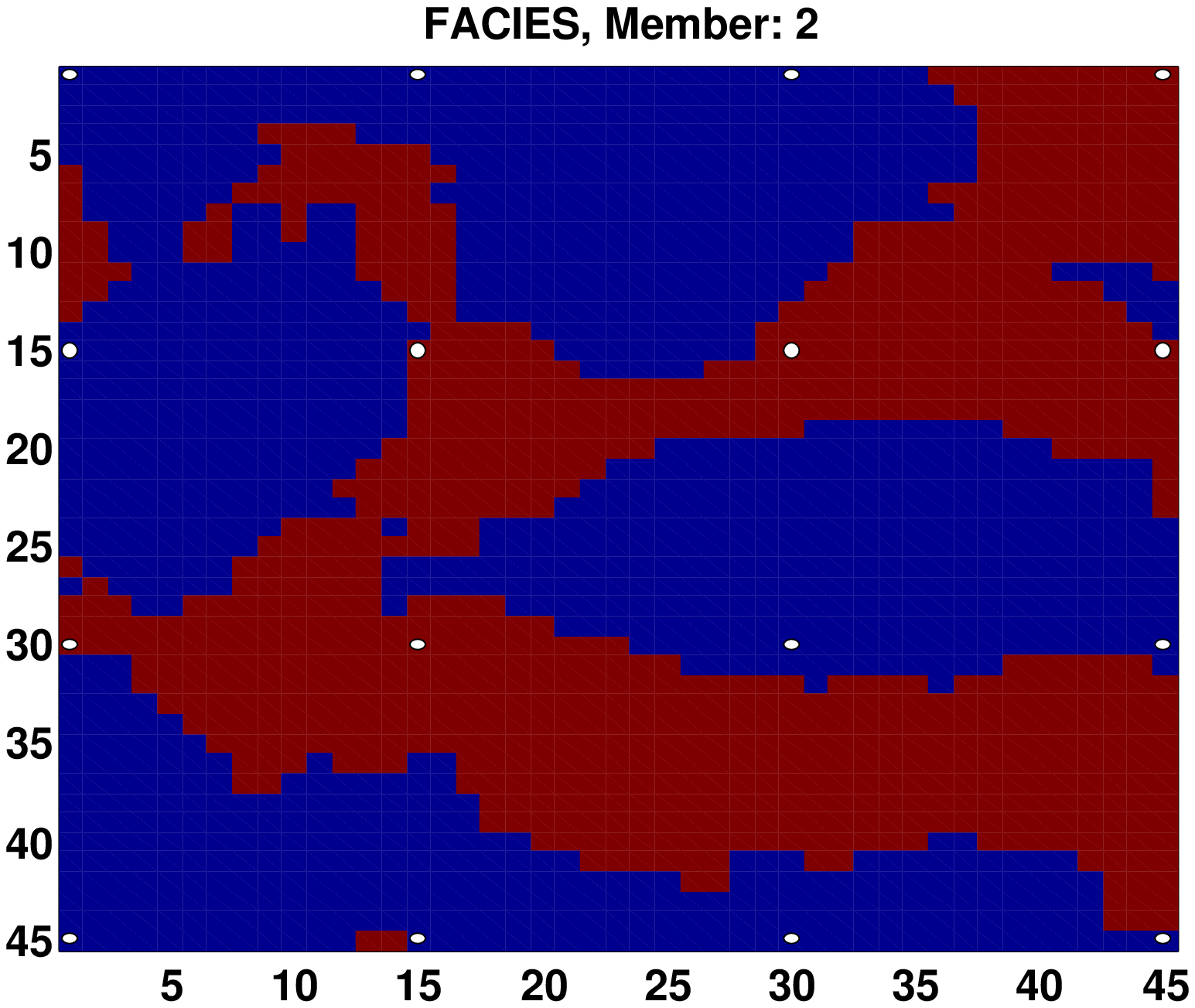}
}
\subfigure[]{ \label{subfig:RLM-MAC_field_PERMX_1_2_ensemble10}
\includegraphics[scale=\nScale]{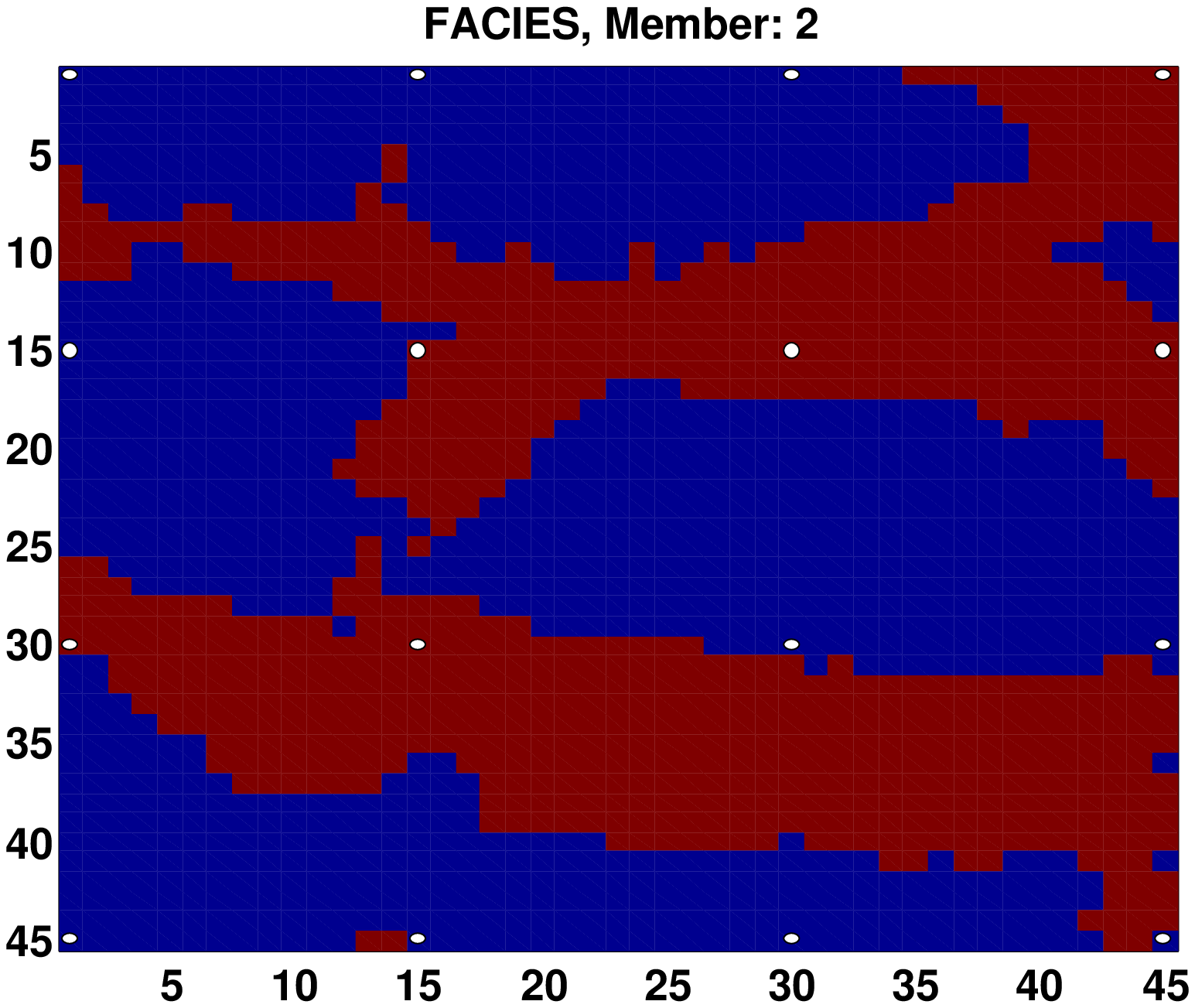}
}

\subfigure[Initial ensemble]{ \label{subfig:field_PERMX_1_3_iniEns}
\includegraphics[scale=\nScale]{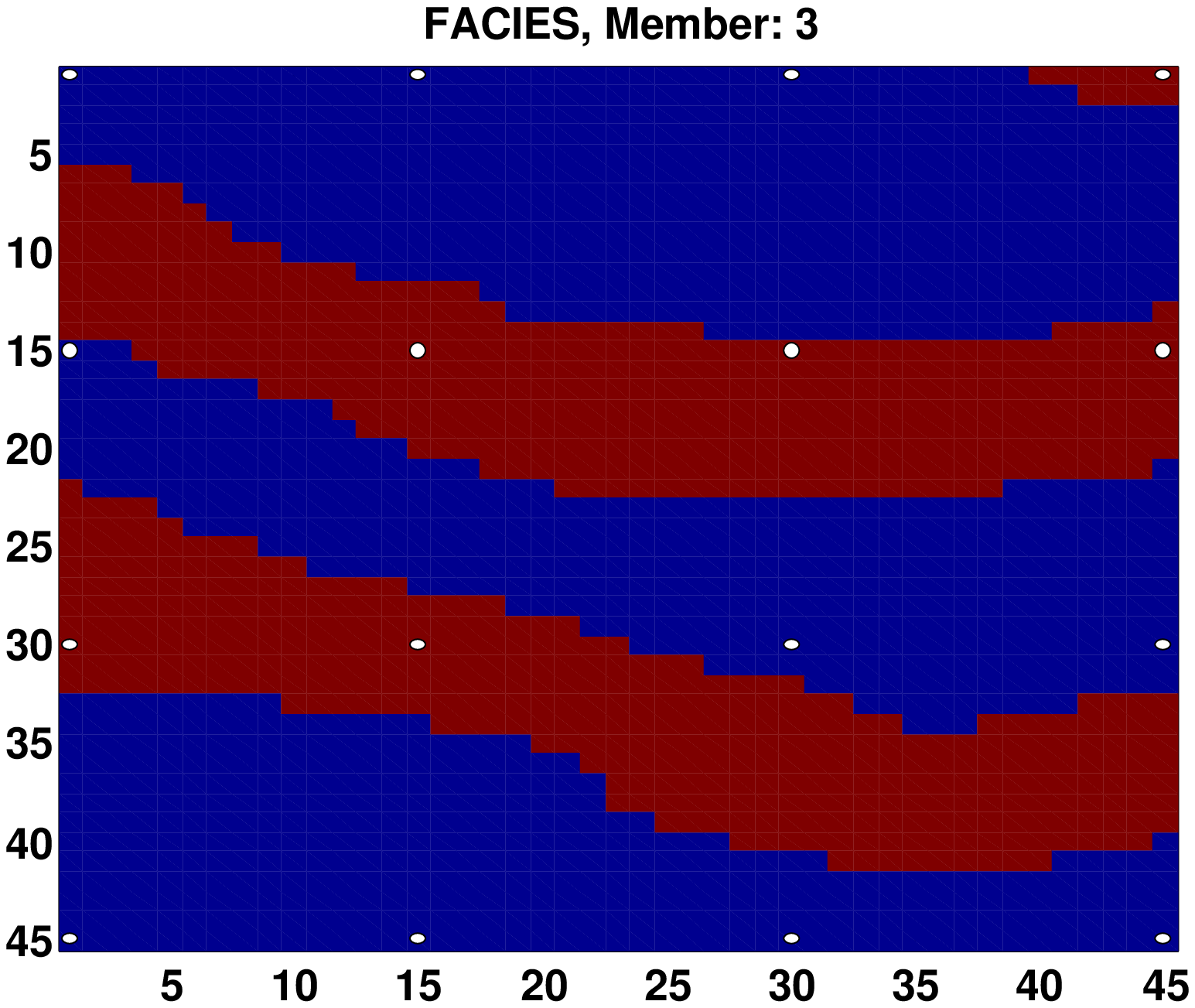}
}
\subfigure[aLM-EnRML]{ \label{subfig:aLM-EnRML_field_PERMX_1_3_ensemble10}
\includegraphics[scale=\nScale]{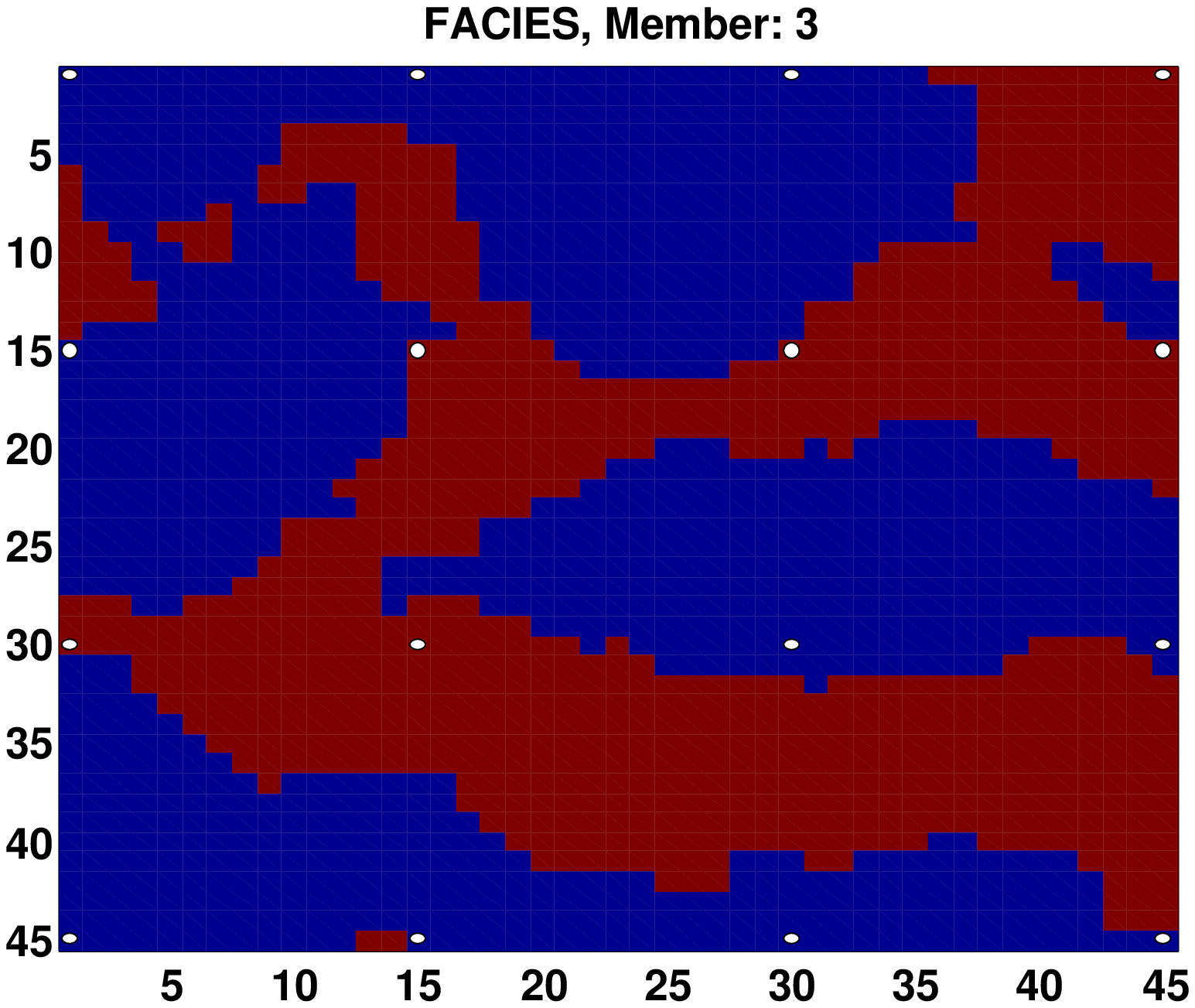}
}
\subfigure[RLM-MAC]{ \label{subfig:RLM-MAC_field_PERMX_1_3_ensemble10}
\includegraphics[scale=\nScale]{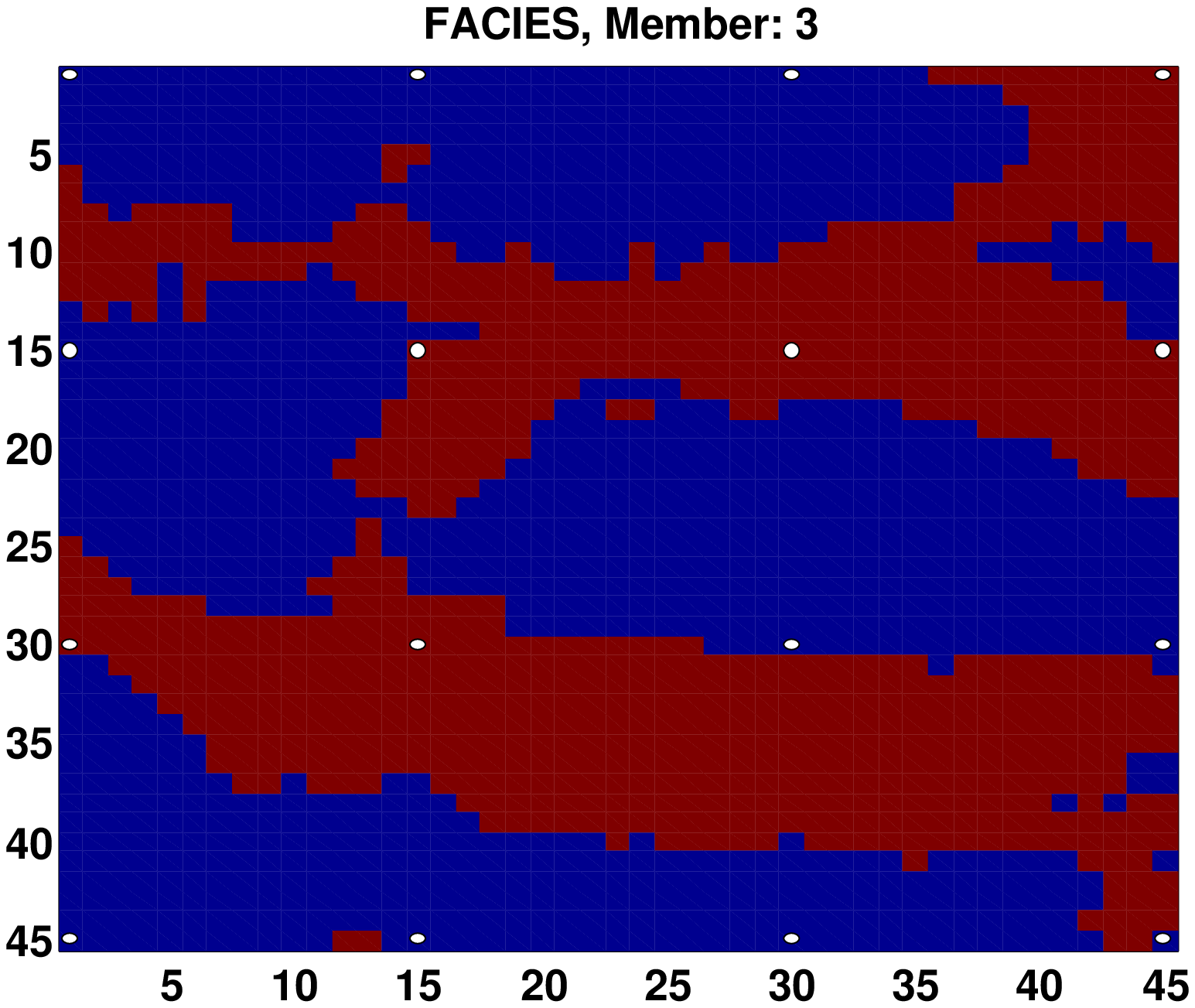}
}
\caption{\label{fig:facies_models} A few models (members 1 - 3) in the initial ensemble (1st column) and the ensembles of the aLM-EnRML (2nd column) and RLM-MAC (3rd column) at the final iteration steps of the facies estimation problem.} 
\end{figure*}  

\clearpage
\renewcommand{\nScale}{0.28}
\begin{figure*}
\centering
\subfigure[Initial ensemble]{ \label{subfig:fieldScore0_1}
\includegraphics[scale=\nScale]{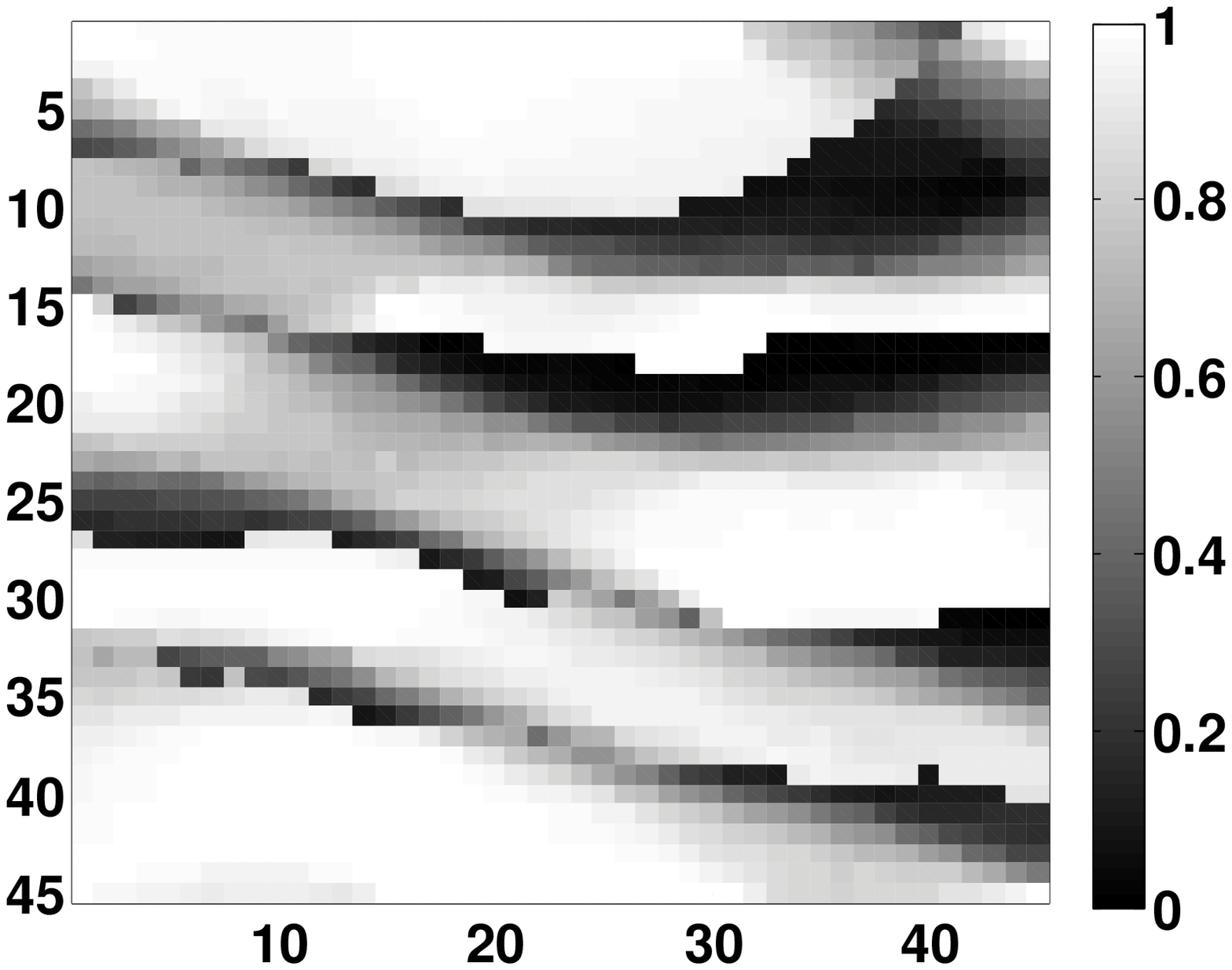}
}
\subfigure[aLM-EnRML]{ \label{subfig:aLM-EnRML_fieldScore11_1}
\includegraphics[scale=\nScale]{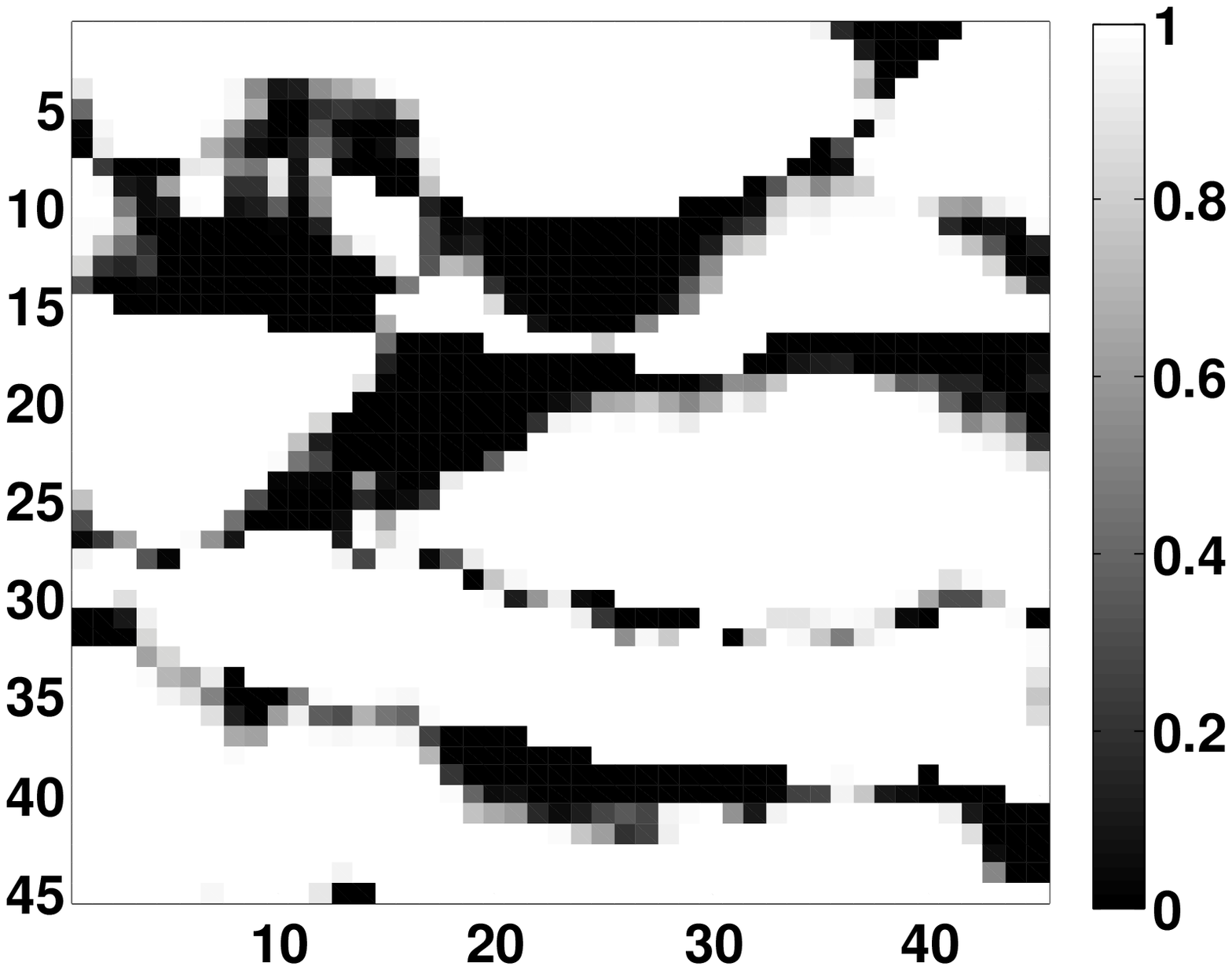}
}
\subfigure[RLM-MAC]{ \label{subfig:RLM-MAC_fieldScore11_1}
\includegraphics[scale=\nScale]{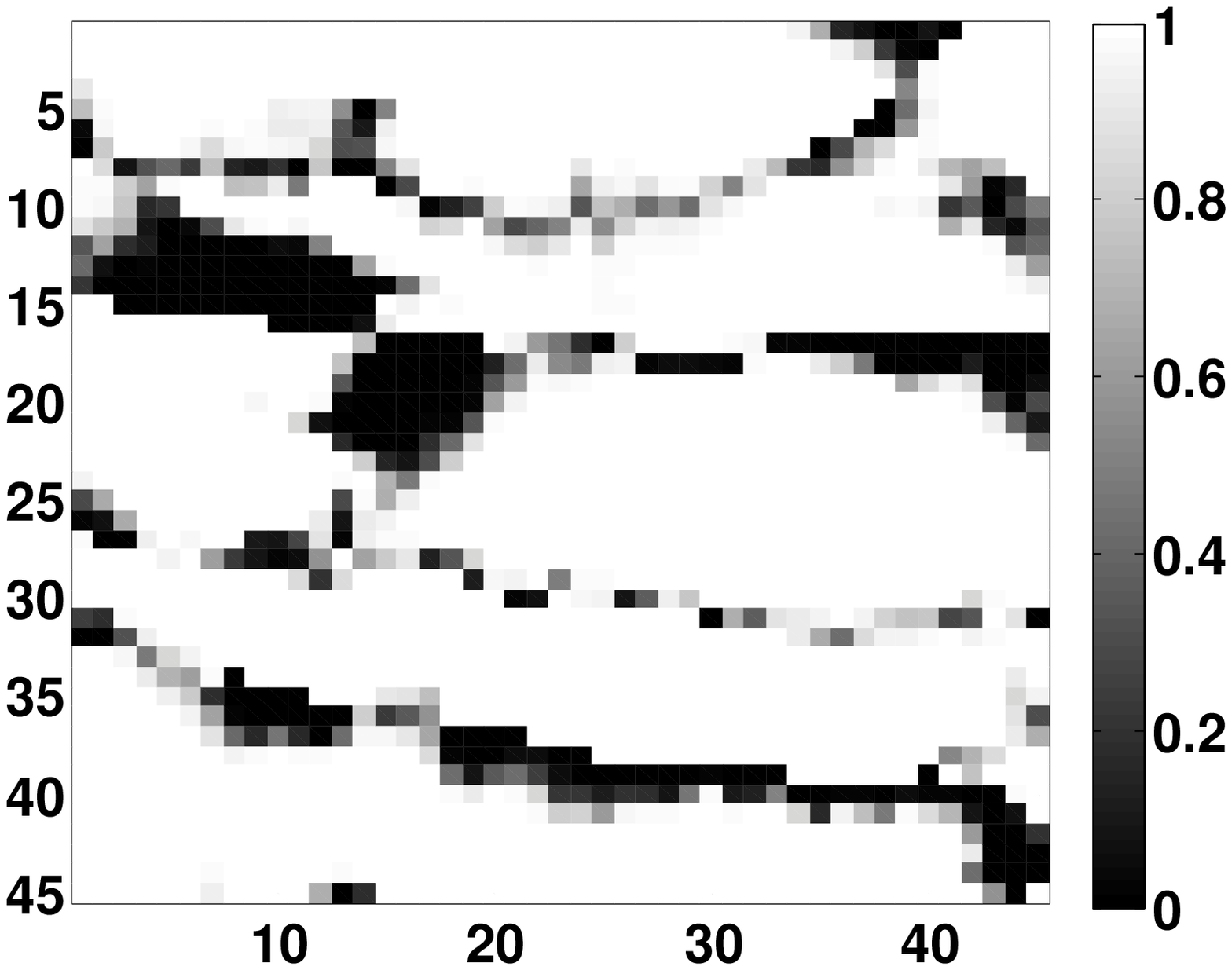}
}
\caption{\label{fig:facies_scores} Average facies matching scores of (a) the initial ensemble, (b) the final ensemble of the aLM-EnRML and (c) the final ensemble of the RLM-MAC. A higher score means that more models have the correct facies.} 
\end{figure*}  

\clearpage
\renewcommand{\nScale}{0.4}
\begin{figure*}
\centering
\subfigure{ \label{subfig:FS1_ensize100_aLM-EnRML_boxplot_objRealIter}
\includegraphics[scale=\nScale]{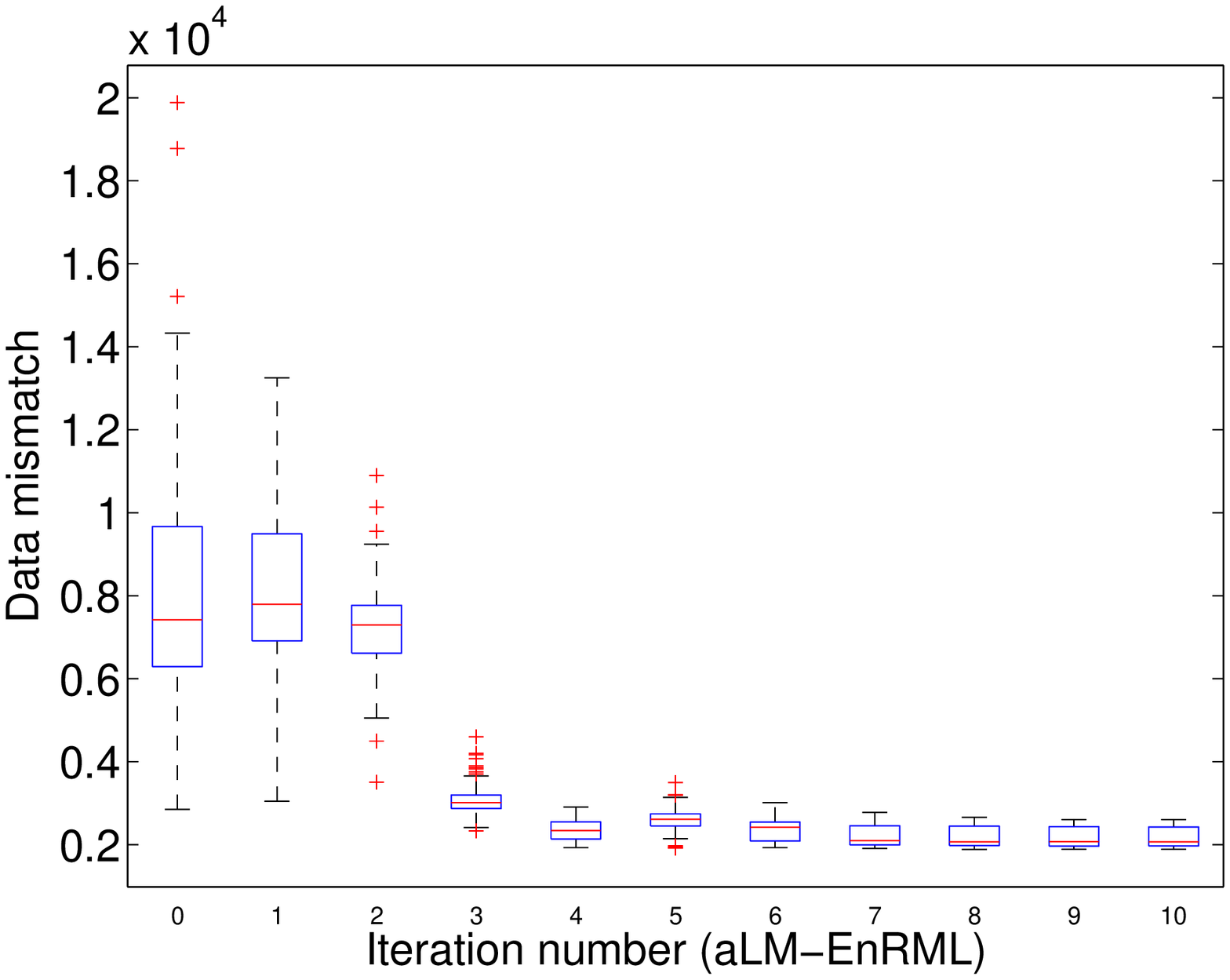}
}
\subfigure{ \label{subfig:FS1_ensize99_RLM-MAC_boxplot_objRealIter}
\includegraphics[scale=\nScale]{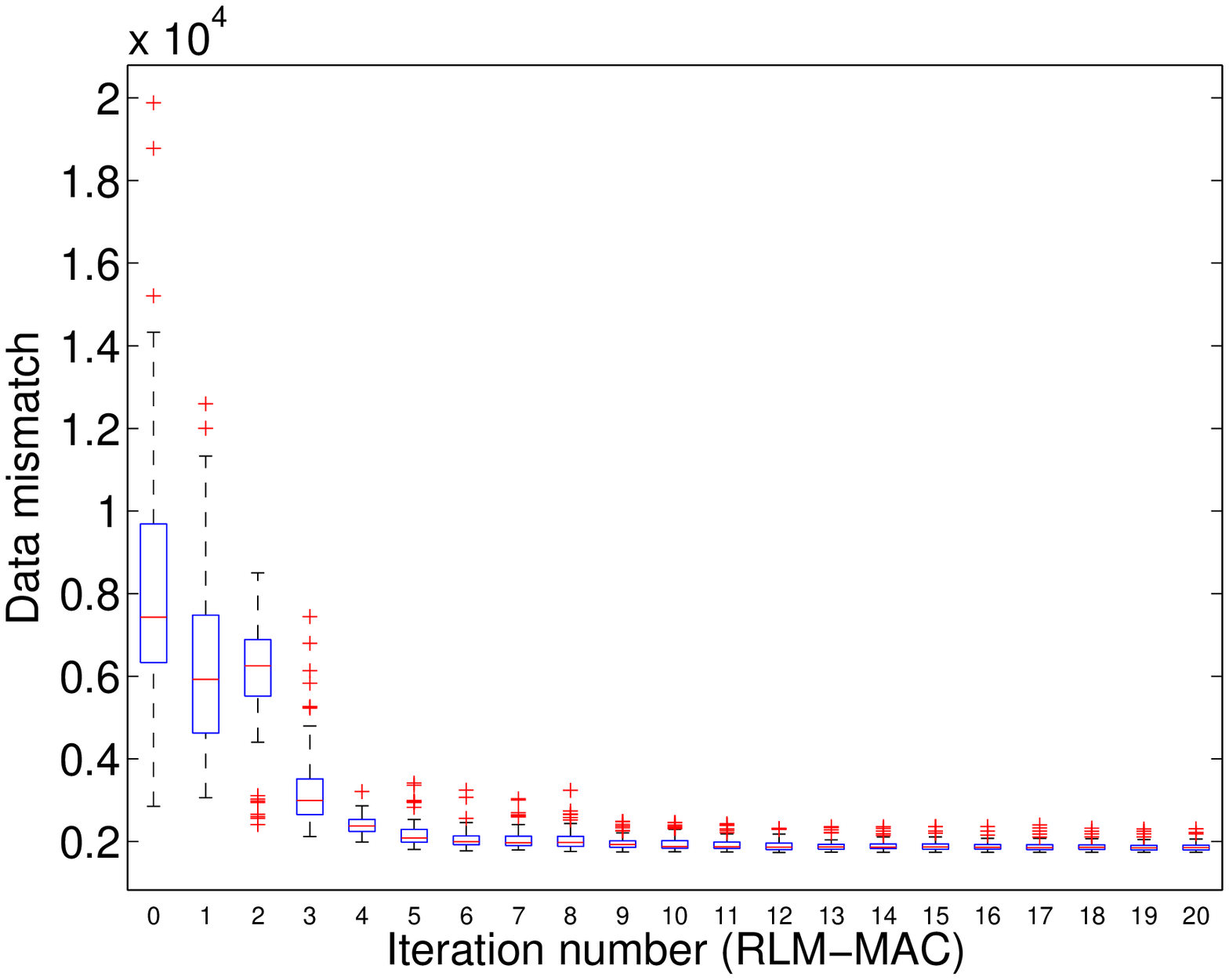}
}
\caption{\label{fig:facies_obj} Box plots of data mismatch at different iteration steps of the facies estimation problem. Left: aLM-EnRML; Right: RLM-MAC.} 
\end{figure*} 

\clearpage
\begin{figure*}
\renewcommand{\nScale}{0.28}
\centering
\subfigure[]{ \label{subfig:init_forecast_0_WOPR_P5}
\includegraphics[scale=\nScale]{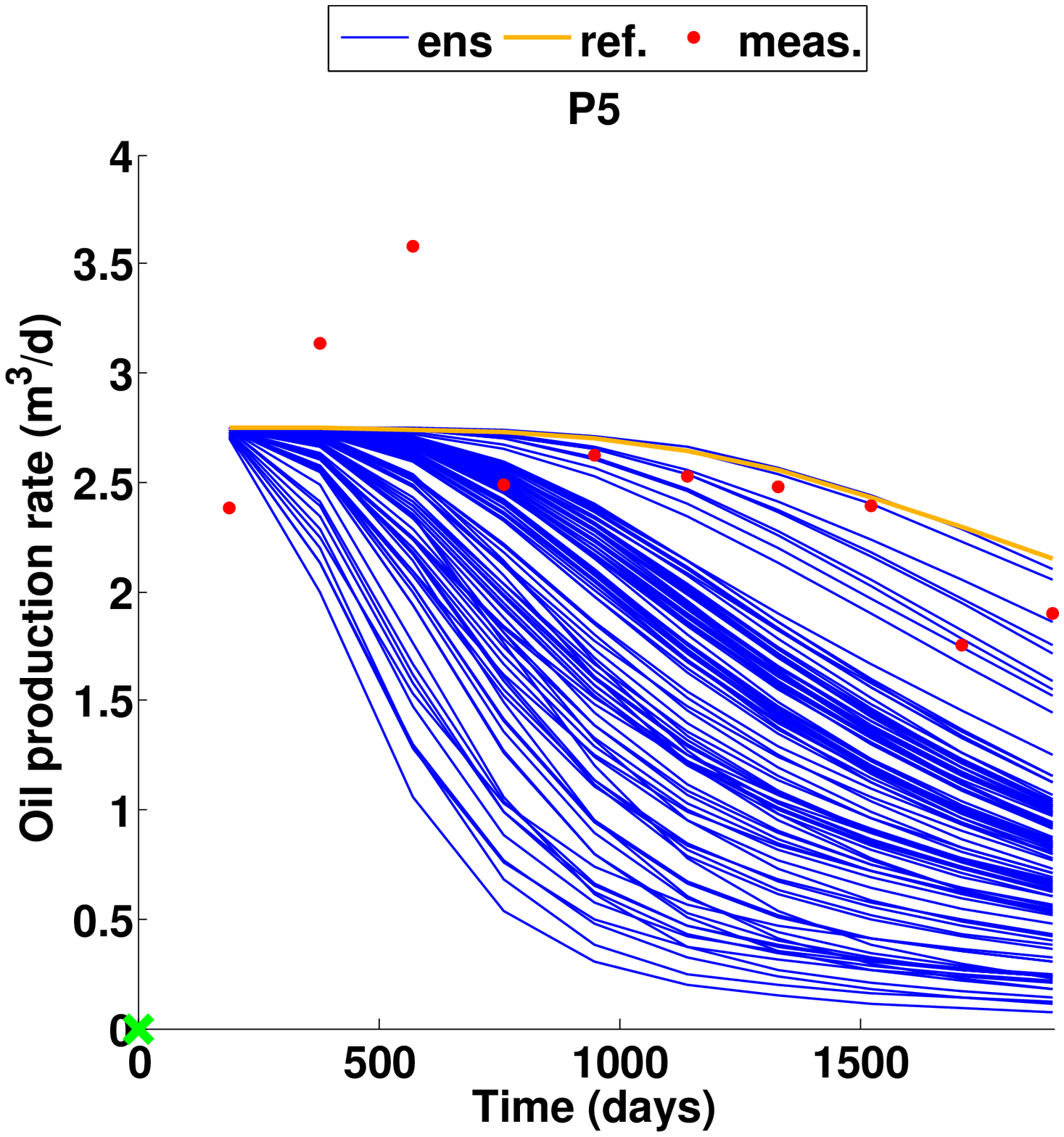}
}
\subfigure[]{ \label{subfig:aLM-EnRML_forecast(ES)__WOPR_P5}
\includegraphics[scale=\nScale]{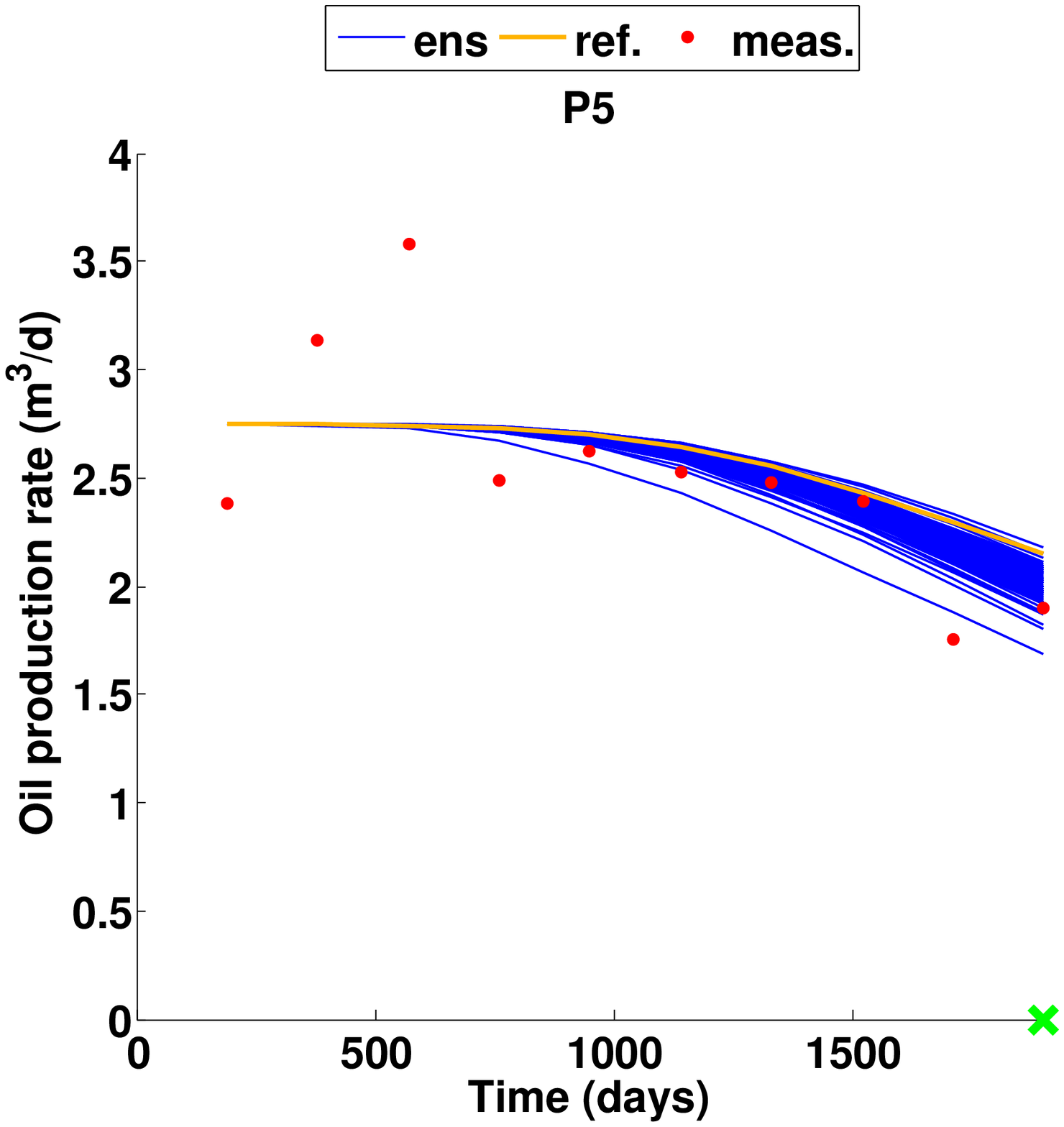}
}
\subfigure[]{ \label{subfig:RLM-MAC_forecast(ES)__WOPR_P5}
\includegraphics[scale=\nScale]{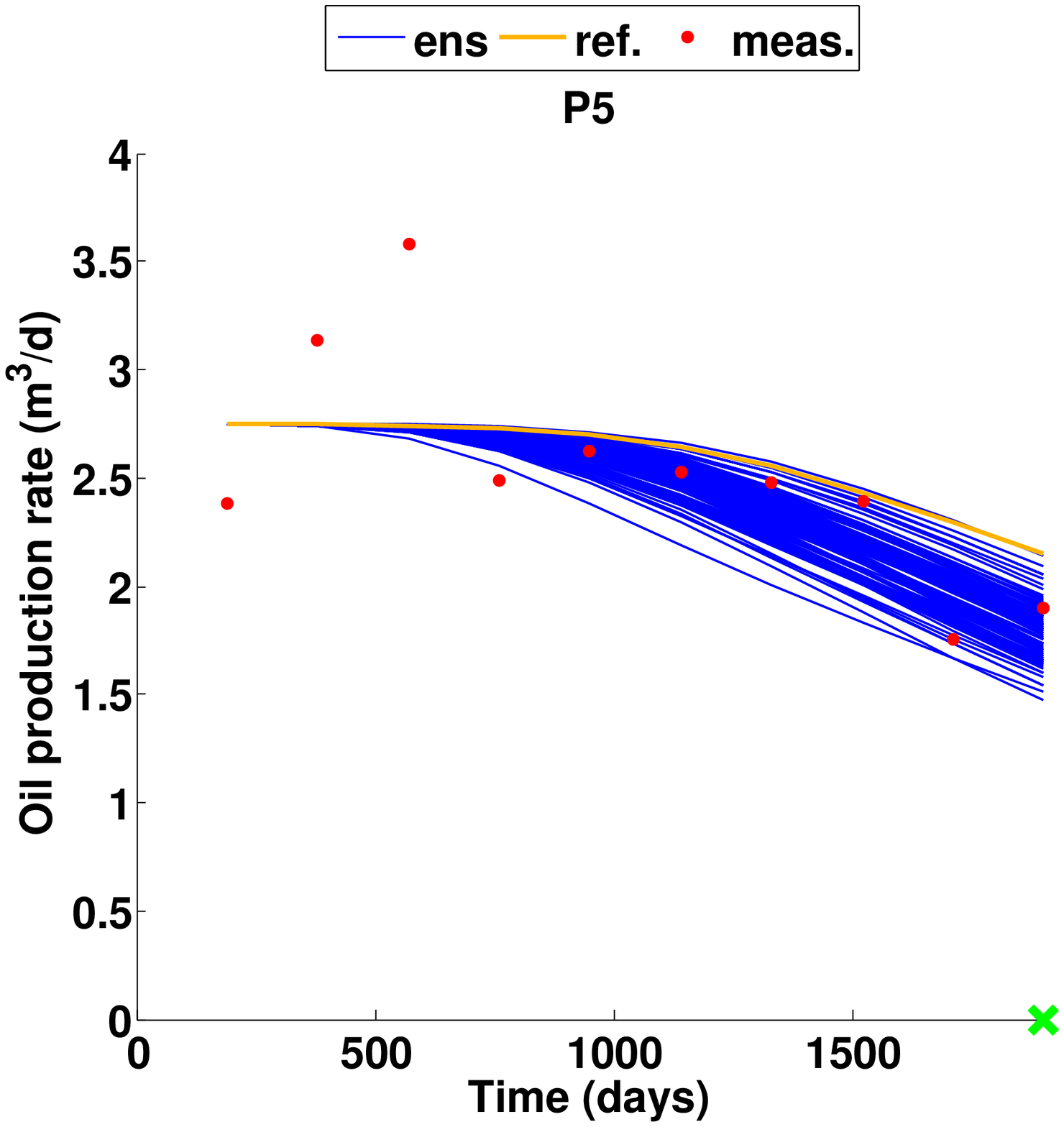}
}

\subfigure[Initial ensemble]{ \label{subfig:init_forecast_0_WOPR_P7}
\includegraphics[scale=\nScale]{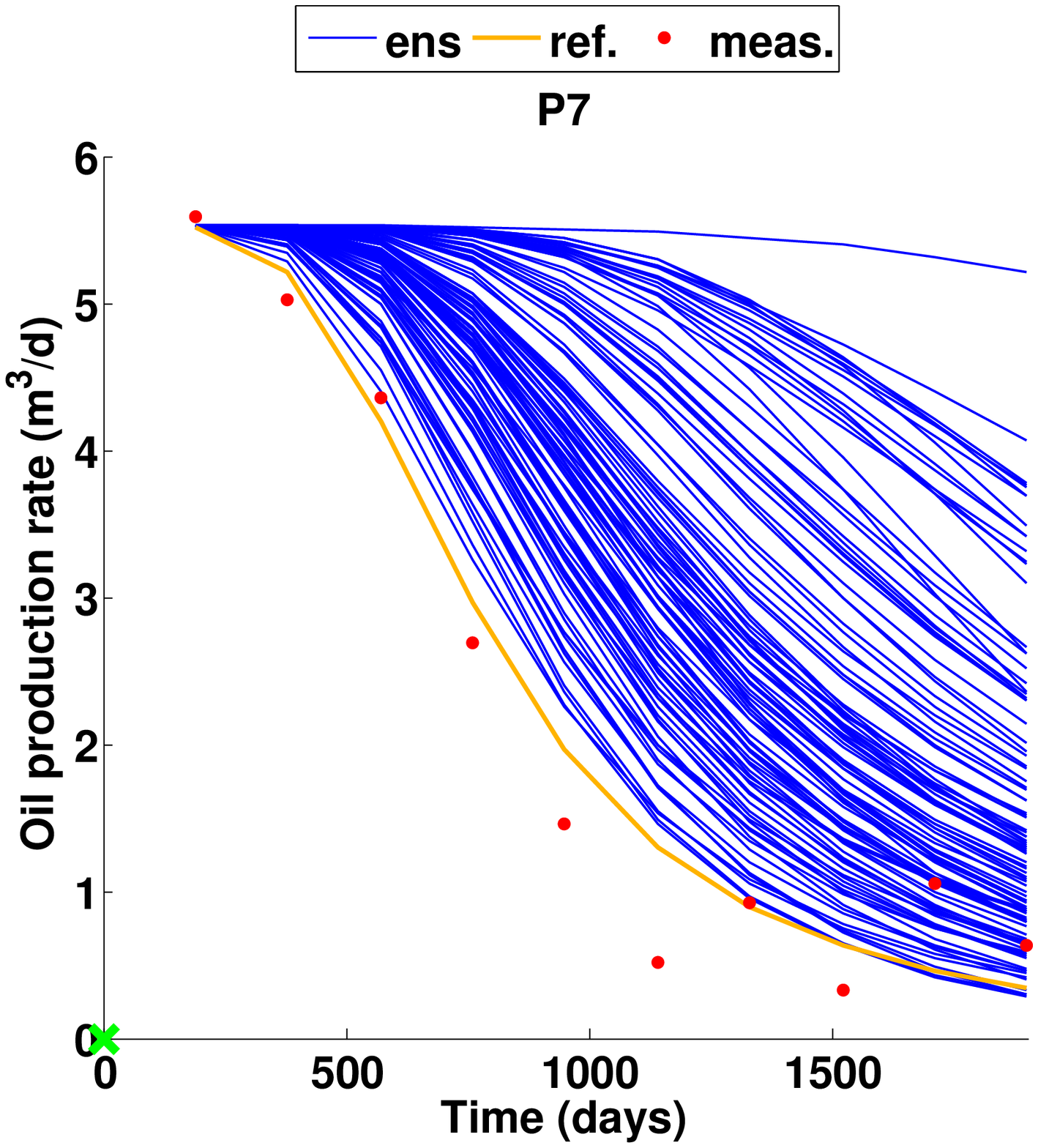}
}
\subfigure[aLM-EnRML]{ \label{subfig:aLM-EnRML_forecast(ES)__WOPR_P7}
\includegraphics[scale=\nScale]{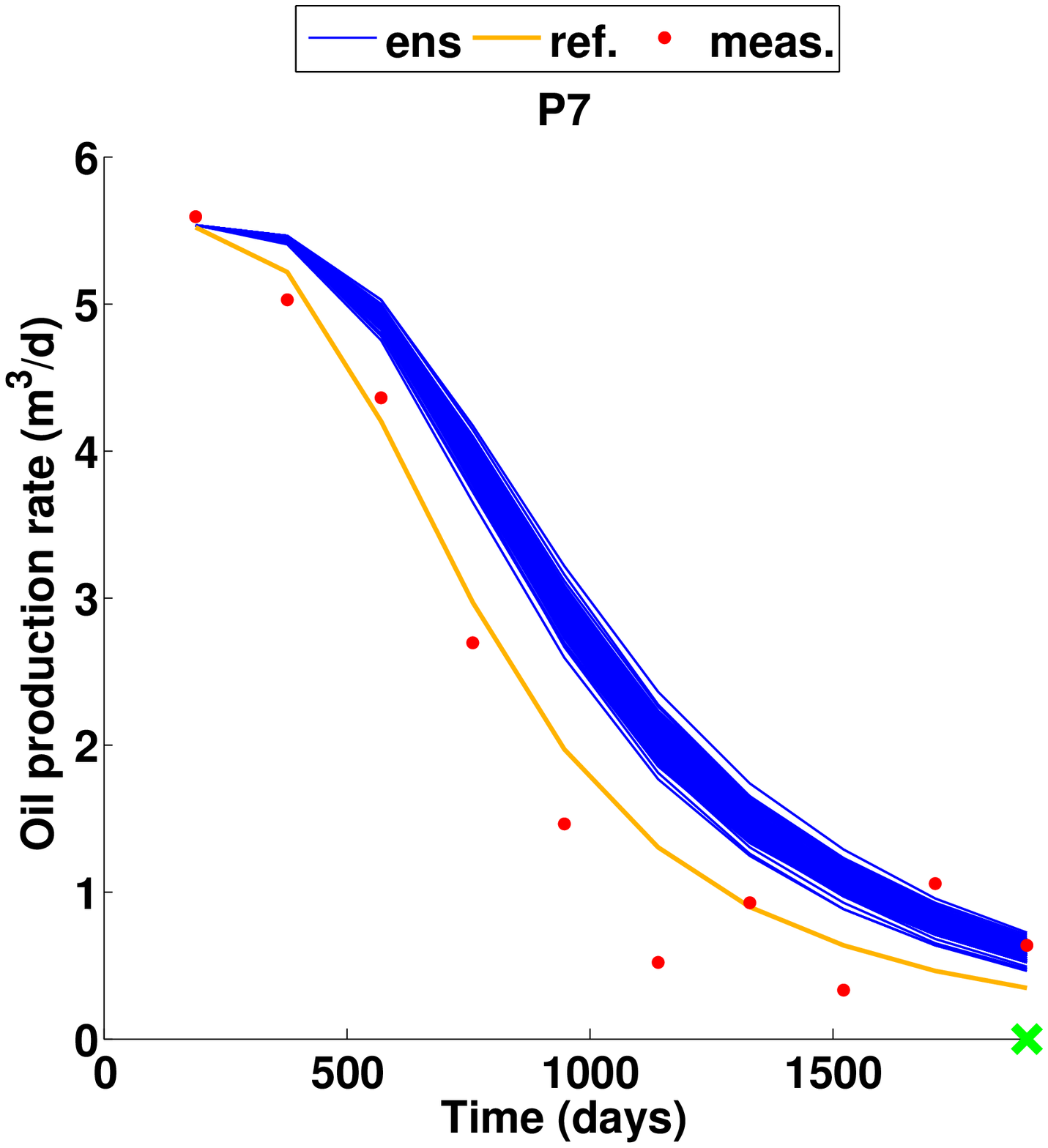}
}
\subfigure[RLM-MAC]{ \label{subfig:RLM-MAC_forecast(ES)__WOPR_P7}
\includegraphics[scale=\nScale]{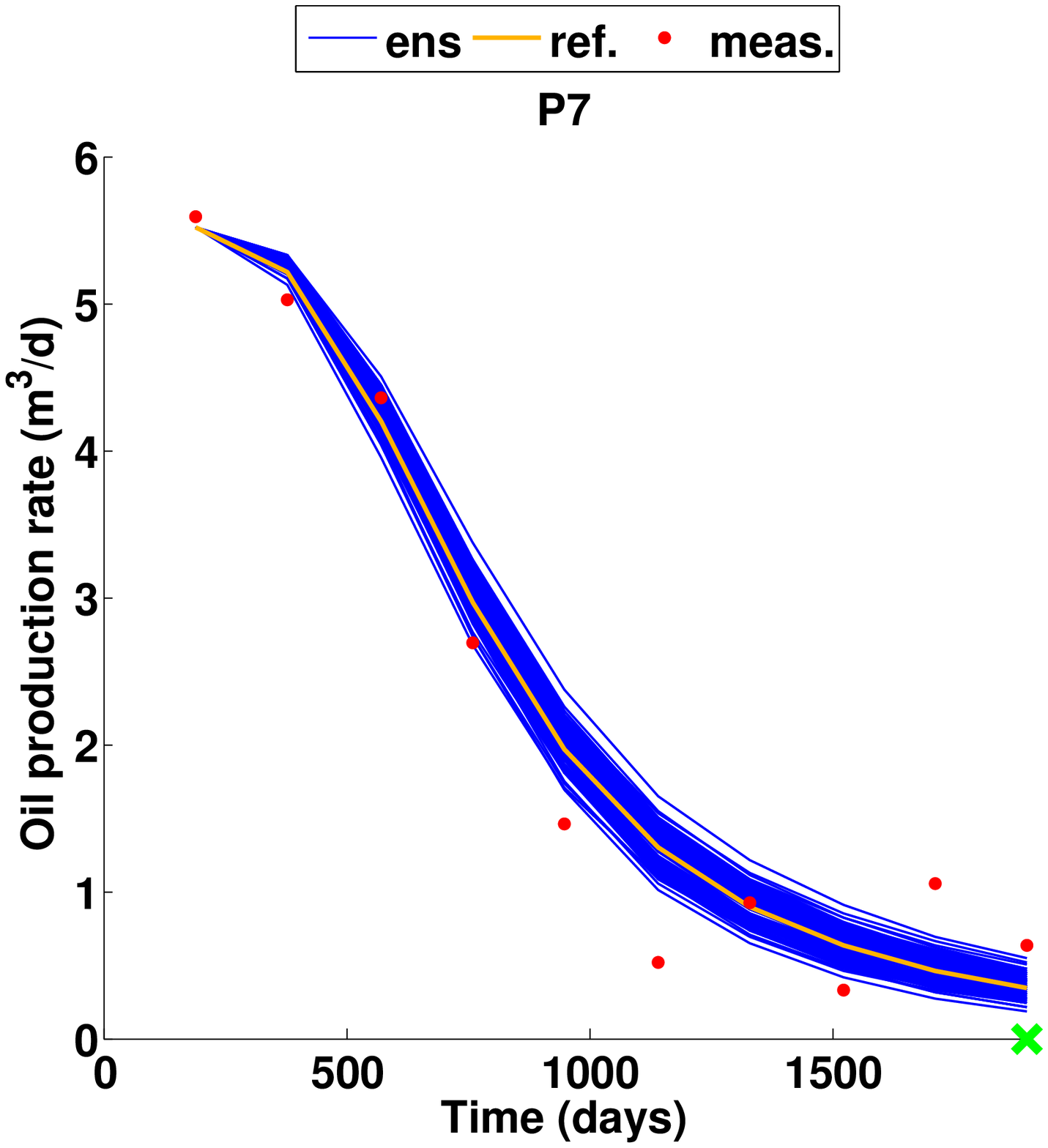}
}
\caption{\label{fig:facies_forecasts_WOPR} History matching of oil production rates at P5 (top row) and P7 (bottom row) of the facies estimation problem, using the initial ensemble (1st column) and the ensembles of the aLM-EnRML (2nd column) and RLM-MAC (3rd column) at the final iteration steps. In each image, the red dots represent the synthetic measurements and the blue curves are the forecasts with respect to the initial ensemble (1st column) and history matching profiles (2nd and 3rd columns). For comparison, the production data of the reference model are also shown in orange.} 
\end{figure*}  

\clearpage
\renewcommand{\nScale}{0.28}
\begin{figure*}
\centering
\subfigure[]{ \label{subfig:init_forecast_0_WWPR_P5}
\includegraphics[scale=\nScale]{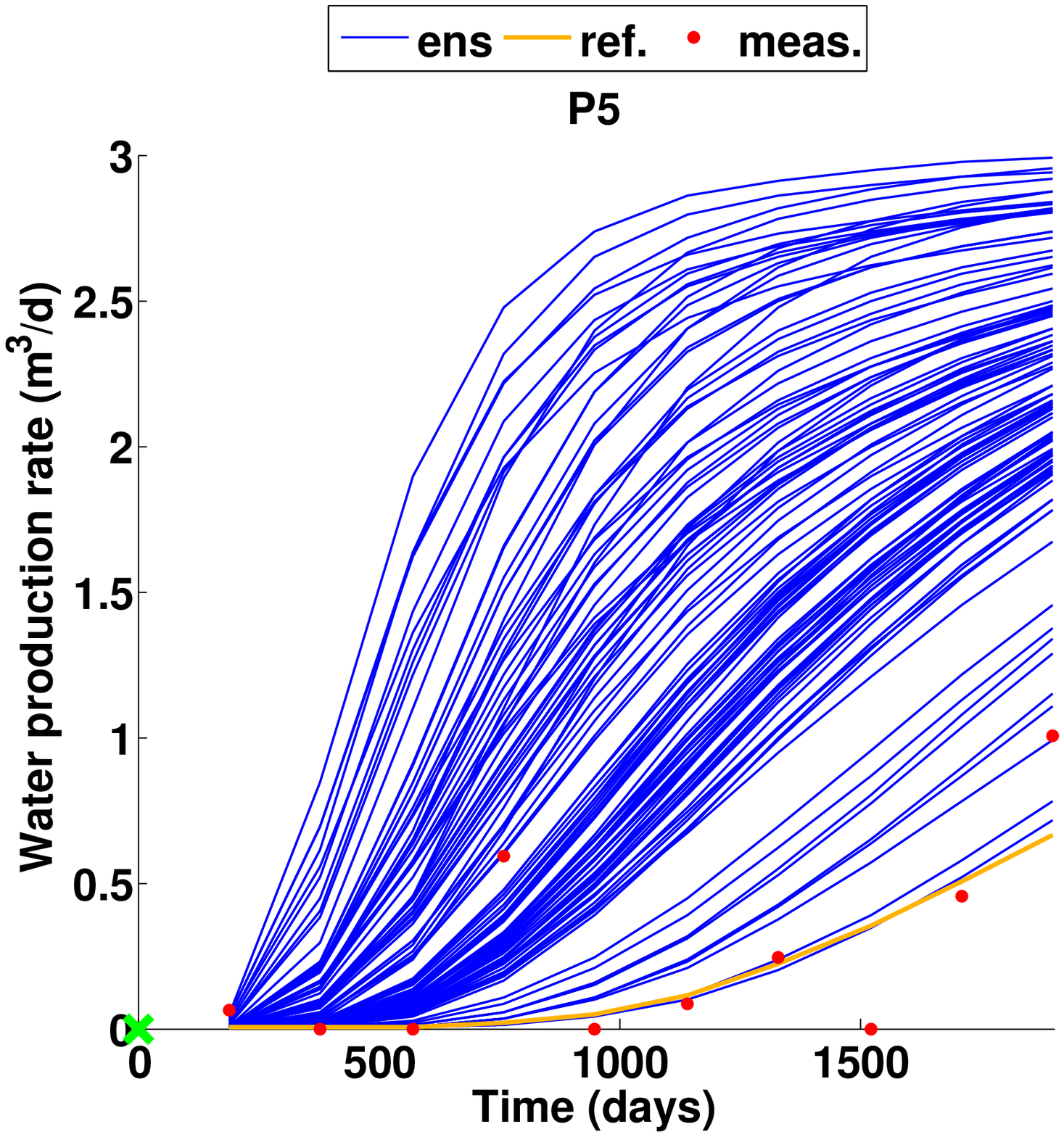}
}
\subfigure[]{ \label{subfig:aLM-EnRML_forecast(ES)__WWPR_P5}
\includegraphics[scale=\nScale]{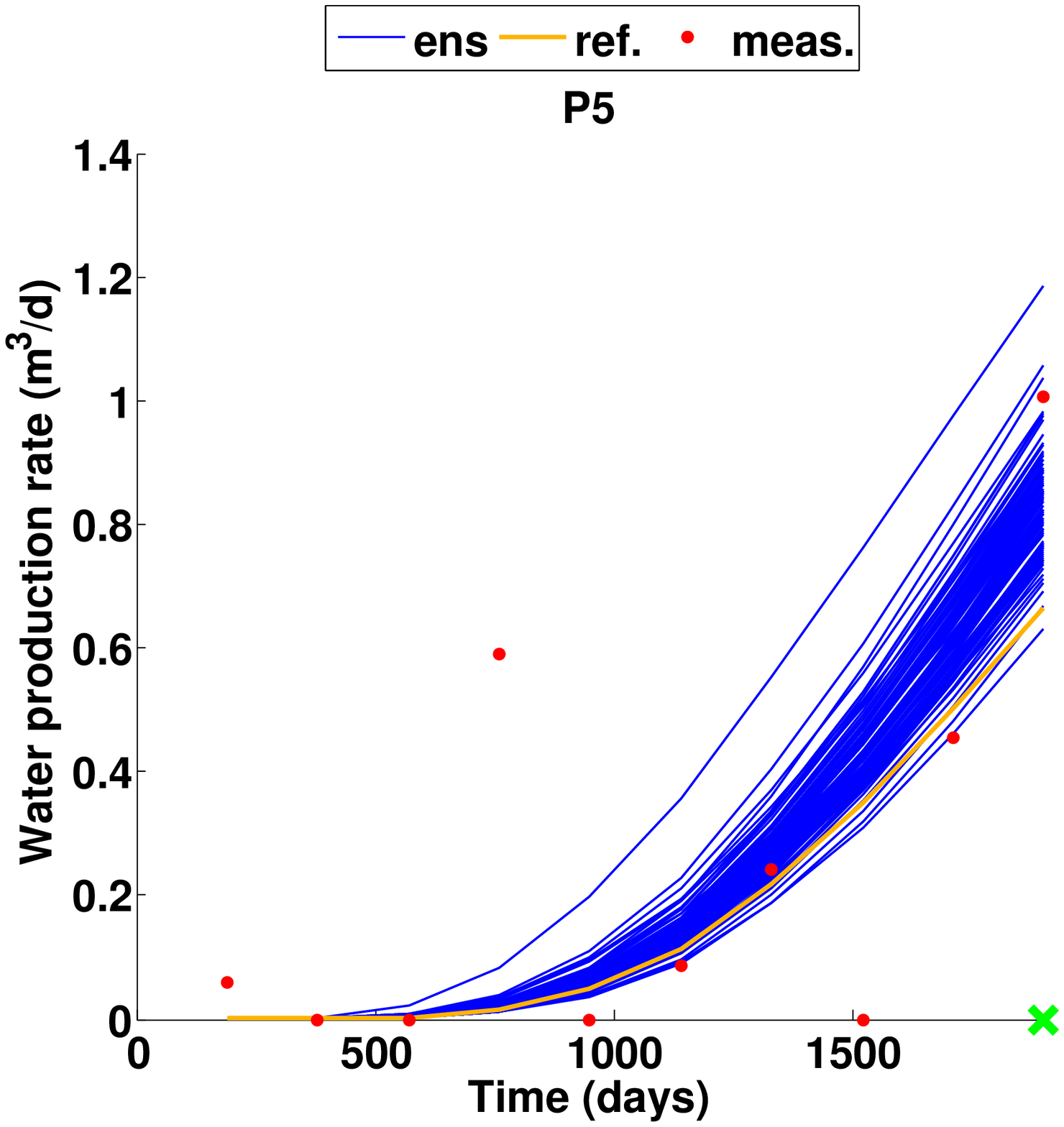}
}
\subfigure[]{ \label{subfig:RLM-MAC_forecast(ES)__WWPR_P5}
\includegraphics[scale=\nScale]{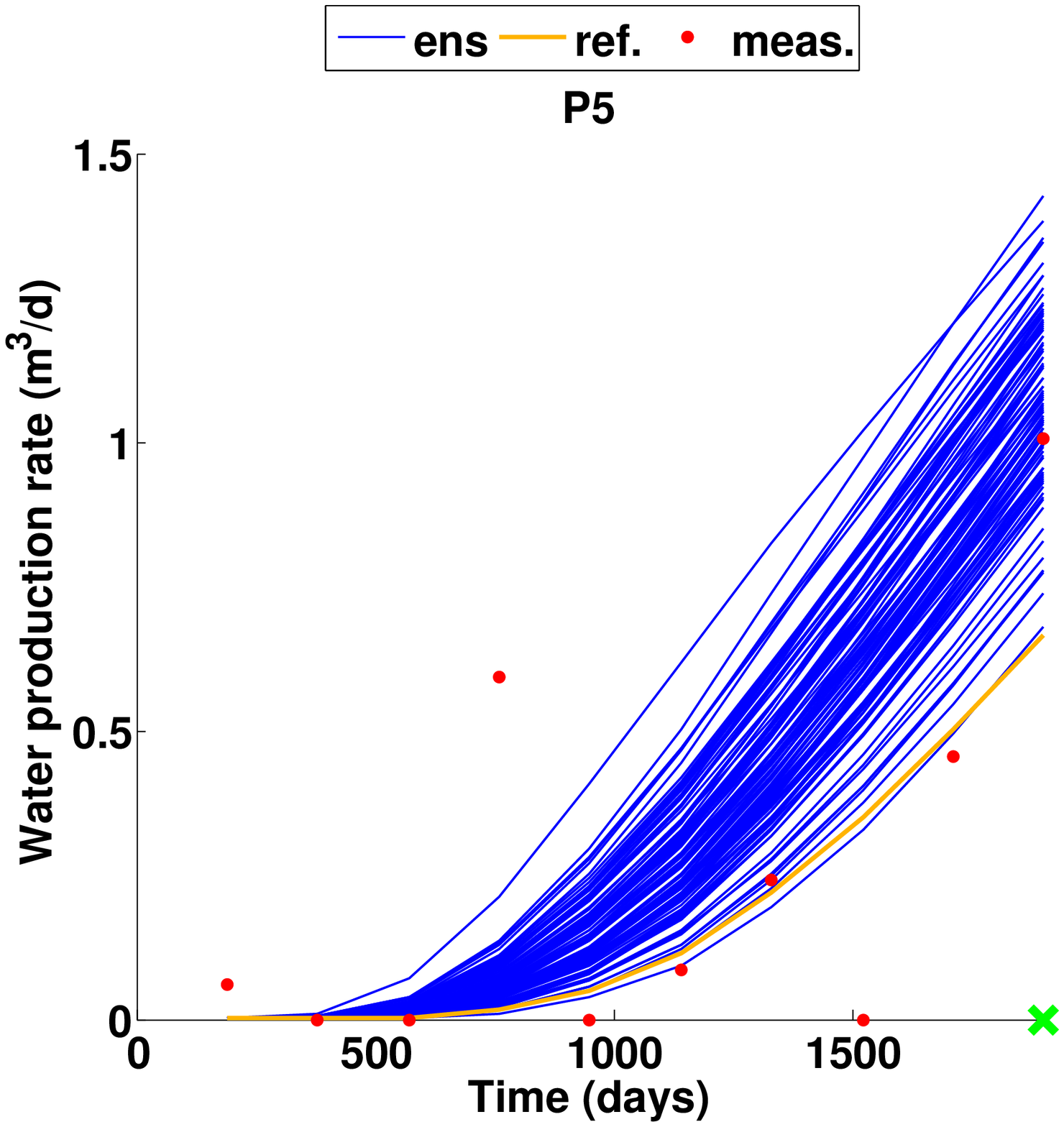}
}

\subfigure[Initial ensemble]{ \label{subfig:init_forecast_0_WWPR_P7}
\includegraphics[scale=\nScale]{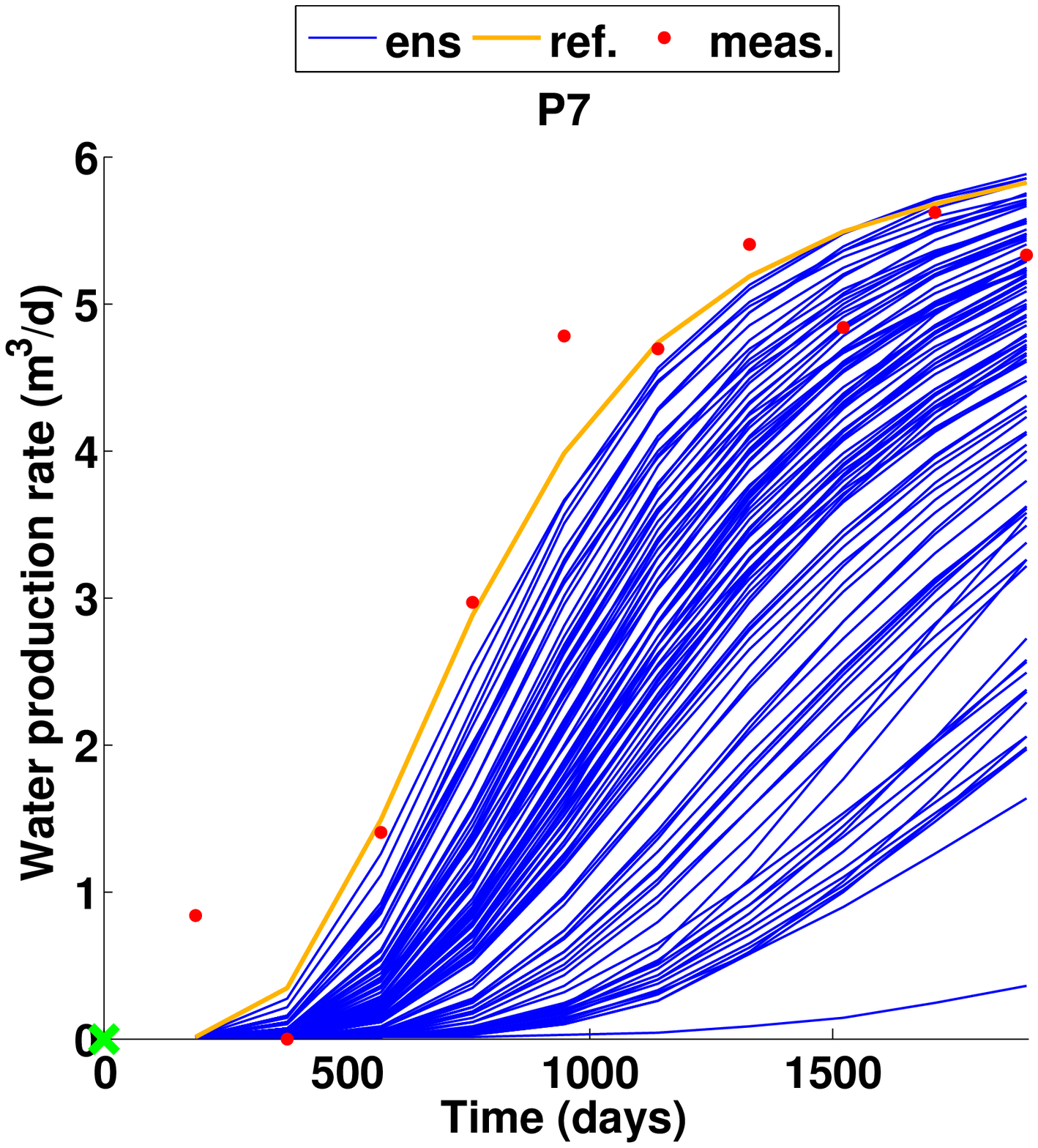}
}
\subfigure[aLM-EnRML]{ \label{subfig:aLM-EnRML_forecast(ES)__WWPR_P7}
\includegraphics[scale=\nScale]{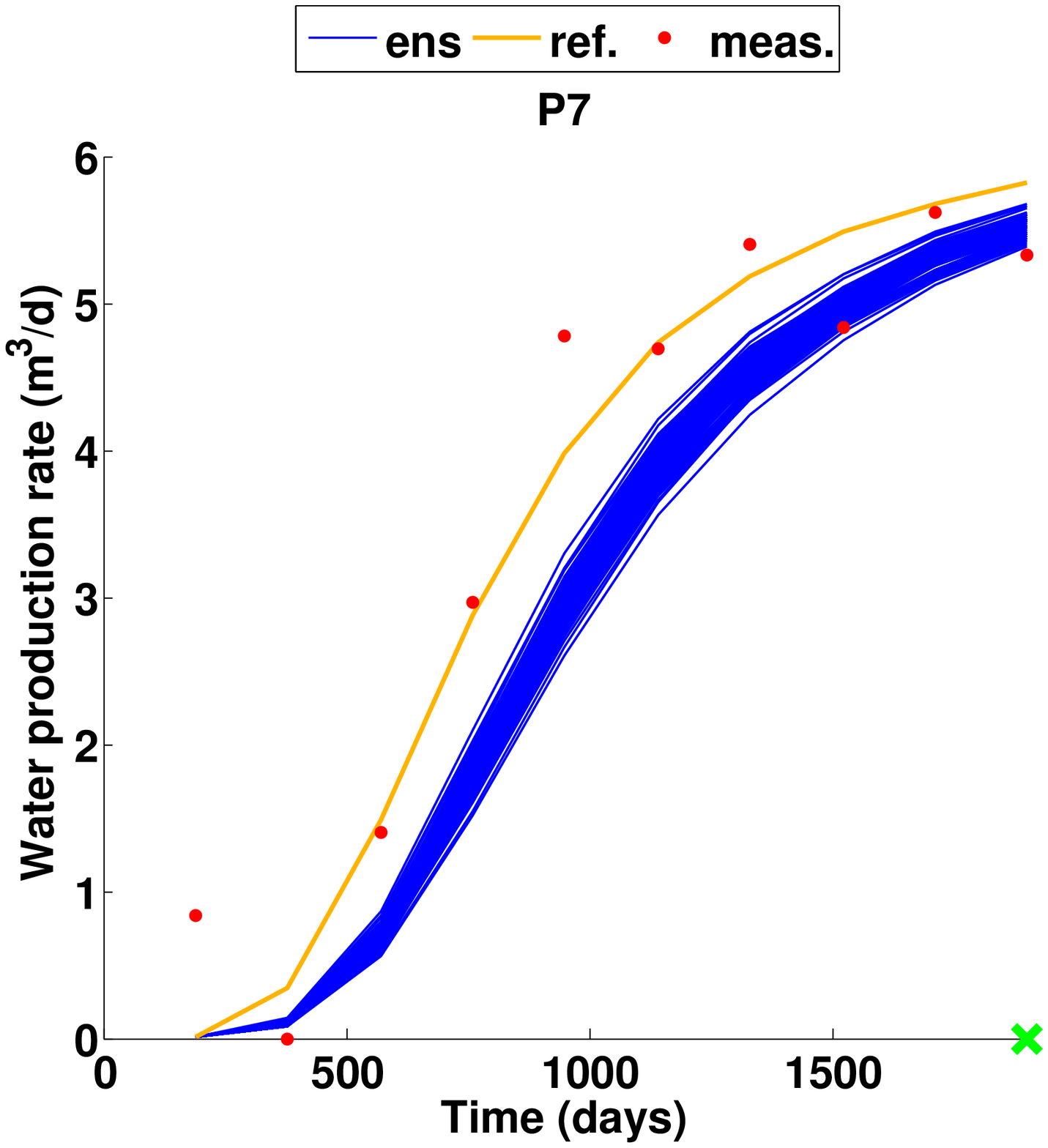}
}
\subfigure[RLM-MAC]{ \label{subfig:RLM-MAC_forecast(ES)__WWPR_P7}
\includegraphics[scale=\nScale]{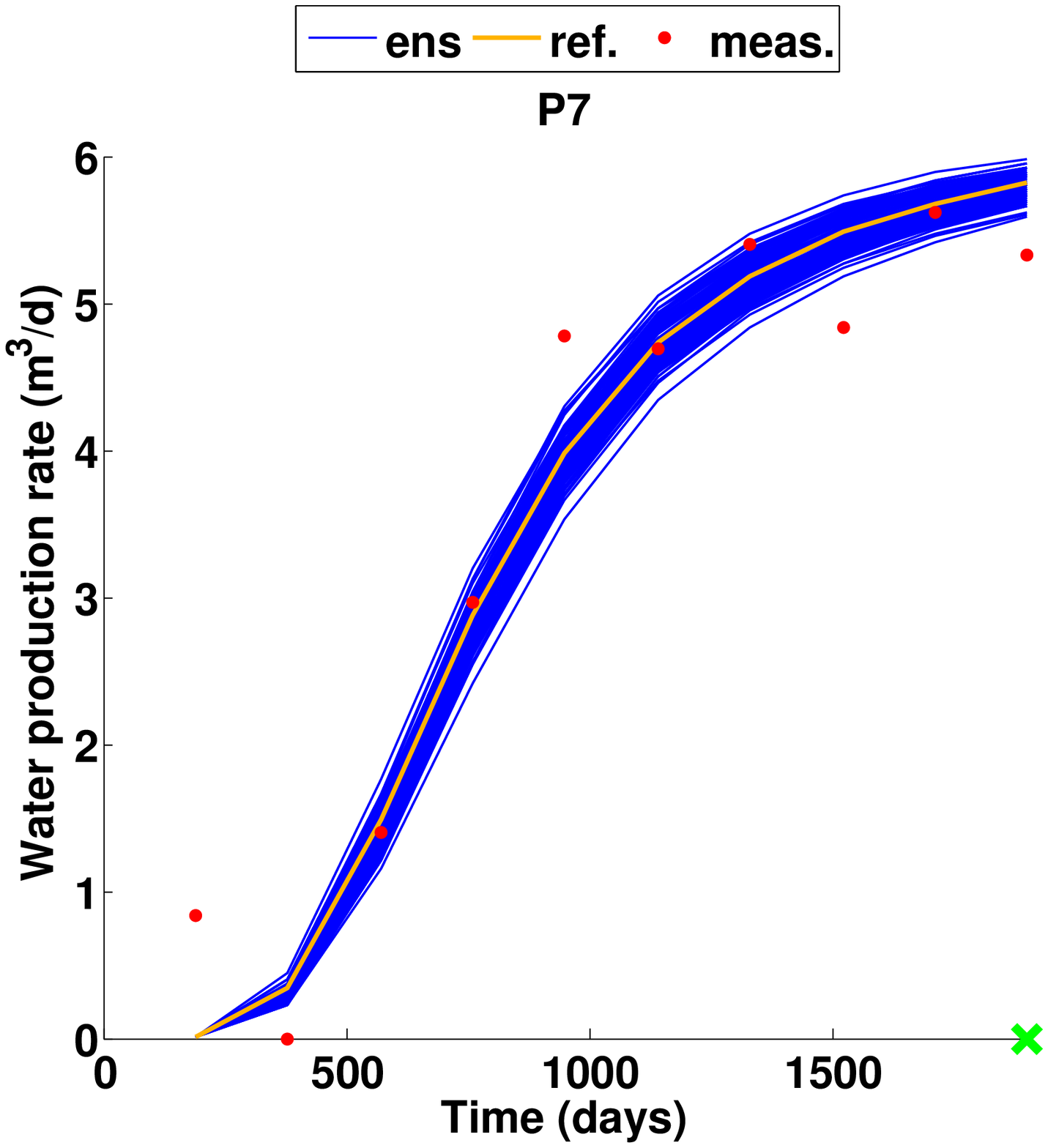}
}
\caption{\label{fig:facies_forecasts_WWPR} As in Figure \ref{fig:facies_forecasts_WOPR}, but for history matching of water production rates in the facies estimation problem.} 
\end{figure*}  

\clearpage
\renewcommand{\nScale}{0.4}
\begin{figure*}
\centering
\subfigure{ \label{subfig:rmse_aLM-EnRML_boxplot_ensemble}
\includegraphics[scale=\nScale]{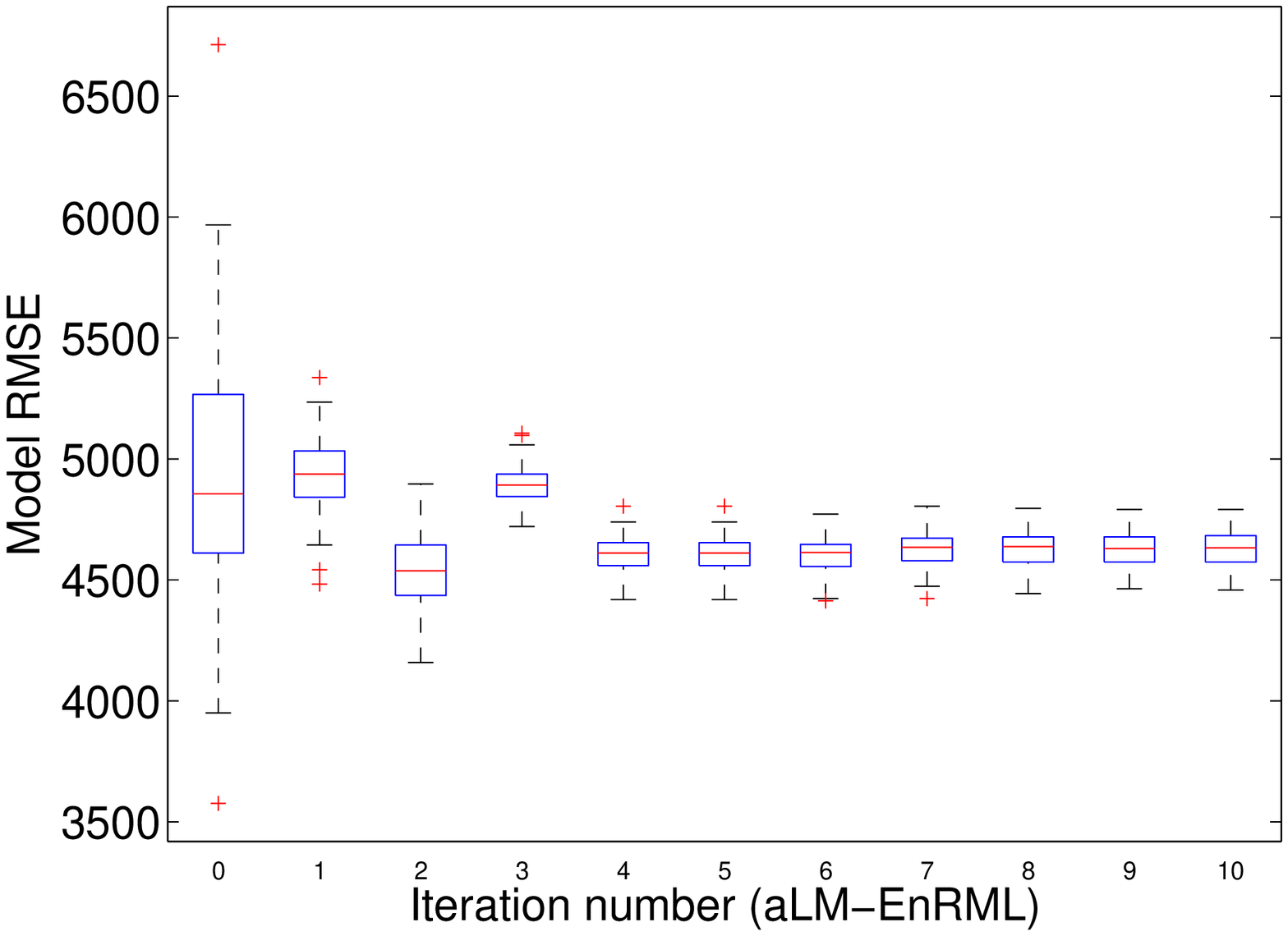}
}
\subfigure{ \label{subfig:rmse_RLM-MAC_boxplot_ensemble}
\includegraphics[scale=\nScale]{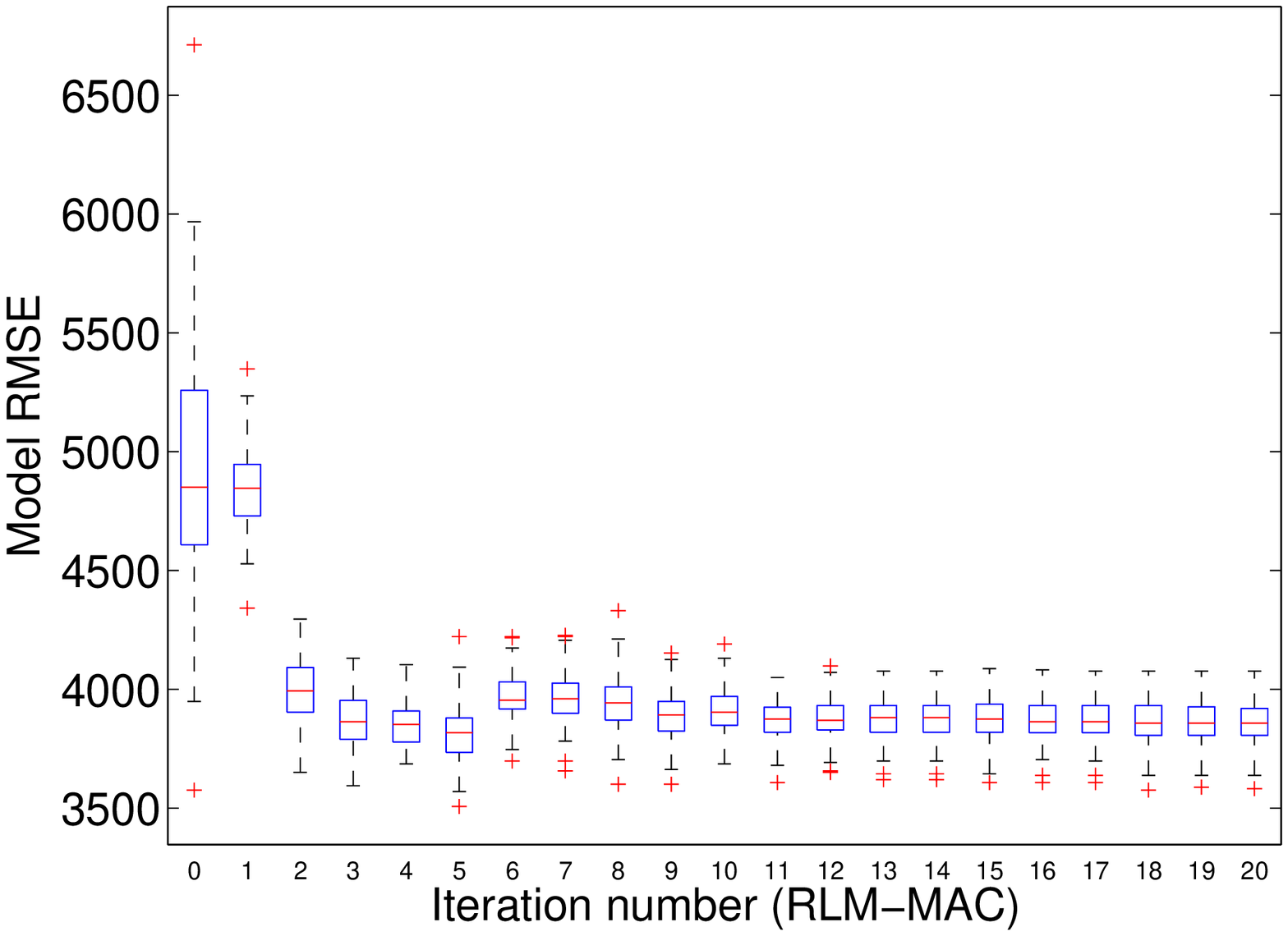}
}

\subfigure{ \label{subfig:spread_aLM-EnRML_boxplot_ensemble}
\includegraphics[scale=\nScale]{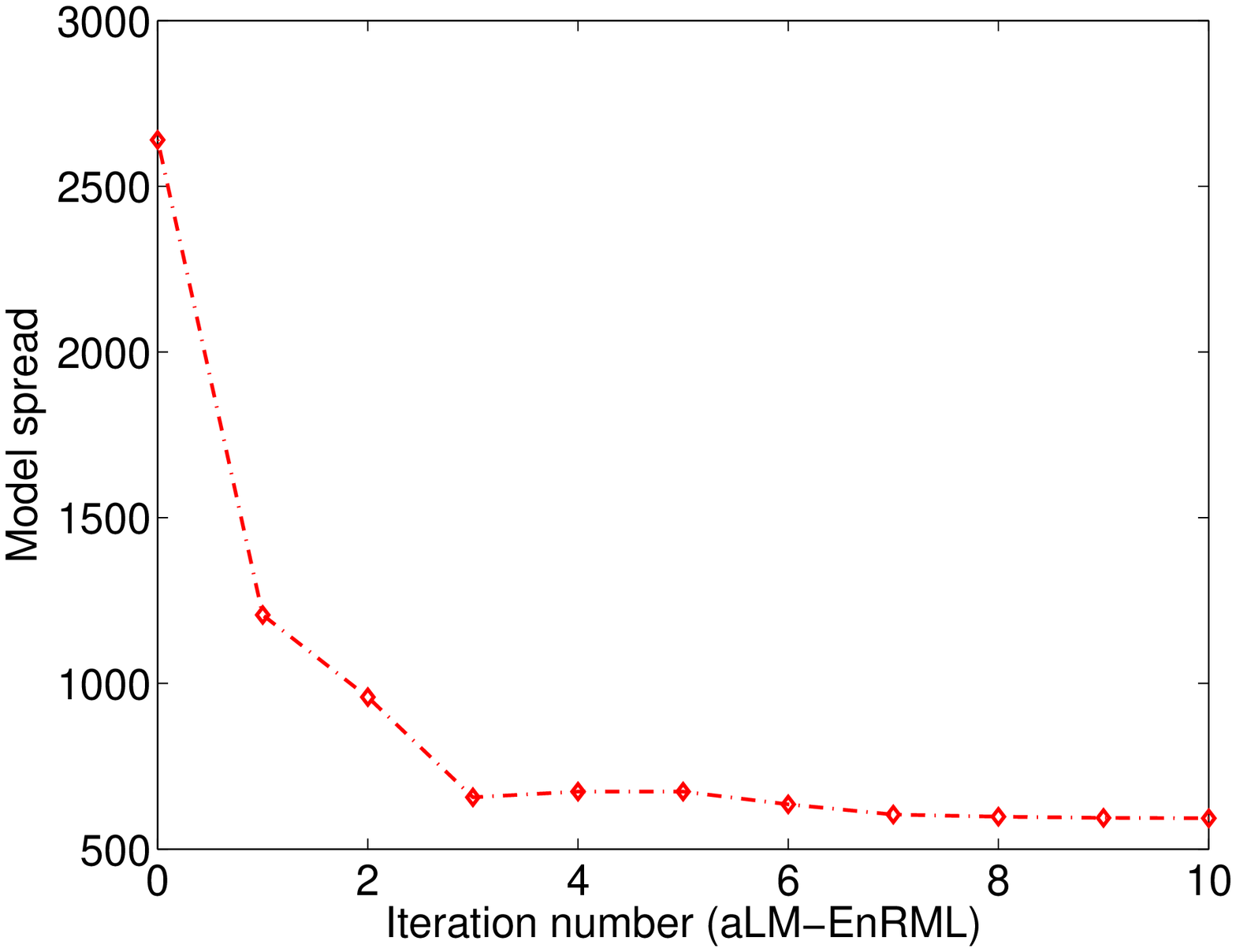}
}
\subfigure{ \label{subfig:spread_RLM-MAC_boxplot_ensemble}
\includegraphics[scale=\nScale]{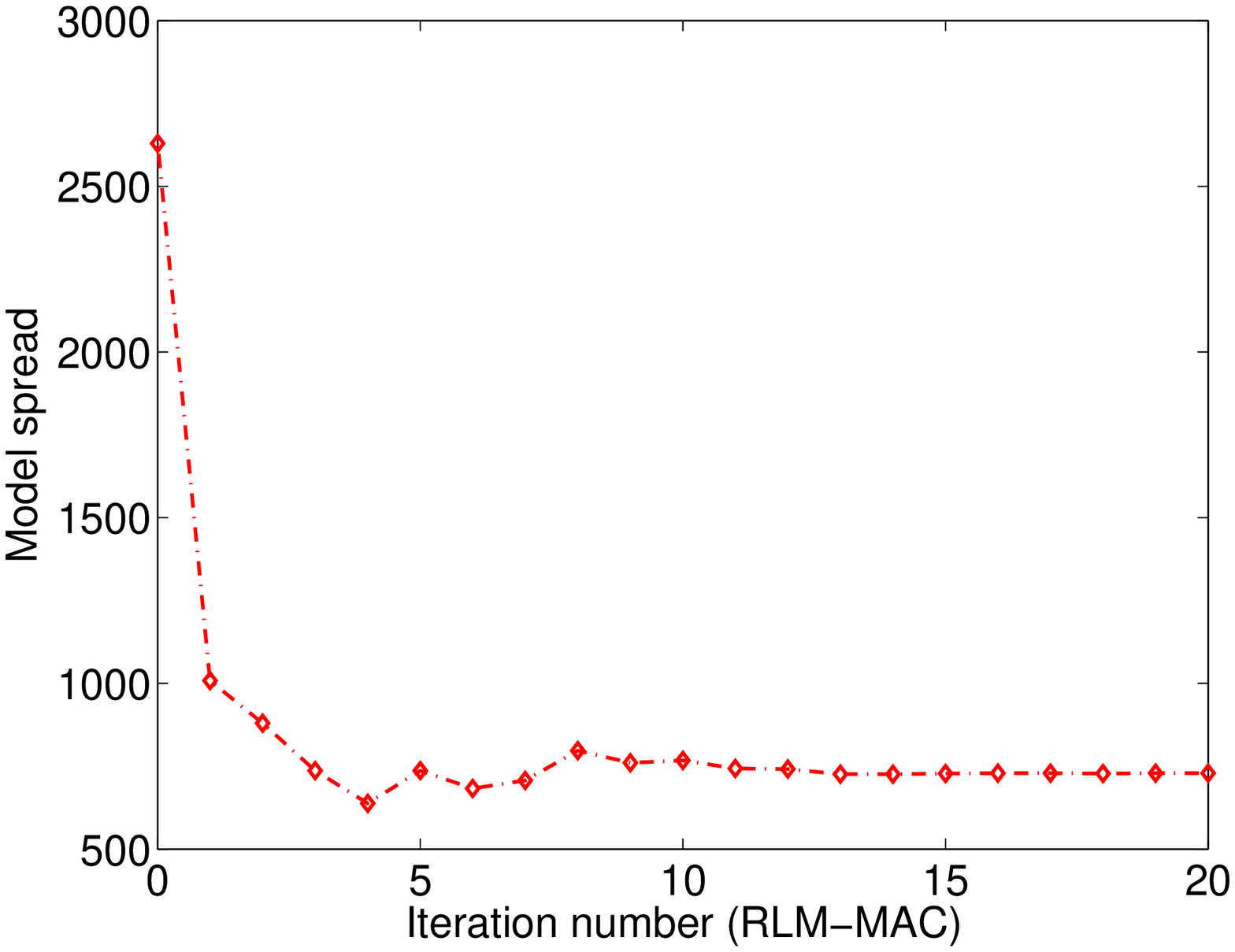}
}
\caption{\label{fig:facies_rmse_spread} Top panels: Box plots of RMSEs of the ensemble models--with respect to the reference model of the facies estimation problem--at different iteration steps in the aLM-EnRML (left) and RLM-MAC (right). Bottom panels: Ensemble spreads of the aLM-EnRML (left) and RLM-MAC (right) at different iteration steps.} 
\end{figure*}  

\clearpage
\renewcommand{\nScale}{0.6}
\begin{figure*}
\centering
\includegraphics[scale=\nScale]{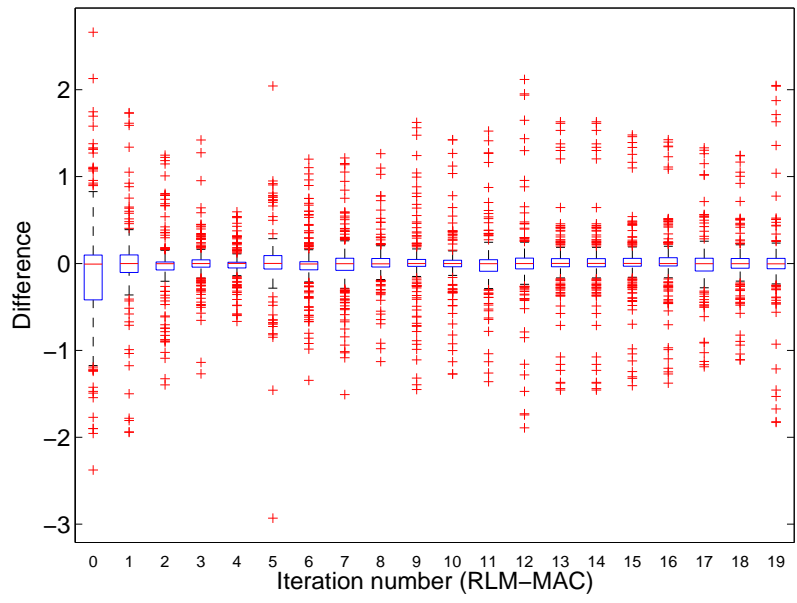}
\caption{\label{fig:facies_obsDiff_boxPlot} Box plots of the normalized differences $\mathbf{C}_d^{-1/2} \left( \mathbf{g} \left(\bar{\mathbf{m}}^{i} \right) - \overline{\mathbf{g}\left(\mathbf{m}_j^{i}\right)} \right)$ at different iteration steps of the facies estimation problem.} 
\end{figure*}  

\clearpage
\renewcommand{\nScale}{0.5}
\begin{figure*}
	\centering
	\subfigure[Initial ensemble]{ \label{subfig:field_PERMX_1_1_brSmallEnsemble_all_active_init_layer1}
		\includegraphics[scale=\nScale]{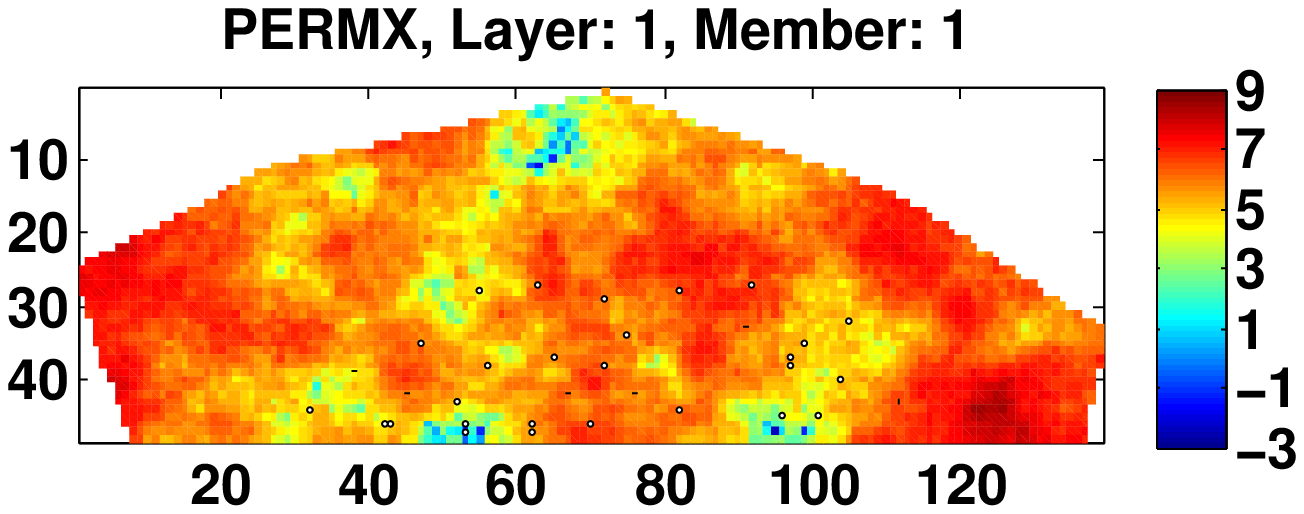}
	}
	\subfigure[Initial ensemble]{ \label{subfig:field_PERMX_1_1_brSmallEnsemble_all_active_init_layer2}
		\includegraphics[scale=\nScale]{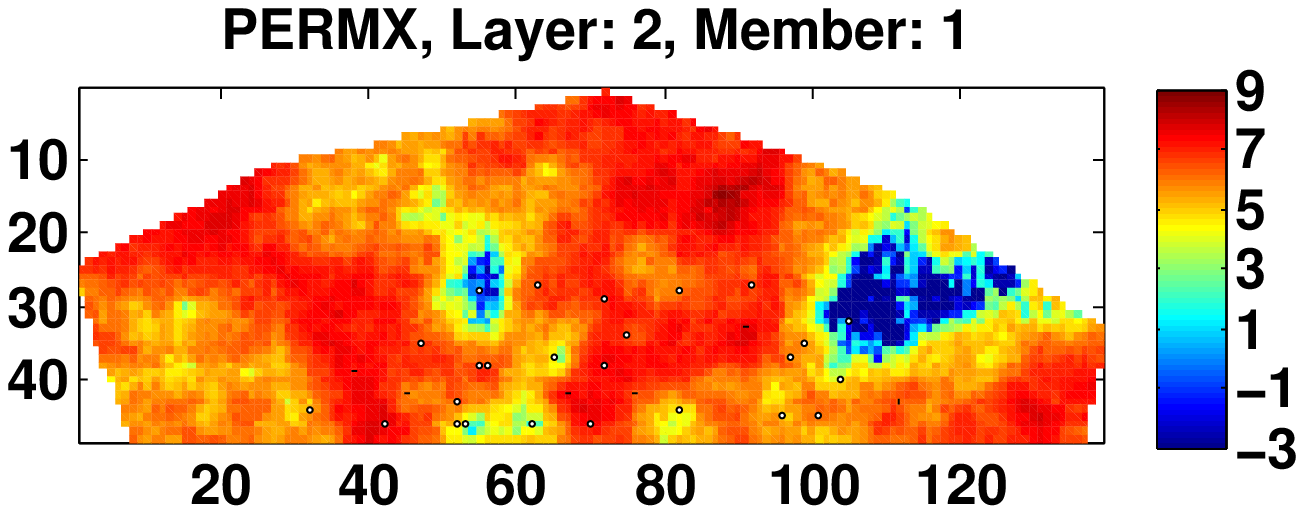}
	}
	
	\subfigure[aLM-EnRML]{ \label{subfig:field_PERMX_1_1_ensemble15_aLMEnRML_layer1}
		\includegraphics[scale=\nScale]{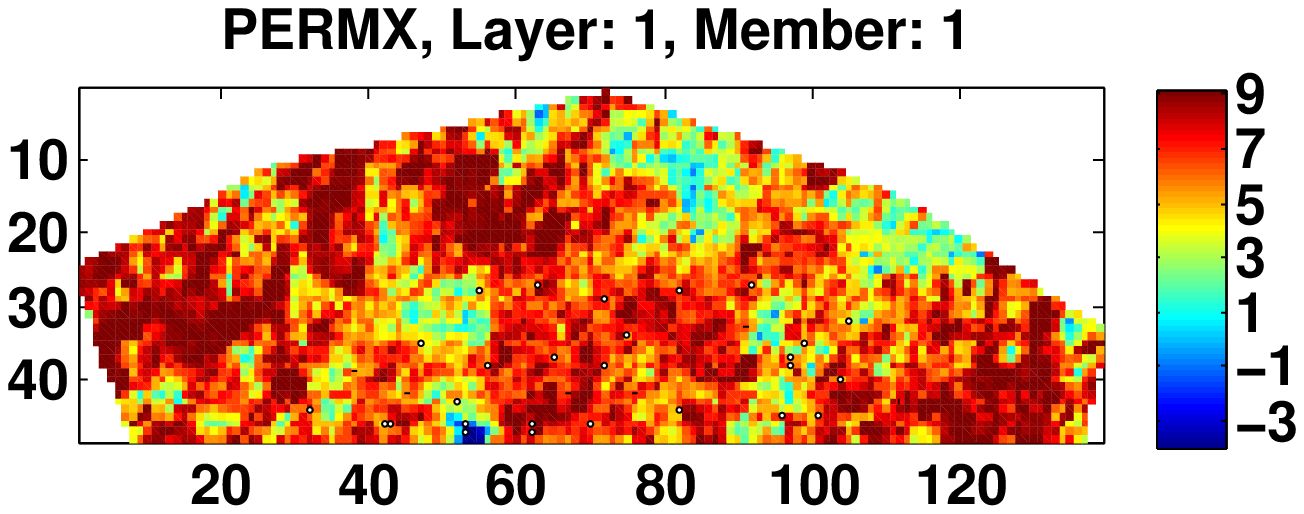}
	}
	\subfigure[aLM-EnRML]{ \label{subfig:field_PERMX_1_1_ensemble15_aLMEnRML_layer2}
		\includegraphics[scale=\nScale]{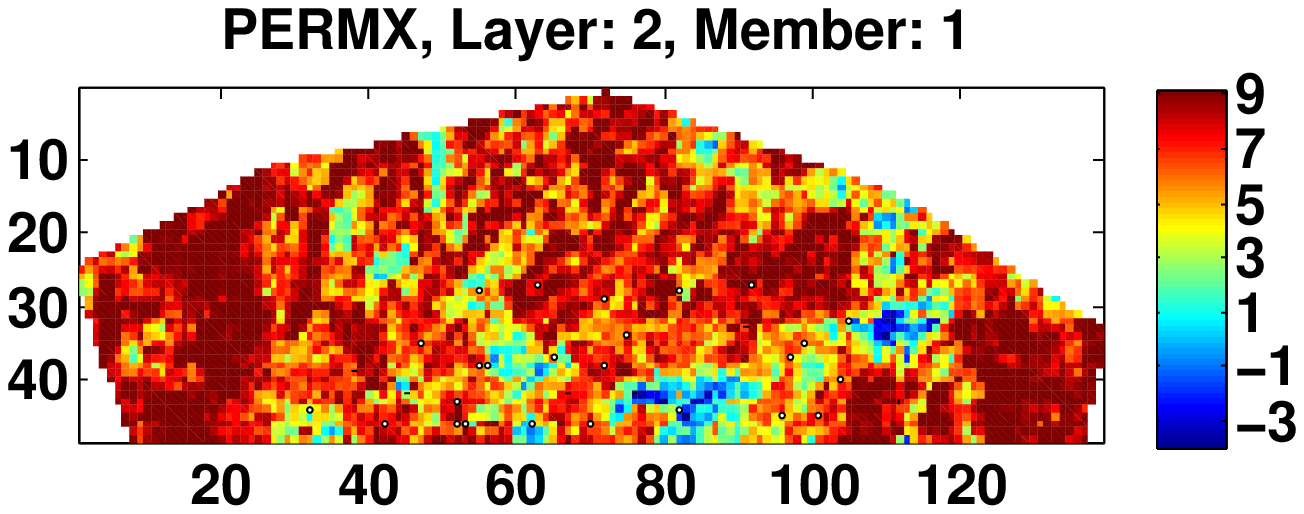}
	}
	
	\subfigure[RLM-MAC]{ \label{subfig:field_PERMX_1_1_ensemble20_RLMMAC_layer1}
		\includegraphics[scale=\nScale]{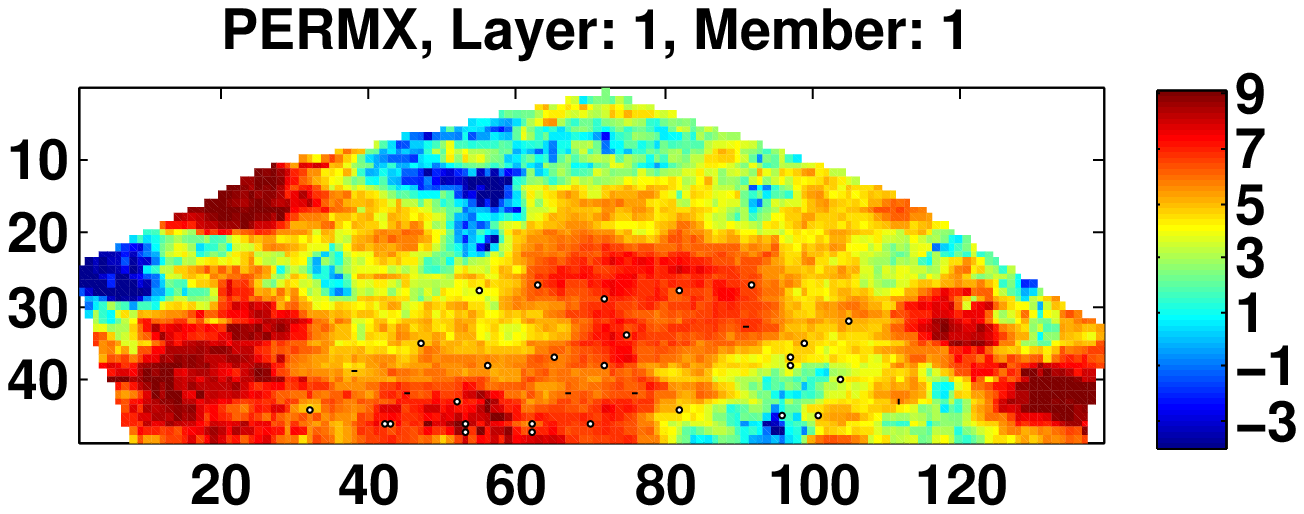}
	}
	\subfigure[RLM-MAC]{ \label{subfig:field_PERMX_1_1_ensemble20_RLMMAC_layer2}
		\includegraphics[scale=\nScale]{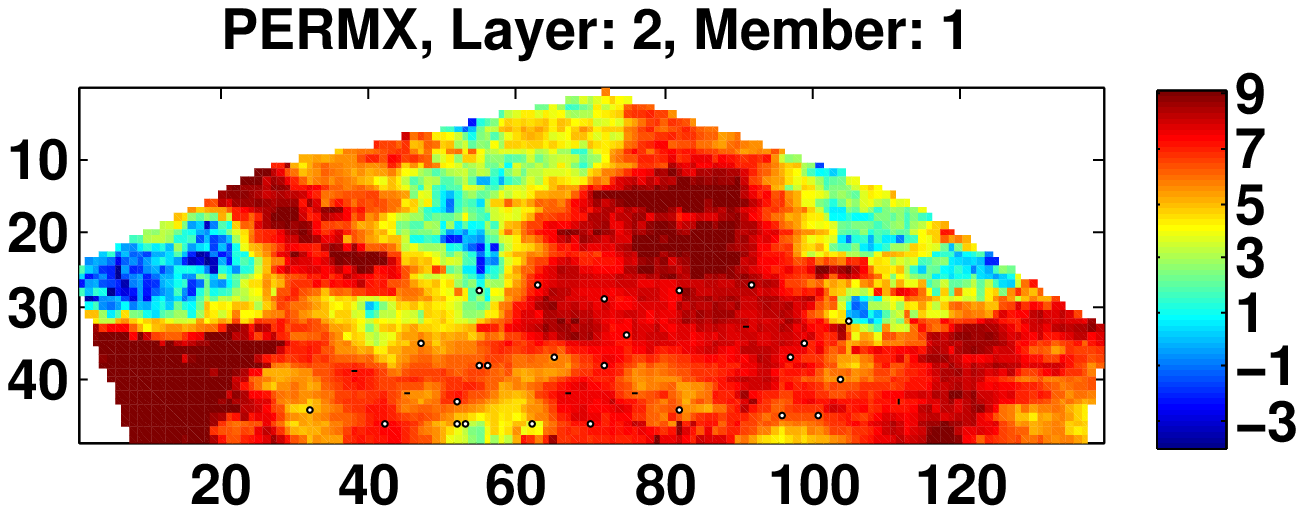}
	}
	\caption{\label{fig:brugge_permx_layers12}  Distributions of PERMX on layers 1 and 2 from a reservoir model of the initial ensemble (a,b), the corresponding reservoir model updated by the aLM-EnRML (c,d), and the corresponding reservoir model updated by the RLM-MAC (e,f). The dots in the figures indicate the locations of the producers and injectors in the Brugge field.} 
\end{figure*}  

\clearpage
\renewcommand{\nScale}{0.4}
\begin{figure*}
	\centering
	\subfigure[aLM-EnRML]{ \label{subfig:Brugge_aLM-EnRML_boxplot_objRealIter}
		\includegraphics[scale=\nScale]{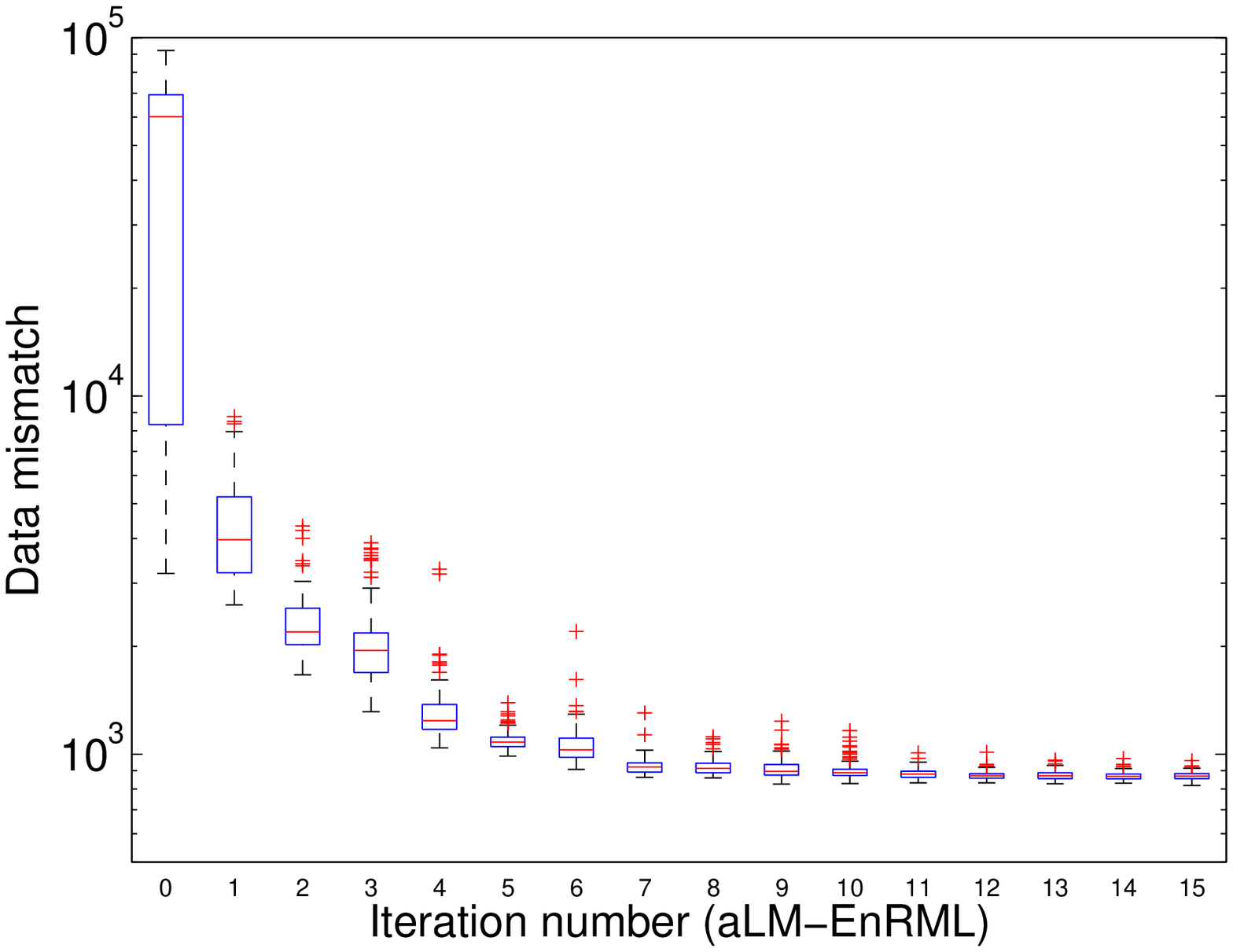}
	}	
	\subfigure[RLM-MAC]{ \label{subfig:Brugge_RLM-MAC_boxplot_objRealIter}
		\includegraphics[scale=\nScale]{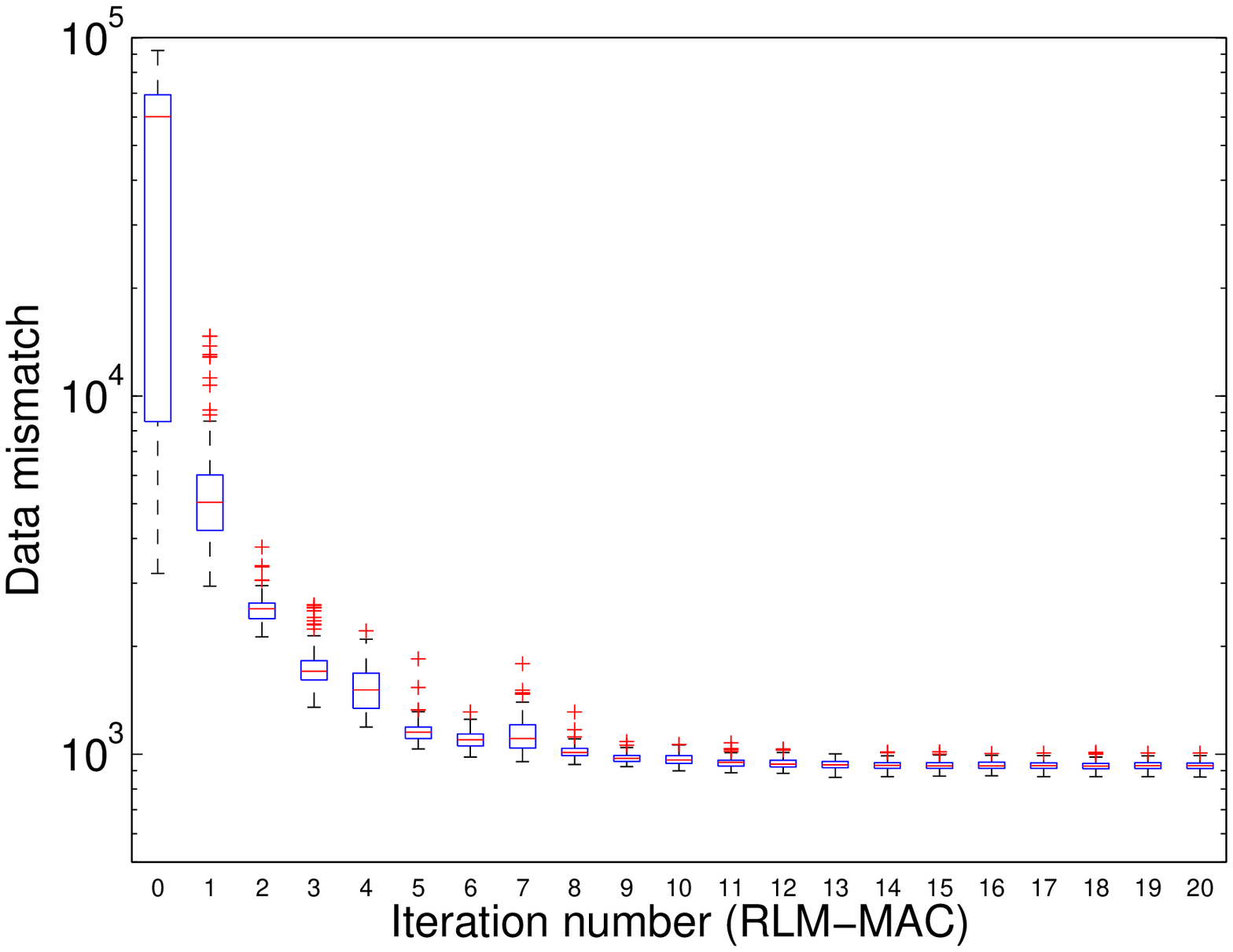}
	}
	\caption{\label{fig:brugge_data_mismatch_hm} Box plots of data mismatch of the aLM-EnRML (left) and the RLM-MAC (right) at different iteration steps in the Brugge field case, using production data at 20 out of 127 time instants in the first 10 years.} 
\end{figure*}  
  
\clearpage
\renewcommand{\nScale}{0.3}
\begin{figure*}
	\centering
	\subfigure[]{ \label{subfig:Brugge_WOPR_BR-P-9_init}
		\includegraphics[scale=\nScale]{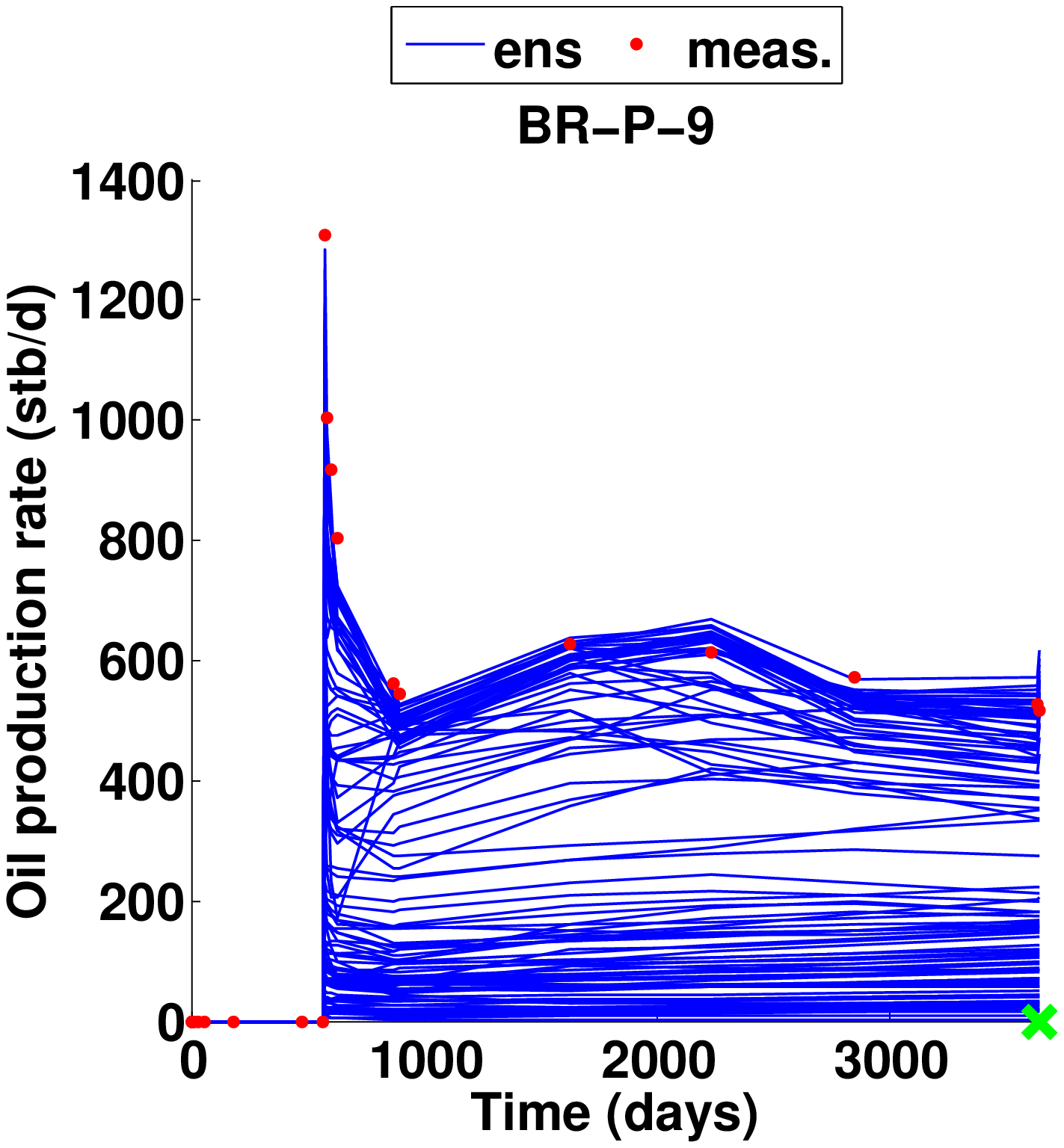}
	}
	\subfigure[]{ \label{subfig:Brugge_WOPR_BR-P-9_aLMEnRML}
		\includegraphics[scale=\nScale]{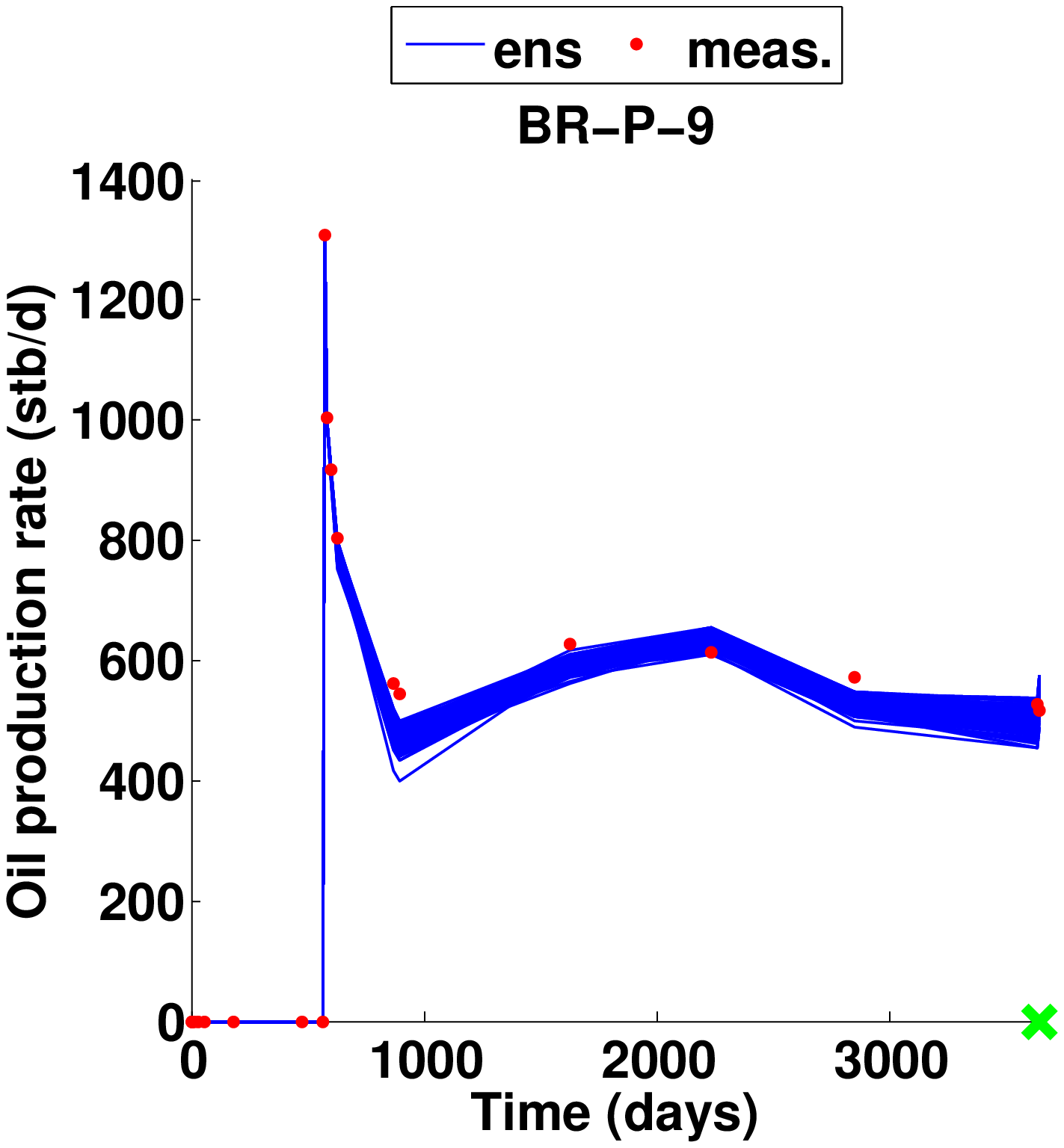}
	}
	\subfigure[]{ \label{subfig:Brugge_WOPR_BR-P-9_RLMMAC}
		\includegraphics[scale=\nScale]{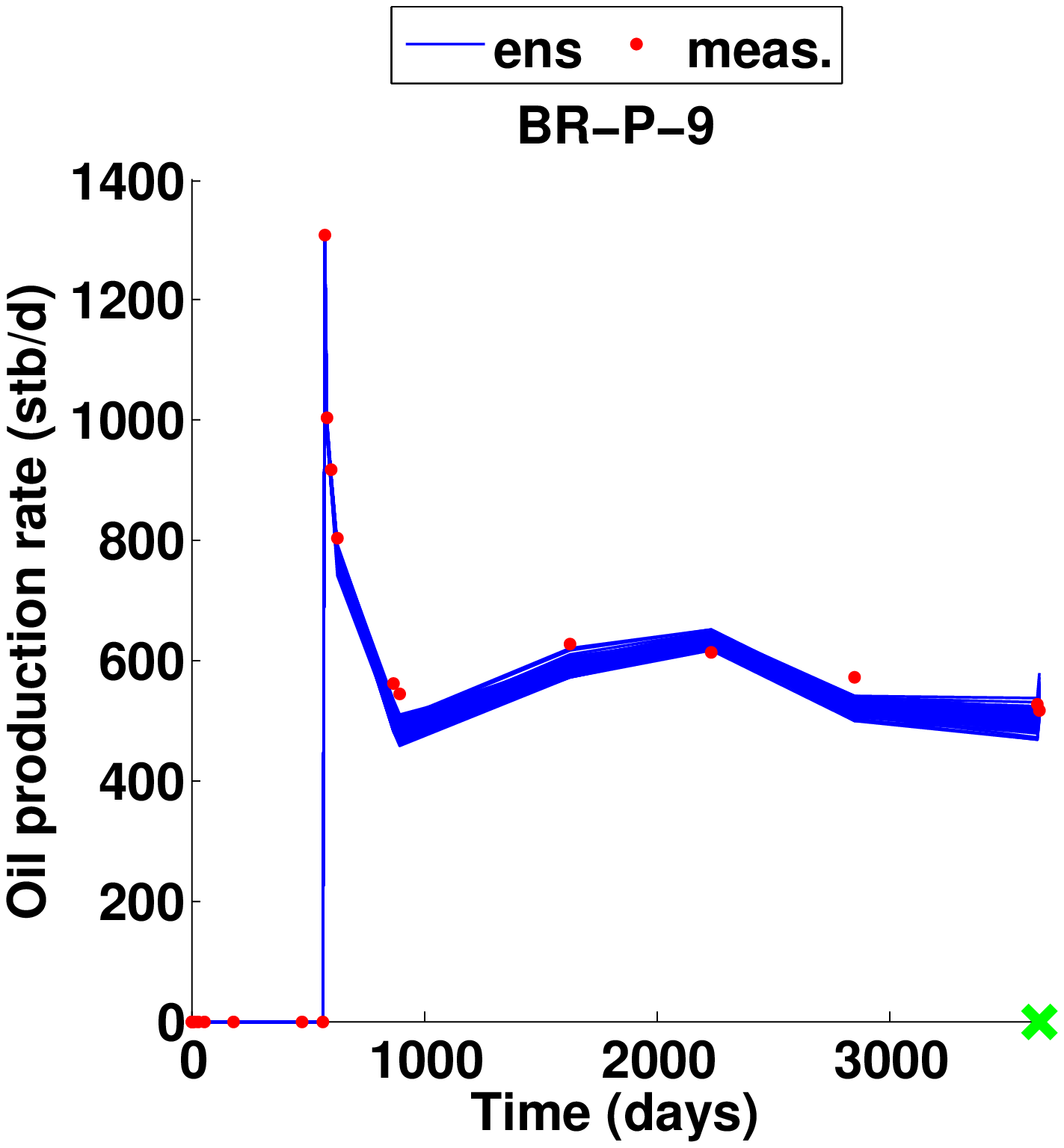}
	}
	
	\subfigure[]{ \label{subfig:Brugge_WOPR_BR-P-13_init}
			\includegraphics[scale=\nScale]{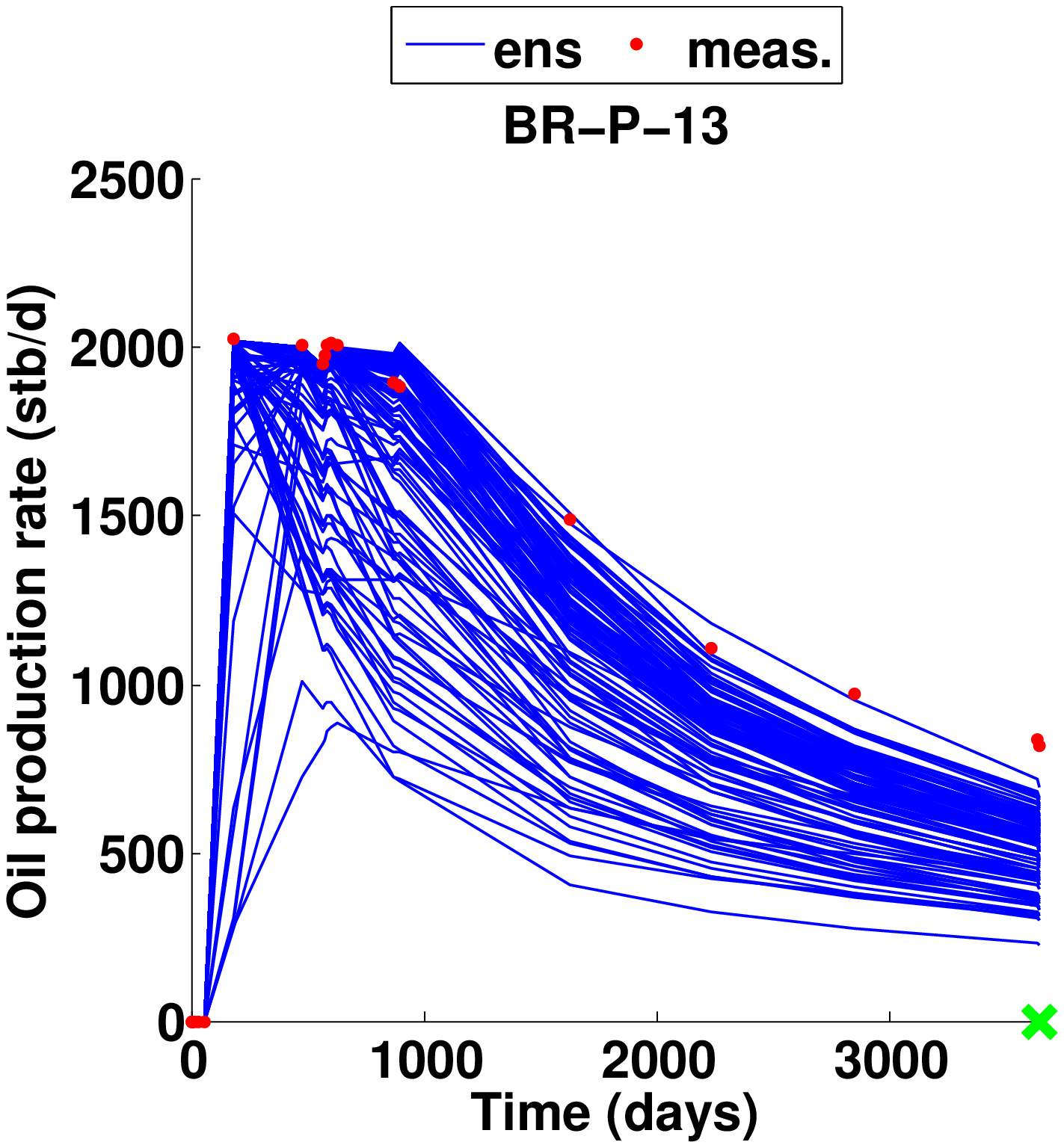}
	}
	\subfigure[]{ \label{subfig:Brugge_WOPR_BR-P-13_aLMEnRML}
			\includegraphics[scale=\nScale]{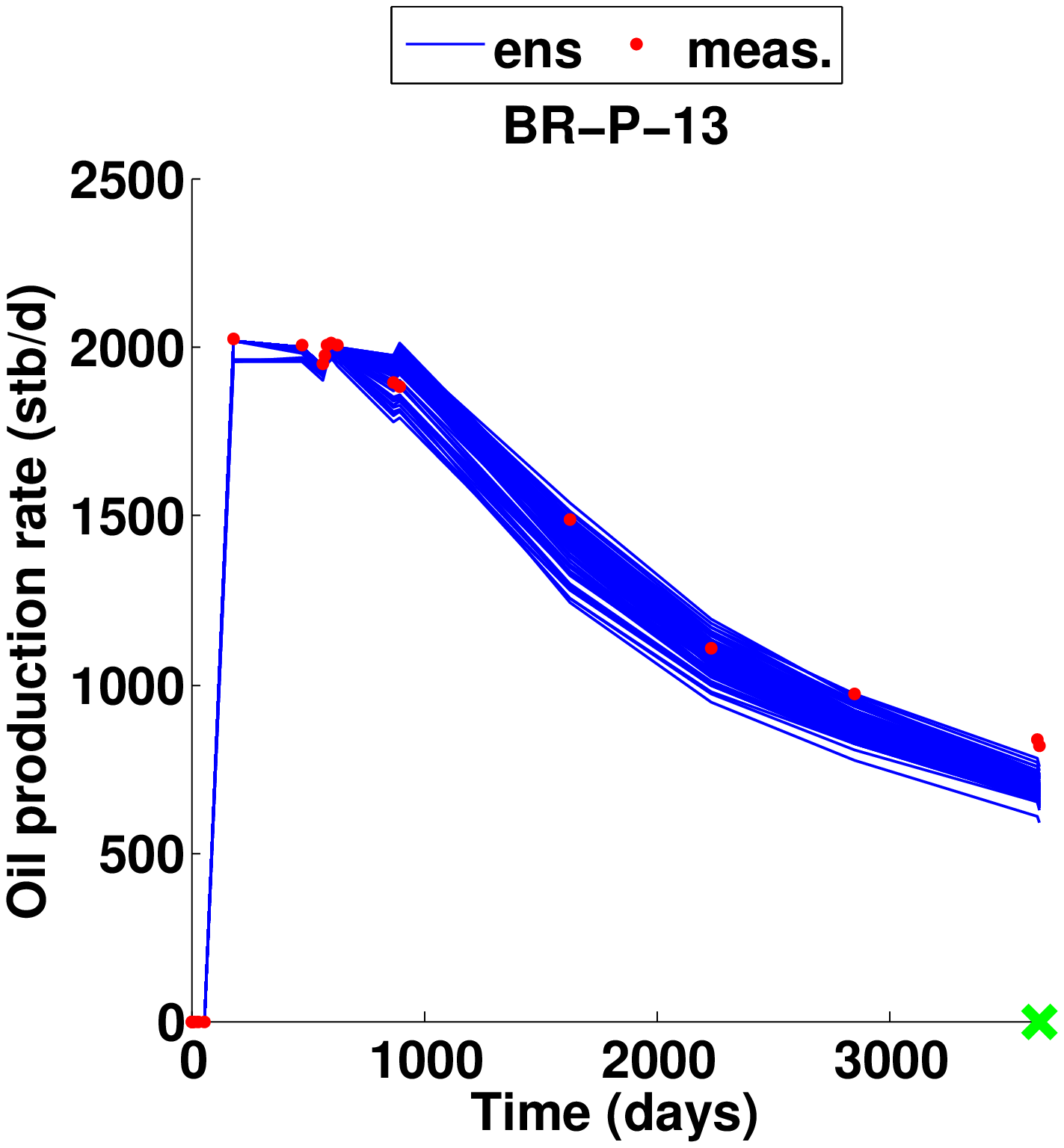}
	}
	\subfigure[]{ \label{subfig:Brugge_WOPR_BR-P-13_RLMMAC}
			\includegraphics[scale=\nScale]{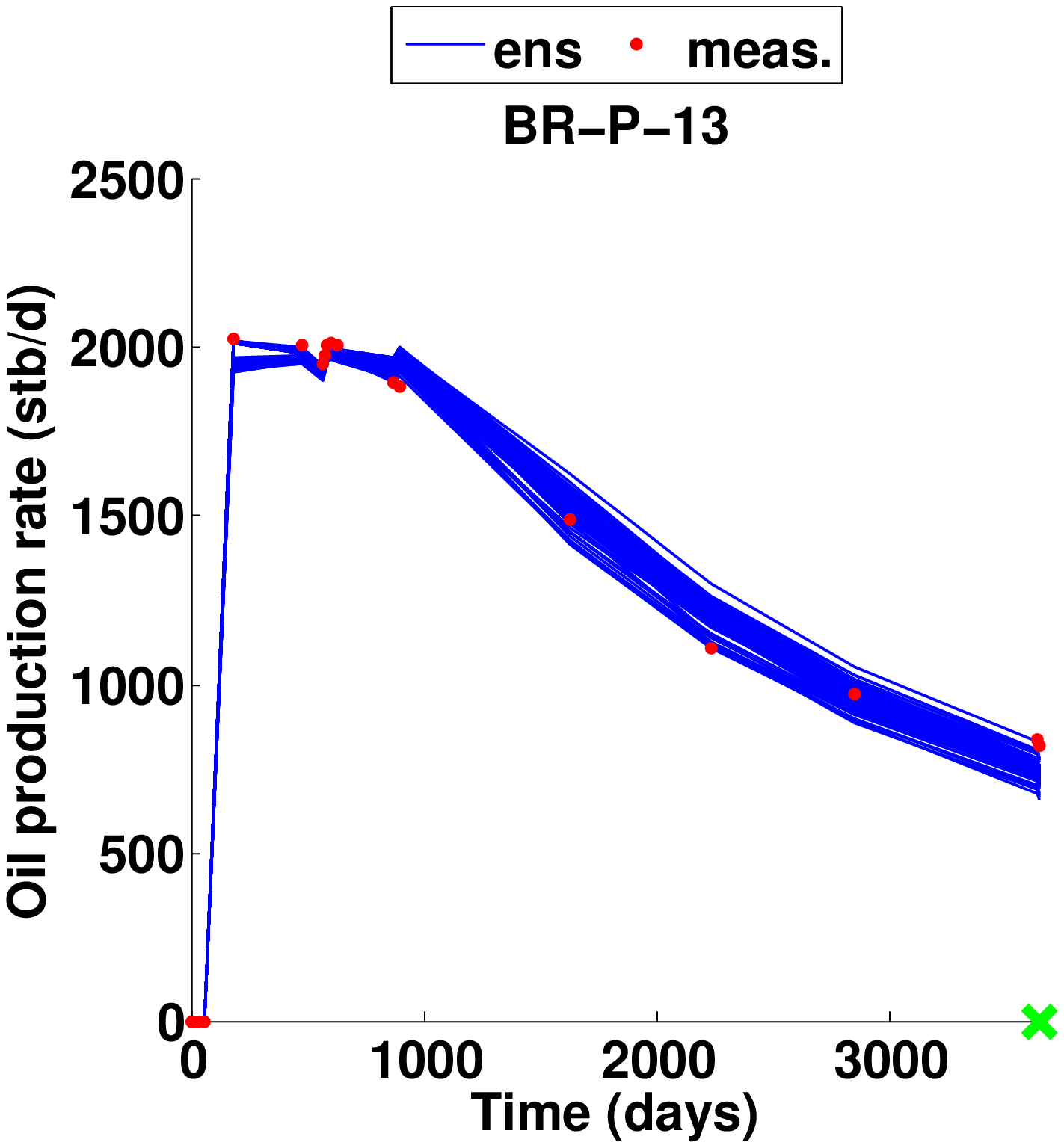}
	}
	
	\subfigure[Initial ensemble]{ \label{subfig:Brugge_WOPR_BR-P-19_init}
		\includegraphics[scale=\nScale]{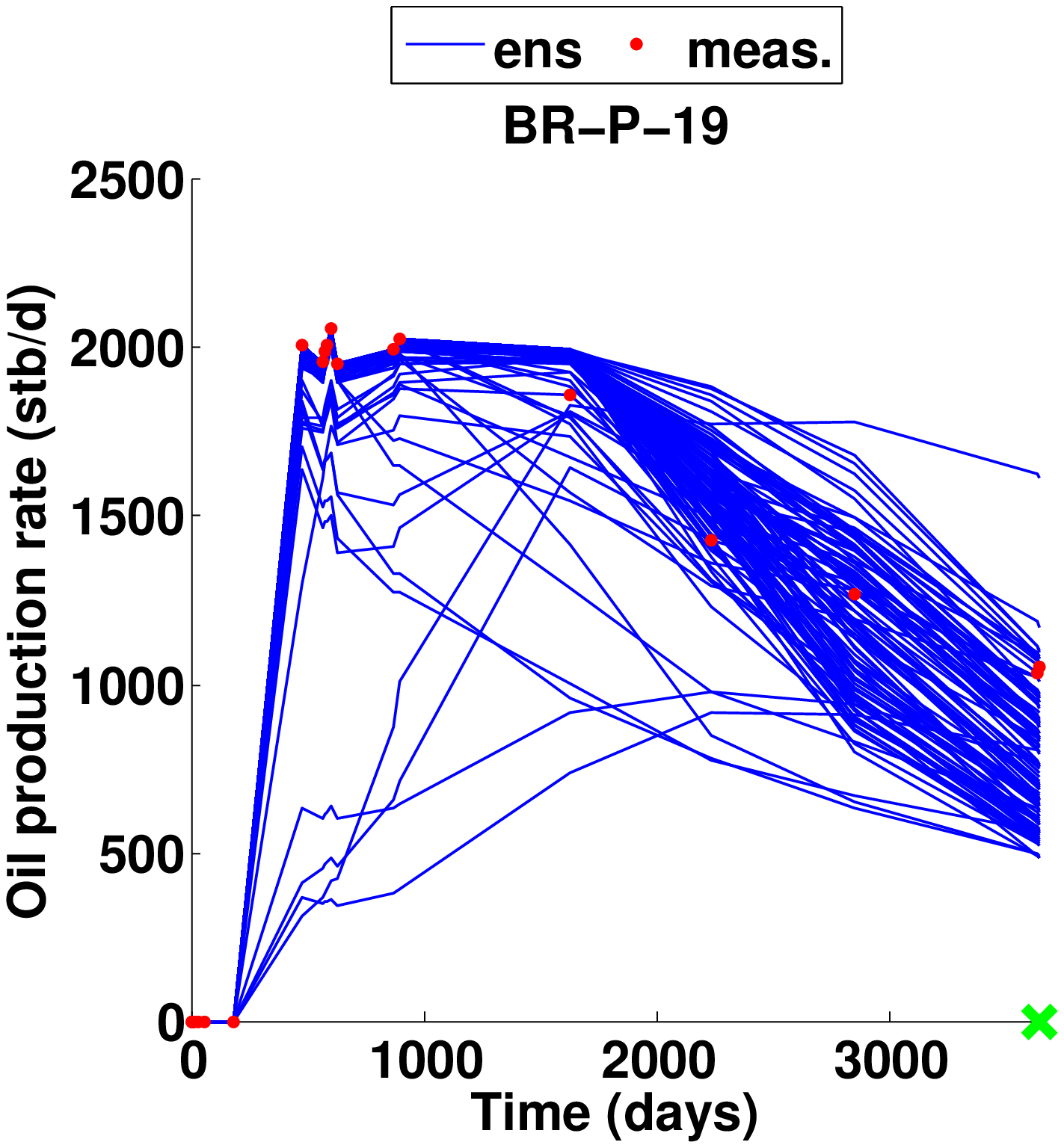}
	}
	\subfigure[aLm-EnRML]{ \label{subfig:Brugge_WOPR_BR-P-19_aLMEnRML}
		\includegraphics[scale=\nScale]{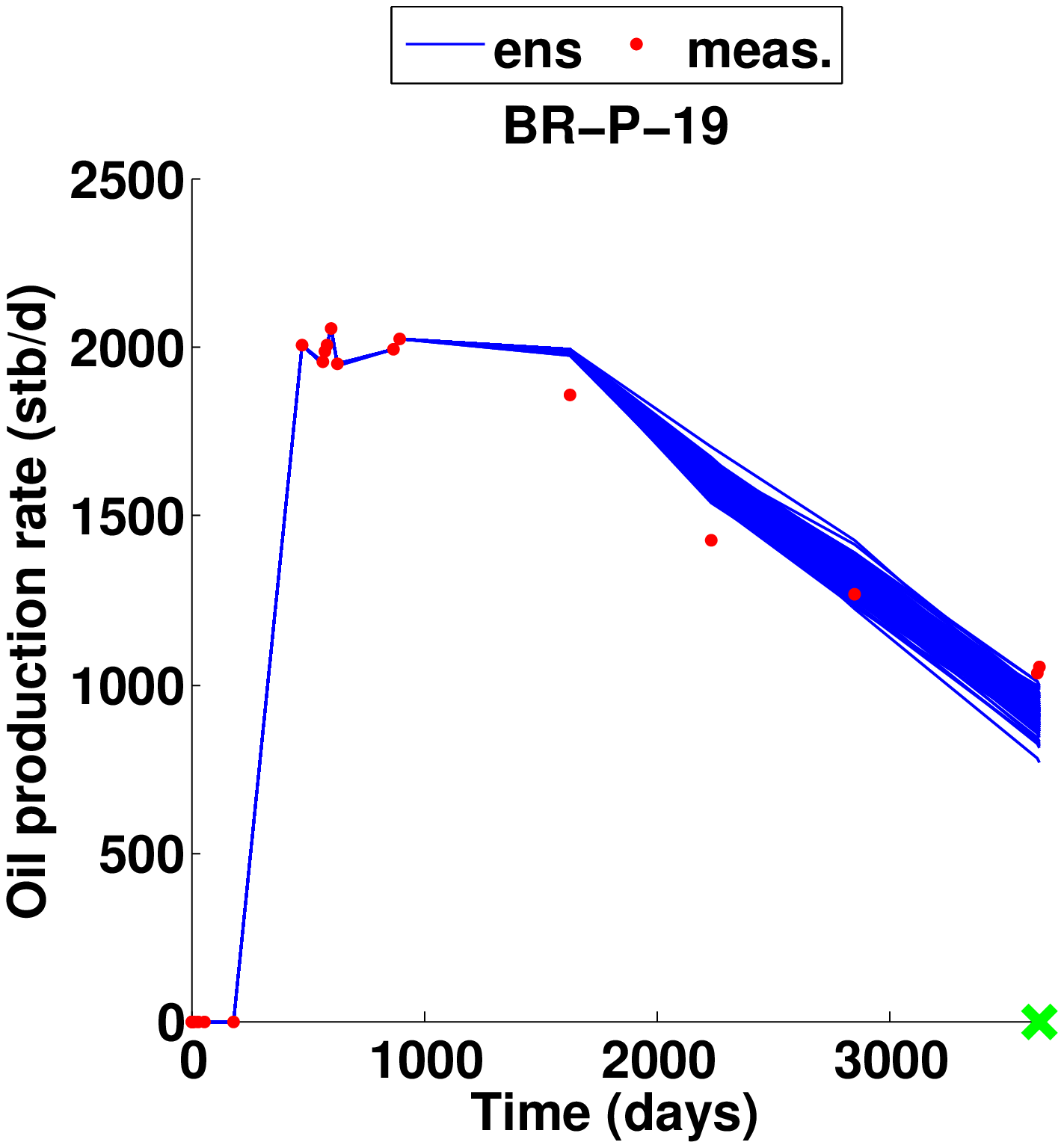}
	}
	\subfigure[RLM-MAC]{ \label{subfig:Brugge_WOPR_BR-P-19_RLMMAC}
		\includegraphics[scale=\nScale]{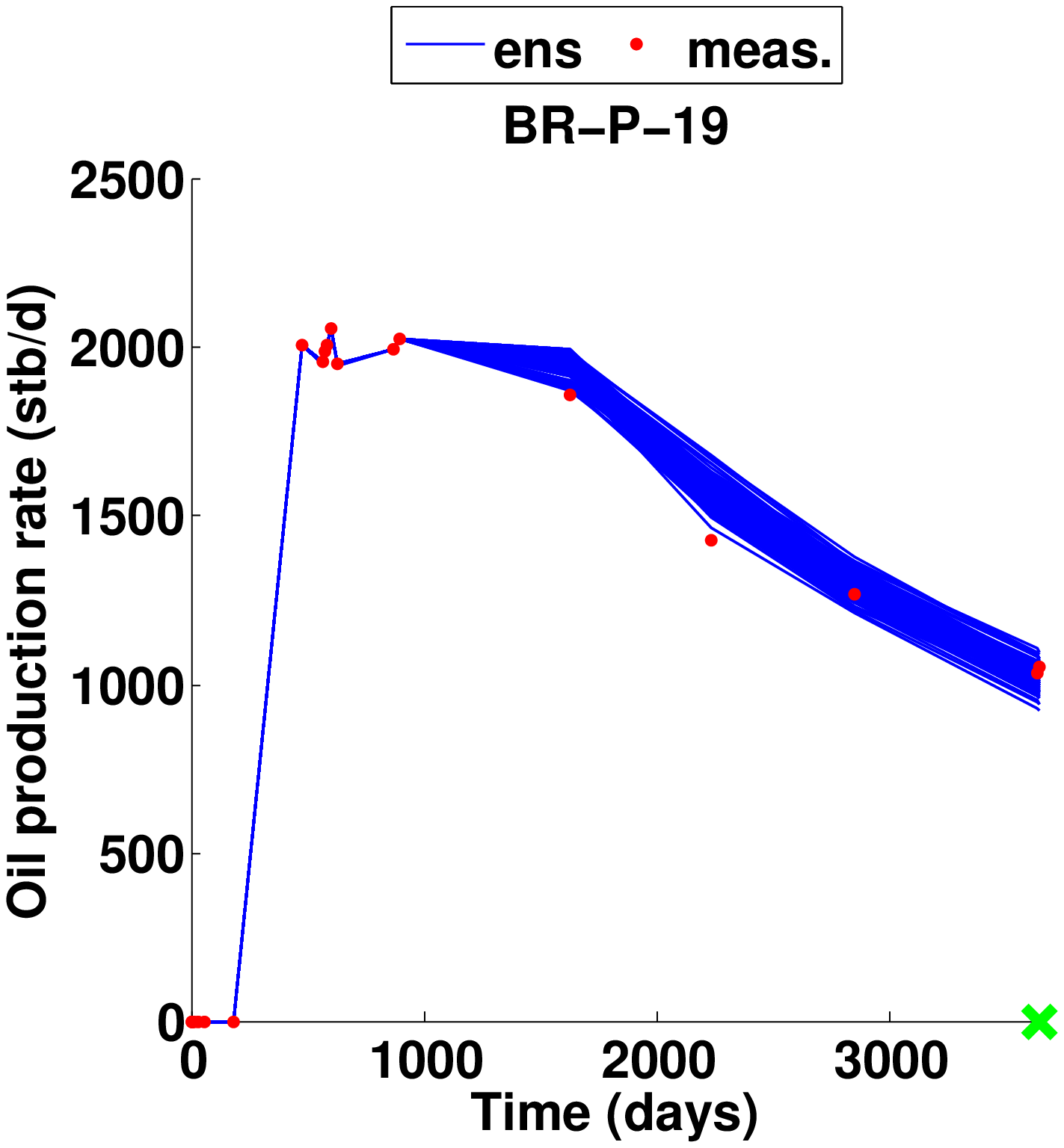}
	}
	\caption{\label{fig:brugge_wopr_hm}  History matching of oil production rates (WOPR) at BR-P-9 (top), BR-P-13 (middle) and BR-P-19 (bottom) in the Brugge field case, using the initial ensemble (1st column) and the ensembles of the aLM-EnRML (2nd column) and RLM-MAC (3rd column) at the final iteration steps. In each image, the red dots represent the historical WOPRH data and the blue curves are the forecasts with respect to the initial ensemble (1st column), and history matching profiles (2nd and 3rd columns).} 
\end{figure*}  

\clearpage
\renewcommand{\nScale}{0.3}
\begin{figure*}
	\centering
	\subfigure[]{ \label{subfig:Brugge_WWCT_BR-P-9_init}
		\includegraphics[scale=\nScale]{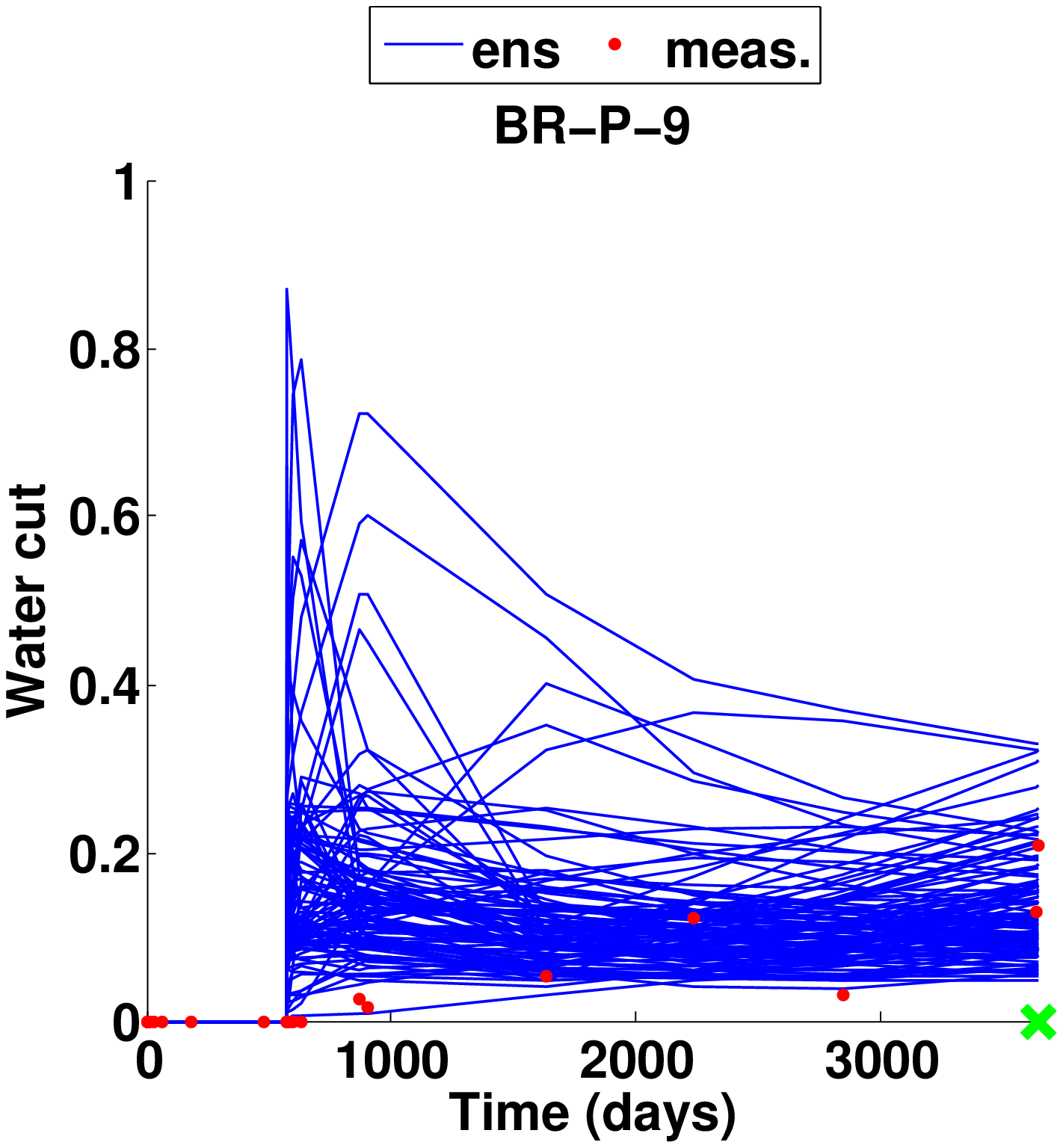}
	}
	\subfigure[]{ \label{subfig:Brugge_WWCT_BR-P-9_aLMEnRML}
		\includegraphics[scale=\nScale]{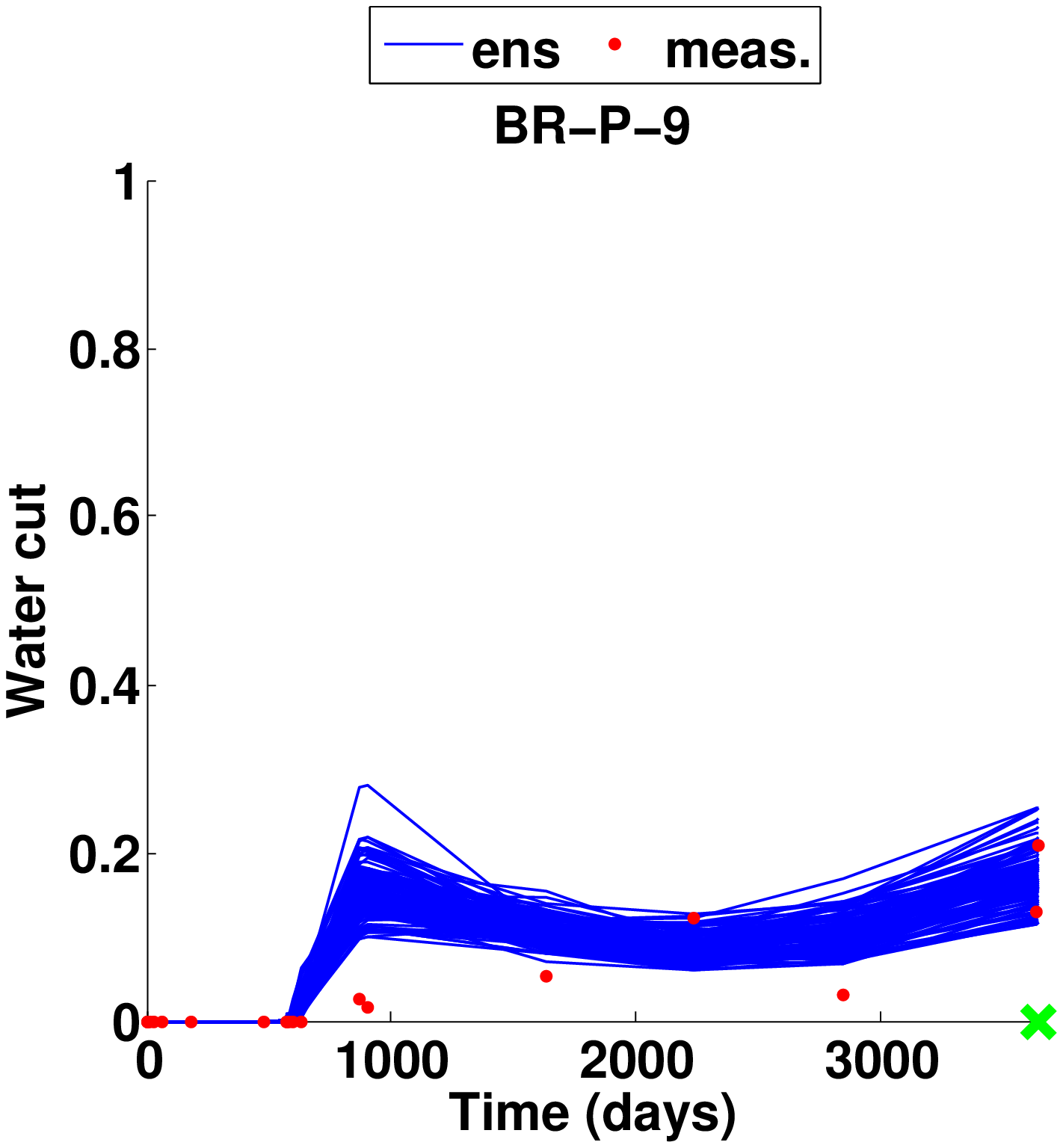}
	}
	\subfigure[]{ \label{subfig:Brugge_WWCT_BR-P-9_RLMMAC}
		\includegraphics[scale=\nScale]{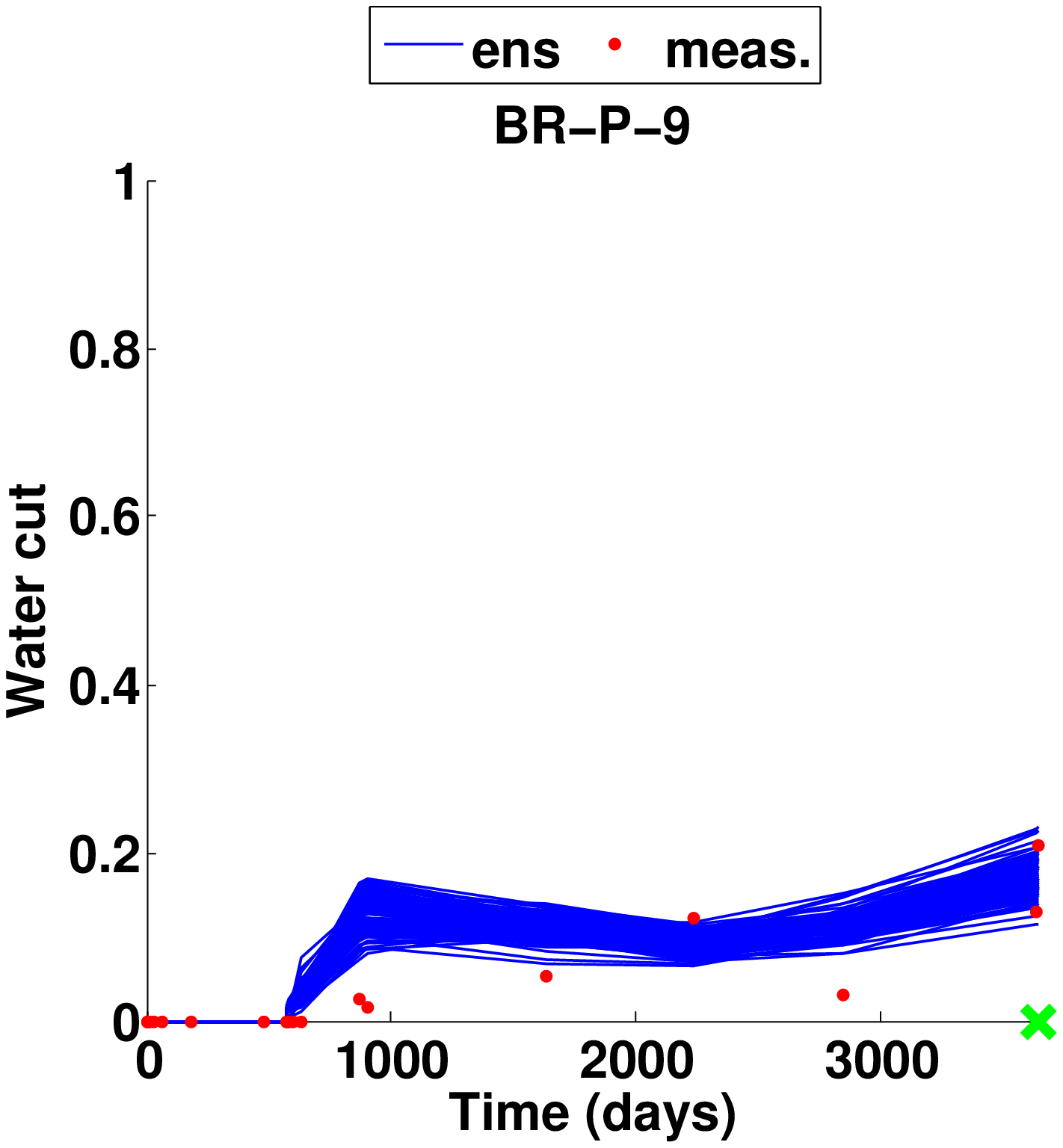}
	}

	\subfigure[]{ \label{subfig:Brugge_WWCT_BR-P-13_init}
		\includegraphics[scale=\nScale]{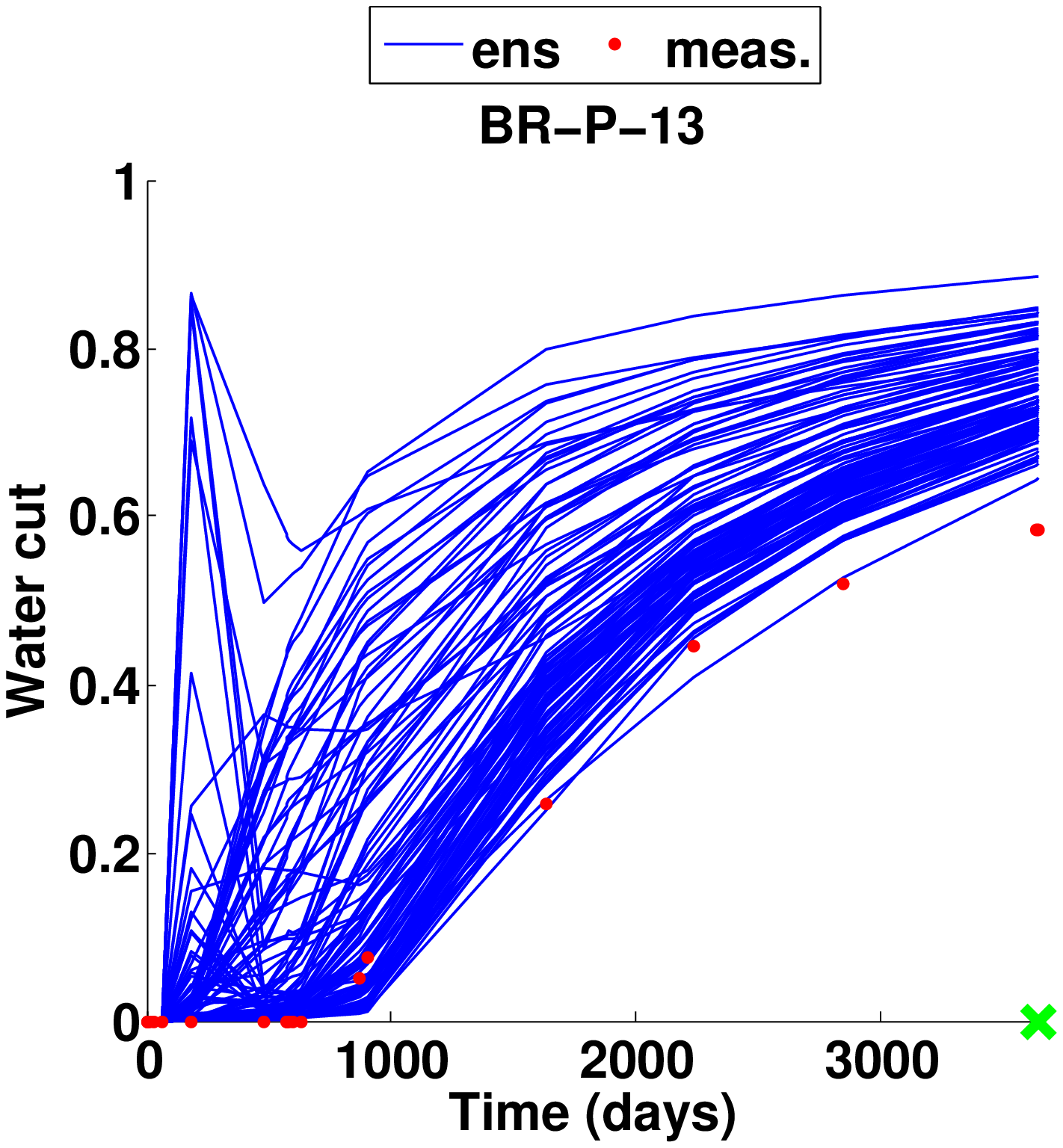}
	}
	\subfigure[]{ \label{subfig:Brugge_WWCT_BR-P-13_aLMEnRML}
		\includegraphics[scale=\nScale]{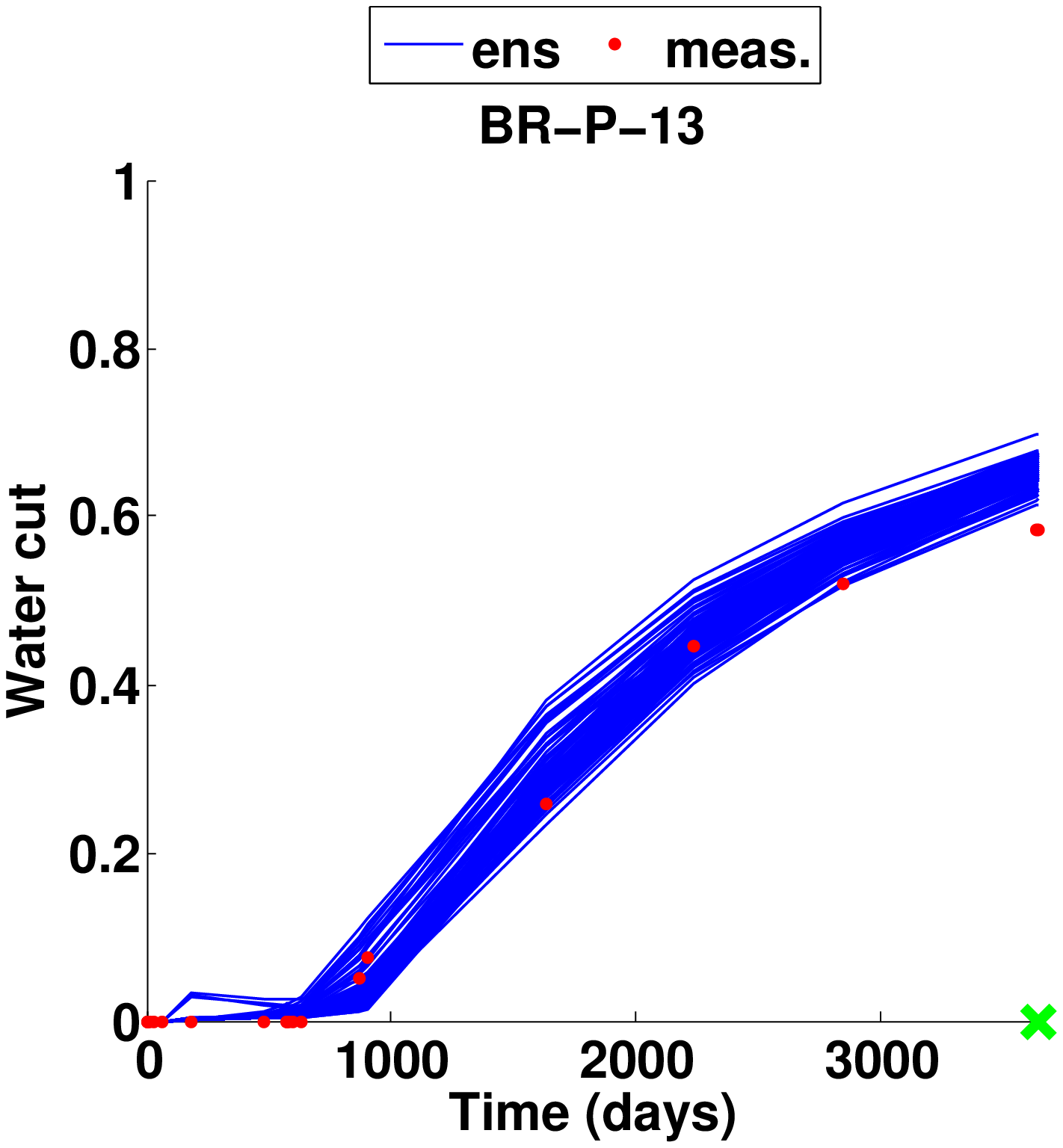}
	}
	\subfigure[]{ \label{subfig:Brugge_WWCT_BR-P-13_RLMMAC}
		\includegraphics[scale=\nScale]{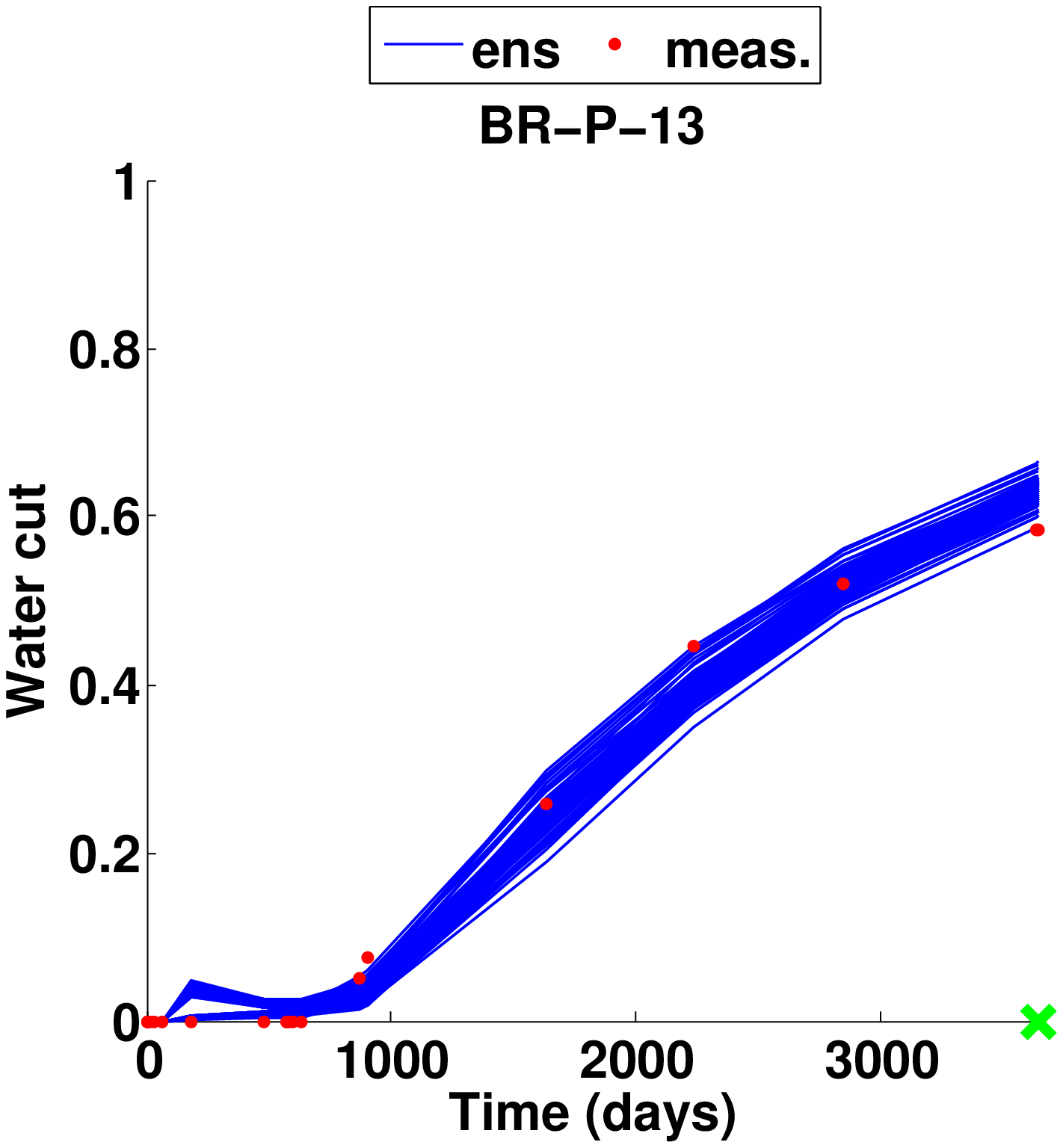}
	}	
	
	\subfigure[Initial ensemble]{ \label{subfig:Brugge_WWCT_BR-P-19_init}
		\includegraphics[scale=\nScale]{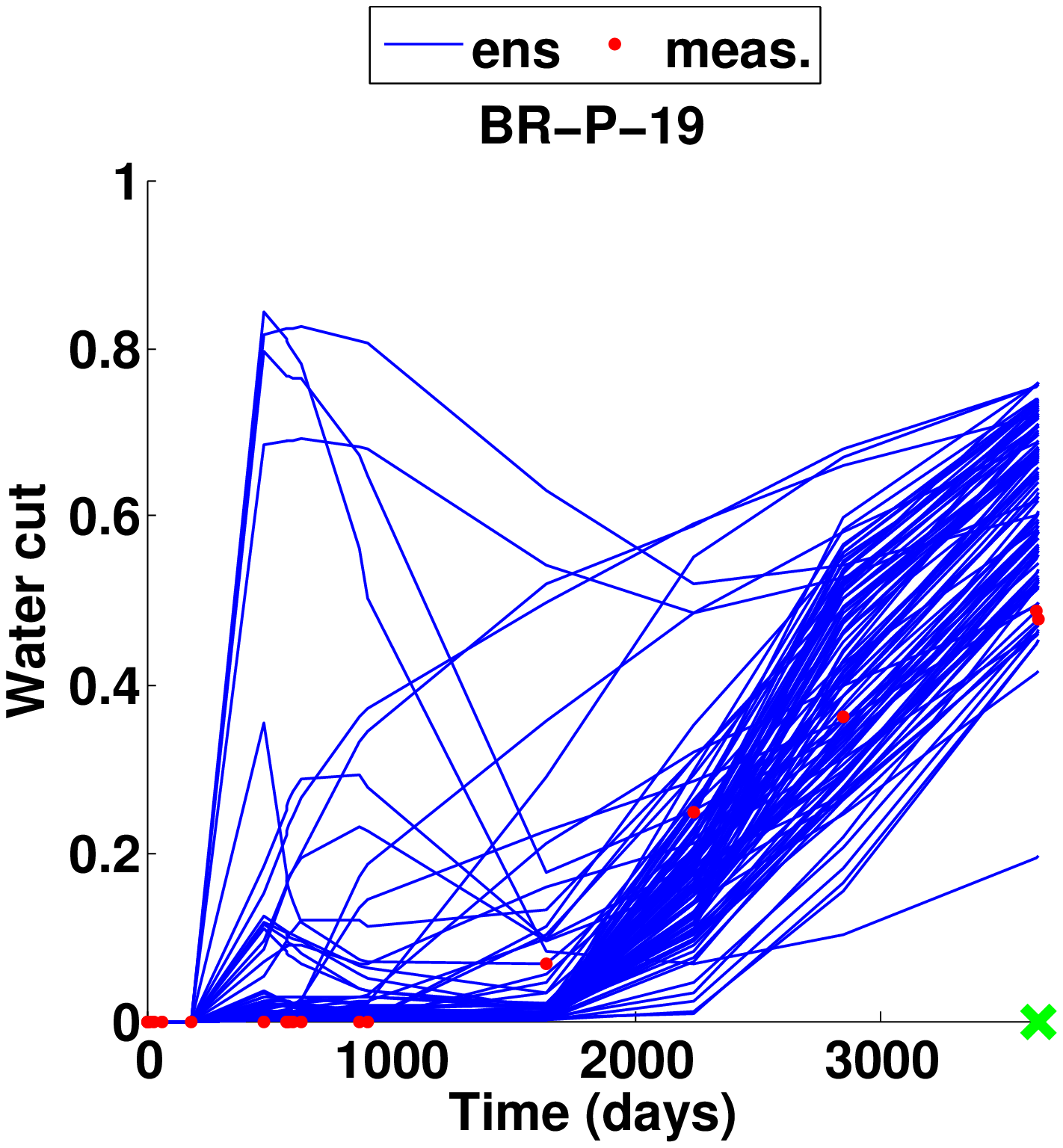}
	}
	\subfigure[aLm-EnRML]{ \label{subfig:Brugge_WWCT_BR-P-19_aLMEnRML}
		\includegraphics[scale=\nScale]{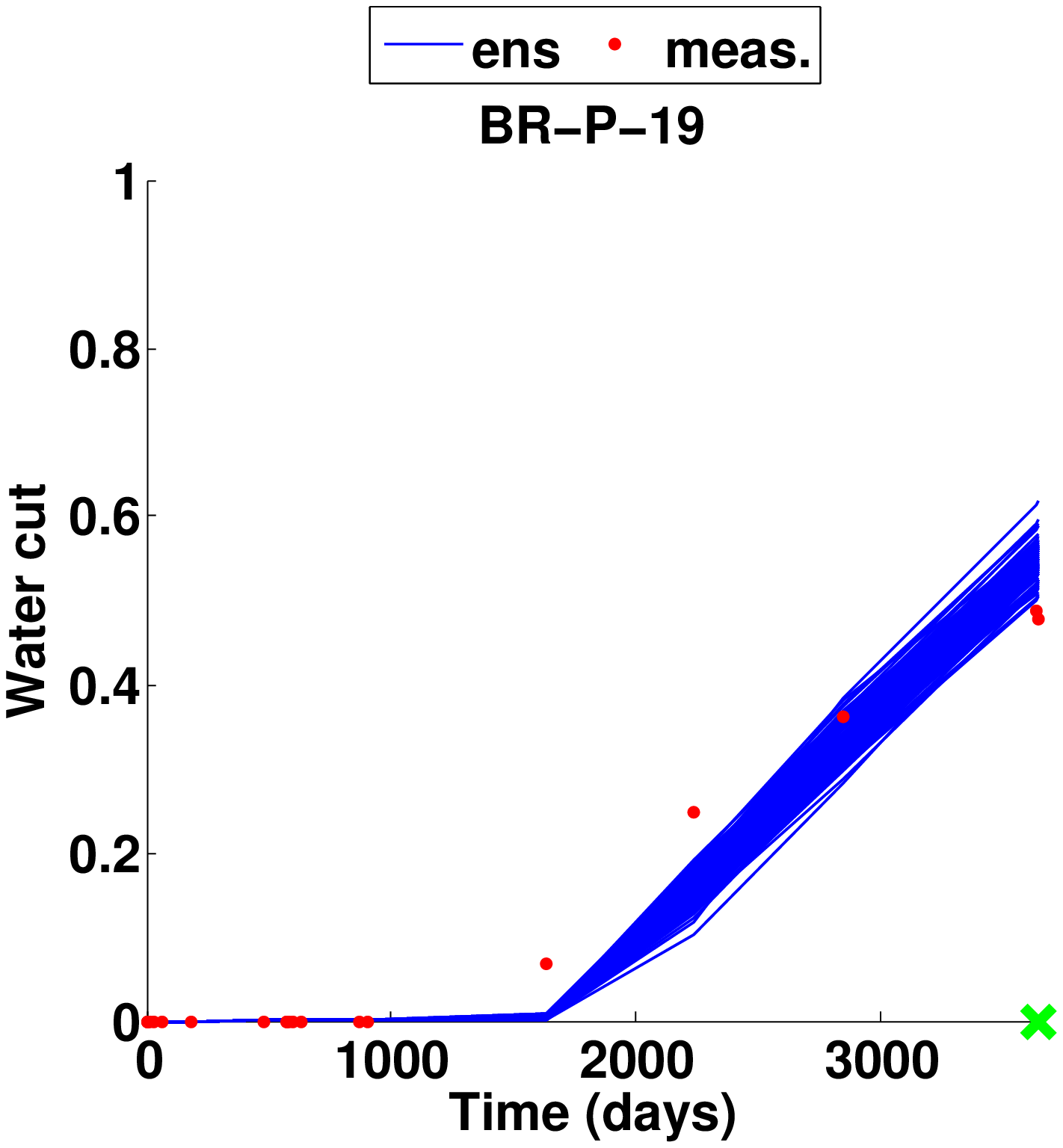}
	}
	\subfigure[RLM-MAC]{ \label{subfig:Brugge_WWCT_BR-P-19_RLMMAC}
		\includegraphics[scale=\nScale]{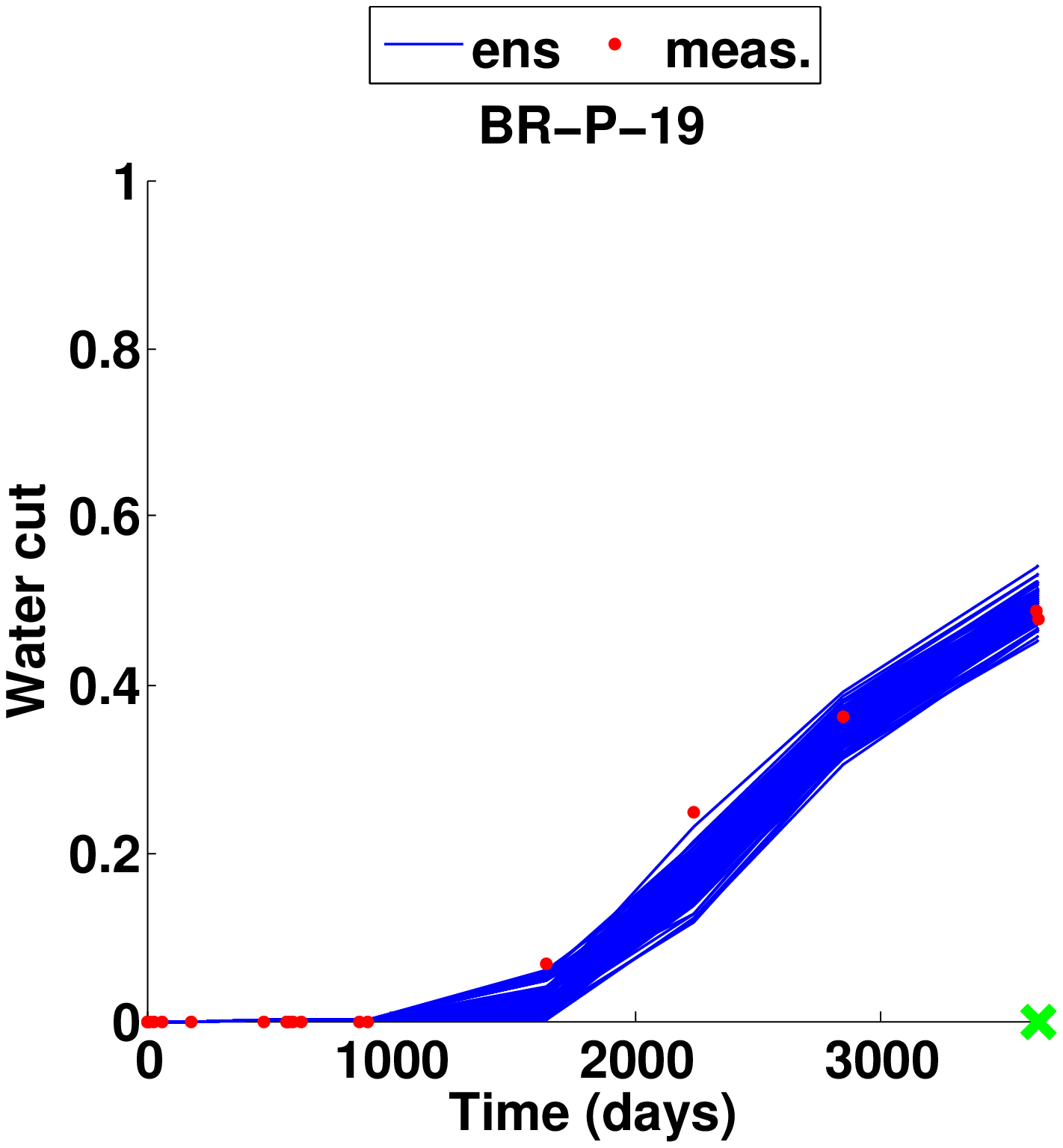}
	}
	\caption{\label{fig:brugge_WWCT_hm} As in Figure \ref{fig:brugge_wopr_hm}, but for water cut (WWCT) at BR-P-9, BR-P-13 and BR-P-19 in the Brugge field case.} 
\end{figure*}  

\clearpage
\renewcommand{\nScale}{0.6}
\begin{figure*}
	\centering
	\includegraphics[scale=\nScale]{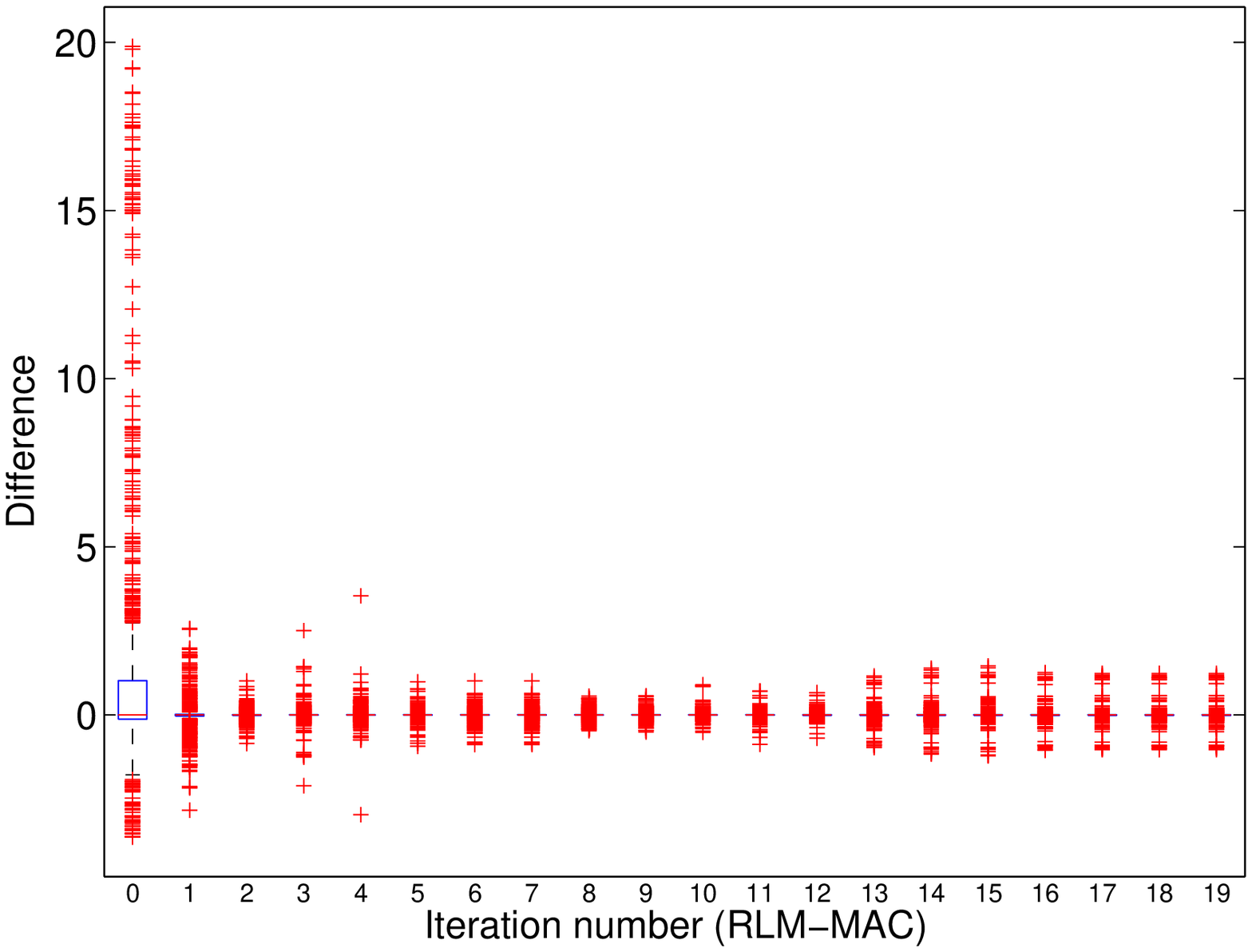}
	\caption{\label{fig:Brugge_obsDiff_boxPlot} Box plots of the normalized differences $\mathbf{C}_d^{-1/2} \left( \mathbf{g} \left(\bar{\mathbf{m}}^{i} \right) - \overline{\mathbf{g}\left(\mathbf{m}_j^{i}\right)} \right)$ at different iteration steps in the Brugge field case.} 
\end{figure*} 

\clearpage
\renewcommand{\nScale}{0.6}
\begin{figure*}
	\centering
	 \label{subfig:comparison_results_allHistory2}
		\includegraphics[scale=\nScale]{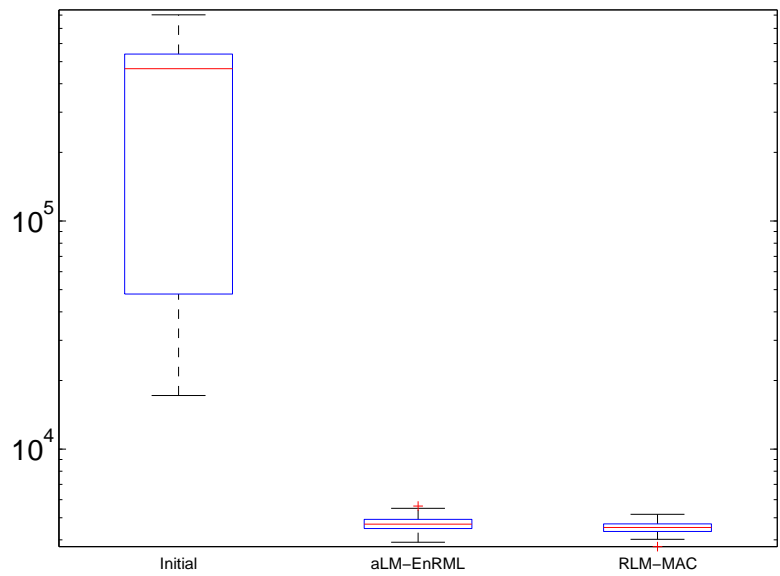}
	\caption{\label{fig:brugge_data_mismatch_validation} Box plots of data mismatch with respect to the initial ensemble (left), the aLM-EnRML (middle) and the RLM-MAC (right) in the Brugge field case. Here, all the production data in the first 20 years are used to cross-validate the history-matched reservoir models obtained by aLM-EnRML and RLM-MAC using production data at 20 out of 127 time instants in the first 10 years.} 
\end{figure*}

\clearpage
\renewcommand{\nScale}{0.3}
\begin{figure*}
	\centering
	\subfigure[]{ \label{subfig:Brugge_WOPR_BR-P-9_init_20yr}
		\includegraphics[scale=\nScale]{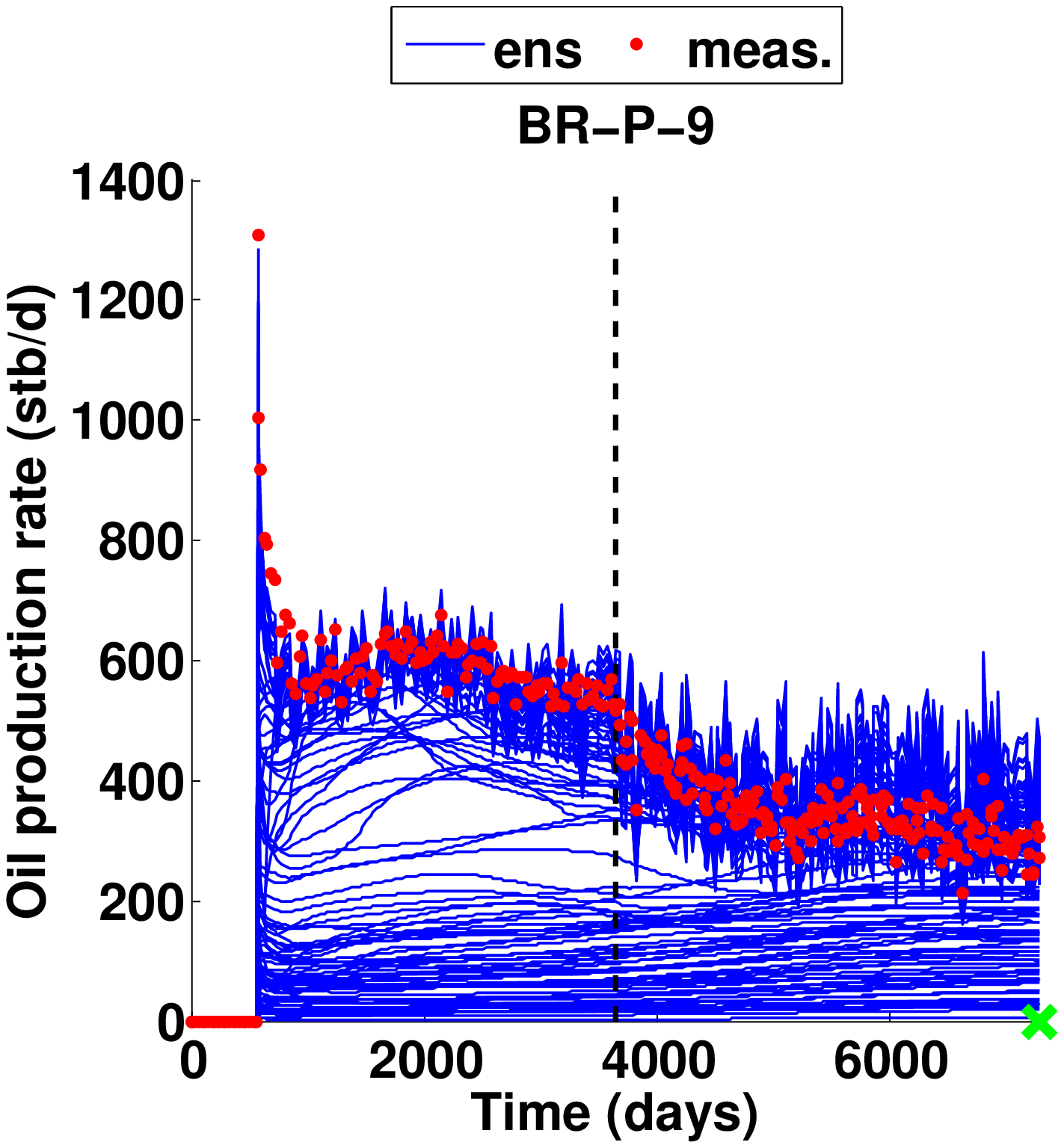}
	}
	\subfigure[]{ \label{subfig:Brugge_WOPR_BR-P-9_aLMEnRML_20yr}
		\includegraphics[scale=\nScale]{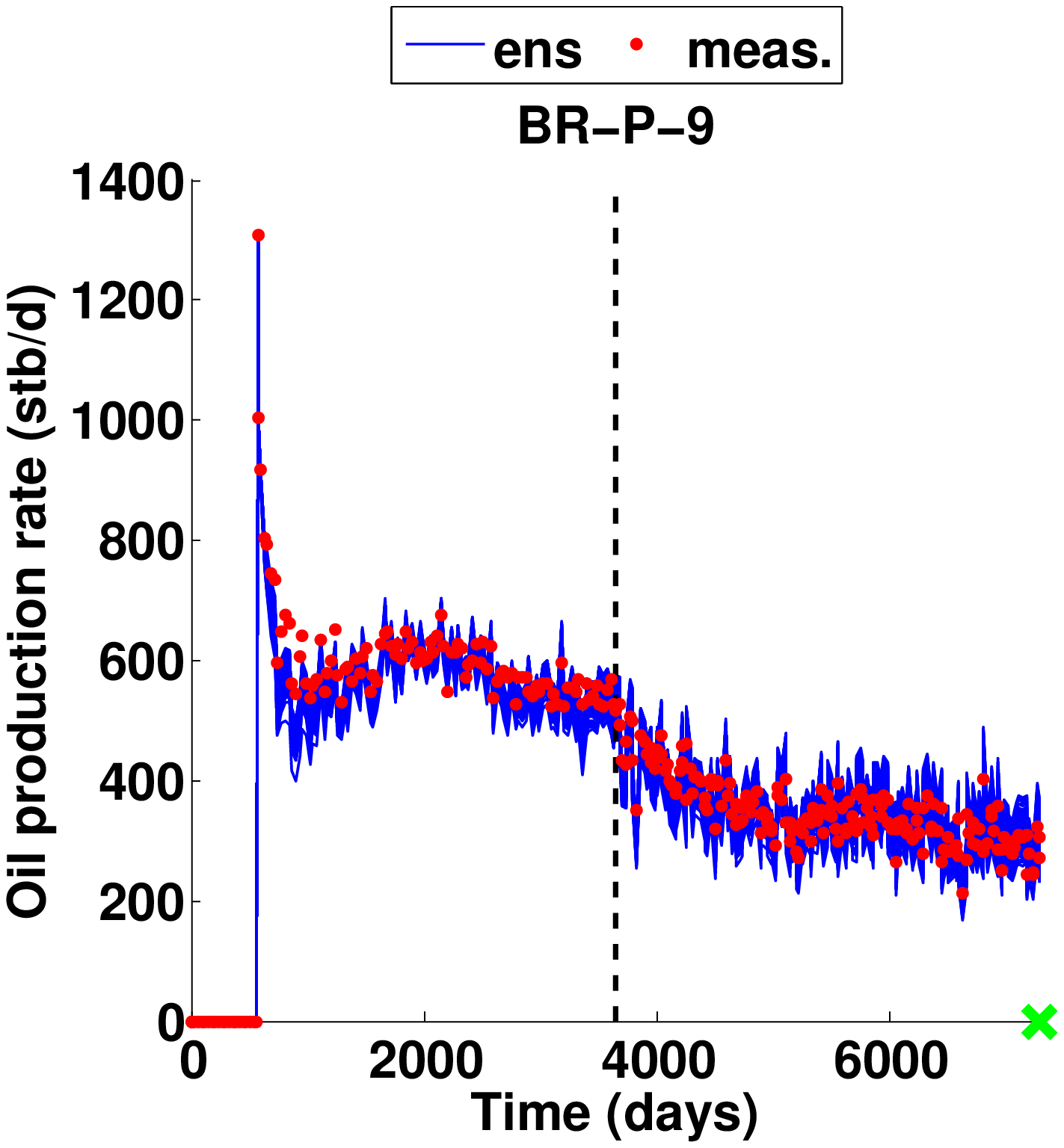}
	}
	\subfigure[]{ \label{subfig:Brugge_WOPR_BR-P-9_RLMMAC_20yr}
		\includegraphics[scale=\nScale]{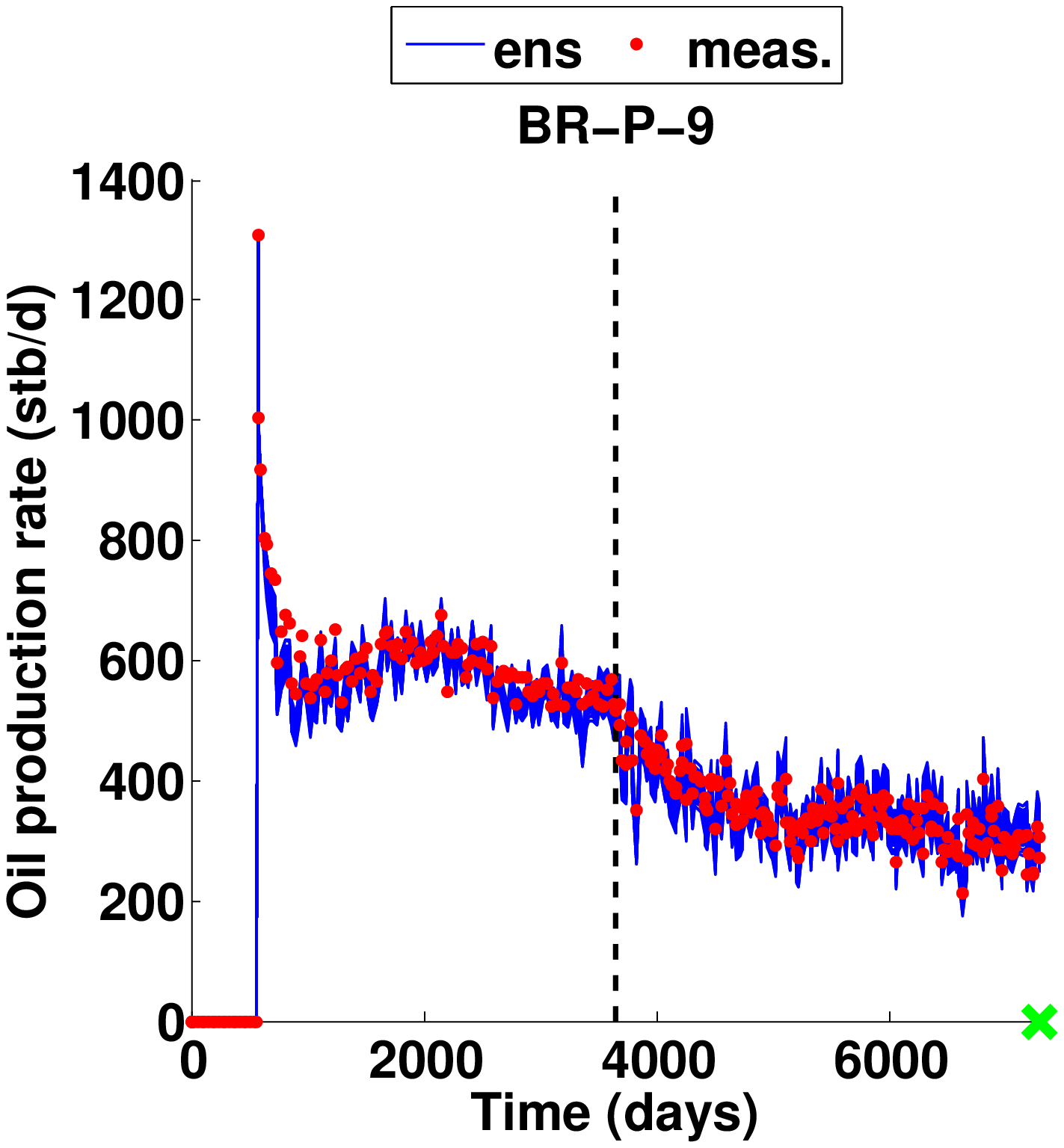}
	}
	
	\subfigure[]{ \label{subfig:Brugge_WOPR_BR-P-13_init_20yr}
		\includegraphics[scale=\nScale]{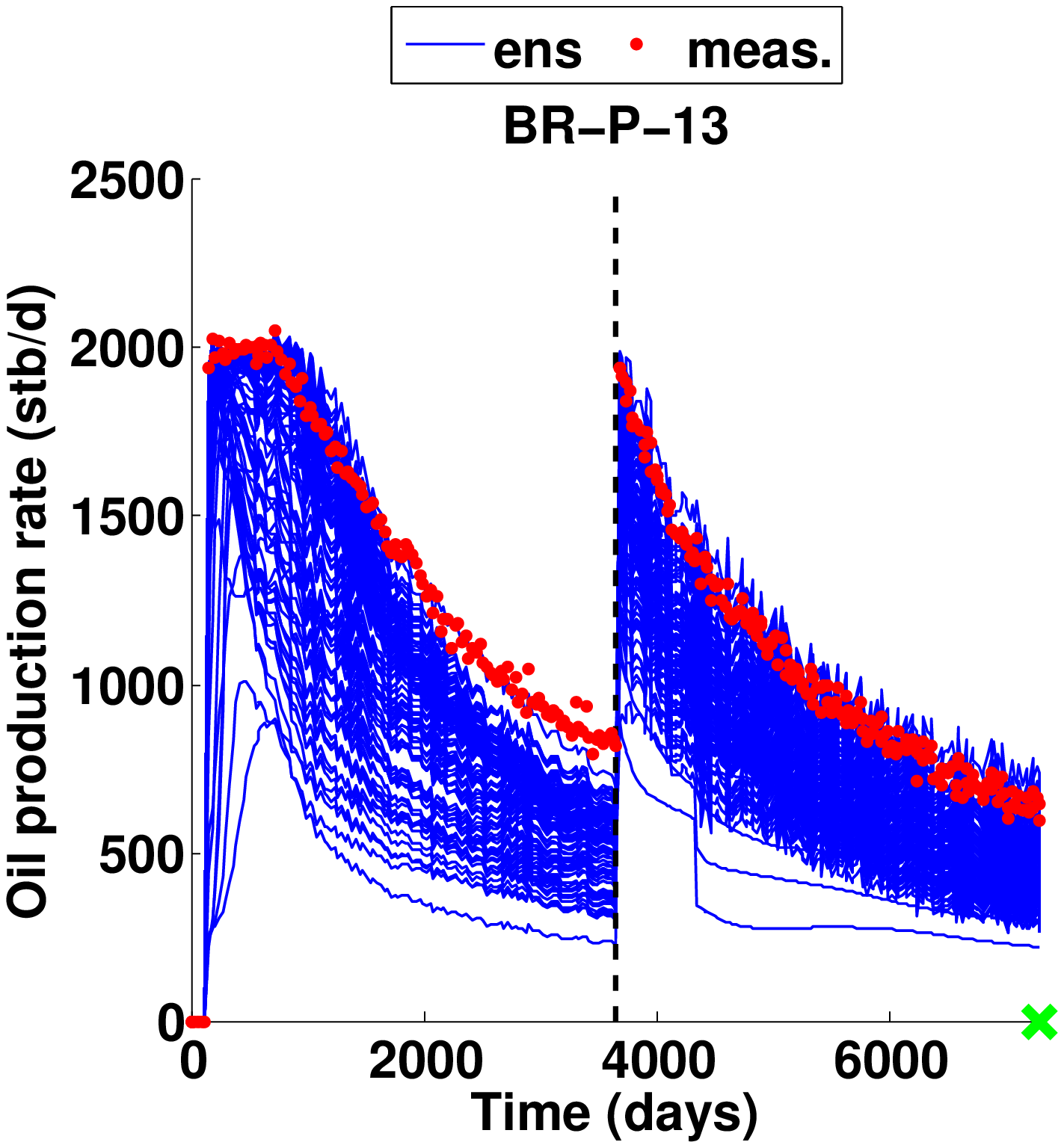}
	}
	\subfigure[]{ \label{subfig:Brugge_WOPR_BR-P-13_aLMEnRML_20yr}
		\includegraphics[scale=\nScale]{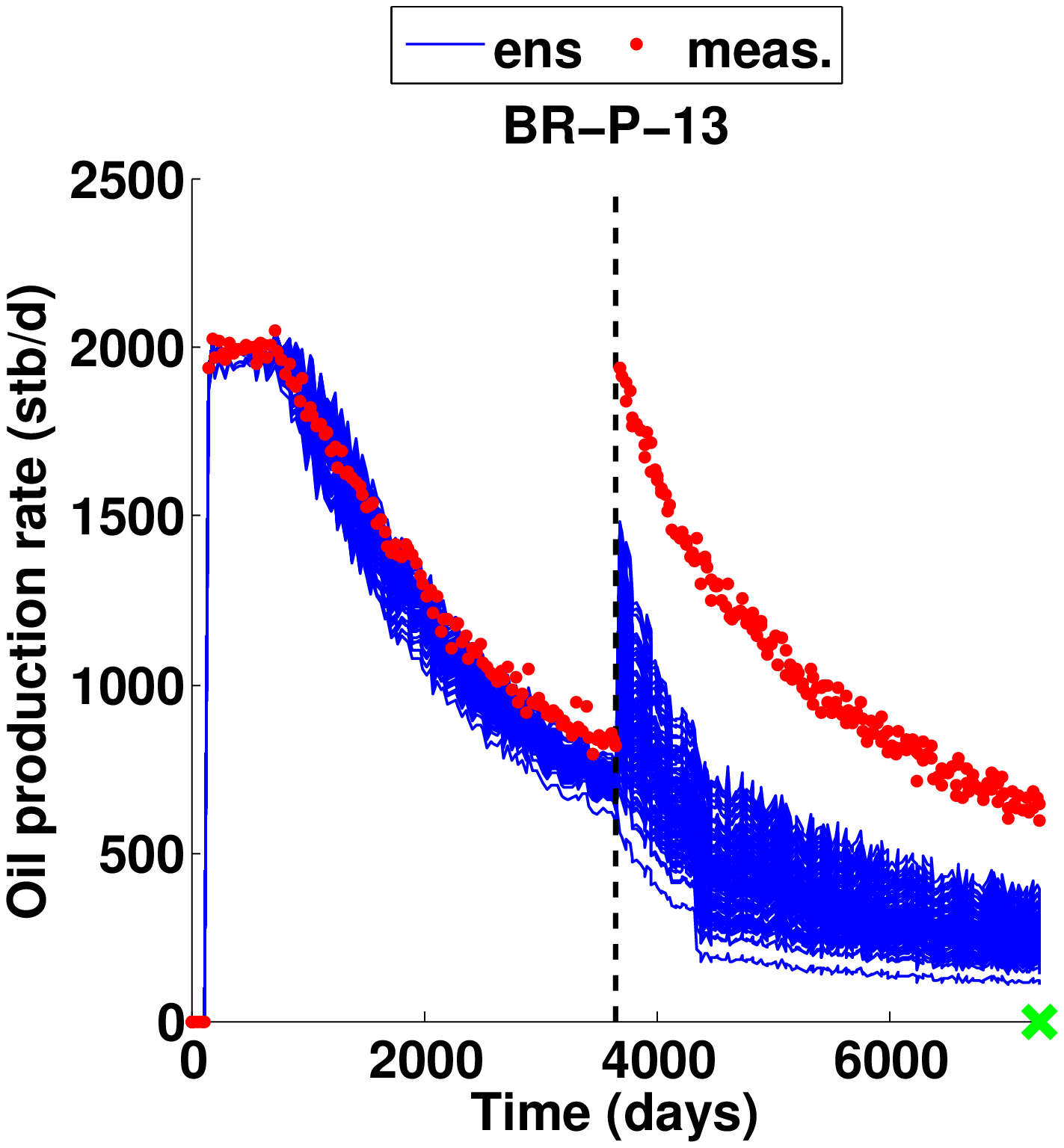}
	}
	\subfigure[]{ \label{subfig:Brugge_WOPR_BR-P-13_RLMMAC_20yr}
		\includegraphics[scale=\nScale]{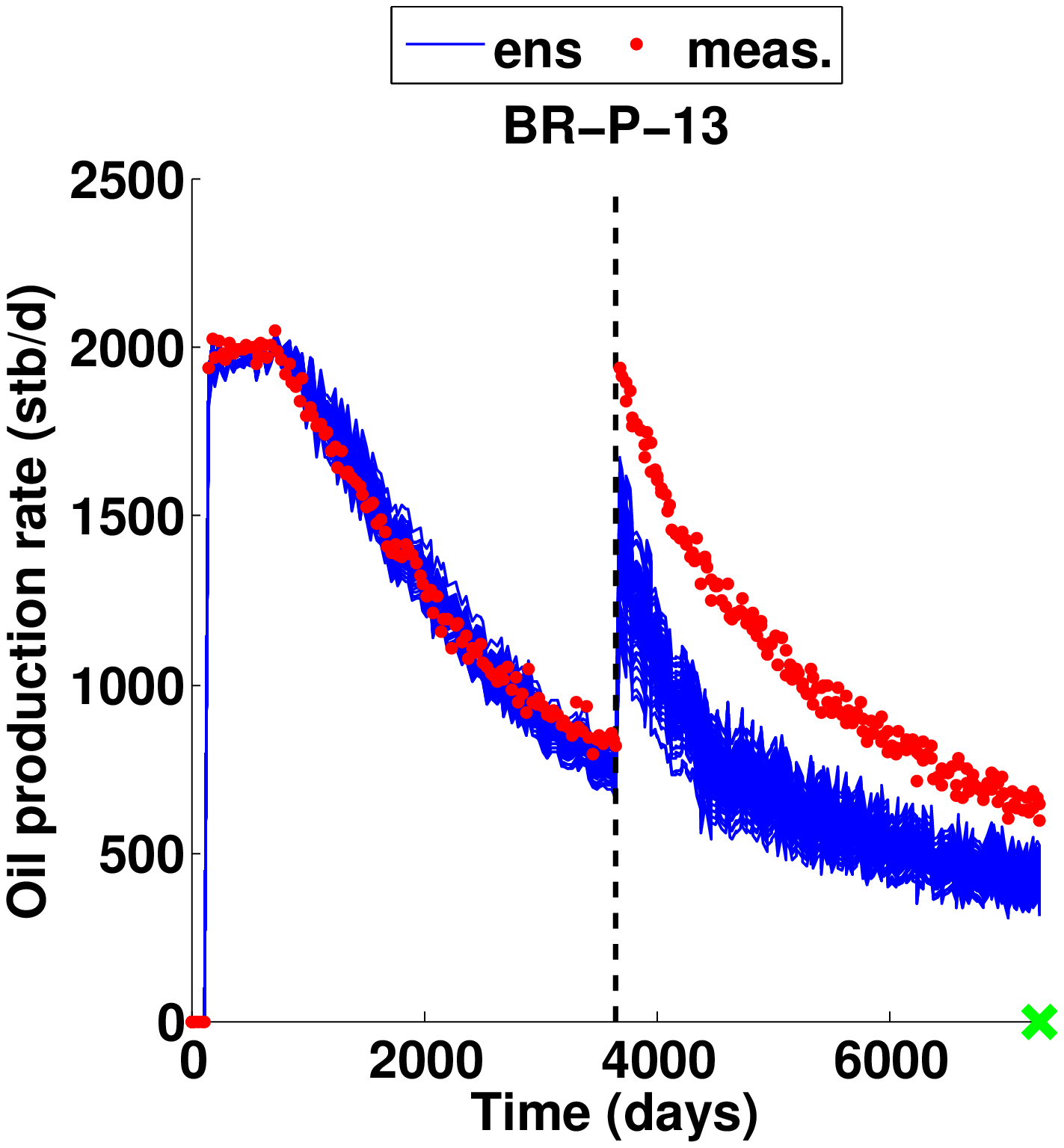}
	}	
	
	\subfigure[Initial ensemble]{ \label{subfig:Brugge_WOPR_BR-P-19_init_20yr}
		\includegraphics[scale=\nScale]{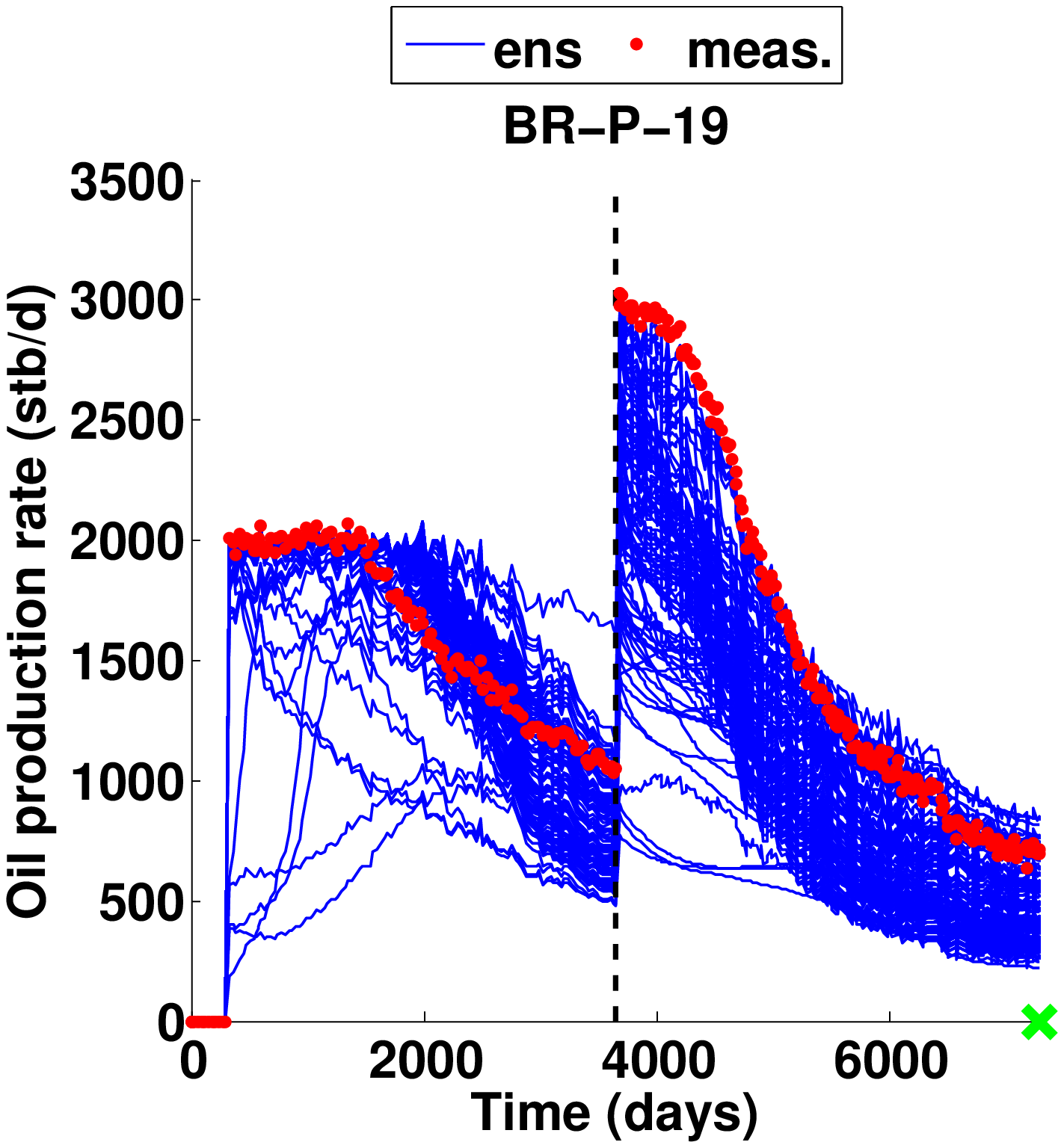}
	}
	\subfigure[aLm-EnRML]{ \label{subfig:Brugge_WOPR_BR-P-19_aLMEnRML_20yr}
		\includegraphics[scale=\nScale]{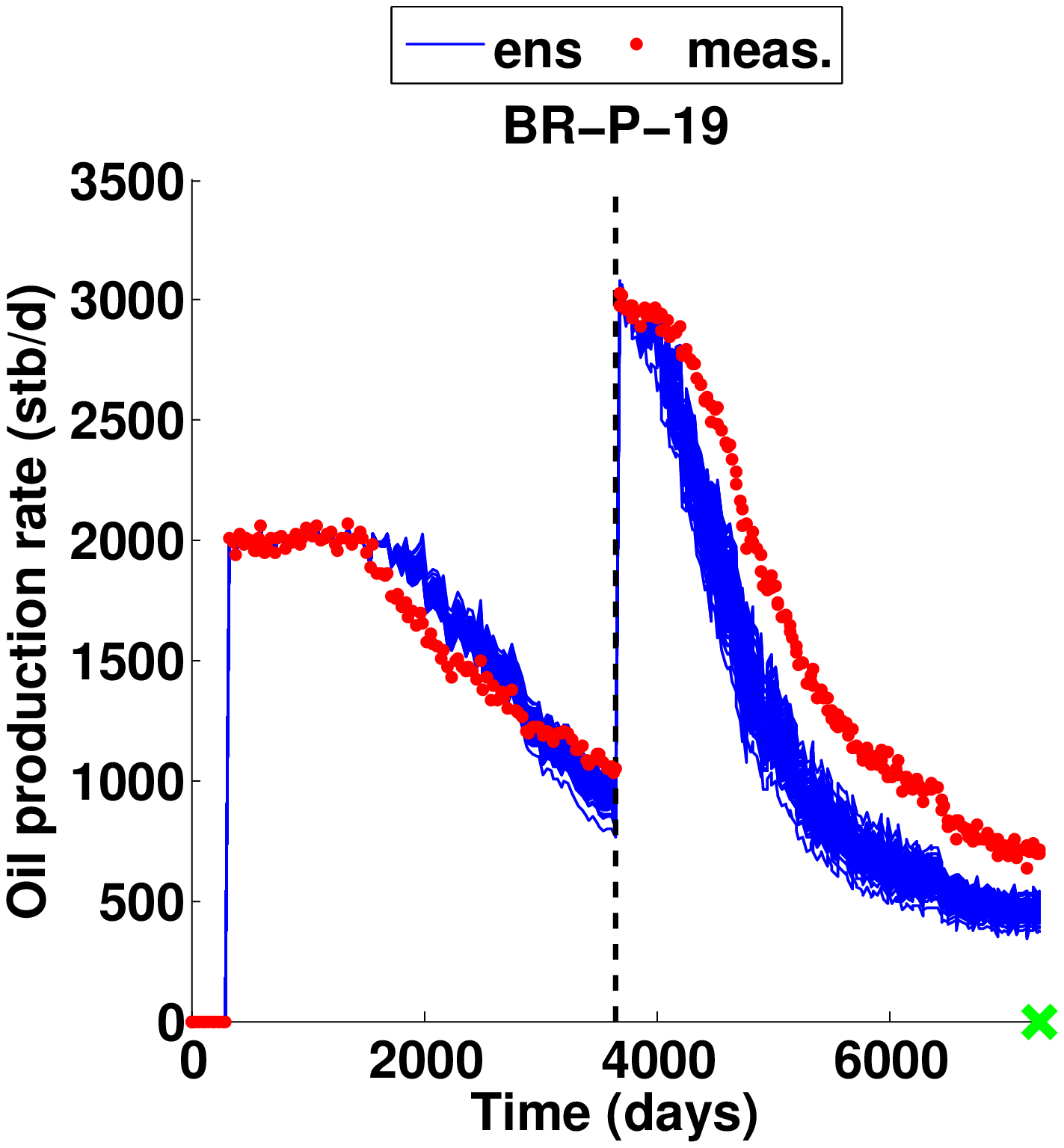}
	}
	\subfigure[RLM-MAC]{ \label{subfig:Brugge_WOPR_BR-P-19_RLMMAC_20yr}
		\includegraphics[scale=\nScale]{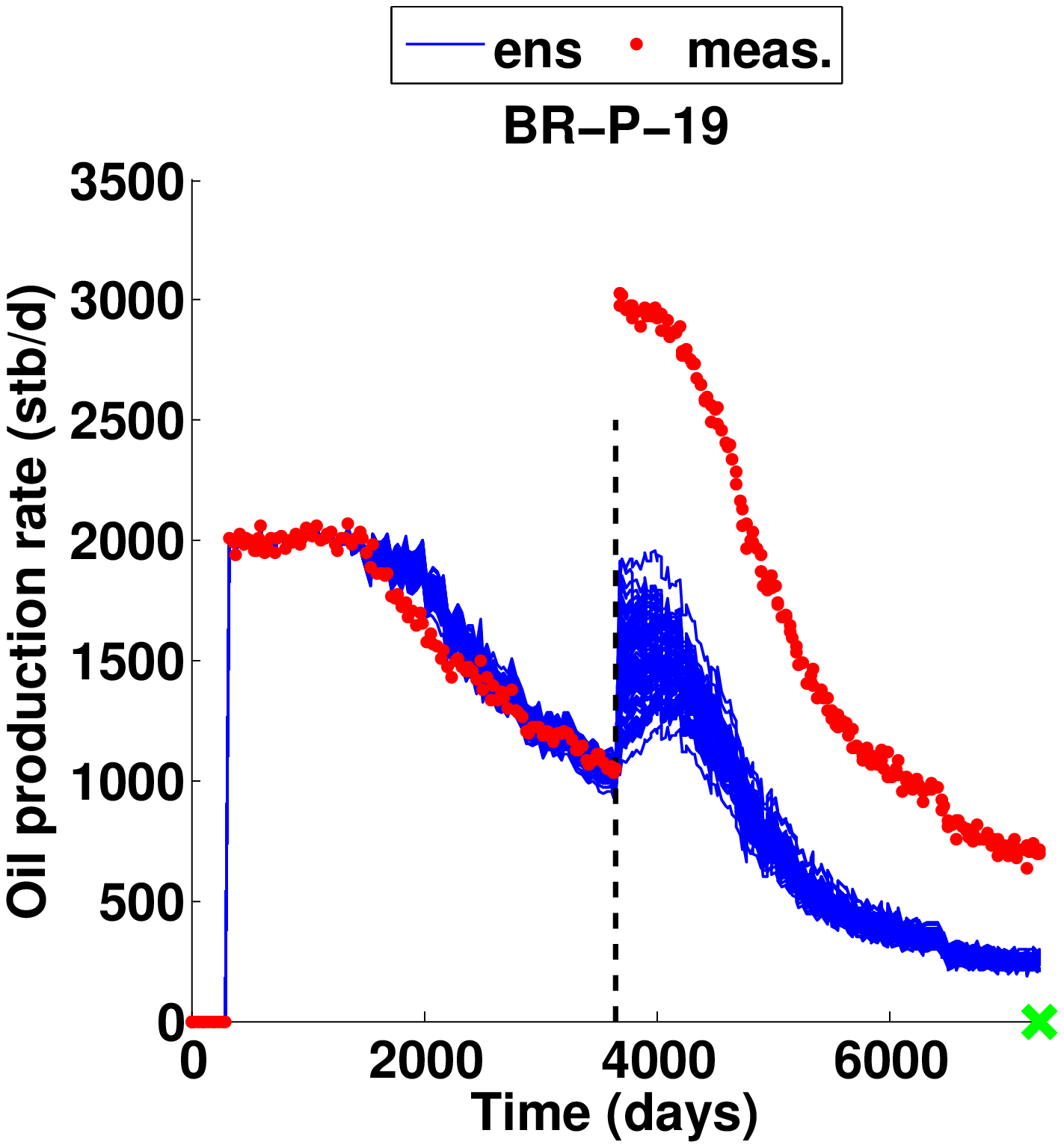}
	}
	\caption{\label{fig:brugge_wopr_validation}  Cross validation of oil production rates (WOPR) at BR-P-9 (top), BR-P-13 (middle) and BR-P-19 (bottom) in the Brugge field case, using the initial ensemble (1st column) and the ensembles of the aLM-EnRML (2nd column) and RLM-MAC (3rd column) at the final iteration steps. In each image, the red dots represent the historical WOPRH data and the blue curves are the forecasts with respect to the initial ensemble (1st column) and the final ensembles obtained by both iES (2nd and 3rd columns). The vertical dashed lines in the figures of the 2nd and 3rd columns separate the first and second decades.} 
\end{figure*}  

\clearpage
\renewcommand{\nScale}{0.3}
	\begin{figure*}
		\centering
		\subfigure[]{ \label{subfig:Brugge_WWCT_BR-P-9_init_20yr}
			\includegraphics[scale=\nScale]{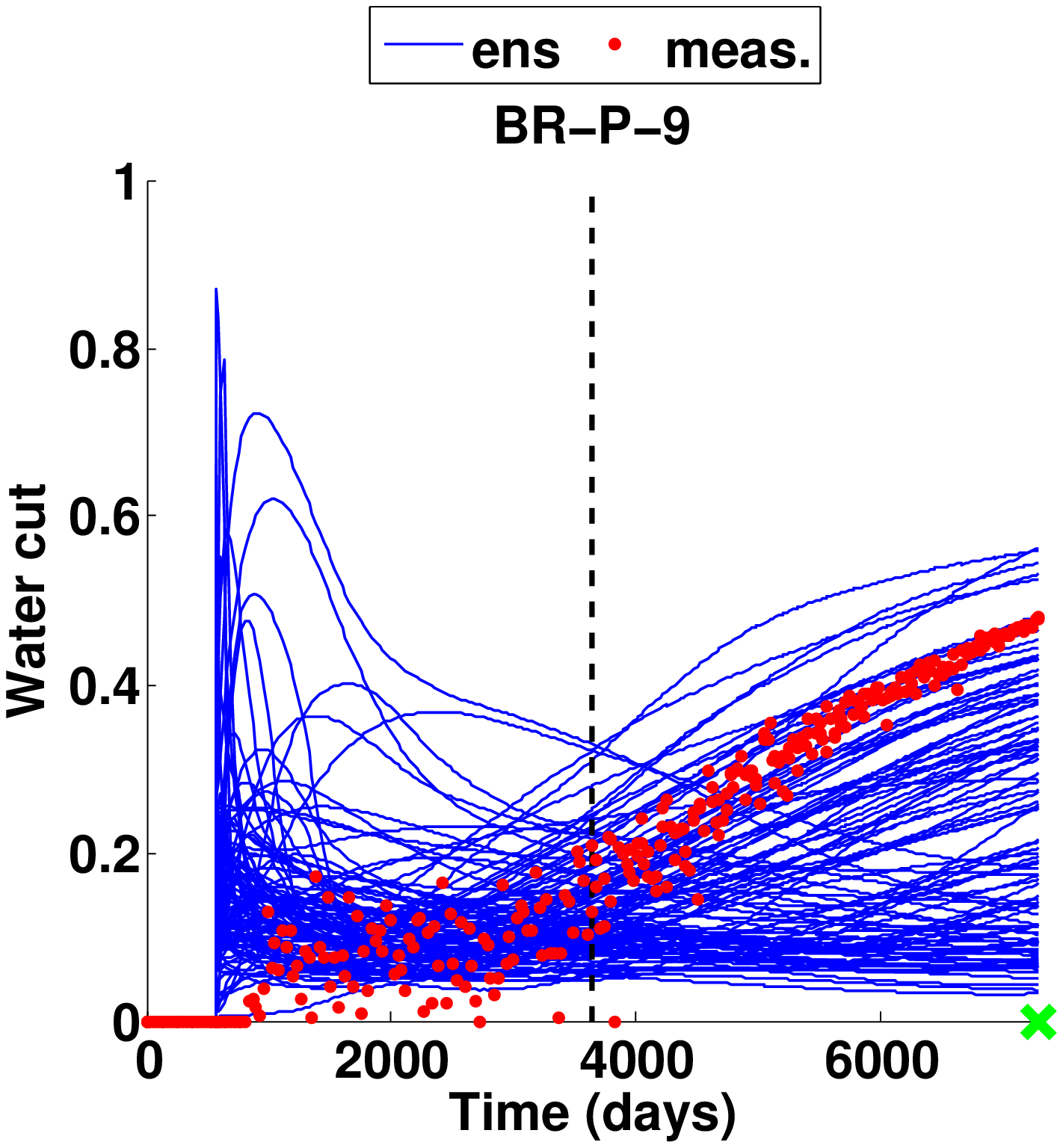}
		}
		\subfigure[]{ \label{subfig:Brugge_WWCT_BR-P-9_aLMEnRML_20yr}
			\includegraphics[scale=\nScale]{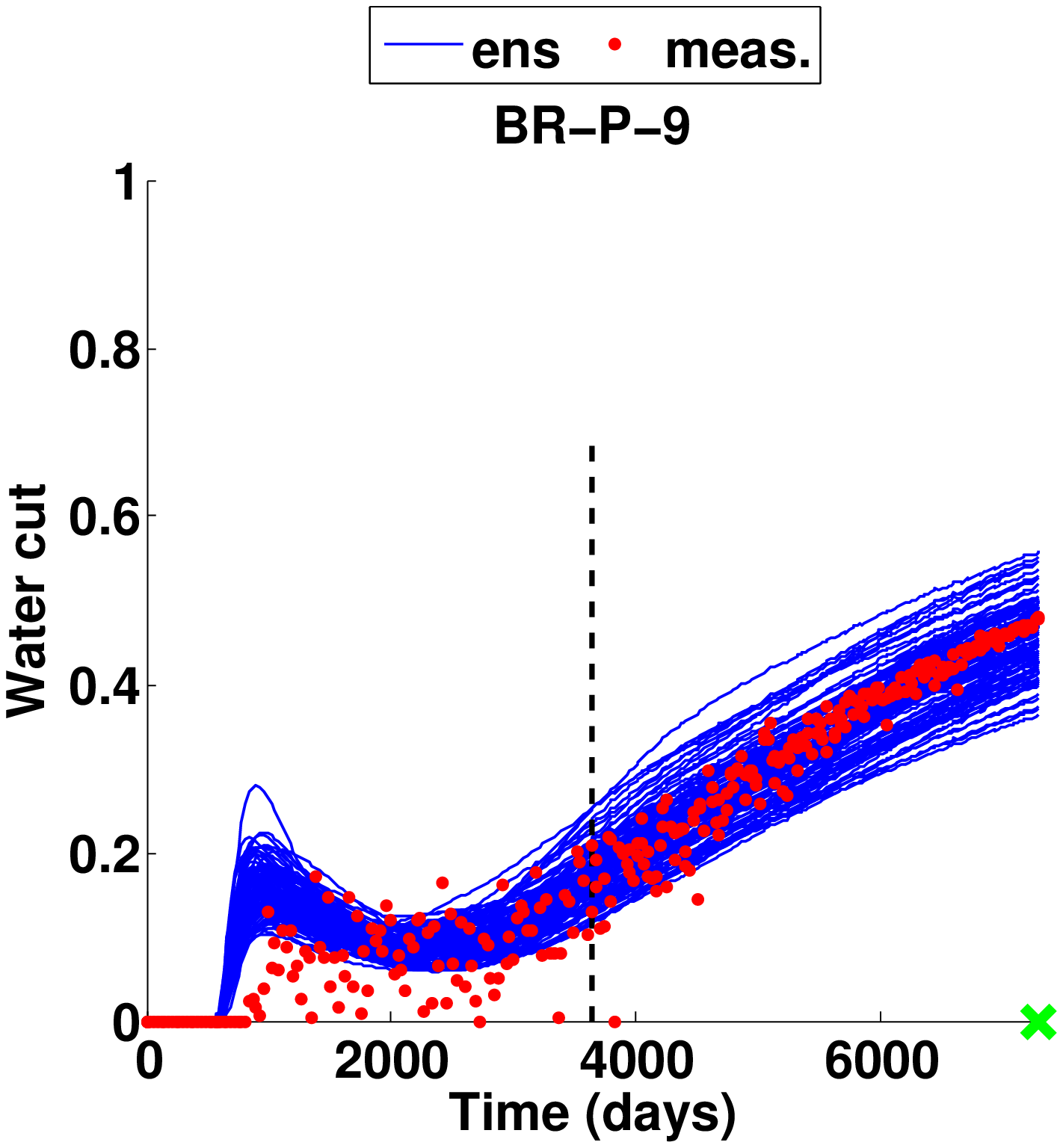}
		}
		\subfigure[]{ \label{subfig:Brugge_WWCT_BR-P-9_RLMMAC_20yr}
			\includegraphics[scale=\nScale]{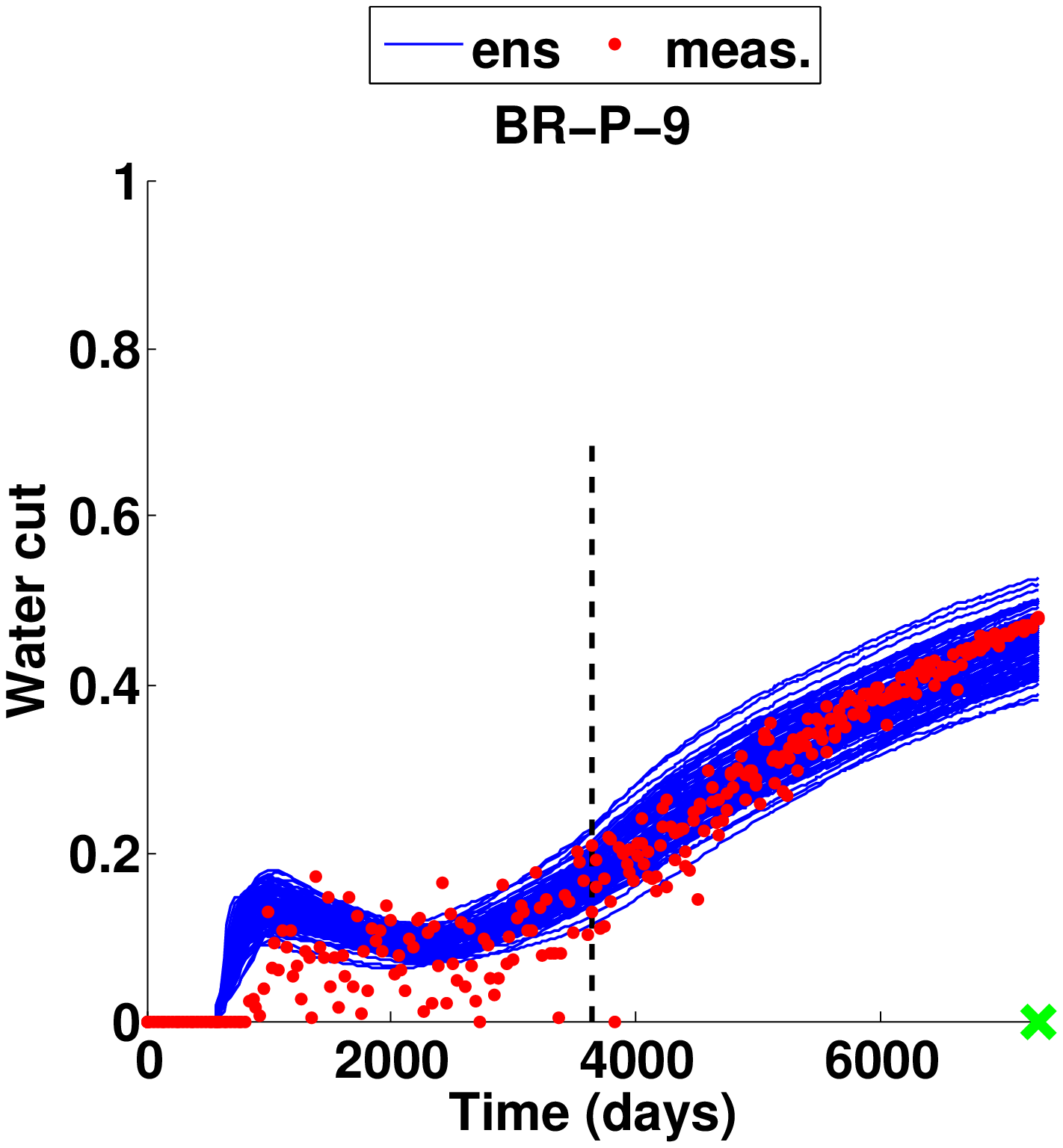}
		}
		
	    \subfigure[]{ \label{subfig:Brugge_WWCT_BR-P-13_init_20yr}
			\includegraphics[scale=\nScale]{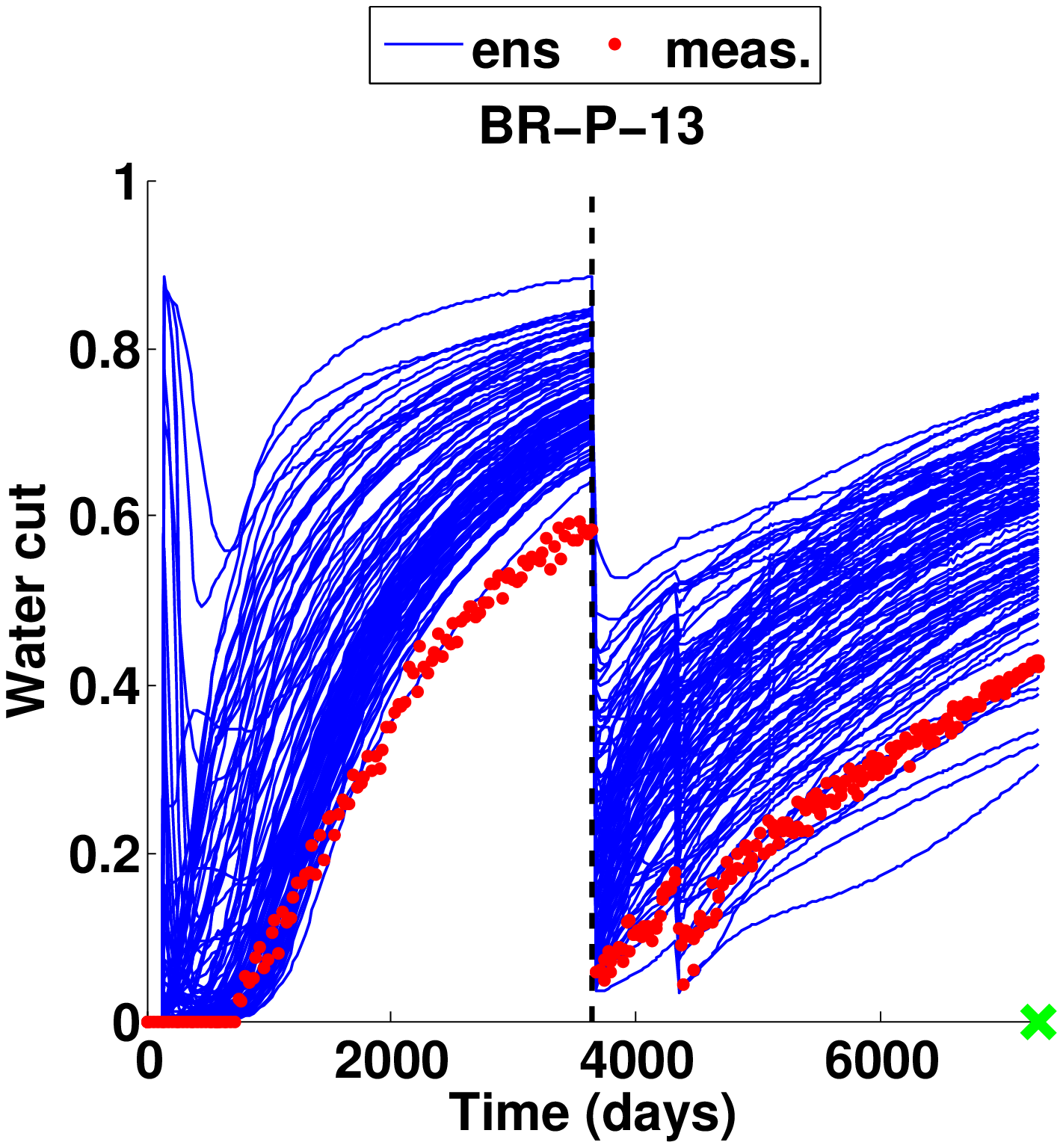}
		}
		\subfigure[]{ \label{subfig:Brugge_WWCT_BR-P-13_aLMEnRML_20yr}
			\includegraphics[scale=\nScale]{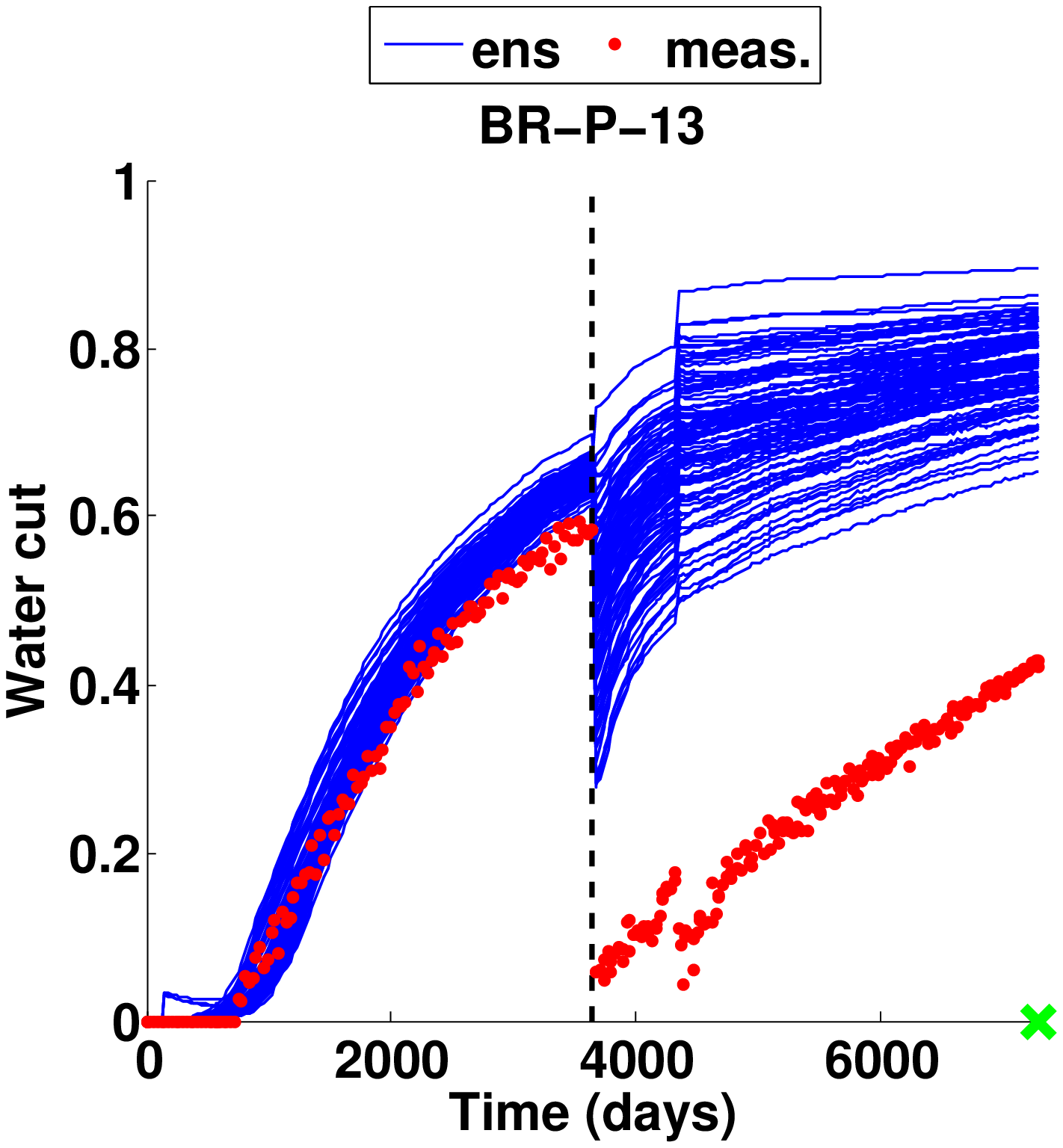}
		}
		\subfigure[]{ \label{subfig:Brugge_WWCT_BR-P-13_RLMMAC_20yr}
			\includegraphics[scale=\nScale]{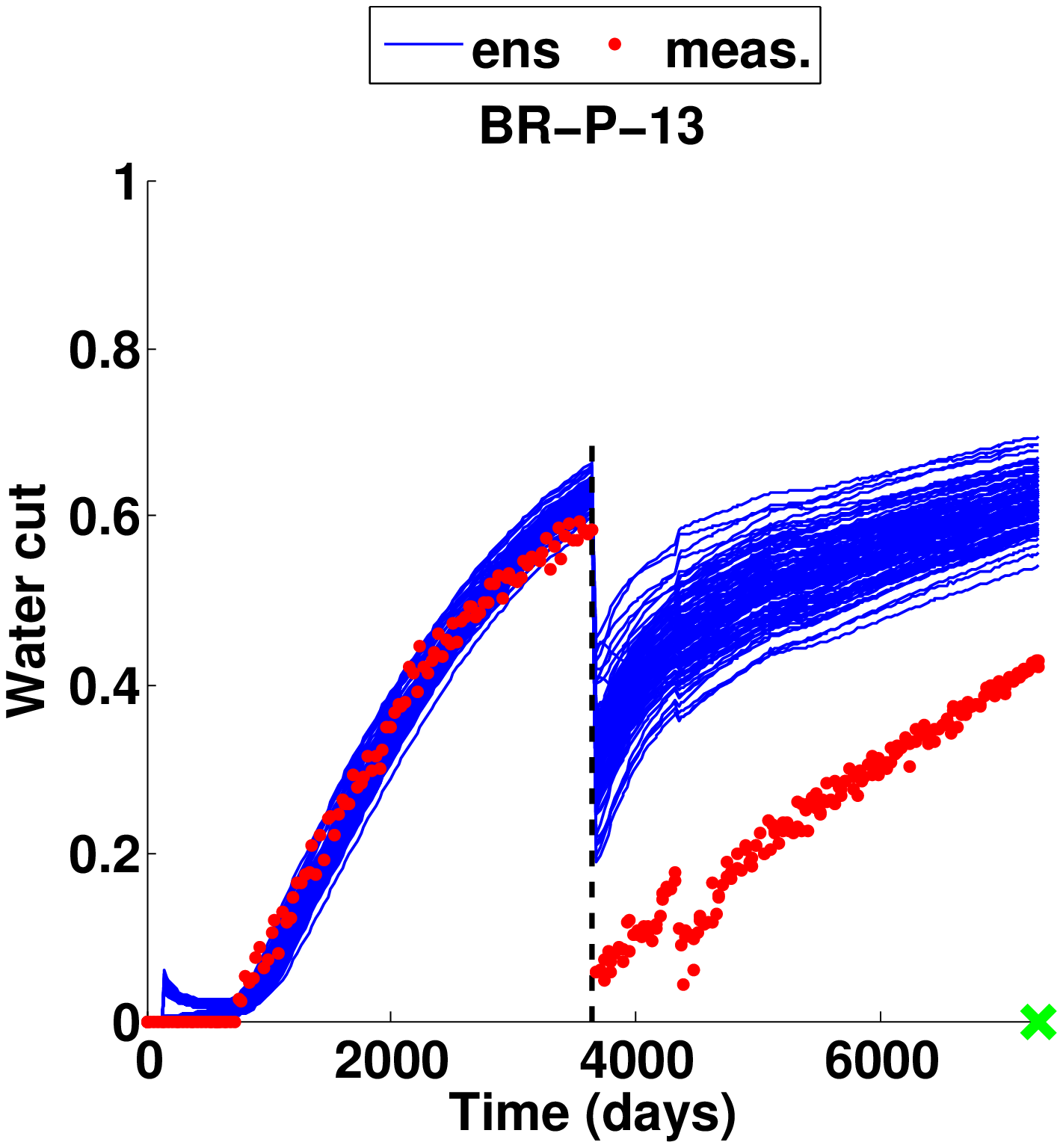}
		}		
		
		\subfigure[Initial ensemble]{ \label{subfig:Brugge_WWCT_BR-P-19_init_20yr}
			\includegraphics[scale=\nScale]{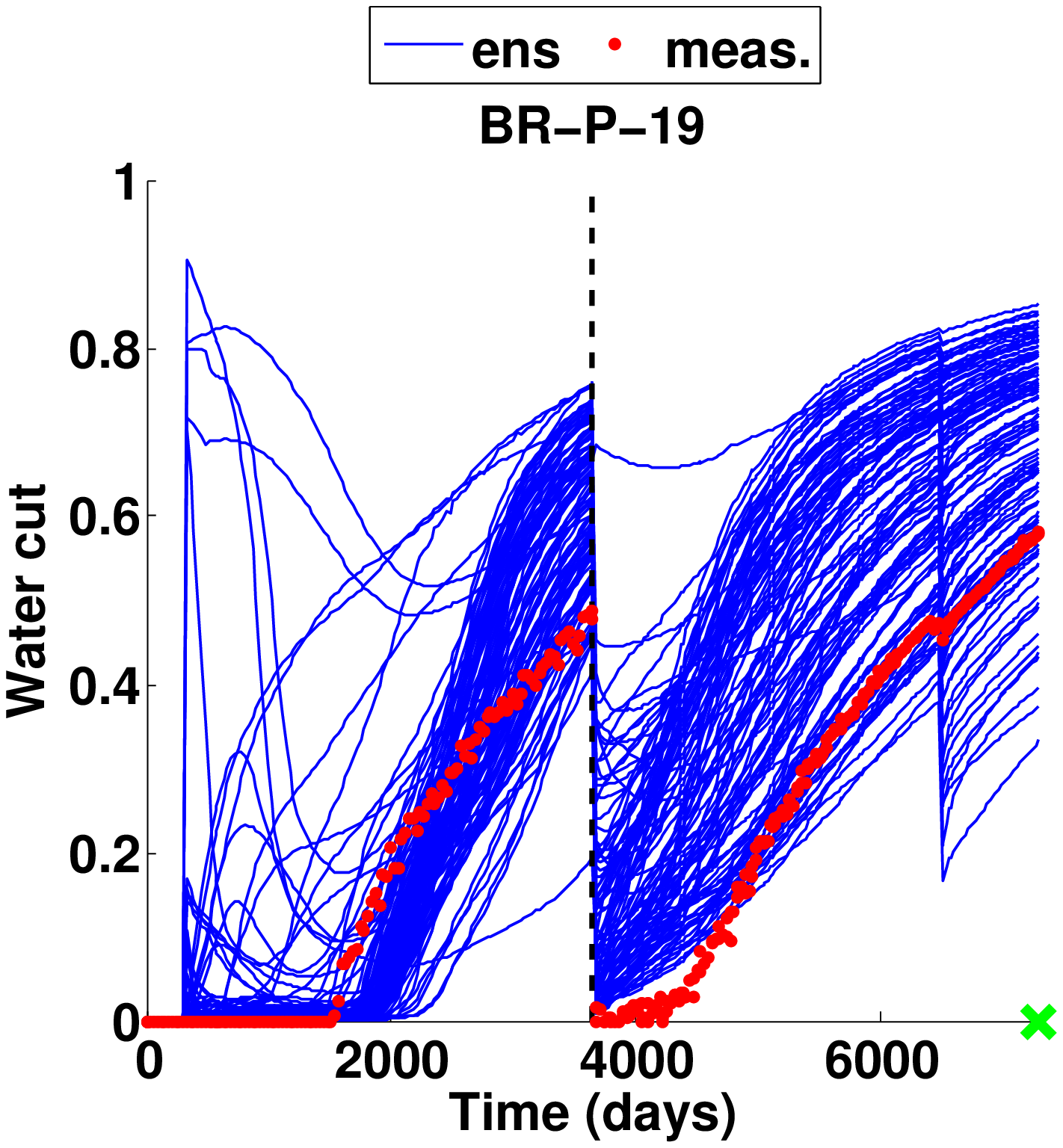}
		}
		\subfigure[aLm-EnRML]{ \label{subfig:Brugge_WWCT_BR-P-19_aLMEnRML_20yr}
			\includegraphics[scale=\nScale]{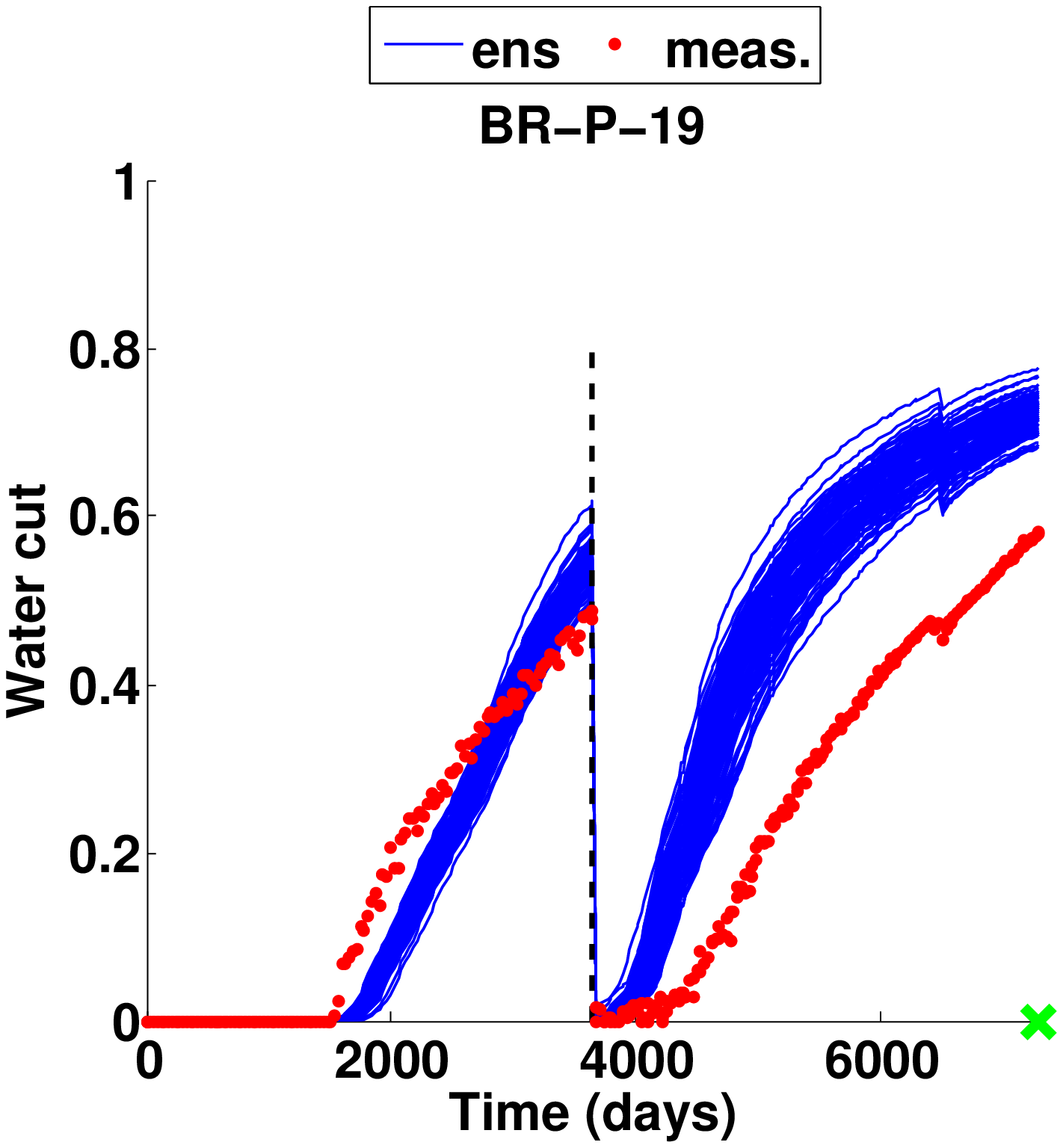}
		}
		\subfigure[RLM-MAC]{ \label{subfig:Brugge_WWCT_BR-P-19_RLMMAC_20yr}
			\includegraphics[scale=\nScale]{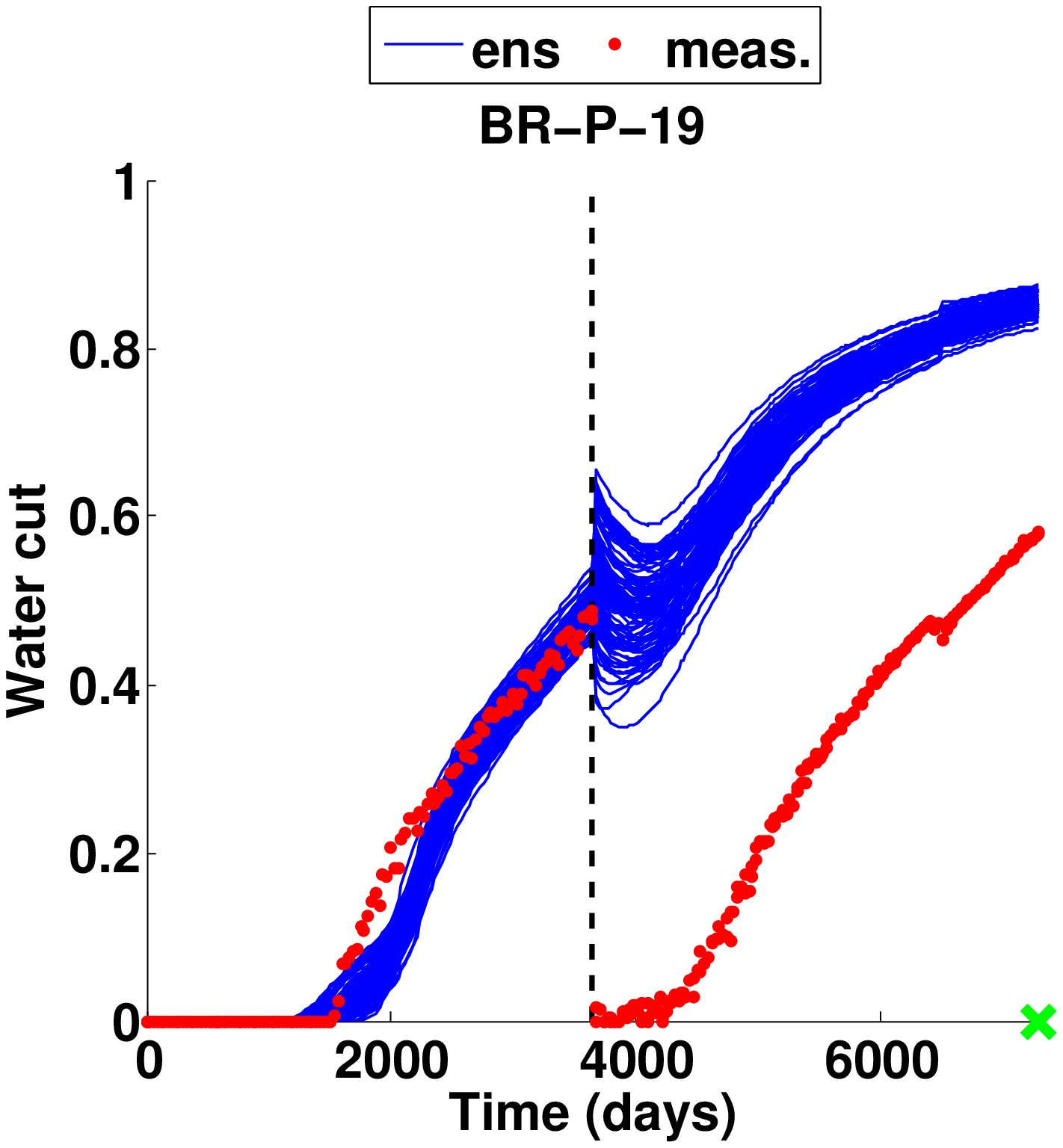}
		}
	\caption{\label{fig:brugge_WWCT_validation} As in Figure \ref{fig:brugge_wopr_hm}, but for water cut (WWCT) at BR-P-9, BR-P-13 and BR-P-19 in the Brugge field case.} 
\end{figure*}

\section*{}
\end{document}